\documentclass[10pt,twoside]{arxiv_article}

\usepackage[table]{xcolor}
\usepackage[latin9]{inputenc}
\usepackage{amssymb}
\usepackage{amsfonts}
\usepackage{amsmath}
\usepackage{fancybox}
\usepackage{fancyhdr}
\usepackage{graphicx}
\usepackage{subcaption}
\usepackage[hyphens]{url}
\usepackage{arydshln}
\usepackage{multirow}
\usepackage{pdflscape}
\usepackage{tikz}
\usepackage{comment}
\usepackage{nicefrac}
\usepackage{subcaption}
\usepackage{dsfont}
\usepackage{natbib}
\usepackage[colorlinks = true, linkcolor = blue, urlcolor = blue,
            citecolor = blue, anchorcolor = blue]{hyperref}
\usepackage{adjustbox}
\usepackage{array}
\usepackage{lipsum}
\usepackage{eurosym}
\usepackage{morefloats}

\newcolumntype{R}[2]{%
    >{\adjustbox{angle=#1,lap=\width-(#2)}\bgroup}%
    l%
    <{\egroup}%
}
\newcommand*\tabr{\multicolumn{1}{R{45}{1em}}}

\newlength{\figurewidth}
\newlength{\figureheight}
\newtheorem{remark}{Remark}

\def\limfunc#1{\mathop{\rm #1}}
\def\func#1{\mathop{\rm #1}}
\newcommand{\mr}[1]{\multirow{2}{*}{#1}}

\newcommand{\TsIII}{\hspace{3pt}}
\newcommand{\TsV}{\hspace{5pt}}

\newcommand{\TsVII}{\hspace{7pt}}
\newcommand{\TsVIII}{\hspace{8pt}}
\newcommand{\TsX}{\hspace{10pt}}
\newcommand{\TsXIII}{\hspace{13pt}}
\newcommand{\TsXV}{\hspace{15pt}}

\DeclareMathAlphabet{\mathpzc}{OT1}{pzc}{m}{it}
\newcommand{\redemption}{\ensuremath{\boldsymbol{\mathpzc{R}}}}
\newcommand{\frequency}{\ensuremath{\boldsymbol{\mathpzc{F}}}}

\usetikzlibrary{positioning}
\usetikzlibrary{patterns,arrows,decorations.pathreplacing}
\usetikzlibrary{backgrounds}
\usetikzlibrary{shapes.symbols}

\newcommand{\bZZZ}{$\circ \circ \circ$}
\newcommand{\bZZ}{$\circ\, \circ$}
\newcommand{\bZ}{$\circ$}
\newcommand{\bI}{$\bullet$}
\newcommand{\bII}{$\bullet\, \bullet$}
\newcommand{\bIII}{$\bullet \bullet \bullet$}

\newif\ifResearchVersion
\ResearchVersiontrue

\begin{document}

\ifResearchVersion

\title{\textbf{\color{amundi_blue}Liquidity Stress Testing in Asset Management\\Part 1. Modeling the Liability Liquidity Risk}%
\footnote{We are grateful to Pascal Blanqu\'e, Bernard de Wit, Vincent Mortier and Eric Vandamme for their helpful comments,
and Th\'eo Roncalli for his research assistance on zero-inflated models.
This research has also benefited from the support of Amundi Asset Management, which has
provided the data. However, the opinions expressed in this article are those
of the authors and are not meant to represent the opinions or official
positions of Amundi Asset Management.}}
\author{
{\color{amundi_dark_blue} Thierry Roncalli} \\
Quantitative Research \\
Amundi Asset Management, Paris \\
\texttt{thierry.roncalli@amundi.com} \and
{\color{amundi_dark_blue} Fatma Karray-Meziou} \\
Risk Management \\
Amundi Asset Management, Paris \\
\texttt{fatma.karraymeziou@amundi.com} \and
{\color{amundi_dark_blue} Fran\c{c}ois Pan} \\
Risk Management \\
Amundi Asset Management, Paris \\
\texttt{francois.pan@amundi.com} \and
{\color{amundi_dark_blue} Margaux Regnault} \\
Quantitative Research \\
Amundi Asset Management, Paris \\
\texttt{margaux.regnault@amundi.com}}

\date{\color{amundi_dark_blue}December 2020}

\maketitle

\begin{abstract}
This article is part of a comprehensive research project on liquidity risk in
asset management, which can be divided into three dimensions. The first
dimension covers liability liquidity risk (or funding
liquidity) modeling, the second dimension focuses on asset
liquidity risk (or market liquidity) modeling, and the third dimension considers
asset-liability liquidity risk management (or asset-liability
matching). The purpose of this research is to propose a methodological and
practical framework in order to perform liquidity stress testing programs,
which comply with regulatory guidelines \citep{ESMA-2019} and are useful
for fund managers. The review of the academic literature and professional
research studies shows that there is a lack of standardized and analytical
models. The aim of this research project is then to fill the gap with the
goal to develop mathematical and statistical approaches, and provide
appropriate answers.\smallskip

In this first part that focuses on liability liquidity risk modeling,
we propose several statistical models for estimating redemption shocks. The
historical approach must be complemented by an analytical approach based on
zero-inflated models if we want to understand the true parameters that
influence the redemption shocks. Moreover, we must also distinguish aggregate
population models and individual-based models if we want to develop
behavioral approaches. Once these different statistical models are
calibrated, the second big issue is the risk measure to assess normal and
stressed redemption shocks. Finally, the last issue is to develop a factor
model that can translate stress scenarios on market risk factors into stress
scenarios on fund liabilities.

\end{abstract}

\medskip


\clearpage

\else

\setcounter{page}{7}

\fi

\section{Introduction}

Liquidity stress testing in asset management is a complex topic because it is
related to three dimensions --- liquidity risk, asset management and stress
testing, whose linkages have been little studied and are hard to capture. First,
liquidity is certainly the risk that is the most difficult to model with the
systemic risk. If we consider market, credit, operational and counterparty
risks, there is a huge amount of academic literature on these topics in terms of models,
statistical inference and analysis. In terms of liquidity risk, the number
of practical studies and applied approaches is limited. Even though a great deal of
research has been completed on this subject, much of it is overly focused
on descriptive analyses of
liquidity, or its impact on systemic risk, or policy rules
for financial regulation. Moreover, this research generally focuses on the banking
sector \citep{Grillet-Aubert-2018}. For instance, the liquidity coverage ratio (LCR) and the net stable
funding ratio (NSFR) of the Basel III regulatory framework are of no help when
measuring the liquidity risk in asset management. In fact, the concept of
liquidity risk in asset management is not well defined. More generally, it is a
recent subject and we must admit that the tools and models used in asset
management are very much lagging those developed in the banking sector.
This is why the culture of asset-liability management (ALM) is poor in
investment management, both on the side of asset managers and asset owners.
Therefore, if we add the third dimension, stress testing, we obtain an
unknown and obscure topic, because the combination of liquidity risk and stress
testing applied to asset management is a new and difficult task.\smallskip

This is not the first time that the regulatory environment has sped up the
development of a risk management framework. Previous occurrences include the case of
market risk with the Amendment of the first Basel Accord \citep{BCBS-1996},
credit risk with the second consultative paper on Basel II
\citep{BCBS-2001}, credit valuation adjustment with the publication of the
Basel III Accord \citep{BCBS-2010}, interest rate risk in the banking book
with the IRRBB guidelines \citep{BCBS-2016}, etc. However, the measurement of
these risks had already benefited from the existence of analytical models
developed by academics and professionals. One exception was operational
risk, since banks started from a blank page when asked to measure it
\citep{BCBS-1999}. Asset managers now face a similar situation at this moment.
Between 2015 and 2018, the US Securities and Exchange Commission established
several rules governing liquidity management \citep{SEC-2015, SEC-2016,
SEC-2018a, SEC-2018b}. In particular, Rule 22e-4 requires investment funds
to classify their positions in one of four liquidity buckets (highly liquid
investments, moderately liquid investments, less liquid investments and
illiquid investments), establish a liquid investment minimum, and develop
policies and procedures on redemptions in kind. From September 2020, European
asset managers must also comply with new guidelines on liquidity stress testing
(LST) published by the European Securities and Markets Authority
\citep{ESMA-2019}. These different regulations are rooted in the agenda proposed
by the Financial Stability Board to monitor and manage systemic risk of
non-bank non-insurer systemically important financial institutions
\citep{FSB-2010}. Even if the original works of the FSB were biased, the idea
that the asset management industry can contribute to systemic risk has gained
ground and is now widely accepted. Indeed, \citet{FSB-2015} confused systemic
risk and systematic market risk \citep{Roncalli-2015a}. However,
\citet{Roncalli-2015b} showed that \textquotedblleft\textsl{the liquidation
channel is the main component of systemic risk to which the asset management
industry contributes}\textquotedblright. In this context, liquidity is the
major risk posed by the asset management industry that regulators must control.
But liquidity risk is not only a concern for regulators. It must also be a
priority for asset managers. The crisis of money market funds in the fourth
quarter of 2008 demonstrated the fragility of some fund managers
\citep{Schmidt-2016}. Market liquidity deteriorated in March and April
2020, triggering a liquidity shock on some investment funds and strategies. However,
aside from the 2008 Global Financial Crisis and 2020 coronavirus pandemic, which
have put all asset managers under pressure, the last ten years have
demonstrated that liquidity is also an individual risk for fund managers. It
was especially true during episodes of flash crash\footnote{For instance,
during the US stock market flash crash (May 6, 2010), the US Treasury flash
crash (October 15, 2014), the US ETF flash crash (August 24, 2015), etc.},
where fund managers reacted differently. In a similar way, idiosyncratic
liquidity events may affect asset managers at the individual level
\citep{Thompson-2019}. Following some high-profile fund suspensions in
mid-2019, asset managers received requests from asset owners to
describe their liquidity policies and conduct a liquidity review of their
portfolios. Therefore, we notice that liquidity is increasingly becoming
a priority for asset managers for three main reasons, because it is a
reputational risk, they are challenged by asset owners and it can be a
vulnerability factor for financial performance.\smallskip

However, even though liquidity stress testing in asset management has
become one of the hot topics in finance, it has attracted few
academics and professionals, implying that the research on this
subject is not as dynamic as one might expect. In fact, it is at the
same stage as operational risk was in the early 2000s, when there was no
academic research on this topic. And it is also at the stage of
ALM banking risk, where the most significant contributions have come
from professionals. Since liquidity stress testing in asset
management is an asset-liability management exercise, modeling
progress mainly comes from professionals, because the subject is
so specific, requires business expertise and must be underpinned by
industry-level data. This is obviously an enormous hurdle for
academics, and this explains the lack of modeling and scientific
approach that asset managers encounter when they want to develop a
liquidity stress testing framework. Therefore, the objective of this
research is twofold. First, the idea is to provide a mathematical
and statistical formalization to professionals in order to go beyond
expert qualitative judgement. Second, the aim is to assist academics
in understanding this topic. This is important, because academic
research generally boosts the development of analytical models,
which are essential for implementing liquidity stress testing
programs in asset management.\smallskip

Liquidity stress testing in asset management involves so many dimensions that
we have decided to split this research into three parts:
\begin{enumerate}
\item liability liquidity risk modeling;
\item asset liquidity risk modeling;
\item asset-liability liquidity risk management.
\end{enumerate}
Indeed, managing liquidity risk consists of three steps. First, we have to
model the liability liquidity of the investment fund, especially the redemption
shocks. By construction, this step must incorporate client behavior. Second, we
have to develop a liquidity model for assets. For that, we must specify a
transaction cost model beyond the traditional bid-ask spread measure. In
particular, the model must incorporate two dimensions: price impact and trading
limits. These first two steps make the distinction between funding liquidity
and market liquidity. As noticed by \citet{Brunnermeier-2009}, these two
types of liquidity may be correlated. However, we suppose that they are
independent at this level of analysis. While the first step gives the liquidity
amount of the investment fund that can be required by the investors, the second
step gives the liquidity amount of the investment fund that can be available in
the market. Therefore, the third step corresponds to the asset-liability
management in terms of liquidity, that is the matching process between required
liquidity and available liquidity. This implies defining the part of the
redemption shock that can be managed by asset liquidation and the associated
liquidity costs. It also implies defining the liquidity tools that can be
put in place in order to manage the non-covered part of the redemption shock or
the liquidation shortfall. For instance, a liquidity buffer is an example of one
of these tools, but this is not the only solution. Redemption gates, side pockets and
redemptions in kind are alternative methods, but they are extreme solutions
that may break the fiduciary duties and liquidity promises of asset managers.
Swing pricing is also an important ALM tool, and is a challenging question when
we consider the fair calibration of swing prices.\smallskip

\begin{figure}[tbh]
\centering
\caption{The sequential approach of liquidity stress testing}
\label{fig:alm1}
\begin{tikzpicture}

\fill[color=gray!30] (2.0,0) rectangle (8.5,1);
\draw[very thick,red] (2.0,0) rectangle (8.5,1);

\fill[color=gray!30] (2.0,2) rectangle (8.5,3);
\draw[very thick, red] (2.0,2) rectangle (8.5,3);

\fill[color=gray!30] (2.0,4) rectangle (8.5,5);
\draw[very thick, red] (2.0,4) rectangle (8.5,5);

\draw (5.25,0.7) node{\textbf{Asset-liability management}};
\draw (5.25,0.2) node{\textbf{(liquidity matching)}};

\draw (5.25,2.5) node{\textbf{Asset (or market) liquidity}};
\draw (5.25,4.5) node{\textbf{Liability (or funding) liquidity}};

\draw[thick, <-, red] (5.25,1.1) -- (5.25,1.9);
\draw[thick, <-, red] (5.25,3.1) -- (5.25,3.9);

\end{tikzpicture}
\end{figure}
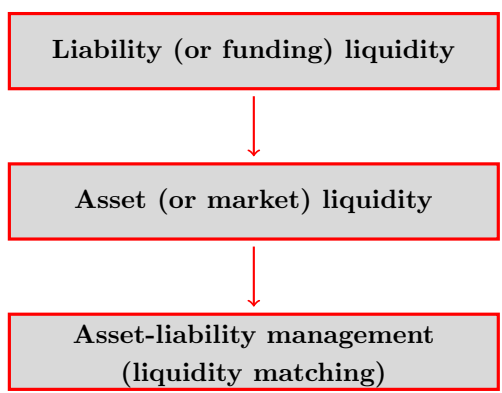

The three-stage process has many advantages in terms of modeling. First, it
splits a complex question into three independent and more manageable problems. This is
particularly the case of liability and asset modeling. Second, managing
liquidity risk becomes a sequential process, where the starting point is
clearly identified. As shown in Figure \ref{fig:alm1}, we should begin with the
liability risk. Indeed, if we observe no inflows or outflows, the process stops
here. As such, the first stage determines the amount to sell in the market and it
is measured with respect to the investor behavior. The liquidity risk has its
roots in the severity of the redemption shock. Market liquidity is part of
the second phase. Depending on the redemption size and the liquidity of the
market, the fund manager will decide the best solution to adopt.
And the sequential process will conclude with the action of the fund manager%
\footnote{Of course, the fund manager's action is not uniquely determined, because it depends
on several parameters. This means that two fund managers can take two different
decisions even if they face the same situation in terms of redemption and
market liquidity.}. Finally, the third advantage concerns the feasibility of
stress testing programs. In this approach, stress testing concerns the two
independent dimensions. We can stress the liquidity on the liability side, or
we can stress the liquidity on the asset side, or both, but the rule is
simple.\smallskip

\begin{figure}[tbh]
\centering
\caption{The network risk of liquidity}
\label{fig:alm2}
\begin{tikzpicture}[scale=0.90, every node/.style={transform shape}]

\draw[very thick, black, outer color=green!50!gray, inner color=white] (6,6) circle (1.3);
\draw[very thick, black, outer color=magenta!50!gray, inner color=white] (2,2.5) circle (1.2);
\draw[very thick, black, outer color=red!50!gray, inner color=white] (10,2.5) circle (1.2);

\draw[very thick, ->, red] (3.0,3.25) -- (5.0,5.05);
\draw[very thick, <-, red] (2.9,3.50) -- (4.9,5.25);

\draw[very thick, ->, red] (9.0,3.25) -- (7.00,5.05);
\draw[very thick, <-, red] (9.1,3.50) -- (7.11,5.25);

\draw[very thick, ->, red] (3.3,2.65) -- (8.7,2.65);
\draw[very thick, <-, red] (3.3,2.4) -- (8.7,2.4);

\draw (6,6) node{\begin{tabular}{c}
                            \textbf{Asset-liability} \\
                            \textbf{management}
                          \end{tabular}};

\draw (2,2.5) node{\begin{tabular}{c}
                            \textbf{Market} \\
                            \textbf{liquidity}     \\
                            \textbf{risk}
                          \end{tabular}};

\draw (10,2.5) node{\begin{tabular}{c}
                            \textbf{Funding} \\
                            \textbf{liquidity}     \\
                            \textbf{risk}
                          \end{tabular}};
\end{tikzpicture}
\end{figure}
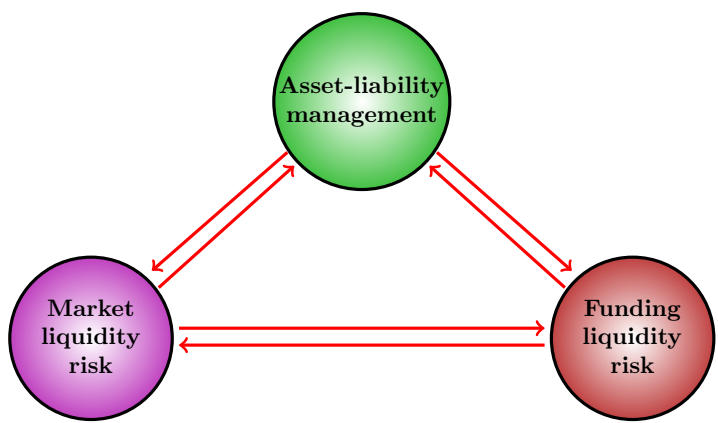

In the sequential approach, the liability of the investment fund is
the central node of the liquidity risk, and the vertex of the
liquidity network. However, it is not so simple, because the three
nodes can be interconnected (Figure \ref{fig:alm2}). If market
liquidity deteriorates sharply, investors may be incited to redeem
in order to avoid a liquidity trap. In this case, funding liquidity
is impacted by market liquidity, reinforcing the feedback loop
between funding and market liquidity, which is described by
\citet{Brunnermeier-2009}. But this is not the only loop. For
instance, the choice of an ALM decision may also influence funding
liquidity. If one asset manager decides to suspend
redemptions, it may be a signal for the investors of the other asset
managers if they continue to meet redemptions. Again, we may observe
a feedback loop between funding liquidity and asset-liability
management. Finally, it is also obvious that market liquidity is
related to ALM decisions, because of many factors such as trading
policies, the first-mover advantage and crowding effects
\citep{Roncalli-2015a}. It follows that the liquidity risk given in
Figure \ref{fig:alm1} is best described by the dense and fully
connected network given in Figure \ref{fig:alm2}. Nevertheless,
developing a statistical model that takes into account the three
reinforcing loops is not straightforward and certainly too complex
for professional use. Therefore, it is more realistic to adjust and
update the sequential models with second-round effects than to have an integrated dynamic
model.\smallskip

Liquidity is a long-standing issue and also an elusive concept. This is
particularly true in asset management, where liquidity covers several
interpretations. For example, some asset classes are considered as highly liquid
whereas other asset classes are illiquid. In the first category, we generally
find government bonds and large cap stocks. The last category includes real
estate and private equities. However, categorizing liquidity of a security is
not easy and there is no consensus. Let us consider for example Rule 22e-4(b)
that is applied in the US. The proposed rule was based on the ability to
convert the security to cash within a given period and distinguished six
buckets: (a) convertible to cash within 1 business day, (b) convertible to cash
within 2-3 business days, (c) convertible to cash within 4-7 calendar days, (d)
convertible to cash within 8-15 calendar days, (e) convertible to cash within
16-30 calendar days (f) convertible to cash in more than one month. Finally,
the adopted rule is the following:
\begin{enumerate}
\item highly liquid investments (convertible to cash within three business
    days);
\item moderately liquid investments (convertible to cash within four to seven
    calendar days);
\item less liquid investments (expected to be sold in more than seven
    calendar days);
\item illiquid investments (cannot be sold in seven calendar days or less
    without significant price impact).
\end{enumerate}
Classifying a security into a bucket may be different from one fund manager to
another. Moreover, the previous categories depend on the market conditions.
Nevertheless, even if the current market liquidity is abundant, securities that
can be categorized in the first bucket must also face episodes of liquidity
shortage \citep{Blanque-2019a}. A typical example concerns government bonds
facing idiosyncratic risks. \citet{Blanque-2019a} gave the case of Italian
bonds in 2018 during the discussion on the budget deficit. However, most
of the time, when we consider the liquidity of an asset class, we assume that
it is static. Certainly, this way of thinking reflects the practice of
portfolio management. Indeed, it is common to include a constant illiquidity
premium when estimating the expected returns of illiquid assets. But investors
should stick to their investments without rebalancing and trading if they want
to capture this illiquidity premium. The split between liquid and illiquid
investments does not help, because it is related to the absolute level of
asset illiquidity, and not liquidity dynamics. However, the issue
is more complex:
\begin{quote}
\textquotedblleft \textsl{[...] there is also broad belief among users of
financial liquidity -- traders, investors and central bankers -- that the
principal challenge is not the average level of financial liquidity... but
its variability and uncertainty}\textquotedblright\ \citep{Persaud-2003}.
\end{quote}
This observation is important because it is related to the liquidity question
from a regulatory point of view. The liquidity risk of private equities or real
assets is not a big concern for regulators, because one knows that these asset
classes are illiquid. Even if they become more illiquid at some point, this
should not dramatically influence investors (asset managers and owners).
Regulators and investors are more concerned by securities that are liquid under
some market conditions and illiquid under other market conditions. At first
sight, it is therefore a paradox that liquidity stress testing programs must mainly
focus on highly or moderately liquid instruments than on illiquid instruments.
In fact, liquidity does not like surprises and changes. This is why the liquidity
issue is related to the cross-section of the expected illiquidity premium for illiquid
assets, but to the time-series illiquidity variance for liquid
assets.\smallskip

This is all the more important that the liquidity risk must be measured and
managed in a stress testing framework, which adds another layer of complexity.
Indeed, stress scenarios are always difficult to interpret, and
calibrating them is a balancing act, because they must correspond to extreme but
also plausible events \citep{Roncalli-2020}. This is why the historical method
is the most used approach when performing stress testing. However, it is very
poor and not flexible in terms of risk management. Parametric approaches must
be preferred since stress periods are very heterogenous and outcomes are
uncertain. Therefore, it makes more sense to estimate and use stressed
liquidity parameters than directly estimate a stressed liquidity outcome. In
this approach, the normal model is the baseline model on which we could apply
scenario analysis on the different parameters that define the liquidity model.
This is certainly the best way to proceed if we want to develop a factor-based
liquidity stress testing program, which is an important issue for fund
management. Otherwise, liquidity stress testing would be likely to remain a
regulatory constraint or a pure exercise of risk measurement, but certainly not
a risk management process supporting investment policies and fund
management.\smallskip

This paper is organized as follows. Section Two introduces the
concept of redemption rates and defines the historical approach of
liquidity stress testing. In Section Three, we consider parametric
models that can be used to estimate redemption shocks. This
implies making the distinction between the redemption event and the
redemption amount. From a statistical point of view, this is
equivalent to modeling the redemption frequency and the redemption
severity. After having developed an aggregate population model, we
consider an individual-based model. It can be considered as a first
attempt to develop a behavioral model, which is the central theme of
Section Four. We analyze the simple case where redemptions between
investors are independent and then extend the model where
redemptions are correlated to take into account spillover effects
and contagion risk. Then, we develop factor-based models of
liquidity stress testing in Section Five. Finally, Section Six
offers some concluding remarks.

\section{Understanding the liability side of liquidity risk}

In order to assess the liquidity risk of an investment fund, we must model its
\textquoteleft \textit{funding}\textquoteright\ liquidity. Therefore, managing
the liquidity in asset management looks like a banking asset-liability
management process \citep{Roncalli-2020}. However, there is a major difference
since banking ALM concerns both balance sheet and income statement. This is not
the case of an investment fund, because we only focus on its balance sheet and
the objective is to model the redemption flows.

\subsection{Balance sheet of an investment fund}

\begin{figure}[h]
\centering
\caption{Balance sheet of mutual funds}
\label{fig:balance-sheet}
\begin{tikzpicture}

\draw[very thick, black, fill=blue!20] (1,0) rectangle (4,4.0);

\draw[very thick, black, fill=magenta!50] (4,0) rectangle (7,1.0);
\draw[very thick, black, fill=magenta!20] (4,1.0) rectangle (7,4.0);

\draw[very thick] (2.5,2.00) node{$A\left(t\right)$};
\draw[very thick] (5.5,2.50) node{$C\left(t\right)$};
\draw[very thick] (5.5,0.50) node{$D\left(t\right)$};

\draw [decorate,decoration={brace,amplitude=5pt,raise=5pt},line width=1pt] (0.75,0)--(0.75,4) node[left,midway,xshift=-10pt] {};

\draw [decorate,decoration={brace,mirror,amplitude=5pt,raise=5pt},line width=1pt] (7.25,1.02)--(7.25,4) node[right,midway,xshift=10pt] {\textbf{Unit capital}};
\draw [decorate,decoration={brace,mirror,amplitude=5pt,raise=5pt},line width=1pt] (7.25,0.0)--(7.25,0.98) node[right,midway,xshift=10pt] {\textbf{Debits}};

\draw[very thick] (0.00,2.25) node[left,xshift=10pt]{\textbf{Total}};
\draw[very thick] (0.00,1.75) node[left,xshift=10pt]{\textbf{Assets}};

\draw[very thick] (-1.50,2.00) node[left]{$\quad$};

\draw[very thick,color = blue!100] (2.50,4.5) node{\textbf{Assets}};
\draw[very thick,color = magenta!100] (5.50,4.5) node{\textbf{Liabilities}};

\end{tikzpicture}

\end{figure}
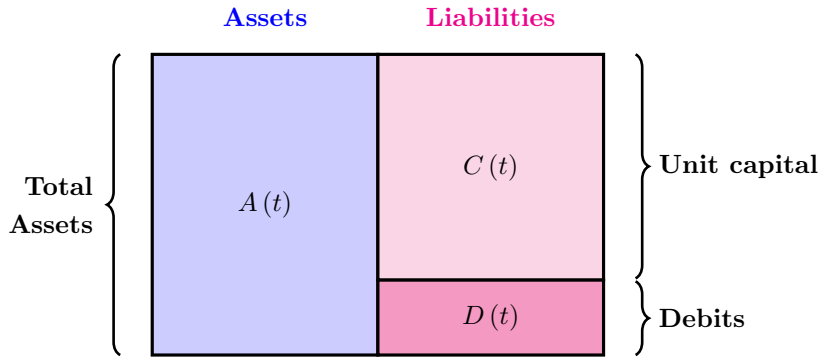

In order to define the liability risk, we first have to understand
the balance sheet of a collective investment fund. A simplified
illustration is given in Figure \ref{fig:balance-sheet} for a mutual
fund. The total (gross) assets $A\left( t\right) $ correspond to the
market value of the investment portfolio. They include stocks, bonds
and all financial instruments that are invested. On the liability
side, we have two main balance sheet items. The first one
corresponds to the debits $D\left( t\right) $, which are also called
current or accrued liabilities. They are all the expenses incurred
by the mutual fund. For instance, the current liabilities include
money owed to lending banks, fees owed to the fund manager and the
custodian, etc. The second liability item is the unit capital
$C\left( t\right) $, which is owned by the investors. Each investor
owns a number of units (or shares) and is referred to as a
\textquoteleft \textit{unitholder}\textquoteright. This unit capital
is equivalent to the equity concept of a financial institution. A
unitholder is then also called a shareholder in reference to capital
markets.

\subsubsection{Definition of net asset value}

The total net assets (TNA) equal the total value of assets less the
current or accrued liabilities:
\begin{equation*}
\limfunc{TNA}\left( t\right) =A\left( t\right) -D\left( t\right)
\end{equation*}%
The net asset value (NAV) represents the share price or the unit price. It is
equal to:
\begin{equation}
\limfunc{NAV}\left( t\right) =\frac{\limfunc{TNA}\left( t\right) }{N\left(
t\right) } \label{eq:nav1}
\end{equation}%
where the total number $N\left( t\right) $ of shares or units in
issue is the sum of all units owned by all unitholders. The previous
accounting rules show that the capital is exactly equal to the total
net assets, which is also called the assets under management (AUM).
The investment fund's capital is therefore an endogenous variable
and depends on the performance of the total net assets:
\begin{eqnarray*}
C\left( t\right) &=&N\left( t\right) \cdot \limfunc{NAV}\left( t\right)  \\
&=&\limfunc{TNA}\left( t\right)
\end{eqnarray*}%
At time $t+1$, we assume that the portfolio's return is equal to $R\left(
t+1\right) $. Since $D\left( t\right) \ll A\left( t\right) $, it follows
that:%
\begin{eqnarray*}
\limfunc{TNA}\left( t+1\right)  &=&A\left( t+1\right) -D\left( t+1\right)  \\
&=&\left( 1+R\left( t+1\right) \right) A\left( t\right) -D\left( t+1\right) \\
&\approx &\left( 1+R\left( t+1\right) \right) \cdot \limfunc{TNA}\left(
t\right)
\end{eqnarray*}%
meaning that:
\begin{equation*}
\limfunc{NAV}\left( t+1\right) \approx \left( 1+R\left( t+1\right) \right) \cdot
\limfunc{NAV}\left( t\right)
\end{equation*}%
The investment fund's capital is therefore time-varying. It
increases when the performance of the asset is positive, and it
decreases otherwise.

\begin{remark}
In the sequel, we assume that the mutual fund is priced daily,
meaning that the NAV of the mutual fund is calculated at the end of the market
day. Therefore, the time $t$ represents the current market day, whereas the
time $t+1$ corresponds to the next market day.
\end{remark}

\subsubsection{The effect of subscriptions and redemptions}

Let us now introduce the impact of subscriptions and redemptions. In this case,
new and current investors may purchase new mutual fund units, while
existing investors may redeem all or part of their shares. Subscription and
redemption orders must be known by the fund manager before $t+1$ in order to be
executed. In this case, the number of units becomes:
\begin{equation*}
N\left( t+1\right) =N\left( t\right) +N^{+}\left( t+1\right) -N^{-}\left(
t+1\right)
\end{equation*}%
where $N^{+}\left( t+1\right) $ is the number of units to be created and
$N^{-}\left( t+1\right) $ is the number of units to be redeemed. At time $t+1$,
we have:
\begin{eqnarray*}
\limfunc{NAV}\left( t+1\right)  &=&\frac{\limfunc{TNA}\left( t+1\right) }{%
N\left( t+1\right) } \\
&=&\frac{\limfunc{TNA}\left( t+1\right) }{N\left( t\right) +N^{+}\left(
t+1\right) -N^{-}\left( t+1\right) }
\end{eqnarray*}%
We deduce that:%
\begin{equation}
\limfunc{TNA}\left( t+1\right) =N\left( t\right) \cdot \limfunc{NAV}\left(
t+1\right) +\mathcal{F}^{+}\left( t+1\right) -\mathcal{F}^{-}\left(
t+1\right)   \label{eq:nav2}
\end{equation}%
where $\mathcal{F}^{+}\left( t+1\right) =N^{+}\left( t+1\right) \cdot
\limfunc{NAV}\left( t+1\right) $ and $\mathcal{F}^{-}\left( t+1\right)
=N^{-}\left( t+1\right) \cdot \limfunc{NAV}\left( t+1\right) $ are the
investment inflows and outflows. Again, we notice that the
investment fund's capital is time-varying and depends on the fund flows.\smallskip

From Equation (\ref{eq:nav2}), we deduce that:
\begin{eqnarray*}
\limfunc{TNA}\left( t+1\right)  &=&N\left( t\right) \cdot \limfunc{NAV}%
\left( t+1\right) +\mathcal{F}^{+}\left( t+1\right) -\mathcal{F}^{-}\left(
t+1\right)  \\
&\approx &N\left( t\right) \cdot \left( 1+R\left( t+1\right) \right) \cdot
\limfunc{NAV}\left( t\right) +\mathcal{F}^{+}\left( t+1\right) -\mathcal{F}%
^{-}\left( t+1\right)  \\
&=&\left( 1+R\left( t+1\right) \right) \cdot \limfunc{TNA}\left( t\right) +%
\mathcal{F}^{+}\left( t+1\right) -\mathcal{F}^{-}\left( t+1\right)
\end{eqnarray*}%
The current net assets are approximatively equal to the previous net
assets plus the performance value plus the net flow. We retrieve the
famous formula of \citet{Sirri-1998} when we want to estimate
the net flow from the NAV and TNA of the fund:
\begin{eqnarray}
\mathcal{F}\left( t+1\right)  &=&\mathcal{F}^{+}\left( t+1\right) -\mathcal{F%
}^{-}\left( t+1\right)   \notag \\
&=&\limfunc{TNA}\left( t+1\right) -\left( 1+R\left( t+1\right) \right) \cdot
\limfunc{TNA}\left( t\right)   \notag \\
&=&\limfunc{TNA}\left( t+1\right) -\left( \frac{\limfunc{NAV}\left(
t+1\right) }{\limfunc{NAV}\left( t\right) }\right) \limfunc{TNA}\left(
t\right)   \label{eq:nav3}
\end{eqnarray}

\subsubsection{Liability risks}

Since the capital is a residual, we face three liability risks. The first one
deals with the accrued liabilities $D\left( t\right) $. Generally, the debits
are a very small part of the liabilities. However, we can potentially face some
situations where the debits are larger than the assets, implying that the net
asset value becomes negative. In particular, this type of situation occurs when
the fund is highly leveraged. The second risk concerns the inflows. If the
investment fund has a big subscription, it may have some difficulties buying
the financial instruments. For instance, this type of situation may occur when
the fund must buy fixed-income securities in a bond bull market and it is
difficult to find investors who are looking to sell bonds. The third liability
risk is produced by the outflows. In this case, the fund manager must sell
assets, which could be difficult in illiquid and stressed market conditions.
The last two situations are produced when supply and demand dynamics are
totally unbalanced (higher supply for buying assets or higher demand for
selling assets). In this article, we focus on the third liability risk, which
is also called redemption risk.

\subsection{Measuring redemption risk}

In order to assess an investment fund's redemption risk, we
need an objective measurement system, which is well scaled. For
instance, the outflows $\mathcal{F}^{-}\left( t\right) $ are not
very interesting, because they depend on the investment fund's
assets under management. In fact, they must be scaled in order to be
a homogeneous measure that can be used to compare the redemption
behavior across time, across funds and across investors.

\subsubsection{Gross redemption rate}

The (gross) redemption rate is defined as the ratio between the
fund's redemption flows and total net assets:
\begin{equation}
\redemption\left( t\right) =\frac{\mathcal{F}^{-}\left( t\right) }{\limfunc{TNA}\left( t\right) }
\label{eq:redemption1}
\end{equation}
We verify the property that $\redemption\left( t\right) \in \left[ 0,1\right]
$. For example, if we observe an outflow of $\$100$ mn for a fund of $\$5$ bn,
we have $\redemption\left( t\right) =100/5\,000=2\%$. In the case where the
outflow is $\$10$ mn and the fund size is $\$100$ mn, the redemption rate is
equal to $10\%$. The redemption is more severe for the small fund than for the
large fund.\smallskip

We notice that Equation (\ref{eq:redemption1}) is used to calculate the
ex-post redemption rate, meaning that the value of outflows is known.
Therefore, Equation (\ref{eq:redemption1}) corresponds to the definition of
the redemption rate, but it can also be used to estimate or predict the
redemption flows. Indeed, we have:
\begin{equation}
\mathcal{\hat{F}}^{-}\left( t+1\right) =\redemption\left( t+1\right) \cdot
\limfunc{TNA}\left( t\right)   \label{eq:redemption2}
\end{equation}%
In this case, $\redemption\left( t+1\right) $ is a random variable
and is not known at the current time $t$. By assuming that
redemption rates are stationary, the challenge is then to model the
associated probability distribution $\mathbf{F}$.

\subsubsection{Net redemption rate}

The guidelines on the liquidity stress testing published by \citet{ESMA-2019}
refer to both gross and net redemptions:
\begin{quote}
\textquotedblleft \textsl{LST should be adapted appropriately to each fund,
including by adapting: [...] the assumptions regarding investor behaviour
(gross and net redemptions)}\textquotedblright\ \citep[page 36]{ESMA-2019}.
\end{quote}
Following this remark, we can also define the net flow rate by
considering both inflows and outflows:%
\begin{equation}
\redemption^{\pm }\left( t\right) =\frac{\mathcal{F}\left( t\right) }{%
\limfunc{TNA}\left( t\right) }  \label{eq:redemption3}
\end{equation}%
This quantity is more complex than the previous one, because it cannot be
used from an ex-ante point of view:
\begin{equation*}
\mathcal{\hat{F}}\left( t+1\right) \neq \redemption^{\pm }\left( t+1\right)
\cdot \limfunc{TNA}\left( t\right)
\end{equation*}%
The reason is that the outflows are bounded and cannot exceed the
current assets under management. This is not the case for the inflows.
For example, we consider a fund with a size of $\$100$ mn. By
construction, we have\footnote{In order to simplify the calculus, we
do not take into account the daily performance of the fund.}
$\mathcal{\hat{F}}^{-}\left( t+1\right) \leq 100$, but we can
imagine that $\mathcal{\hat{F}}^{+}\left( t+1\right) >100$. The fund
size can double or triple, in particular when the investment fund is
young and small.\smallskip

Nevertheless, the use of net flows is not foolish since the true liability risk
of the fund is on the net flows. If the fund manager faces a large redemption,
which is offset by a large subscription, there is no liquidity risk. The
issue is that the use of net flows\ is difficult to justify in stress periods.
In these cases, inflows generally disappear and the probability distribution of
$\redemption^{\pm }\left( t\right) $ may not reflect the liability risk in a
stress testing framework. For example, let us consider an asset class that has
experienced a bull market over the last three years. Certainly, we will
mainly observe positive net flows and a very small number of observations with
negative net flows. We may think that these data are not relevant for
building stress scenarios. More generally, if an asset manager uses net flow
rates for stress testing purposes, only the observations during historical
stress periods are relevant, meaning that the calibration is based on a small
fraction of the dataset.\smallskip

In fact, the use of net flows is motivated by other considerations. Indeed,
the computation of $\redemption\left( t\right) $ requires us to know the
outflows $\mathcal{F}^{-}\left( t\right) $ exactly. Moreover, as we will see later,
$\redemption\left( t\right) $ must be computed for all the investor
categories that are present in the fund (retail, private banking,
institutional, etc.). This implies in-depth knowledge of the fund's balance sheet
liability, meaning that the asset manager must have a database
with all the flows of all the investors on a daily basis. From an industrial
point of view, this is a big challenge in terms of IT systems between the asset
manager and the custodian. This is why many asset managers don't have the
disaggregated information on the liability flows. An alternative measure is
to compute the net redemption rate, which corresponds to the negative part of
the net flow rate:
\begin{equation*}
\redemption^{-}\left( t\right) =\max \left( 0,-\frac{\mathcal{F}\left(
t\right) }{\limfunc{TNA}\left( t\right) }\right)
\end{equation*}%
It has the good mathematical property that $\redemption^{-}\left( t\right) \in \left[ 0,1%
\right] $. Indeed, we have:%
\begin{equation}
\redemption^{-}\left( t\right) =\max \left( 0,\frac{\mathcal{F}^{-}\left(
t\right) -\mathcal{F}^{+}\left( t\right) }{\limfunc{TNA}\left( t\right) }%
\right)   \label{eq:redemption4}
\end{equation}%
and its maximum value is reached when $\mathcal{F}^{-}\left( t\right) =%
\limfunc{TNA}\left( t\right) $ and $\mathcal{F}^{+}\left( t\right) =0$.
Moreover, we notice that the net redemption rate is equal to the gross
redemption rate when there are no inflows:
\begin{equation*}
\redemption^{-}\left( t\right) =\max \left( 0,\frac{\mathcal{F}^{-}\left(
t\right) }{\limfunc{TNA}\left( t\right) }\right) =\redemption\left( t\right)
\end{equation*}%
Otherwise, we have:%
\begin{equation*}
\redemption^{-}\left( t\right) <\redemption\left( t\right)
\end{equation*}%
From a risk management point of view, it follows that redemption
shocks based on net redemptions may be underestimated compared to
redemption shocks based on gross redemptions. However, we will see
later that the approximation $\redemption\left( t\right) \approx
\redemption^{-}\left( t\right)$ may be empirically valid under some
conditions.

\subsubsection{Liability classification}

The computation of redemption rates only makes sense if they are
homogeneous, coherent and comparable. Let us assume that we compute
the redemption rate $\redemption\left( t\right) $ at the level of
the asset management company, and we have the historical data for the
last ten years. By assuming that there are $260$ market days per
year, we have a sample of $2\,600$ redemption rates. We can compute
the mean, the standard deviation, different quantiles, etc. Does it
help with building a stress scenario for a mutual fund? Certainly
not, because redemptions depend on the specific investor behavior at
the fund level and not on the overall investor behavior at the asset
manager level. For instance, we can assume that an investor does not
have the same behavior if he is invested in an equity fund or a
money market fund. We can also assume that the redemption behavior is
not the same for a central bank, a retail investor, or a pension
fund. Therefore, we must build categories that correspond to
homogenous behaviors. Otherwise, we will obtain categories, whose
behavior is non-stationary. But, without the stationarity property,
risk measurement is impossible and stress testing is a hazardous
exercise.\smallskip

Therefore, liability categorization is an important step before computing
redemption rates. For instance, \citet{ESMA-2019} considers four factors
regarding investor behavior: investor category, investor concentration,
investor location and investor strategy. Even though the last three factors are
significant, the most important factor remains the investor type. For
instance, \citet[page 12]{AMF-2017} gives an example with the following
investor types: large institutional (tier one), small institutional (tier
two), investment (or mutual) fund, private banking network and retail
investor. Other categories can be added: central bank, sovereign, corporate,
third-party distributor, employee savings plan, wealth management, etc.
Moreover, it is also important to classify funds into homogeneous buckets\
such as balanced funds, bond funds, equity funds, etc. An example of
an investor/fund categorization matrix is given in Table \ref{tab:liability1}.

\begin{table}[tbph]
\center
\caption{An example of two-dimensional categorization matrix}
\label{tab:liability1}
\begin{tabular}{l|c|c|c|c|c|c|c|c|c}
\multicolumn{1}{l}{Investor category} & \tabr{Absolute return} & \tabr{Balanced} & \tabr{Bond} & \tabr{Commodity} &
           \tabr{Enhanced treasury} & \tabr{Equity} &
           \tabr{Money market} & \tabr{Real asset} & \tabr{Structured} \\ \hline
Central bank             & & & & & & & & & \\
Corporate                & & & & & & & & & \\
Institutional            & & & & & & & & & \\
Insurance                & & & & & & & & & \\
Internal                 & & & & & & & & & \\
Pension fund             & & & & & & & & & \\
Retail                   & & & & & & & & & \\
Sovereign                & & & & & & & & & \\
Third-party distributor  & & & & & & & & & \\
Wealth management        & & & & & & & & & \\ \hline
\end{tabular}
\end{table}

\begin{remark}
The granularity of the investor/fund classification is an important issue. It
is important to have a very detailed classification at the level of the
database in order to group categories together from a computational point of view. In
order to calibrate stress scenarios, we must have a sufficient number of
observations in each cell of the classification matrix. Let us for instance
consider the case of central banks. We can suppose that their behavior is very
different to the other investors. Therefore, it is important for an asset
manager to be aware of the liabilities with respect to central banks. Nevertheless,
there are few central banks in the world, meaning we may not have enough
observations for calibrating some cells (e.g. central bank/equity or central
bank/real asset), and we have to merge some cells (across investor and fund
categories).
\end{remark}

\subsubsection{The arithmetic of redemption rates}

We consider a fund. We note $\limfunc{TNA}\nolimits_{i}\left(
t\right) $ the assets under management of the investor $i$ for this
fund. By definition, we have:
\begin{equation*}
\limfunc{TNA}\nolimits_{i}\left( t\right) =N_{i}\left( t\right) \cdot
\limfunc{NAV}\left( t\right)
\end{equation*}%
where $\limfunc{NAV}\left( t\right) $ is the net asset value of the
fund and $N_{i}\left( t\right) $ is the number of units held by the
investor $i$ for the fund. The fund's assets under management are
equal to:
\begin{equation*}
\limfunc{TNA}\left( t\right) =\sum_{k}\limfunc{TNA}\nolimits_{\left(
k\right) }\left( t\right)
\end{equation*}%
where $\limfunc{TNA}\nolimits_{\left( k\right) }\left( t\right) =\sum_{i\in
\mathcal{IC}_{\left( k\right) }}\limfunc{TNA}\nolimits_{i}\left( t\right) $,
and $\mathcal{IC}_{\left( k\right) }$ is the \textit{k}$^\mathrm{th}$ investor
category. It follows that:%
\begin{eqnarray*}
\limfunc{TNA}\left( t\right)  &=&\sum_{k}\sum_{i\in \mathcal{IC}_{\left(
k\right) }}\limfunc{TNA}\nolimits_{i}\left( t\right)  \\
&=&\sum_{k}\sum_{i\in \mathcal{IC}_{\left( k\right) }}N_{i}\left( t\right)
\cdot \limfunc{NAV}\left( t\right)  \\
&=&N\left( t\right) \cdot \limfunc{NAV}\left( t\right)
\end{eqnarray*}%
where $N\left( t\right) =\sum_{k}\sum_{i\in \mathcal{IC}_{\left(
k\right) }}N_{i}\left( t\right) $ is the total number of units in
issue. We retrieve the definition of the assets under management (or
total net assets) at the fund level. We can obtain a similar
breakdown for the outflows\footnote{We have
$\mathcal{F}_{k}^{-}\left( t\right) =\sum_{i\in \mathcal{IC}_{\left(
k\right) }}\mathcal{F}_{i}^{-}\left( t\right) $.}:
\begin{equation*}
\mathcal{F}^{-}\left( t\right) =\sum_{k}\sum_{i\in \mathcal{IC}_{\left(
k\right) }}\mathcal{F}_{i}^{-}\left( t\right) =\sum_{k}\mathcal{F}_{\left(
k\right) }^{-}\left( t\right)
\end{equation*}%
The redemption rate for the investor category $\mathcal{IC}_{\left( k\right)
}$ is then equal to:%
\begin{equation}
\redemption_{\left( k\right) }\left( t\right) =\frac{\mathcal{F}_{\left(
k\right) }^{-}\left( t\right) }{\limfunc{TNA}\nolimits_{\left( k\right)
}\left( t\right) }  \label{eq:redemption5}
\end{equation}%
We deduce that the relationship between the investor-based redemption rates
and the fund-based redemption rate is:
\begin{eqnarray}
\redemption\left( t\right)  &=&\frac{\mathcal{F}^{-}\left( t\right) }{%
\limfunc{TNA}\left( t\right) }  \notag \\
&=&\frac{\sum_{k}\mathcal{F}_{\left( k\right) }^{-}\left( t\right) }{%
\limfunc{TNA}\left( t\right) }  \notag \\
&=&\frac{\sum_{k}\limfunc{TNA}\nolimits_{\left( k\right) }\left( t\right)
\cdot \redemption_{\left( k\right) }\left( t\right) }{\limfunc{TNA}\left(
t\right) }  \notag \\
&=&\sum_{k}\omega _{\left( k\right) }\left( t\right) \cdot \redemption%
_{\left( k\right) }\left( t\right)   \label{eq:redemption6}
\end{eqnarray}%
where $\omega _{\left( k\right) }\left( t\right) $ represents the weights of
the investor category $\mathcal{IC}_{\left( k\right) }$ in the fund:%
\begin{equation*}
\omega _{\left( k\right) }\left( t\right) =\frac{\limfunc{TNA}%
\nolimits_{\left( k\right) }\left( t\right) }{\limfunc{TNA}\left( t\right) }
\end{equation*}%
Equation (\ref{eq:redemption6}) is very important, because it shows that the
redemption rate at the fund level is a weighted-average of the redemption
rates of the different investor categories.\smallskip

Let us now consider different funds. We note $\redemption_{\left( f,k\right)
}\left( t\right) $ as the redemption rate of the investor category
$\mathcal{IC}_{\left( k\right) }$ for the fund $f$ at time $t$. By relating
the fund $f$ to its fund category $\mathcal{FC}_{\left( j\right) }$, we
obtain a database of redemption rates by investor category
$\mathcal{IC}_{\left( k\right) }$ and fund category $\mathcal{FC}_{\left(
j\right) }$:
\begin{equation*}
\mathcal{DB}_{\left( j,k\right) }\left( T\right) =\left\{
\redemption_{\left( f,k\right) }\left( t\right) :f\in \mathcal{FC}_{\left( j\right)
},t\in T\right\}
\end{equation*}%
$\mathcal{DB}_{\left( j,k\right) }\left( T\right) $ is then the sample of all
redemption rates of the investor category $\mathcal{IC}_{\left( k\right) }$ for
all the funds that fall into the fund category $\mathcal{FC}_{\left( j\right)
}$ during the observation period $T$. We notice that $\mathcal{DB}_{\left(
j,k\right) }\left( t\right) $ does not have a unique element for a given date $t$
because we generally observe several redemptions at the same date for different
funds and the same investor category.

\subsection{Calibration of historical redemption scenarios}

The key parameter for computing the redemption flows is the redemption rate,
which is defined for an investor category and a fund category. It is
not calibrated at the fund level, because past redemption data
for a given specific fund are generally not enough to obtain a robust
estimation. This is why we have pooled redemption data as described in the
previous paragraph. Using these data, we can estimate the probability
distribution $\mathbf{F}$ of the redemption rate and define several
statistics that can help to build stress scenarios.

\subsubsection{Data}

In what follows, we consider the liability data provided by Amundi
Asset Management from January, $1^{\mathrm{st}}$ 2019 to August,
$19^{\mathrm{th}}$ 2020. The database is called \textquoteleft
\textit{Amundi Cube Database}\textquoteright\ and contains
$1\,617\,403$ observations if we filter based on funds with assets under
management greater than $\text{\euro}5$ mn. The breakdown by
investor categories\footnote{The Amundi database contains $13$
investor and $13$ fund categories.} is given in Table
\ref{tab:data1-2} on page \pageref{tab:data1-2}. The number of
observations is $464\,399$ for retail investors, $310\,452$ for
third-party distributors, $267\,600$ for institutionals, etc. The
investor category which is the smallest is central banks with
$15\,523$ observations. In terms of fund categories, bond, equity
and balanced funds dominate with respectively $452\,942$, $436\,401$
and $361\,488$ observations. The smallest categories are private
loan funds and real estate funds. In terms of classification matrix,
the largest matrix cells are retail/balanced, third-party
distributor/equity, retail/equity, institutional/bond, retail/bond,
third-party distributor/bond, retail/structured, etc.

\begin{remark}
In what follows, we apply a filter that consists in removing observations that
corresponds to dedicated mutual funds (FCP and SICAV) and mandates
(see Table \ref{tab:data3-4} on page \pageref{tab:data3-4}). The motivation is
to focus on mutual funds with several investors, and this issue will be extensively
discussed in Section \ref{section:mandates} on page \pageref{section:mandates}.
\end{remark}

\subsubsection{Net flow, net redemption and gross redemption rates}

\begin{figure}[tbph]
\centering
\caption{Retail investor}
\label{fig:rate1}
\includegraphics[width = \figurewidth, height = \figureheight]{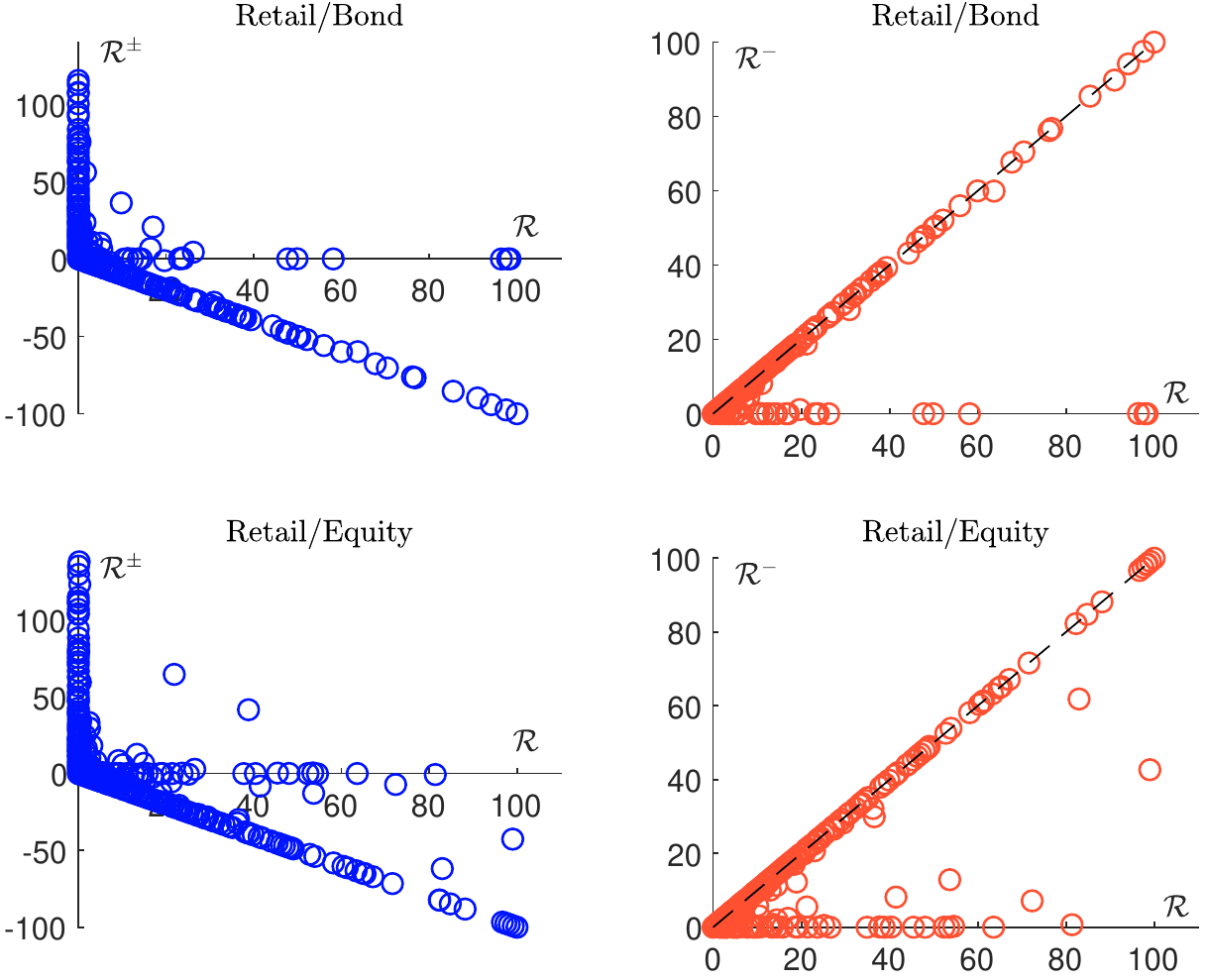}
\end{figure}

\begin{figure}[tbph]
\centering
\caption{Insurance}
\label{fig:rate2}
\includegraphics[width = \figurewidth, height = \figureheight]{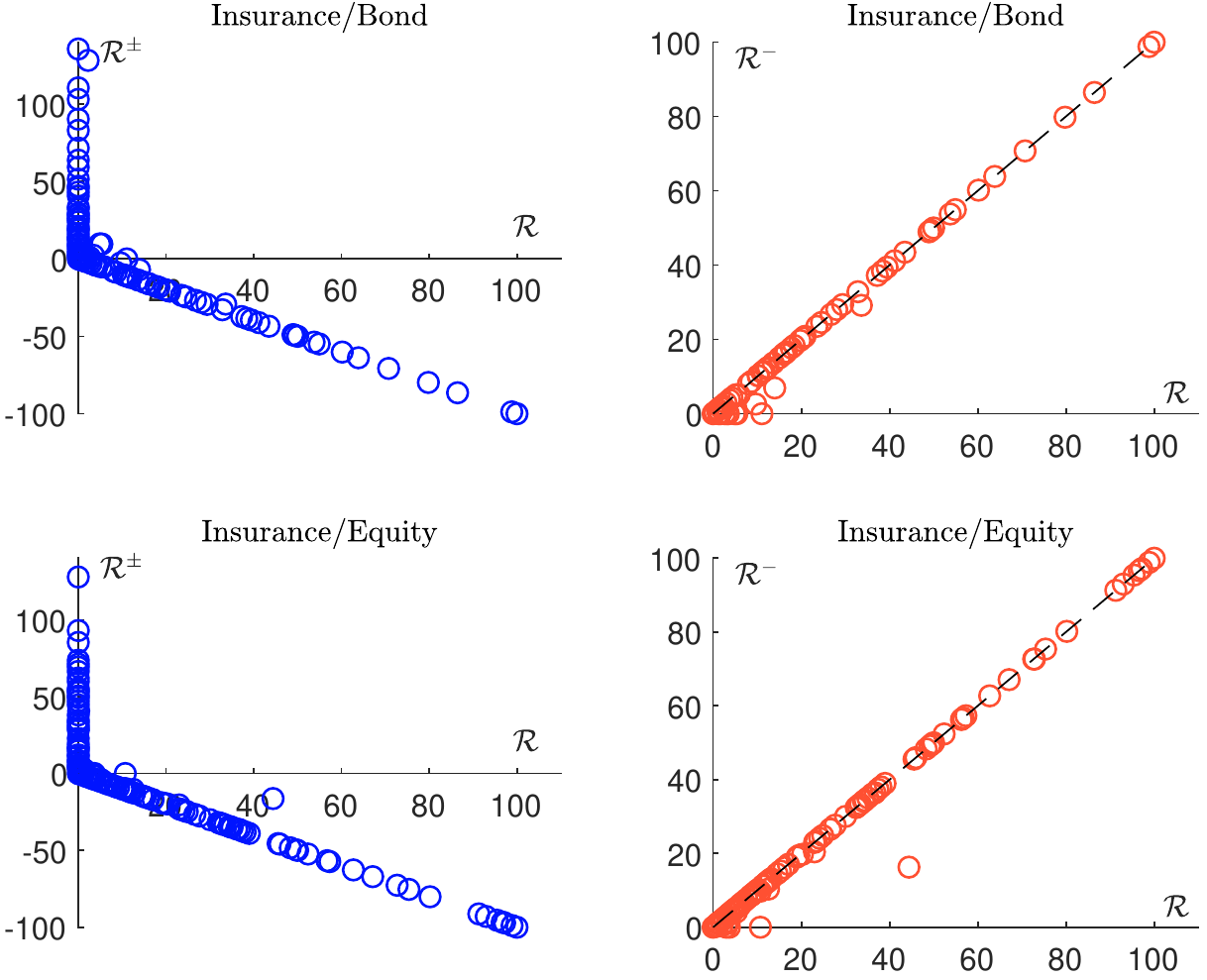}
\end{figure}

\begin{figure}[tbph]
\centering
\caption{Money market fund}
\label{fig:rate4}
\includegraphics[width = \figurewidth, height = \figureheight]{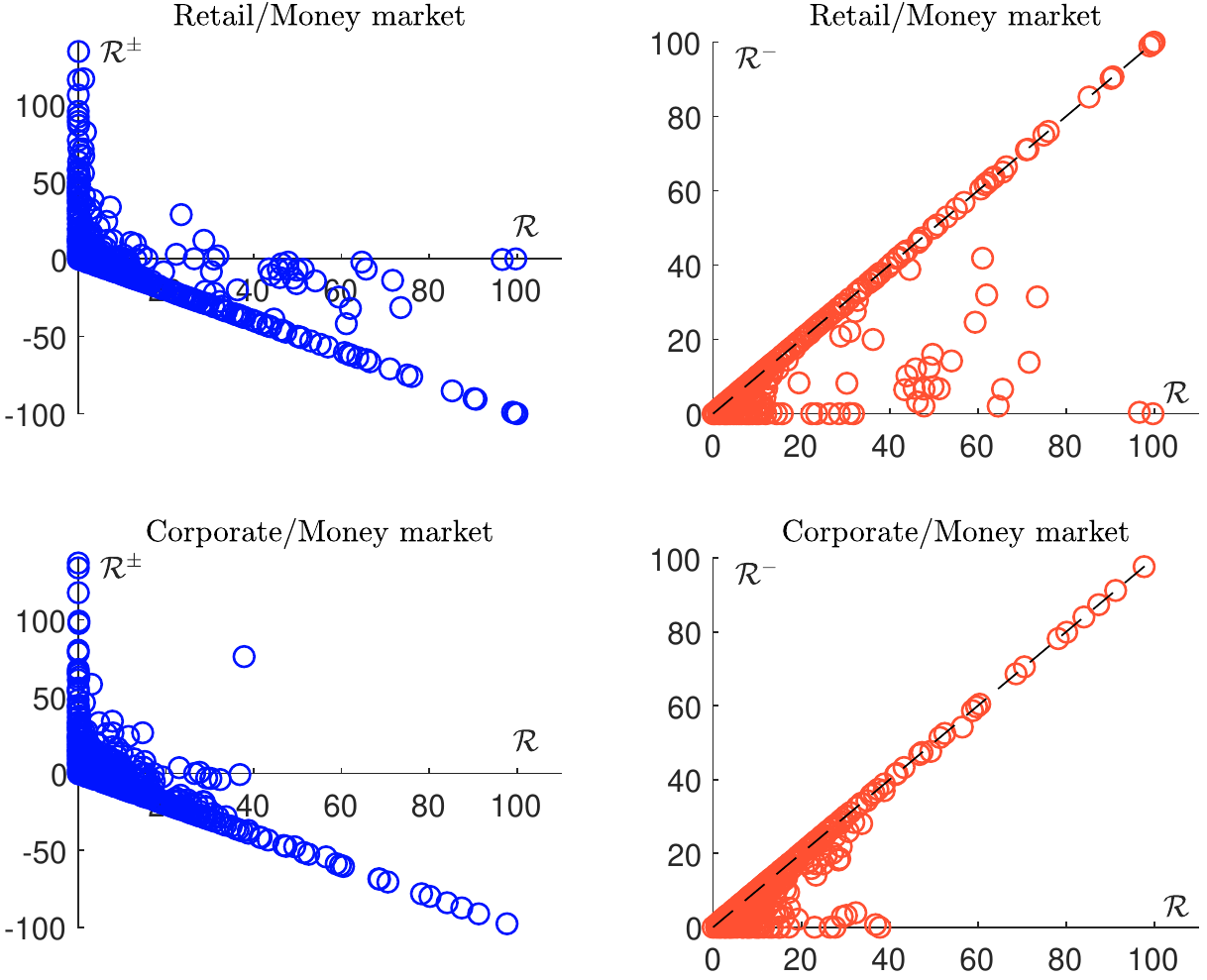}
\end{figure}

We first begin by comparing the gross redemption rate $\redemption$, the net
flow rate $\redemption^{\pm }$ and the net redemption rate $\redemption^{-}$.
Some results are given in Figures \ref{fig:rate1} and \ref{fig:rate2} for
retail and insurance investors and bond and equity funds. In the case of
insurance companies, we notice that the approximation $\redemption\approx
\redemption^{-}\approx -\redemption^{\pm }$ is valid when the redemption rate
is greater than 20\%. This is not the case for retail investors, where we
observe that some large redemptions may be offset by large
subscriptions\footnote{We observe the same phenomenon when we consider the data
of third-party distributors (see Figure \ref{fig:rate3} on page
\pageref{fig:rate3}).}. The difference between retail and insurance categories
lies in the investor concentration. When an investor category is concentrated,
there is a low probability that this offsetting effect will be observed. This is not the
case when the granularity of the investor category is high. We also observe
that the approximation $\redemption\approx \redemption^{-}\approx
-\redemption^{\pm }$ depends on the fund category. For instance, it is not
valid for money market funds. The reason is that we generally observe
subscriptions in a bull market and redemptions in a bear market
when the investment decision mainly depends on the performance of the
asset class. This is why large redemptions and subscriptions tend to be
mutually exclusive (in the mathematical sense) in equity or bond funds. The
mutual exclusivity property is more difficult to observe for money market,
enhanced treasury or balanced funds, because their inflows are less dependent
on market conditions. We conclude that net redemption rates may be used in
order to perform stress scenarios under some conditions regarding the
concentration of the investor category and the type of mutual
fund.\smallskip

\subsubsection{Statistical risk measures}

For a given investor/fund category, we note $\mathbf{F}$ as the probability
distribution of the redemption rate. We can define several risk measures
\citep[pages 62-63]{Roncalli-2020}:
\begin{itemize}
\item the mean:%
\begin{equation*}
\mathbb{M}=\int_{0}^{1}x\,\mathrm{d}\mathbf{F}\left( x\right)
\end{equation*}

\item the standard deviation-based risk measure:%
\begin{equation*}
\mathbb{SD}\left( c\right) =\mathbb{M}+c\int_{0}^{1}\left( x-\mathbb{M}%
^{2}\right) \,\mathrm{d}\mathbf{F}\left( x\right)
\label{eq:SD-measure}
\end{equation*}

\item the quantile (or the value-at-risk) at the confidence level $\alpha $:%
\begin{equation*}
\mathbb{Q}\left( \alpha \right) =\mathbf{F}^{-1}\left( \alpha \right)
\end{equation*}

\item the average beyond the quantile (or the conditional value-at-risk):%
\begin{equation*}
\mathbb{C}\left( \alpha \right) =\mathbb{E}\left[ \redemption\mid \redemption%
\geq \mathbf{F}^{-1}\left( \alpha \right) \right]
\end{equation*}
\end{itemize}
The choice of a risk measure depends on its use. For instance, $\mathbb{M}$
can be used by the fund manager daily, because it is the expected
value of the daily redemption rate. If the fund manager prefers to have a
more conservative measure, he can use $\mathbb{SD}\left( 1\right) $.
$\mathbb{M}$ and $\mathbb{SD}\left( c\right) $ make sense in normal
periods from a portfolio management perspective, but they are less relevant
in a stress period. This is why it is better to use $\mathbb{Q}\left( \alpha
\right) $ and $\mathbb{C}\left( \alpha \right) $ from a risk management point
of view. In the asset management industry, the consensus is to set $\alpha
=99\%$.\smallskip

In Table \ref{tab:historical1}, we have reported the values of $\mathbb{M}$,
$\mathbb{Q}\left(99\%\right)$ and $\mathbb{C}\left(99\%\right)$ by considering
the empirical distribution of gross redemption rates by client category. We do
not consider the $\mathbb{SD}$-measure because we will see later that there is
an issue when it is directly computed from a sample of historical redemption
rates. On average, the expected redemption rate is roughly equal to $20$ bps. It
differs from one client category to another, since the lowest value of
$\mathbb{M}$ is observed for central banks whereas the highest value
of $\mathbb{M}$ is observed for corporates. The $99\%$ value-at-risk is
equal to $3.5\%$. This means that we observe a redemption rate of $3.5\%$ every
$100$ days, that is every five months. Again, there are some big differences
between the client categories. The riskiest category is corporate followed by
sovereign and auto-consumption. If we focus on the
conditional value-at-risk, we are surprised by the high values taken by
$\mathbb{C}\left(99\%\right)$. If we consider all investor categories,
$\mathbb{C}\left(99\%\right)$ is more than $15\%$, and the ratio $\mathbb{R}\left(99\%\right)$ between
$\mathbb{C}\left(99\%\right)$ and $\mathbb{Q}\left(99\%\right)$ is equal to
$4.51$. This is a very high figure since this ratio is generally less than $2$
for market and credit risks. For example, in the case of a Gaussian
distribution $\mathcal{N}\left(0,\sigma^2\right)$, the ratio $\mathbb{R}\left(\alpha\right)$ between the
conditional value-at-risk and the value-at-risk is equal to:
\begin{equation*}
\mathbb{R}\left(\alpha\right) = \frac{\mathbb{C}\left( \alpha \right) }{\mathbb{Q}\left( \alpha \right) }=%
\frac{\phi \left( \Phi ^{-1}\left( \alpha \right) \right) }{\left(
1-\alpha \right) \Phi ^{-1}\left( \alpha \right) }
\end{equation*}%
This ratio is respectively equal to $1.37$ and $1.15$ when $\alpha =90\%$ and
$\alpha =99\%$. Moreover, \citet[page 118]{Roncalli-2020} showed that it is a
decreasing function of $\alpha $ and:
\begin{equation*}
\lim_{\alpha \rightarrow 1^{-}}\mathbb{R}\left(\alpha\right) = 1
\end{equation*}%
We deduce that the ratio is lower than $1.5$ for reasonable values
of the confidence level $\alpha $. Therefore, the previous figure
$\mathbb{R}\left(99\%\right) = 4.51$ indicates
that redemption risk is more skewed than market and credit risks.

\begin{table}[tbph]
\centering
\caption{Redemption statistical measures in \% by investor category}
\label{tab:historical1}
\begin{tabular}{lcccc}
\hline
Client                   & $\mathbb{M}$ &
                           $\mathbb{Q}\left(99\%\right)$ &
                           $\mathbb{C}\left(99\%\right)$ &
                           $\mathbb{R}\left(99\%\right)$ \\ \hline
Auto-consumption          & $0.38$ & ${\TsV}7.44$ &      $24.86$ & ${\TsV}3.34$  \\
Central bank              & $0.04$ & ${\TsV}0.00$ & ${\TsV}4.38$ &  ${\TsX}\infty$ \\
Corporate                 & $0.54$ &      $12.71$ &      $28.21$ & ${\TsV}2.22$  \\
Corporate pension fund    & $0.13$ & ${\TsV}0.50$ &      $13.06$ &      $26.22$  \\
Employee savings plan     & $0.06$ & ${\TsV}1.13$ & ${\TsV}4.86$ & ${\TsV}4.30$  \\
Institutional             & $0.27$ & ${\TsV}5.11$ &      $22.79$ & ${\TsV}4.46$  \\
Insurance                 & $0.26$ & ${\TsV}5.25$ &      $21.24$ & ${\TsV}4.05$  \\
Other                     & $0.23$ & ${\TsV}3.41$ &      $20.22$ & ${\TsV}5.92$  \\
Retail                    & $0.15$ & ${\TsV}1.92$ & ${\TsV}9.18$ & ${\TsV}4.77$  \\
Sovereign                 & $0.45$ & ${\TsV}8.28$ &      $39.85$ & ${\TsV}4.81$  \\
Third-party distributor   & $0.23$ & ${\TsV}3.90$ &      $13.72$ & ${\TsV}3.52$  \\ \hline
Total                     & $0.22$ & ${\TsV}3.50$ &      $15.79$ & ${\TsV}4.51$  \\ \hline
\end{tabular}
\end{table}

Table \ref{tab:historical2} reports the statistical measures by fund category.
Again, we observe some big differences. Money market and enhanced treasury
funds face a high redemption risk followed by bond and equity funds. This is
normal because treasury funds can be converted to cash very quickly, investors
are motivated to redeem these funds when they need cash, and their holding
period is short. At the global level, we also notice that the redemption
behavior is similar between bond and equity funds. For instance, their $99\%$
value-at-risk is close to $3\%$ (compared to $6\%$ for the enhanced treasury
category and $22\%$ for money market funds). Another interesting result is
the lower redemption rate of balanced funds compared to bond and equity funds.
This result is normal because balanced funds are more diversified. Therefore,
investors in balanced funds are more or less protected by a bond crisis or an
equity crisis. Finally, structured funds are the least exposed category to
redemption risk, because they generally include a capital guarantee
or protection option.\smallskip

\begin{table}[tbph]
\centering
\caption{Redemption statistical measures in \% by fund category}
\label{tab:historical2}
\begin{tabular}{lcccc}
\hline
Fund                     & $\mathbb{M}$ &
                           $\mathbb{Q}\left(99\%\right)$ &
                           $\mathbb{C}\left(99\%\right)$ &
                           $\mathbb{R}\left(99\%\right)$ \\ \hline
Balanced           & $0.14$ & ${\TsV}1.77$ & ${\TsV}8.14$ & $4.61$  \\
Bond               & $0.20$ & ${\TsV}3.18$ &      $14.23$ & $4.47$  \\
Enhanced treasury  & $0.40$ & ${\TsV}6.30$ &      $31.15$ & $4.94$  \\
Equity             & $0.18$ & ${\TsV}2.68$ &      $12.94$ & $4.84$  \\
Money market       & $1.06$ &      $21.76$ &      $46.13$ & $2.12$  \\
Other              & $0.11$ & ${\TsV}1.19$ & ${\TsV}9.32$ & $7.84$  \\
Structured         & $0.04$ & ${\TsV}0.45$ & ${\TsV}3.52$ & $7.88$  \\ \hline
Total              & $0.22$ & ${\TsV}3.50$ &      $15.79$ & $4.51$  \\ \hline
\end{tabular}
\end{table}

\begin{table}[tbph]
\centering
\caption{Historical $\mathbb{M}$-statistic in \% by investor/fund category}
\label{tab:historical4a}
\begin{tabular}{lcccccccc}
\hline
                          &    (1) &         (2) &         (3) &         (4) &         (5) &         (6) &         (7) &         (8)  \\ \hline
Auto-consumption          &  $0.27$ & $0.36$ &      $0.65$ & $0.30$ &      $1.58$ &      $0.18$ & ${\TsV}   $ & $0.38$  \\
Central bank              &  $0.01$ & $0.06$ & ${\TsV}   $ & $0.11$ & ${\TsV}   $ & ${\TsV}   $ & ${\TsV}   $ & $0.04$  \\
Corporate                 &  $0.08$ & $0.15$ &      $0.27$ & $0.25$ &      $1.52$ &      $0.07$ & ${\TsV}   $ & $0.54$  \\
Corporate pension fund    &  $0.17$ & $0.05$ &      $0.10$ & $0.10$ &      $0.55$ &      $0.00$ & ${\TsV}   $ & $0.13$  \\
Employee savings plan     &  $0.03$ & $0.05$ &      $0.13$ & $0.06$ &      $0.06$ & ${\TsV}   $ &      $0.08$ & $0.06$  \\
Institutional             &  $0.13$ & $0.16$ &      $0.64$ & $0.18$ &      $1.47$ &      $0.06$ & ${\TsV}   $ & $0.27$  \\
Insurance                 &  $0.17$ & $0.15$ &      $0.12$ & $0.16$ &      $0.90$ &      $0.08$ & ${\TsV}   $ & $0.26$  \\
Other                     &  $0.08$ & $0.10$ &      $0.33$ & $0.21$ &      $0.76$ &      $0.02$ & ${\TsV}   $ & $0.23$  \\
Retail                    &  $0.15$ & $0.14$ &      $0.26$ & $0.16$ &      $0.91$ &      $0.07$ &      $0.04$ & $0.15$  \\
Sovereign                 &  $0.01$ & $0.01$ &      $0.16$ & $0.19$ &      $1.91$ &      $0.06$ & ${\TsV}   $ & $0.45$  \\
Third-party distributor   &  $0.12$ & $0.24$ &      $0.67$ & $0.19$ &      $0.92$ &      $0.28$ &      $0.08$ & $0.23$  \\ \hline
Total                     &  $0.14$ & $0.20$ &      $0.40$ & $0.18$ &      $1.06$ &      $0.11$ &      $0.04$ & $0.22$  \\ \hline
\end{tabular}
\end{table}

\begin{table}[tbph]
\centering
\caption{Historical $\mathbb{Q}$-statistic in \% by investor/fund category}
\label{tab:historical4b}
\begin{tabular}{lcccccccc}
\hline
                          &    (1) &         (2) &          (3) &         (4) &          (5) &         (6) &         (7) &         (8)  \\ \hline
Auto-consumption          & $2.93$ & $7.57$ &      $12.62$ & $5.46$ &      $25.98$ &      $3.23$ & ${\TsV}   $ & ${\TsV}7.44$  \\
Central bank              & $0.00$ & $0.00$ &  ${\TsX}   $ & $0.12$ &  ${\TsX}   $ & ${\TsV}   $ & ${\TsV}   $ & ${\TsV}0.00$  \\
Corporate                 & $0.30$ & $1.58$ & ${\TsV}4.90$ & $3.88$ &      $24.14$ &      $0.00$ & ${\TsV}   $ &      $12.71$  \\
Corporate pension fund    & $0.39$ & $0.05$ & ${\TsV}1.30$ & $0.03$ &      $13.09$ &      $0.00$ & ${\TsV}   $ & ${\TsV}0.50$  \\
Employee savings plan     & $1.06$ & $1.70$ & ${\TsV}2.35$ & $1.08$ & ${\TsV}2.51$ & ${\TsV}   $ &      $0.25$ & ${\TsV}1.13$  \\
Institutional             & $0.84$ & $1.94$ & ${\TsV}8.68$ & $3.10$ &      $34.82$ &      $0.00$ & ${\TsV}   $ & ${\TsV}5.11$  \\
Insurance                 & $0.32$ & $0.21$ & ${\TsV}3.87$ & $0.50$ &      $18.39$ &      $0.00$ & ${\TsV}   $ & ${\TsV}5.25$  \\
Other                     & $0.73$ & $0.56$ & ${\TsV}2.40$ & $2.20$ &      $14.75$ &      $0.05$ & ${\TsV}   $ & ${\TsV}3.41$  \\
Retail                    & $2.01$ & $1.50$ & ${\TsV}4.72$ & $1.65$ &      $18.36$ &      $1.17$ &      $0.45$ & ${\TsV}1.92$  \\
Sovereign                 & $0.11$ & $0.14$ & ${\TsV}7.98$ & $0.22$ &      $66.36$ &      $0.00$ & ${\TsV}   $ & ${\TsV}8.28$  \\
Third-party distributor   & $1.32$ & $4.59$ &      $11.13$ & $3.38$ &      $14.66$ &      $3.96$ &      $1.11$ & ${\TsV}3.90$  \\ \hline
Total                     & $1.77$ & $3.18$ & ${\TsV}6.30$ & $2.68$ &      $21.76$ &      $1.19$ &      $0.45$ & ${\TsV}3.50$  \\ \hline
\end{tabular}
\end{table}

\begin{table}[tbph]
\centering
\caption{Historical $\mathbb{C}$-statistic in \% by investor/fund category}
\label{tab:historical4c}
\begin{tabular}{lcccccccc}
\hline
                          &           (1) &          (2) &          (3) &          (4) &          (5) &          (6) &         (7) &          (8)  \\ \hline
Auto-consumption          &      $21.08$ &      $23.37$ &      $40.73$ &      $21.24$ &      $54.96$ &      $15.50$ & ${\TsV}   $ &      $24.86$  \\
Central bank              & ${\TsV}1.28$ & ${\TsV}6.05$ &  ${\TsX}   $ &      $10.11$ &  ${\TsX}   $ &  ${\TsX}   $ & ${\TsV}   $ & ${\TsV}4.38$  \\
Corporate                 & ${\TsV}7.31$ &      $14.98$ &      $22.80$ &      $22.48$ &      $38.37$ & ${\TsV}6.52$ & ${\TsV}   $ &      $28.21$  \\
Corporate pension fund    &      $17.22$ & ${\TsV}5.14$ & ${\TsV}9.24$ & ${\TsV}9.58$ &      $32.33$ & ${\TsV}0.00$ & ${\TsV}   $ &      $13.06$  \\
Employee savings plan     & ${\TsV}2.48$ & ${\TsV}3.16$ &      $10.60$ & ${\TsV}4.91$ & ${\TsV}4.97$ &  ${\TsX}   $ &      $7.91$ & ${\TsV}4.86$  \\
Institutional             &      $10.99$ &      $15.40$ &      $62.30$ &      $16.27$ &      $58.10$ & ${\TsV}6.26$ & ${\TsV}   $ &      $22.79$  \\
Insurance                 &      $16.35$ &      $14.65$ &      $10.59$ &      $15.32$ &      $37.28$ & ${\TsV}7.62$ & ${\TsV}   $ &      $21.24$  \\
Other                     & ${\TsV}7.45$ & ${\TsV}9.84$ &      $32.56$ &      $18.61$ &      $46.88$ & ${\TsV}2.17$ & ${\TsV}   $ &      $20.22$  \\
Retail                    & ${\TsV}7.02$ & ${\TsV}8.34$ &      $15.99$ & ${\TsV}8.95$ &      $44.38$ & ${\TsV}5.03$ &      $3.03$ & ${\TsV}9.18$  \\
Sovereign                 & ${\TsV}0.39$ & ${\TsV}1.35$ &      $15.20$ &      $17.97$ &      $86.47$ & ${\TsV}5.73$ & ${\TsV}   $ &      $39.85$  \\
Third-party distributor   & ${\TsV}6.69$ &      $14.53$ &      $42.24$ &      $11.22$ &      $32.68$ &      $20.16$ &      $6.85$ &      $13.72$  \\ \hline
Total                     & ${\TsV}8.14$ &      $14.23$ &      $31.15$ &      $12.94$ &      $46.13$ & ${\TsV}9.32$ &      $3.52$ &      $15.79$  \\ \hline
\end{tabular}
\medskip

\begin{flushleft}
\begin{footnotesize}
(1) = balanced, (2) = bond, (3) = enhanced treasury, (4) = equity, (5) = money market,
(6) = other, (7) = structured, (8) = total
\end{footnotesize}
\end{flushleft}
\end{table}

The historical statistical measures\footnote{They are not calculated if the
number of observations is less than $200$.} for the classification matrix are
given in Tables \ref{tab:historical4a}, \ref{tab:historical4b} and
\ref{tab:historical4c}. We notice that the two dimensions are important, since
one dimension does not dominate the other. This means that a low-risk (resp.
high-risk) investor category tends to present the lowest (resp. highest)
redemption statistics whatever the fund category. In addition, the ranking of redemption
statistics between fund categories is similar whatever the investor category.
Nevertheless, we observe some exceptions and new stylized facts. For instance,
we have previously noticed that bond and equity funds have similar redemption rates
on average. This is not the case for the corporate, corporate pension fund and
sovereign categories, for which historical $\mathbb{C}$-statistics are more
important for equity funds than bond funds. For the corporate pension fund category, the
risk is also higher for balanced funds than for bond funds.

\subsubsection{Defining historical stress scenarios}

According to \citet[page 60]{BCBS-2017}, a historical stress scenario
\textquotedblleft \textsl{aims at replicating the changes in risk factor shocks
that took place in an actual past episode}\textquotedblright. If we apply this
definition to the redemption risk, the computation of the historical stress
scenario is simple. First, we have to choose a stress period
$T^{\mathrm{stress}}$ and second, we compute the maximum redemption rate:
\begin{equation*}
\mathbb{X}\left( T^{\mathrm{stress}}\right) =\max_{t\in T^{\mathrm{stress}}}\redemption\left( t\right)
\end{equation*}%
For example, if we apply this definition to our study period, we obtain the results
given in Table \ref{tab:historical4d}. We recall that the study period runs
from January 2019 to August 2020 and includes the Coronavirus pandemic crisis,
which was a redemption stress period. We observe that the
$\mathbb{X} $-statistic is generally equal to $100\%$! This is a big issue,
because it is not helpful to consider that liquidity stress testing of
liabilities leads to a figure of $100\%$. The problem is that the
$\mathbb{X}$-statistic is not adapted to redemption risk. Let us consider an
investor category $\mathcal{IC}_{\left( k\right) }$ and a fund category
$\mathcal{FC}_{\left( j\right) }$. The
$\mathbb{X}$-statistic is computed by taking the maximum of all redemption rates
for all funds that belong to the fund category:
\begin{equation*}
\mathbb{X}_{\left( j,k\right) }\left( T^{\mathrm{stress}}\right)
=\max_{t\in T^{\mathrm{stress}}}\left\{ \redemption_{\left(
f,k\right) }\left( t\right) :f\in \mathcal{FC}_{\left( j\right) }\right\}
\end{equation*}%
If there is one fund with only one investor and if this investor
redeems $100\%$, $\mathbb{X}_{\left( j,k\right) }\left(
T^{\mathrm{stress}}\right) $ is equal to 100\%. However, the asset
manager does not really face a liquidity risk in this situation, because there
is no other investor in this fund. So, the other investors are not penalized.
We have excluded this type of fund. However, we face a similar situation in many
other cases: small funds with a large fund holder, funds with a low number of
unitholders, etc. Moreover, this type of approach penalizes big asset managers,
which have hundreds of funds. Let us consider an example. For a given
investor/fund category, the fund manager $A$ has $100$ funds of $\$100$
million, whereas the fund manager $B$ has one fund of $\$10$ billion. From a
theoretical point of view, $A$ and $B$ face the same redemption risk, since
they both manage $\$10$ billions for the same investor/fund
category. However, it is obvious that $\mathbb{X}_{A}\gg \mathbb{X}_{B}$,
meaning that the historical stress scenario for the fund manager $A$ will be
much higher than the historical stress scenario for the fund manager $B$. This
is just a probabilistic counting principle as shown in Appendix
\ref{appendix:x-statistic} on page \pageref{appendix:x-statistic}. If we
consider the previous example, the historical stress scenario for the fund
manager $A$ is larger than $99.9\%$ when the historical stress scenario for the
fund manager $B$ is larger than $6.68\%$ (see Figure \ref{fig:xstatistic2} on
page \pageref{fig:xstatistic2}). More generally, the two stress scenarios are
related in the following manner:
\begin{equation*}
\mathbb{X}_{n}=1-\left( 1-\mathbb{X}_{1}\right) ^{n}
\end{equation*}%
where $\mathbb{X}_{1}$ is the $\mathbb{X}$-measure for one fund and $\mathbb{%
X}_{n}$ is the $\mathbb{X}$-measure for $n$ funds.

\begin{table}[tbph]
\centering
\caption{Historical $\mathbb{X}$-statistic in \% by investor/fund category}
\label{tab:historical4d}
\begin{tabular}{lccccccc}
\hline
                          &           (1) &           (2) &           (3) &           (4) &          (5) &           (6) &           (7) \\ \hline
Auto-consumption          &       $100.00$ &      $100.00$ &      $100.00$ &      $100.00$ & ${\TsV}99.65$ &      $100.00$ &         $   $ \\
Central bank              &   ${\TsX}9.17$ & ${\TsV}29.60$ &         $   $ & ${\TsV}50.00$ &         $   $ &         $   $ &         $   $ \\
Corporate                 &  ${\TsV}78.64$ & ${\TsV}83.44$ &      $100.00$ & ${\TsV}94.14$ & ${\TsV}97.72$ &      $100.00$ &         $   $ \\
Corporate pension fund    &       $100.00$ &      $100.00$ & ${\TsV}15.79$ &      $100.00$ & ${\TsV}94.78$ &  ${\TsX}0.00$ &         $   $ \\
Employee savings plan     &  ${\TsV}50.79$ & ${\TsV}15.35$ &      $100.00$ &      $100.00$ & ${\TsV}14.71$ &         $   $ &      $100.00$ \\
Institutional             &  ${\TsV}99.09$ &      $100.00$ &      $100.00$ &      $100.00$ &      $100.00$ &      $100.00$ &         $   $ \\
Insurance                 &  ${\TsV}99.99$ &      $100.00$ & ${\TsV}56.96$ &      $100.00$ & ${\TsV}99.93$ & ${\TsV}77.13$ &         $   $ \\
Other                     &  ${\TsV}50.00$ &      $100.00$ &      $100.00$ &      $100.00$ &      $100.00$ &      $100.00$ &         $   $ \\
Retail                    &       $100.00$ &      $100.00$ &      $100.00$ &      $100.00$ &      $100.00$ &      $100.00$ &      $100.00$ \\
Sovereign                 &   ${\TsX}5.44$ & ${\TsV}21.12$ & ${\TsV}24.91$ &      $100.00$ &      $100.00$ &      $100.00$ &         $   $ \\
Third-party distributor   &       $100.00$ &      $100.00$ &      $100.00$ &      $100.00$ & ${\TsV}97.04$ &      $100.00$ & ${\TsV}97.98$ \\ \hline
\end{tabular}

\begin{flushleft}
\begin{footnotesize}
(1) = balanced, (2) = bond, (3) = enhanced treasury, (4) = equity, (5) = money
market, (6) = other, (7) = structured
\end{footnotesize}
\end{flushleft}
\vspace*{-30pt}
\end{table}

\begin{remark}
Another approach consists in computing the average redemption
rate daily:
\begin{equation*}
\redemption_{\left( j,k\right) }\left( t\right) =\sum_{f\in \mathcal{FC}%
_{\left( j\right) }}\frac{\limfunc{TNA}\nolimits_{\left( f\right) }}{%
\sum_{f\in \mathcal{FC}_{\left( j\right) }}\limfunc{TNA}\nolimits_{\left(
f\right) }}\redemption_{\left( f,k\right) }\left( t\right)
\end{equation*}%
where the weights are proportional to the size of funds $f$ that belong to the
$j^{\mathrm{th}}$ fund category $\mathcal{FC}_{\left( j\right) }$. In this
case, we have:
\begin{equation*}
\mathbb{X}_{\left( j,k\right) }\left( T^{\mathrm{stress}}\right) =\max_{t\in
T^{\mathrm{stress}}}\redemption_{\left( j,k\right) }\left( t\right)
\end{equation*}%
This method does not have the previous drawback, but it has other shortcomings such
as an information loss. However, the biggest disadvantage is that the
historical stress scenario is generally based on the largest fund, except when
the funds have similar size.
\end{remark}

Since $\mathbb{X}$-measures can not be used to build redemption shocks, we
propose using $\mathbb{Q}$ or $\mathbb{C}$-measures. $\mathbb{Q}\left(
99\%\right) $ is the daily value-at-risk at the $99\%$ confidence level. This
means that its return period is $100$ days. On average, we must observe that
redemption shocks are greater than $\mathbb{Q}\left( 99\%\right)
$ two and a half times per year. We can also use the conditional value-at-risk
$\mathbb{C}\left( 99\%\right) $ if we want more severe redemption shocks. The
drawback of $\mathbb{C}\left( 99\%\right) $ is that we don't know the
return period of such event. However, it does make sense because it is a very
popular measure in risk management, and it is well received by regulatory bodies
and supervisors \citep{Roncalli-2020}. Nevertheless, we must be cautious about
the computed figures obtained in Tables \ref{tab:historical4b} and
\ref{tab:historical4c} on page \pageref{tab:historical4b}. For example, we
don't have the same confidence level between the matrix cells, because the
estimates are not based on the same number of observations. In the case of
retail investors or third-party distributors, we generally use a huge number of
observations whereas this is not the case with the other categories. In Table
\ref{tab:historical5}, we give an example of confidence level codification. We
see that some cells are not well estimated since the number of observations is
less than $10\,000$. For some of them, the number of observations is very low
(less than $200$), implying that the confidence on these estimates is
very poor.\smallskip

\begin{table}[tbph]
\centering
\caption{Confidence in estimated values with respect to the number of observations}
\label{tab:historical5}
\begin{tabular}{lccccccc}
\hline
                        &   (1) &   (2) &   (3) &   (4) &   (5) &   (6) &   (7)  \\ \hline
Auto-consumption        & \bIII & \bIII & \bII  & \bIII & \bII  & \bIII & \bZZZ \\
Central bank            & \bII  & \bI   & \bZZZ & \bI   & \bZZZ & \bZZZ & \bZZZ \\
Corporate               & \bII  & \bII  & \bII  & \bII  & \bII  & \bII  & \bZZZ \\
Corporate pension fund  & \bII  & \bII  & \bI   & \bII  & \bII  & \bII  & \bZZZ \\
Employee savings plan   & \bII  & \bII  & \bII  & \bIII & \bII  & \bZZZ & \bII  \\
Institutional           & \bII  & \bIII & \bII  & \bIII & \bII  & \bIII & \bZZZ \\
Insurance               & \bII  & \bIII & \bII  & \bIII & \bII  & \bII  & \bZZZ \\
Other                   & \bII  & \bIII & \bII  & \bII  & \bII  & \bIII & \bZZZ \\
Retail                  & \bIII & \bIII & \bII  & \bIII & \bIII & \bIII & \bIII \\
Sovereign               & \bII  & \bII  & \bI   & \bII  & \bII  & \bII  & \bZZZ \\
Third-party distributor & \bIII & \bIII & \bII  & \bIII & \bIII & \bIII & \bII  \\
\hline
\end{tabular}
\begin{flushleft}
\begin{footnotesize}
\bZZZ\ $0-10$, \bZZ\ $11-50$, \bZ\ $51-200$, \bI\ $201-1\,000$, \bII\ $1\,001-10\,000$, \bIII\  $+10\,000$
\end{footnotesize}
\end{flushleft}
\vspace*{-30pt}
\end{table}

Therefore, the estimated values cannot be directly used as redemption shocks.
However, they help risk managers and business experts to build redemption
shocks. Starting from these figures, they can modify them and build a table of
redemption shocks that respect the risk coherency\label{marker:coherency}
$\mathcal{C}_{\mathrm{investor}}$ between investor categories and the risk
coherency $\mathcal{C}_{\mathrm{fund}}$ between fund categories\footnote{For
instance, if we consider the sovereign category, it is difficult to explain the
big difference of $\mathbb{C}\left( 99\%\right) $ between bond and equity
funds}. The risk coherency $\mathcal{C}_{\mathrm{investor}}$ means that if one
investor category is assumed to be riskier than another, the global
redemption shock of the first category must be greater than the global
redemption shock of the second category:
\begin{equation*}
\mathcal{IC}_{\left( k_{1}\right) }\succ \mathcal{IC}_{\left( k_{2}\right)
}\Rightarrow \mathbb{S}_{\left( k_{1}\right) }\geq \mathbb{S}_{\left(
k_{2}\right) } \label{eq:coherency1}
\end{equation*}%
For example, if we consider the $\mathbb{Q}$-measure, we can propose the
following risk ordering:
\begin{enumerate}
\item central bank, corporate pension fund
\item employee savings plan, retail
\item other, third-party distributor
\item institutional, insurance
\item auto-consumption, corporate, sovereign
\end{enumerate}
In this case, the redemption shock $\mathbb{S}_{\left( j,k\right) } $ for the
$\left( j,k\right) $-cell depends on the global redemption shock
$\mathbb{S}_{\left( k\right) }$ for the investor category $\mathcal{IC}_{\left(
k\right) }$. For instance, we can set the following rule of thumb:
\begin{equation}
\mathbb{S}_{\left( j,k\right) }=m_{\left( j\right) }\cdot \mathbb{S}_{\left(
k\right) }  \label{eq:rule-C1}
\end{equation}%
where $m_{\left( j\right) }$ is the multiplicative factor of the fund category
$\mathcal{FC}_{\left( j\right) }$. In a similar way, the risk coherency
$\mathcal{C}_{\mathrm{fund}}$ means that if one fund category is assumed to be
riskier than another, the global redemption shock of the first category must
be greater than the global redemption shock of the second category:
\begin{equation*}
\mathcal{FC}_{\left( j_{1}\right) }\succ \mathcal{FC}_{\left( j_{2}\right)
}\Rightarrow \mathbb{S}_{\left( j_{1}\right) }\geq \mathbb{S}_{\left(
j_{2}\right) } \label{eq:coherency2}
\end{equation*}%
For example, if we consider the $\mathbb{Q}$-measure, we can propose the
following risk ordering:
\begin{enumerate}
\item structured
\item balanced, other
\item bond, equity
\item enhanced treasury
\item money market
\end{enumerate}
The redemption shock $\mathbb{S}_{\left( j,k\right) }$ for the $\left(
j,k\right) $-cell depends then on the redemption shock $\mathbb{S}_{\left(
j\right) }$ for the fund category $\mathcal{IC}_{\left( j\right) }$. Again, we
can set the following rule of thumb:
\begin{equation}
\mathbb{S}_{\left( j,k\right) }=m_{\left( k\right) }\cdot \mathbb{S}_{\left(
j\right) }  \label{eq:rule-C2}
\end{equation}%
where $m_{\left( k\right) }$ is the multiplicative factor of the investor
category $\mathcal{IC}_{\left( k\right) }$. We can also combine the two
rules of thumb and we obtain the mixed rule:
\begin{equation}
\mathbb{S}_{\left( j,k\right) }=\frac{m_{\left( k\right) }\cdot \mathbb{S}%
_{\left( j\right) }+m_{\left( j\right) }\cdot \mathbb{S}_{\left( k\right) }}{%
2}  \label{eq:rule-C3}
\end{equation}
Let us illustrate the previous rules of thumb by considering the
$\mathbb{Q}$-measure. Table \ref{tab:historical6a} gives an example of
$\mathbb{S}_{\left( j,k\right) }$ by considering the risk coherency\footnote{We
use the following values: $\mathbb{S}_{\left( k\right) }=0.5\%$ for central
banks and corporate pension funds, $\mathbb{S}_{\left( k\right) }=2\%$ for
employee savings plans and retail, $\mathbb{S}_{\left( k\right) }=3.5\%$ for
other and third-party distributors, $\mathbb{S}_{\left( k\right) }=5\%$ for
institutionals and insurance companies, and $\mathbb{S}_{\left( k\right) }=8\%$
for auto-consumption, corporates and sovereigns. For the multiplication factor, we assume that
$m_{\left( j\right) }=0.25$ for structured, $m_{\left( j\right) }=0.5$ for
balanced and other, $m_{\left( j\right) }=1$ for bond and equity, $m_{\left(
j\right) }=1.75$ for enhanced treasury, and $m_{\left( j\right) }=6$ for money
market.} $\mathcal{C}_{\mathrm{investor}}$, whereas Table
\ref{tab:historical6b} corresponds to the risk coherency\footnote{We use the
following values: $\mathbb{S}_{\left( j\right) }=0.5\%$ for structured,
$\mathbb{S}_{\left( j\right) }=1.5\%$ for balanced and other,
$\mathbb{S}_{\left( j\right) }=3\%$ for bond and equity, $\mathbb{S}_{\left(
j\right) }=5\%$ for enhanced treasury, and $\mathbb{S}_{\left( j\right)
}=20\%$ for money market. For the multiplication factor, we assume that
$m_{\left( k\right) }=0.25$ for central banks and corporate pension funds,
$m_{\left( k\right) }=0.5$ for employee savings plans and retail, $m_{\left( k\right)
}=1$ for other and third-party distributors, $m_{\left( k\right) }=1.5$ for
institutionals and insurance companies, and $m_{\left( k\right)
}=2$ for auto-consumption, corporates and sovereigns.} $\mathcal{C}_{\mathrm{fund}}$. The
mixed rule is reported in Table \ref{tab:historical6c}. These figures can then
be modified by risk managers and business experts by considering the
specificity of some matrix cells. For instance, it is perhaps not realistic to
have the same redemption shock for balanced funds between auto-consumption and
corporates. Moreover, these redemption shocks can also be modified by taking
into account the $\mathbb{C}$-measure. For instance, the conditional
value-at-risk for bond funds is much higher for third-party
distributors than for sovereigns. Perhaps we can modify the redemption shock of
$3.3\%$ and have a larger value for third-party distributors. It is even
more likely that the estimated values of $\mathbb{Q}$ and $\mathbb{C}$ are
based on $75\,591$ observations for the third-party distributor category, and
$2\,261$ for the sovereign category. Therefore, we can consider that the
estimated value of $4.59\%$ obtained in Table \ref{tab:historical4b} on page
\pageref{tab:historical4b} does make more sense than the proposed value of
$3.3\%$ obtained in Table \ref{tab:historical6c} for the third-party
distributor/bond matrix cell. In a similar way, we can consider that the
estimated value of $0.14\%$ does make less sense than the proposed value of
$7.0\%$ for the sovereign/bond matrix cell.\smallskip

The previous analysis shows that building redemption shocks in a stress testing
framework is more of an art than a science. A pure quantitative approach is
dangerous because it is data-driven and it does not respect some coherency
properties. However, historical statistics are very important because they provide
an anchor point for risk managers and business experts in order to propose
stress scenarios that are satisfactory from regulatory, risk management and
fund management points of view. Historical data are also important because they
help to understand the behavior of clients. It is different from one fund
category to another, it also depends on the granularity of the classification,
it may depend on the time period, etc. In what follows, we complete this pure
historical analysis using more theoretical models. These models are important,
because an historical approach is limited when we want to understand
contagion effects between investors, correlation patterns between funds, time
properties of redemption risk, the impact of the holding period, etc.
The idea is not to substitute one model with
another, but to rely on several approaches, because there is not just one single
solution to the liability stress testing problem.

\begin{table}[p]
\centering
\caption{Redemption shocks in \% computed with the rule of thumb (\ref{eq:rule-C1})}
\label{tab:historical6a}
\begin{tabular}{lcccccccc}
\hline
                        &   (1) &   (2) &         (3) &   (4) &         (5) &   (6) &   (7) &   (8) \\ \hline
Auto-consumption        & $4.0$ & $8.0$ &      $14.0$ & $8.0$ &      $48.0$ & $4.0$ & $2.0$ & $8.0$ \\
Central bank            & $0.3$ & $0.5$ & ${\TsV}0.9$ & $0.5$ & ${\TsV}3.0$ & $0.3$ & $0.1$ & $0.5$ \\
Corporate               & $4.0$ & $8.0$ &      $14.0$ & $8.0$ &      $48.0$ & $4.0$ & $2.0$ & $8.0$ \\
Corporate pension fund  & $0.3$ & $0.5$ & ${\TsV}0.9$ & $0.5$ & ${\TsV}3.0$ & $0.3$ & $0.1$ & $0.5$ \\
Employee savings plan   & $1.0$ & $2.0$ & ${\TsV}3.5$ & $2.0$ &      $12.0$ & $1.0$ & $0.5$ & $2.0$ \\
Institutional           & $2.5$ & $5.0$ & ${\TsV}8.8$ & $5.0$ &      $30.0$ & $2.5$ & $1.3$ & $5.0$ \\
Insurance               & $2.5$ & $5.0$ & ${\TsV}8.8$ & $5.0$ &      $30.0$ & $2.5$ & $1.3$ & $5.0$ \\
Other                   & $1.8$ & $3.5$ & ${\TsV}6.1$ & $3.5$ &      $21.0$ & $1.8$ & $0.9$ & $3.5$ \\
Retail                  & $1.0$ & $2.0$ & ${\TsV}3.5$ & $2.0$ &      $12.0$ & $1.0$ & $0.5$ & $2.0$ \\
Sovereign               & $4.0$ & $8.0$ &      $14.0$ & $8.0$ &      $48.0$ & $4.0$ & $2.0$ & $8.0$ \\
Third-party distributor & $1.8$ & $3.5$ & ${\TsV}6.1$ & $3.5$ &      $21.0$ & $1.8$ & $0.9$ & $3.5$ \\ \hline
Total                   & $1.8$ & $3.5$ & ${\TsV}6.1$ & $3.5$ &      $21.0$ & $1.8$ & $0.9$ & $3.5$ \\
\hline
\end{tabular}
\end{table}

\begin{table}[p]
\centering
\caption{Redemption shocks in \% computed with the rule of thumb (\ref{eq:rule-C2})}
\label{tab:historical6b}
\begin{tabular}{lcccccccc}
\hline
                        &   (1) &   (2) &         (3) &   (4) &         (5) &   (6) &   (7) &   (8) \\ \hline
Auto-consumption        & $3.0$ & $6.0$ &      $10.0$ & $6.0$ &      $40.0$ & $3.0$ & $1.0$ & $7.0$ \\
Central bank            & $0.4$ & $0.8$ & ${\TsV}1.3$ & $0.8$ & ${\TsV}5.0$ & $0.4$ & $0.1$ & $0.9$ \\
Corporate               & $3.0$ & $6.0$ &      $10.0$ & $6.0$ &      $40.0$ & $3.0$ & $1.0$ & $7.0$ \\
Corporate pension fund  & $0.4$ & $0.8$ & ${\TsV}1.3$ & $0.8$ & ${\TsV}5.0$ & $0.4$ & $0.1$ & $0.9$ \\
Employee savings plan   & $0.8$ & $1.5$ & ${\TsV}2.5$ & $1.5$ &      $10.0$ & $0.8$ & $0.3$ & $1.8$ \\
Institutional           & $2.3$ & $4.5$ & ${\TsV}7.5$ & $4.5$ &      $30.0$ & $2.3$ & $0.8$ & $5.3$ \\
Insurance               & $2.3$ & $4.5$ & ${\TsV}7.5$ & $4.5$ &      $30.0$ & $2.3$ & $0.8$ & $5.3$ \\
Other                   & $1.5$ & $3.0$ & ${\TsV}5.0$ & $3.0$ &      $20.0$ & $1.5$ & $0.5$ & $3.5$ \\
Retail                  & $0.8$ & $1.5$ & ${\TsV}2.5$ & $1.5$ &      $10.0$ & $0.8$ & $0.3$ & $1.8$ \\
Sovereign               & $3.0$ & $6.0$ &      $10.0$ & $6.0$ &      $40.0$ & $3.0$ & $1.0$ & $7.0$ \\
Third-party distributor & $1.5$ & $3.0$ & ${\TsV}5.0$ & $3.0$ &      $20.0$ & $1.5$ & $0.5$ & $3.5$ \\ \hline
Total                   & $1.5$ & $3.0$ & ${\TsV}5.0$ & $3.0$ &      $20.0$ & $1.5$ & $0.5$ & $3.5$ \\
\hline
\end{tabular}
\end{table}

\begin{table}[p]
\centering
\caption{Redemption shocks in \% computed with the rule of thumb (\ref{eq:rule-C3})}
\label{tab:historical6c}
\begin{tabular}{lcccccccc}
\hline
                        &   (1) &   (2) &         (3) &   (4) &         (5) &   (6) &   (7) &   (8) \\ \hline
Auto-consumption        & $3.5$ & $7.0$ &      $12.0$ & $7.0$ &      $44.0$ & $3.5$ & $1.5$ & $7.5$ \\
Central bank            & $0.3$ & $0.6$ & ${\TsV}1.1$ & $0.6$ & ${\TsV}4.0$ & $0.3$ & $0.1$ & $0.7$ \\
Corporate               & $3.5$ & $7.0$ &      $12.0$ & $7.0$ &      $44.0$ & $3.5$ & $1.5$ & $7.5$ \\
Corporate pension fund  & $0.3$ & $0.6$ & ${\TsV}1.1$ & $0.6$ & ${\TsV}4.0$ & $0.3$ & $0.1$ & $0.7$ \\
Employee savings plan   & $0.9$ & $1.8$ & ${\TsV}3.0$ & $1.8$ &      $11.0$ & $0.9$ & $0.4$ & $1.9$ \\
Institutional           & $2.4$ & $4.8$ & ${\TsV}8.1$ & $4.8$ &      $30.0$ & $2.4$ & $1.0$ & $5.1$ \\
Insurance               & $2.4$ & $4.8$ & ${\TsV}8.1$ & $4.8$ &      $30.0$ & $2.4$ & $1.0$ & $5.1$ \\
Other                   & $1.6$ & $3.3$ & ${\TsV}5.6$ & $3.3$ &      $20.5$ & $1.6$ & $0.7$ & $3.5$ \\
Retail                  & $0.9$ & $1.8$ & ${\TsV}3.0$ & $1.8$ &      $11.0$ & $0.9$ & $0.4$ & $1.9$ \\
Sovereign               & $3.5$ & $7.0$ &      $12.0$ & $7.0$ &      $44.0$ & $3.5$ & $1.5$ & $7.5$ \\
Third-party distributor & $1.6$ & $3.3$ & ${\TsV}5.6$ & $3.3$ &      $20.5$ & $1.6$ & $0.7$ & $3.5$ \\ \hline
Total                   & $1.6$ & $3.3$ & ${\TsV}5.6$ & $3.3$ &      $20.5$ & $1.6$ & $0.7$ & $3.5$ \\
\hline
\end{tabular}
\medskip

\begin{flushleft}
\begin{footnotesize}
(1) = balanced, (2) = bond, (3) = enhanced treasury, (4) = equity, (5) = money market,
(6) = other, (7) = structured, (8) = total
\end{footnotesize}
\end{flushleft}
\end{table}

\section{The frequency-severity modeling approach}

The direct computation of value-at-risk, conditional value-at-risk and other
statistics from historical redemption rates is particularly problematic. Indeed,
we observe a large proportion of zeros in the redemption rate database. On
average, we have $68.9\%$ of zeros, this proportion reaches $99.5\%$ for some
investors and it is more than $99.9\%$ for some matrix cells. Therefore,
the data of redemption rates are \textquotedblleft
\textit{clumped-at-zero}\textquotedblright, meaning that the redemption rate is
a semi-continuous random variable, and not a continuous random variable
\citep{Min-2002}. This discontinuity is a real problem when estimating the
proba\-bi\-li\-ty distribution $\mathbf{F}$. This is why we consider that the
redemption rate is not the right redemption risk factor. We prefer to
assume that the redemption risk is driven by two dimensions or two risk
factors:
\begin{enumerate}
\item the redemption frequency, which measures the occurrence $\mathcal{E}$
of the redemption;

\item the redemption severity $\redemption^{\star }$, which measures the
amount of the redemption.
\end{enumerate}
It is obvious that this modeling approach finds its root in other risk models
that deal with extreme events or counting processes, such as operational and
insurance risks \citep{Roncalli-2020}.

\subsection{Zero-inflated models}

\begin{figure}[h]
\centering
\caption{Zero-inflated modeling of the redemption risk}
\label{fig:zero-inflated}
\begin{tikzpicture}[->,>=stealth',level/.style={sibling distance = 4.0cm/#1,level distance = 4.0cm},grow = right,scale=0.9, every node/.style={transform shape}]
\tikzset{
  loss_node/.style = {align=center, inner sep=0pt, text centered, circle, text width=5.15em,
  very thick, red, draw=black, fill=white!30}}
\node [loss_node] {$\boldsymbol{\mathcal{E}}$}
    child{  node [loss_node] {$\boldsymbol{\mathcal{E} = 1}$}
            child{ node [loss_node] {$\boldsymbol{\redemption^{\star} = 50\%}$}
                edge from parent node[below] {$5\%$}
                }
            child{ node [loss_node] {$\boldsymbol{\redemption^{\star} = 15\%}$}
                edge from parent node[above] {$10\%$}
                }
            child{ node [loss_node] {$\boldsymbol{\redemption^{\star} = 5\%}$}
                edge from parent node[above] {$25\%$}
                }
            child{ node [loss_node] {$\boldsymbol{\redemption^{\star} = 2\%}$}
                edge from parent node[above] {$60\%$}
                }
    edge from parent node[below] {$\boldsymbol{5\%}\quad$}
    }
    child{ node [loss_node] {$\boldsymbol{\mathcal{E} = 0}$}
    edge from parent node[above] {$\boldsymbol{95\%}\quad$}
    }
;
\node[right] at (8.9, 2.50) {$\boldsymbol{\redemption = 0\%}$ ($\boldsymbol{\Pr = 95\%}$)};
\node[right] at (8.9, 1.00) {$\boldsymbol{\redemption = 2\%}$ ($\boldsymbol{\Pr = 3\%}$)};
\node[right] at (8.9,-1.00) {$\boldsymbol{\redemption = 5\%}$ ($\boldsymbol{\Pr = 1.25\%}$)};
\node[right] at (8.9,-3.00) {$\boldsymbol{\redemption = 15\%}$ ($\boldsymbol{\Pr = 0.5\%}$)};
\node[right] at (8.9,-5.00) {$\boldsymbol{\redemption = 50\%}$ ($\boldsymbol{\Pr = 0.25\%}$)};
\end{tikzpicture}
\end{figure}
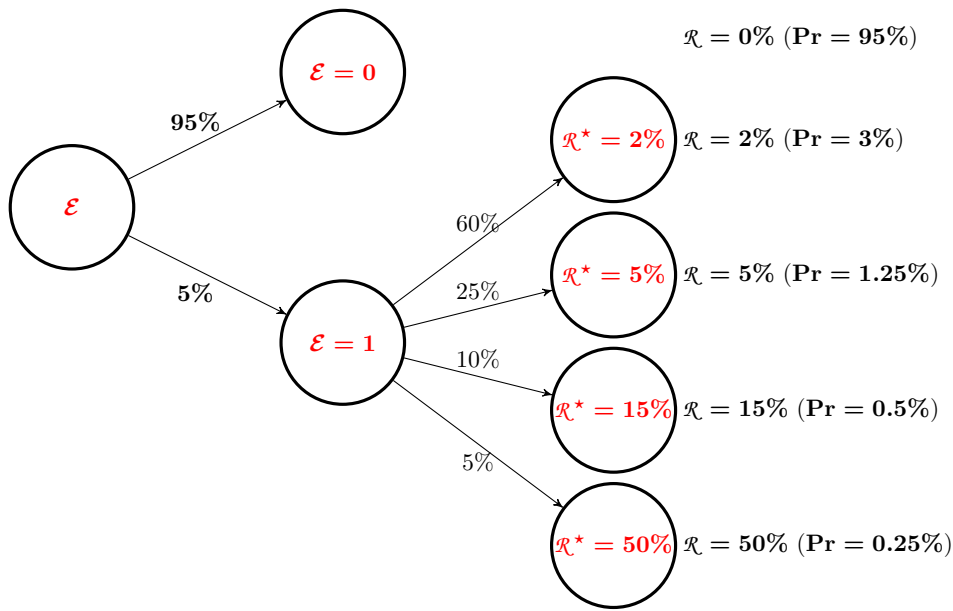

In the frequency-severity approach, we distinguish the redemption event
$\mathcal{E}$ that indicates if there is a redemption, and the redemption
amount $\redemption^{\star}$ that measures the redemption rate in case of a
redemption. An example is provided in Figure \ref{fig:zero-inflated}. The
probability to observe a redemption is equal to $5\%$, and in the case of a
redemption, the amount can be $2\%$, $5\%$, $15\%$ and $50\%$. It follows that
the redemption rate is the convolution of two risk factors.

\begin{figure}[tbph]
\centering
\caption{Zero-inflated probability density function}
\label{fig:inflated2}
\includegraphics[width = \figurewidth, height = \figureheight]{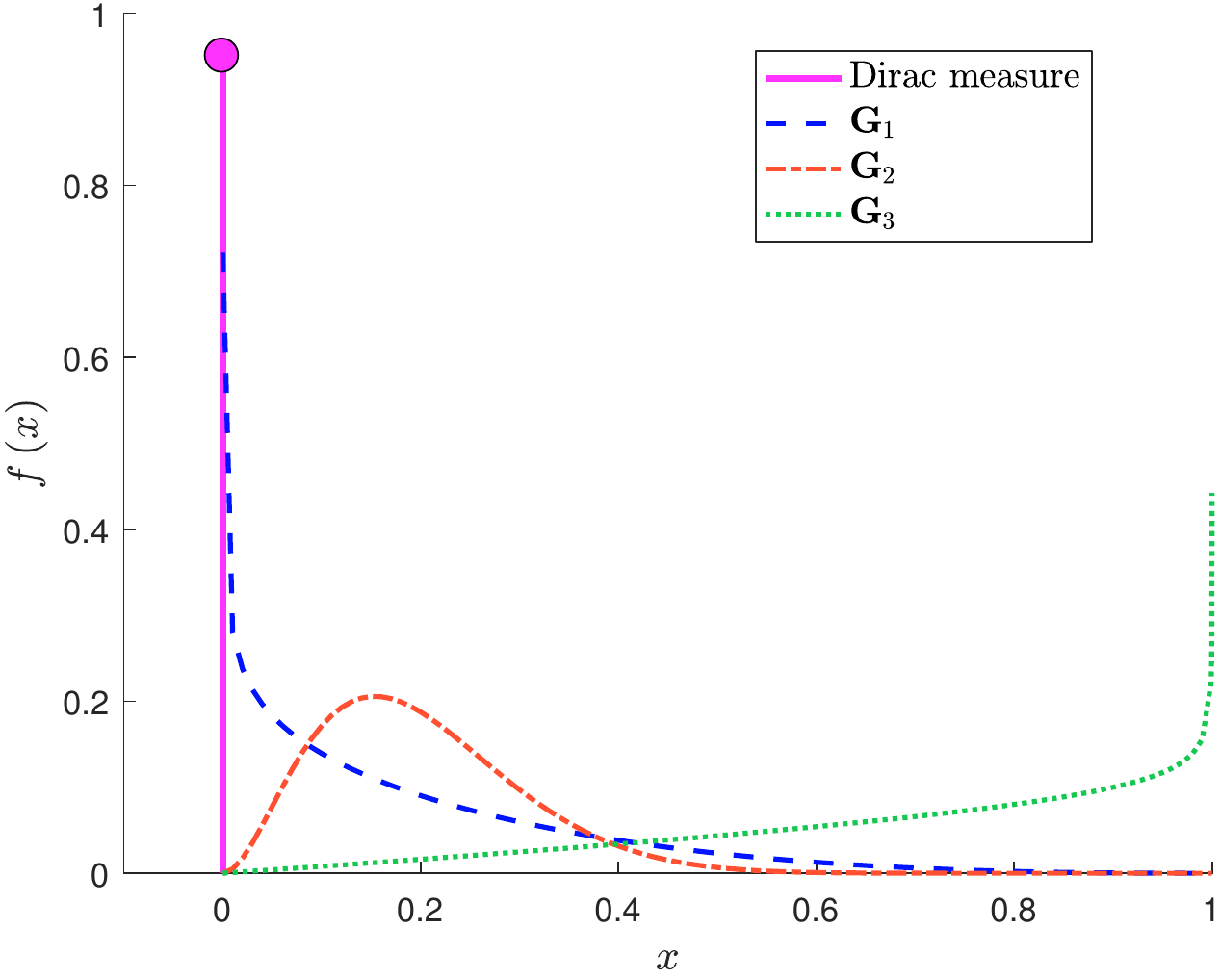}
\vspace*{-10pt}
\end{figure}

\subsubsection{Zero-inflated probability distribution}

We assume that the redemption event $\mathcal{E}$ follows a bernoulli
distribution $\mathcal{B}\left( p\right) $, whereas the redemption
severity\footnote{It is defined as the non-zero redemption rate.}
$\redemption^{\star }$ follows a continuous probability distribution
$\mathbf{G}$. We have:
\begin{equation*}
\Pr \left\{ \mathcal{E}=1\right\} =\Pr \left\{ \redemption>0\right\} = p
\end{equation*}%
and:%
\begin{equation*}
\Pr \left\{ \redemption\leq x\mid \mathcal{E}=1\right\} =\mathbf{G}\left(
x\right)
\end{equation*}%
We deduce that the unconditional probability distribution of the redemption
rate is given by:
\begin{eqnarray*}
\mathbf{F}\left( x\right)  &=&\Pr \left\{ \redemption\leq x\right\}  \\
&=&\mathds{1}\left\{ x \geq 0\right\} \cdot \left( 1-p\right) +\mathds{1}\left\{ x>0\right\} \cdot p\cdot \mathbf{G}%
\left( x\right)
\end{eqnarray*}%
Its density probability function is singular at $x=0$:
\begin{equation*}
f\left( x\right) =\left\{
\begin{array}{ll}
1-p & \text{if }x=0 \\
p\cdot g\left( x\right)  & \text{otherwise}%
\end{array}%
\right.
\end{equation*}
where $g\left( x\right)$ is the density function of $\mathbf{G}$. Some examples
are provided in Figure \ref{fig:inflated2} when $p=5\%$. We observe that the
density function is composed of a dirac measure and a continuous function. In
the case of $\mathbf{G}_1$, the distribution is right-skewed, meaning that the
probability to observe small redemptions is high. In the case of
$\mathbf{G}_2$, we have a bell curve, meaning that the redemption amount is
located around the mean if there is a redemption. Finally, the distribution is
left-skewed in the case of $\mathbf{G}_3$, meaning that the probability to
observe high redemptions is high if there is of course a redemption, because we
recall that the probability to observe a redemption is only equal to
$5\%$.\smallskip

From a probabilistic point of view, the redemption rate is then the product of
the redemption event and the redemption severity:
\begin{equation*}
\redemption=\mathcal{E\cdot }\redemption^{\star }
\end{equation*}%
In Appendix \ref{appendix:zero-inflated-moments} on page
\pageref{appendix:zero-inflated-moments}, we show that:
\begin{equation}
\mathbb{E}\left[ \redemption\right] =p\mathbb{E}\left[ \redemption^{\star }%
\right]
\end{equation}%
and:%
\begin{equation}
\sigma ^{2}\left( \redemption\right) =p\sigma ^{2}\left( \redemption^{\star
}\right) +p\left( 1-p\right) \mathbb{E}^{2}\left[ \redemption^{\star }\right]
\end{equation}%
Moreover, the skewness coefficient is equal to:%
\begin{equation}
\gamma _{1}\left( \redemption\right) =\frac{\vartheta _{1}\left( \redemption%
^{\star }\right) }{\left( p\sigma ^{2}\left( \redemption^{\star }\right)
+p\left( 1-p\right) \mathbb{E}^{2}\left[ \redemption^{\star }\right] \right)
^{3/2}}
\end{equation}%
where:%
\begin{eqnarray*}
\vartheta _{1}\left( \redemption^{\star }\right)  &=&p\gamma _{1}\left( %
\redemption^{\star }\right) \sigma ^{3}\left( \redemption^{\star }\right) +
3p\left( 1-p\right) \sigma ^{2}\left( \redemption^{\star }\right) \mathbb{E%
}\left[ \redemption^{\star }\right] + \\
&&p\left( 1-p\right) \left( 1-2p\right) \mathbb{E}^{3}\left[ \redemption%
^{\star }\right]
\end{eqnarray*}%
For the excess kurtosis coefficient, we obtain:%
\begin{equation}
\gamma _{2}\left( \redemption\right) =\frac{\vartheta _{2}\left( \redemption%
^{\star }\right) }{\left( p\sigma ^{2}\left( \redemption^{\star }\right)
+p\left( 1-p\right) \mathbb{E}^{2}\left[ \redemption^{\star }\right] \right)
^{2}}
\end{equation}%
where:%
\begin{eqnarray*}
\vartheta _{2}\left( \redemption^{\star }\right)  &=&\left( p\gamma _{2}\left( %
\redemption^{\star }\right) +3p\left( 1-p\right) \right) \sigma ^{4}\left( %
\redemption^{\star }\right) +
4p\left( 1-p\right) \gamma _{1}\left( \redemption^{\star }\right) \sigma
^{3}\left( \redemption^{\star }\right) \mathbb{E}\left[ \redemption^{\star }%
\right] + \\
&&6p\left( 1-p\right) \left( 1-2p\right) \sigma ^{2}\left( \redemption%
^{\star }\right) \mathbb{E}^{2}\left[ \redemption^{\star }\right] +
p\left( 1-p\right) \left( 1-6p+6p^{2}\right) \mathbb{E}^{4}\left[ %
\redemption^{\star }\right]
\end{eqnarray*}
\smallskip

\begin{figure}[tbph]
\centering
\caption{Statistical moments of the redemption rate $\redemption$ in zero-inflated models}
\label{fig:inflated3}
\includegraphics[width = \figurewidth, height = \figureheight]{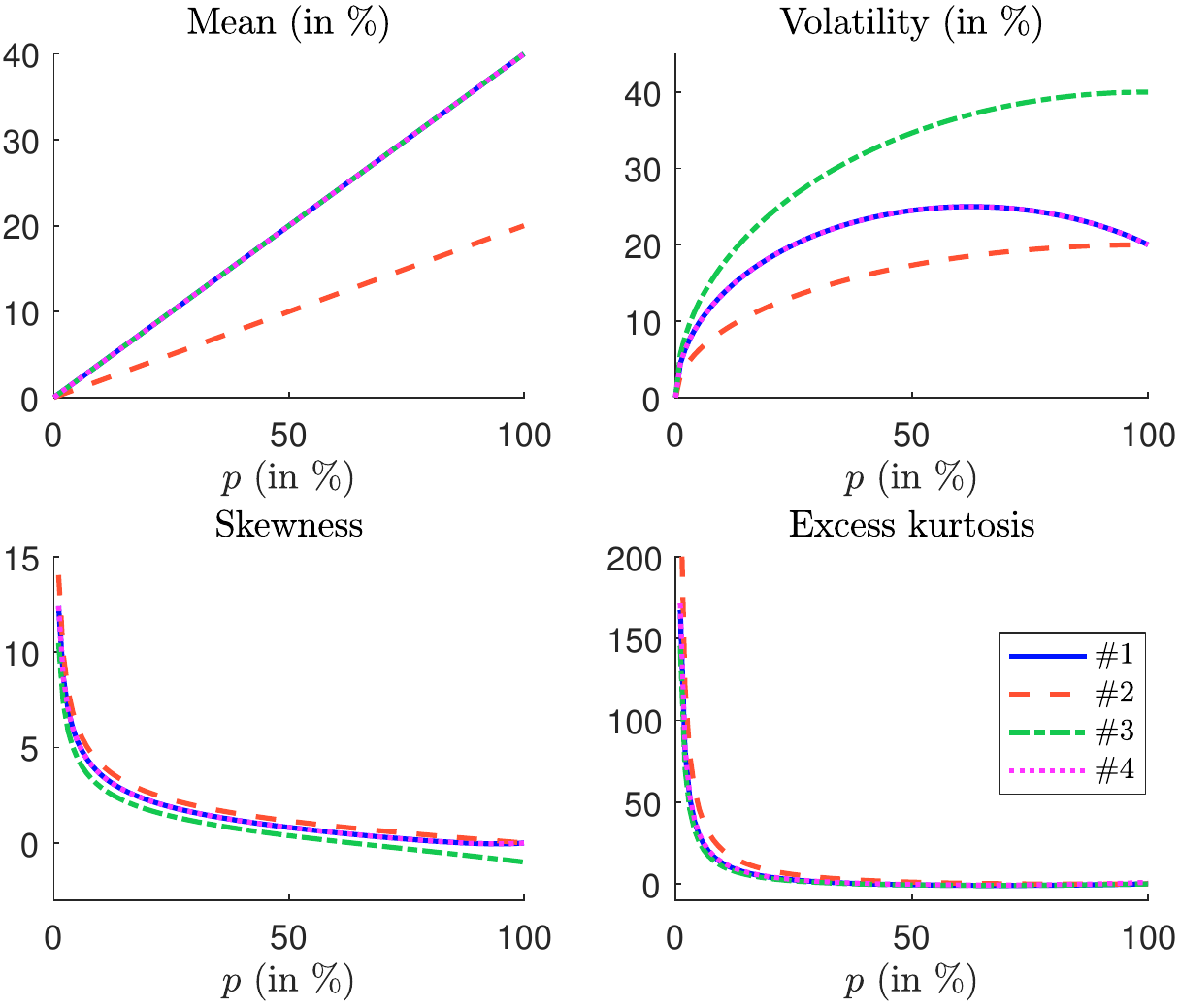}
\end{figure}

In Figure \ref{fig:inflated3} we have reported the moments of the redemption
rate $\redemption$ by considering the following set of parameters:
\begin{itemize}
\item[\#1] $\mathbb{E}\left[ \redemption^{\star }\right] =40\%$, $\sigma
    \left( \redemption^{\star }\right) =20\%$, $\gamma _{1}\left(
    \redemption ^{\star }\right) =0$ and $\gamma _{2}\left(
    \redemption^{\star }\right) =0$;

\item[\#2] $\mathbb{E}\left[ \redemption^{\star }\right] =20\%$, $\sigma
    \left( \redemption^{\star }\right) =20\%$, $\gamma _{1}\left( \redemption
    ^{\star }\right) =0$ and $\gamma _{2}\left( \redemption^{\star }\right)
    =0$;

\item[\#3] $\mathbb{E}\left[ \redemption^{\star }\right] =40\%$, $\sigma
    \left( \redemption^{\star }\right) =40\%$, $\gamma _{1}\left( \redemption
    ^{\star }\right) =-1$ and $\gamma _{2}\left( \redemption^{\star }\right)
    =0$;

\item[\#4] $\mathbb{E}\left[ \redemption^{\star }\right] =40\%$, $\sigma
    \left( \redemption^{\star }\right) =20\%$, $\gamma _{1}\left( \redemption
    ^{\star }\right) =0$ and $\gamma _{2}\left( \redemption^{\star }\right)
    =1$.
\end{itemize}
We notice that the parameter values of $\redemption^{\star }$ have a major
impact on the statistical moments, but the biggest effect comes from the
frequency probability $p$. Indeed, we verify the following properties:
\begin{equation}
\left\{
\begin{array}{l}
\lim_{p\rightarrow 0^{+}}\mathbb{E}\left[ \redemption\right] =\lim_{p\rightarrow
0^{+}}\mathbb{\sigma }\left( \redemption\right) =0 \\
\lim_{p\rightarrow 0^{+}}\mathbb{\gamma }_{1}\left( \redemption\right)
=\lim_{p\rightarrow 0^{+}}\mathbb{\gamma }_{2}\left( \redemption\right) =\infty
\end{array}%
\right. \label{eq:zero-inflated-limits}
\end{equation}%
This means that the redemption risk is very high for small frequency
properties. In this case, the expected redemption rate and its standard
deviation are very low, but skewness and kurtosis risk are very high! This
creates a myopic situation where the asset manager may have the feeling that
redemption risk is not a concern because of historical data. Indeed, when $p$
is low, the probability of observing large redemption rates is small, implying
that they are generally not observed in the database. For instance, let us
consider two categories that have the same redemption severity distribution,
but differ from their redemption frequency probability. One has a probability
of $50\%$, the other has a probability of $1\%$. It is not obvious that the
second category experienced sufficient severe redemption events such that the
historical data are representative of the severity risk.\smallskip

\subsubsection{Statistical risk measures of the zero-inflated model}

\begin{figure}[tbph]
\centering
\caption{Relationship between $p$ and $\alpha_{\mathbf{G}}$ for the $99\%$ value-at-risk}
\label{fig:inflated5}
\includegraphics[width = \figurewidth, height = \figureheight]{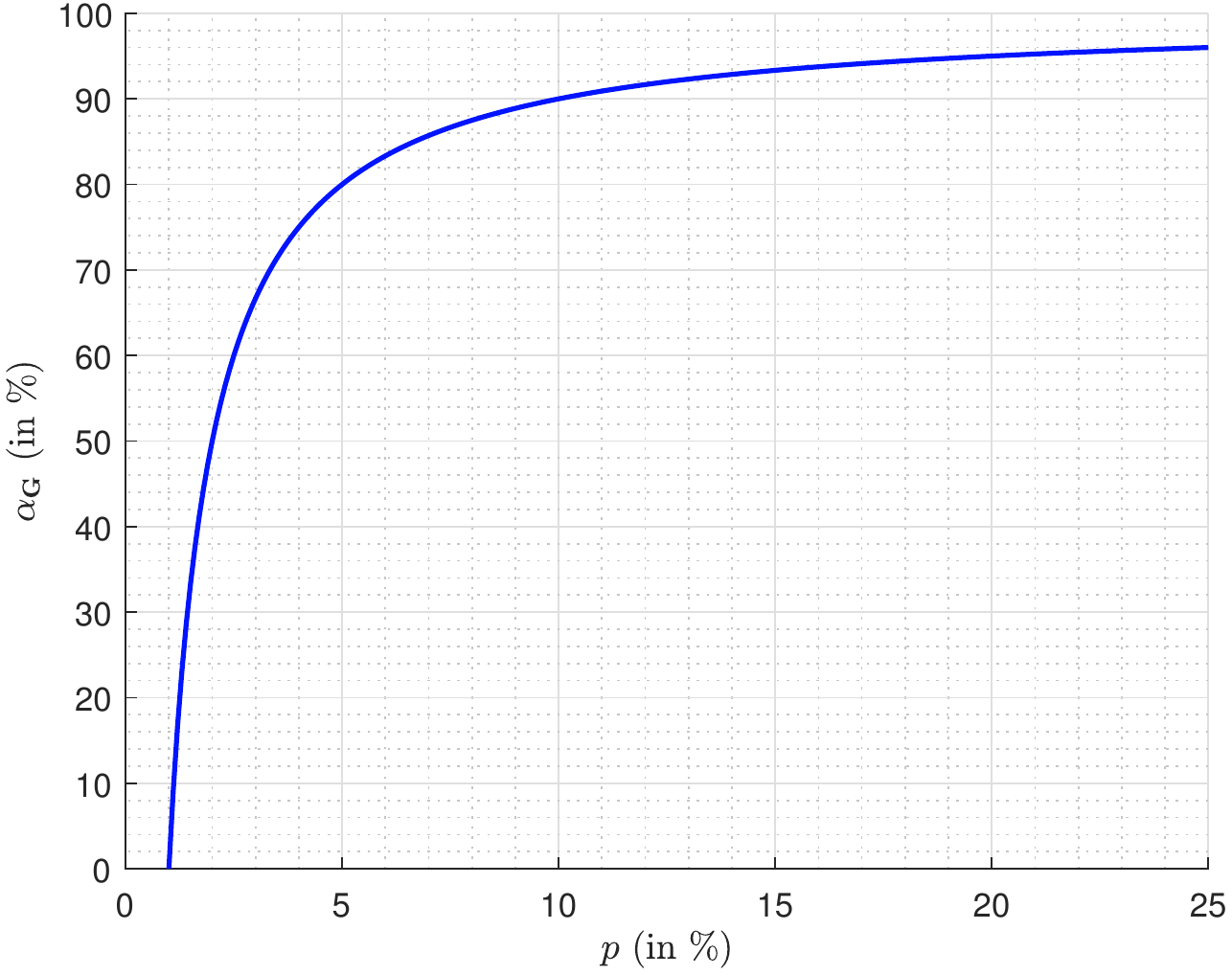}
\end{figure}

For the $\mathbb{M}$-measure, we have:
\begin{equation}
\mathbb{M} = p \mathbb{E}\left[ \redemption^{\star } \right]
\label{eq:inflated-measure1}
\end{equation}
The formula of the value-at-risk is equal to:
\begin{equation}
\mathbb{Q}\left(\alpha\right) =\left\{
\begin{array}{ll}
0 & \text{if }p\leq 1-\alpha \\
\mathbf{G}^{-1}\left( \dfrac{\alpha +p-1}{p}\right) & \text{otherwise}%
\end{array}%
\right.
\label{eq:inflated-measure2}
\end{equation}
We notice that computing the quantile $\alpha$ of the unconditional
distribution $\mathbf{F}$ is equivalent to compute the quantile
$\alpha_\mathbf{G}$ of the severity distribution $\mathbf{G}$:
\begin{equation*}
\alpha_\mathbf{G} = \max\left(0, \frac{\alpha +p-1}{p}\right)
\end{equation*}
The relationship between $p$, $\alpha$ and $\alpha_\mathbf{G}$ is illustrated
in Figure \ref{fig:inflated4} on page \pageref{fig:inflated4}. Let us focus on
the $99\%$ value-at-risk:
\begin{equation*}
\mathbb{Q}\left( 99\%\right) =\left\{
\begin{array}{ll}
0 & \text{if }p\leq 1\% \\
\mathbf{G}^{-1}\left( \dfrac{p-1\%}{p}\right)  & \text{otherwise}%
\end{array}%
\right.
\end{equation*}
If the redemption frequency probability is greater than $1\%$, the
value-at-risk corresponds to the quantile $\left(p-1\%\right)/p$. The
relationship between $p$ and $\alpha_{\mathbf{G}}=\left(p-1\%\right)/p$ is
shown in Figure \ref{fig:inflated5}. If $p$ is greater than $20\%$,
$\alpha_{\mathbf{G}}$ is greater than $95\%$. If $p$ is less than $5\%$, we
observe a high curvature of the relationship, implying that we face a high
estimation risk. For instance, if $p$ is equal to $1.5\%$, the $99\%$
value-at-risk corresponds to the quantile $3.33\%$ of the redemption severity.
If $p$ becomes $2.0\%$, the $99\%$ value-at-risk is then equal to the quantile
$50\%$ of the redemption severity! Therefore, there is a high sensitivity of
the $99\%$ value-at-risk when $p$ is low, implying that a small error in the
estimated value of $p$ leads to a high impact on the value-at-risk.\smallskip

For the conditional value-at-risk, we obtain:
\begin{equation}
\mathbb{C}\left( \alpha \right) =\frac{1}{1-\alpha }\int_{\alpha }^{1}%
\mathbb{Q}\left( u\right) \,\mathrm{d}u
\label{eq:inflated-measure3}
\end{equation}%
where $\mathbb{Q}\left( u\right) $ is the quantile function of $\redemption$
for the confidence level $u$. In the case where $p>1-\alpha $, we obtain:
\begin{equation*}
\mathbb{C}\left( \alpha \right) =\frac{p}{1-\alpha}%
\int_{1-p^{-1}\left( 1-\alpha \right) }^{1}\mathbf{G}^{-1}\left( u\right) \,%
\mathrm{d}u
\end{equation*}%
Another expression of the conditional value-at-risk is:
\begin{equation*}
\mathbb{C}\left( \alpha \right) =\frac{1}{1-\alpha }\int_{\mathbb{Q}\left(
\alpha \right) }^{1}x\,\mathrm{d}\mathbf{F}\left( x\right)
\end{equation*}%
In the case where $p>1-\alpha $, we obtain:%
\begin{equation*}
\mathbb{C}\left( \alpha \right) =\frac{p}{1-\alpha }\int_{\mathbb{Q}\left(
\alpha \right) }^{1}xg\left( x\right) \,\mathrm{d}x
\end{equation*}%
where $g\left( x\right) $ is the probability density function of
$\mathbf{G}\left( x\right) $. All these formulas can be computed numerically
thanks to Gauss-Legendre integration.\smallskip

We now introduce a new risk measure which is very popular when considering
parametric model. \citet{Roncalli-2020} defines the distribution-based (or
parametric-based) stress scenario $\mathbb{S}\left( \mathcal{T}\right) $ for
a given horizon time $\mathcal{T}$ such that the return time of this scenario
is exactly equal to $\mathcal{T}$. From a mathematical point of view, we
have:
\begin{equation*}
\frac{1}{\Pr \left\{ \redemption\geq \mathbb{S}\left( \mathcal{T}\right)
\right\} }=\mathcal{T}
\end{equation*}%
$\Pr \left\{ \redemption\geq \mathbb{S}\left( \mathcal{T}\right) \right\} $
is the exceedance probability of the stress scenario, implying that the
quantity $\Pr\left\{ \redemption\geq \mathbb{S}\left( \mathcal{T}\right)
\right\} ^{-1}$ is the return time of the exceedance event. For example, if
we set $\mathbb{S}\left( \mathcal{T} \right) =\mathbb{Q}\left( \alpha \right)
$, we have $\Pr \left\{\redemption \geq \mathbb{S}\left( \mathcal{T}\right)
\right\} =1-\alpha $ and $\mathcal{T}=\left( 1-\alpha \right) ^{-1}$. The
return time associated to a $99\%$ value-at-risk is then equal to $100$ days,
the return time associated to a $99.9\%$ value-at-risk is equal to $1\,000$
days (or approximately 4 years), etc. This parametric approach of stress
testing is popular among professionals, regulators and academics when they
use the extreme value theory for modeling the risk factors.\smallskip

By combining the two definitions $\mathbb{S}\left( \mathcal{T}\right) =
\mathbb{Q}\left( \alpha \right) $ and $\mathcal{T}=\left( 1-\alpha \right)
^{-1}$, we obtain the mathematical expression of the parametric stress
scenario:
\begin{equation}
\mathbb{S}\left( \mathcal{T}\right) =\mathbb{Q}\left( 1-\frac{1}{\mathcal{T}}%
\right)
\end{equation}%
If we consider the zero-inflated model, we deduce that:%
\begin{equation}
\mathbb{S}\left( \mathcal{T}\right) =\left\{
\begin{array}{ll}
0 & \text{if }p\leq \mathcal{T}^{-1} \\
\mathbf{G}^{-1}\left( 1-\dfrac{1}{p\mathcal{T}}\right)  & \text{otherwise}%
\end{array}%
\right.
\end{equation}%
The magnitude of $\mathcal{T}$ is the year, but the unit of $\mathcal{T}$ is the day.
For example, since one year corresponds to 260 market days, the five-year stress scenario is equal to\footnote{We
assume that the redemption frequency is greater than $1/1300$ or $7.69$ bps.
Otherwise, the quantile is equal to zero.}:
\begin{equation*}
\mathbb{S}\left( 5\right) =\mathbf{G}^{-1}\left( 1-\dfrac{1}{1300\,p}\right)
\end{equation*}

\subsubsection{The zero-inflated beta model}

The choice of the severity distribution is an important issue. Since
$\redemption^{\star }$ is a random variable between $0$ and $1$, it is
natural to use the two-parameter beta distribution $\mathcal{B}\left( a ,b
\right) $. We have:
\begin{equation*}
\mathbf{G}\left( x\right) =\mathfrak{B}\left( x;a,b\right)
\end{equation*}%
where $\mathfrak{B}\left( x;a,b \right) $ is the incomplete beta function.
The corresponding probability density function is equal to:
\begin{equation*}
g\left( x\right) =\frac{x^{a -1}\left( 1-x\right) ^{b -1}}{%
\mathfrak{B}\left( a ,b \right) }
\end{equation*}%
where $\mathfrak{B}\left( a ,b \right) $ is the beta function:
\begin{equation*}
\mathfrak{B}\left( a ,b \right) =\frac{\Gamma \left( a \right)
\Gamma \left( b \right) }{\Gamma \left( a +b \right) }
\end{equation*}%
Concerning the statistical moments, the formulas are given in Appendix
\ref{appendix:zero-inflated-moments-beta} on page
\pageref{appendix:zero-inflated-moments-beta}.
\smallskip

\begin{figure}[tbph]
\centering
\caption{Density function of the beta distribution}
\label{fig:inflated6}
\includegraphics[width = \figurewidth, height = \figureheight]{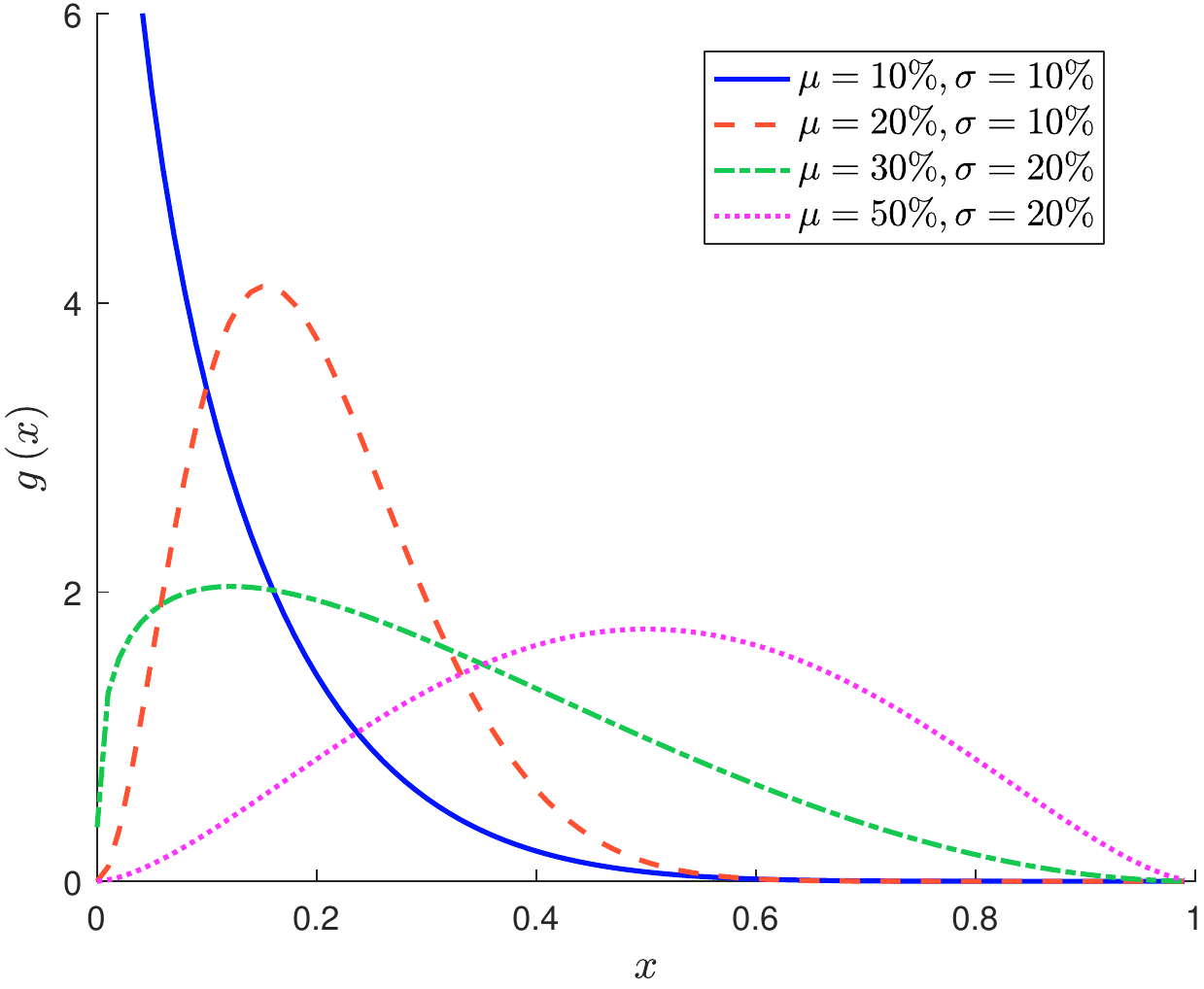}
\end{figure}

\begin{figure}[tbph]
\centering
\caption{Statistical moments of the zero-inflated beta distribution}
\label{fig:inflated7}
\includegraphics[width = \figurewidth, height = \figureheight]{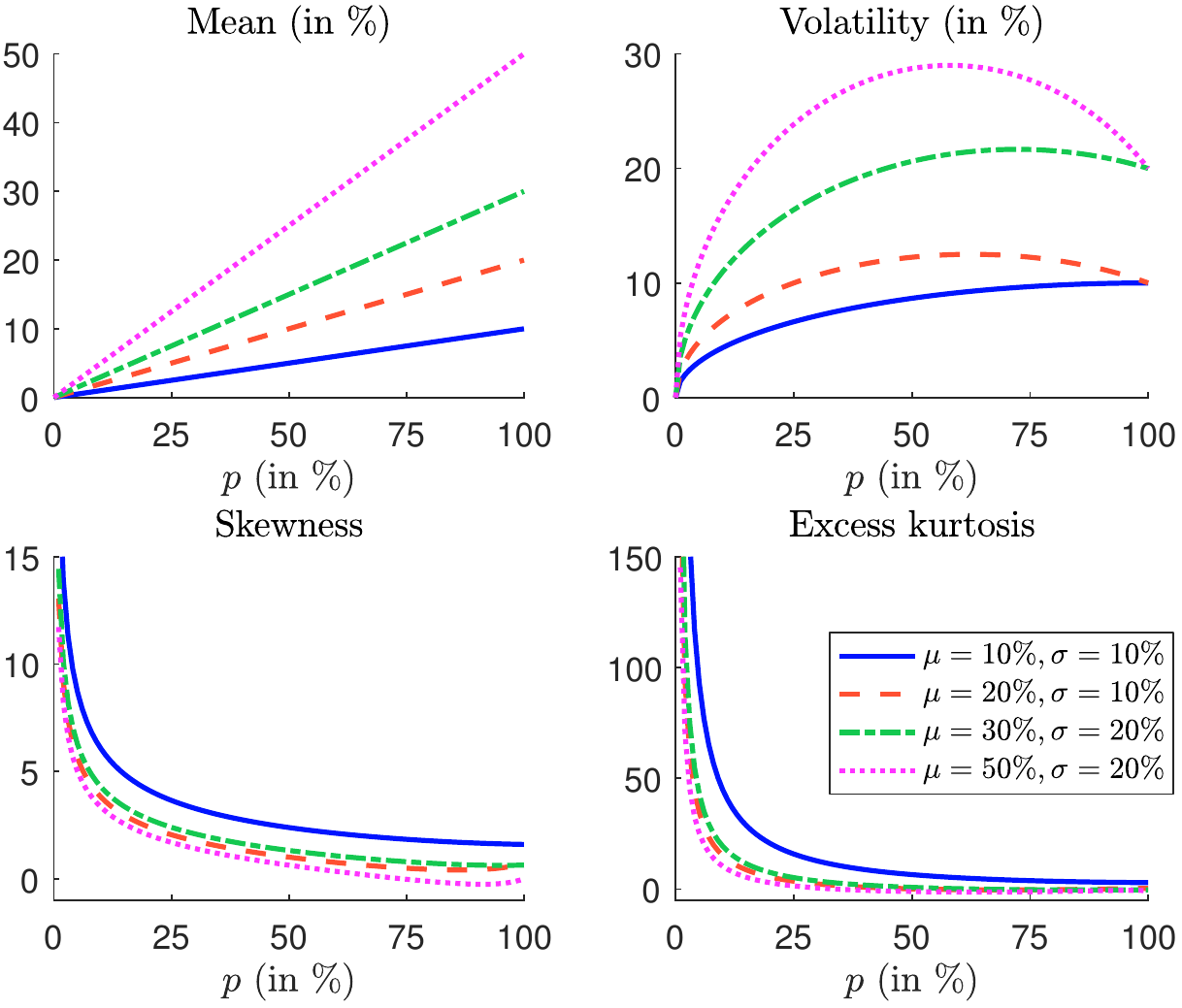}
\end{figure}

We report some examples of density function in Figure \ref{fig:inflated6}.
Instead of providing the parameters $a$ and $b$, we have indicated the value
$\mu$ and $\sigma$ of the mean and the volatility. The first distribution is
skewed, because the volatility is high compared to the mean. The other three
distributions have a mode. Figure \ref{fig:inflated7} shows the corresponding
statistical moments of the associated zero-inflated model. We notice that the
first and third distributions have the largest skewness and kurtosis.\smallskip

\begin{figure}[tbph]
\centering
\caption{$\mathbb{Q}\left(99\%\right)$-measure in \% with respect to the redemption frequency}
\label{fig:inflated8}
\includegraphics[width = \figurewidth, height = \figureheight]{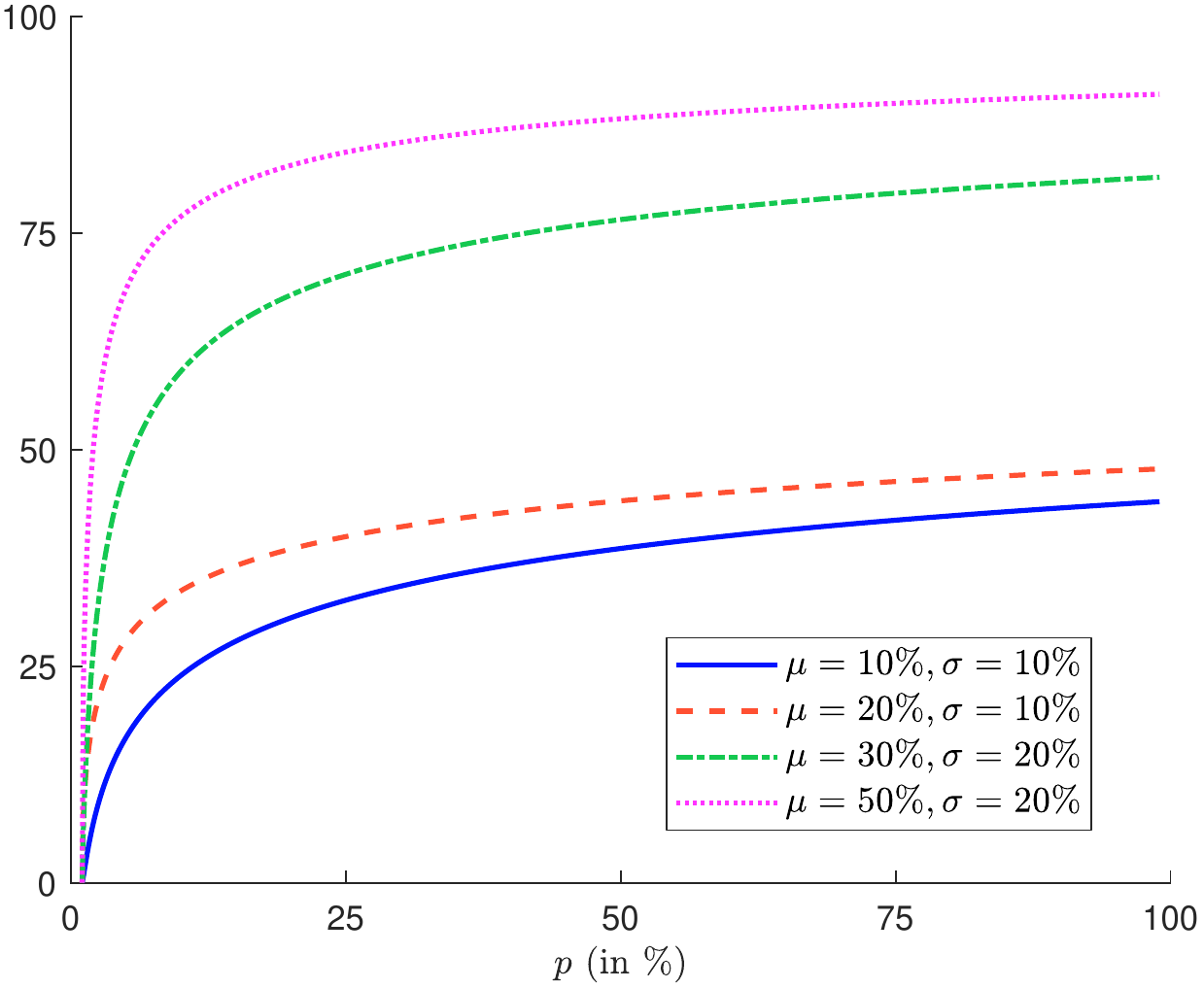}
\end{figure}

\begin{figure}[tbph]
\centering
\caption{Ratio $\mathbb{R}\left(99\%\right)$ with respect to the redemption frequency}
\label{fig:inflated9}
\includegraphics[width = \figurewidth, height = \figureheight]{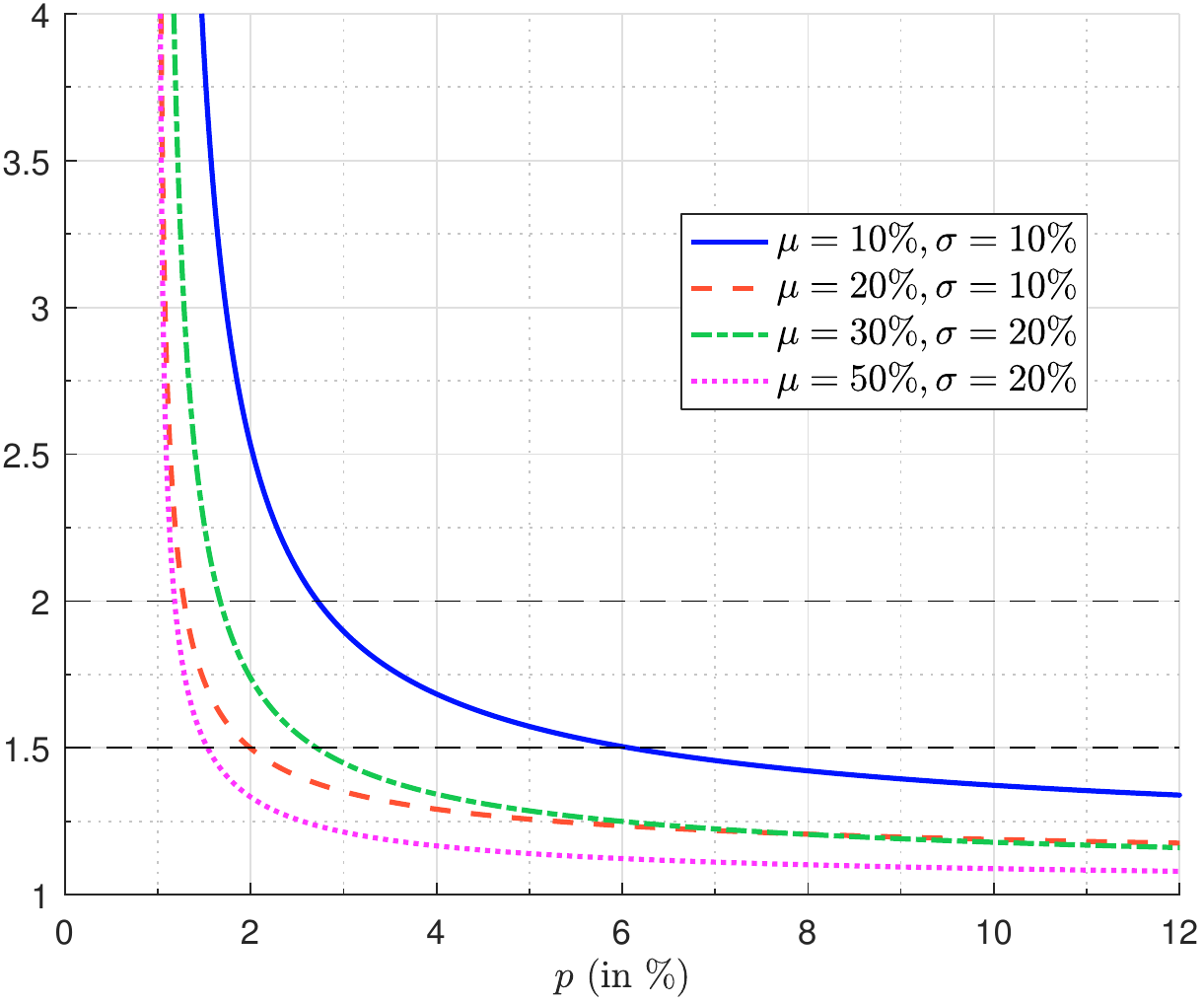}
\end{figure}

\begin{figure}[tbph]
\centering
\caption{Stress scenario $\mathbb{S}\left(\mathcal{T}\right)$ in \% ($p = 1\%$)}
\label{fig:inflated10}
\includegraphics[width = \figurewidth, height = \figureheight]{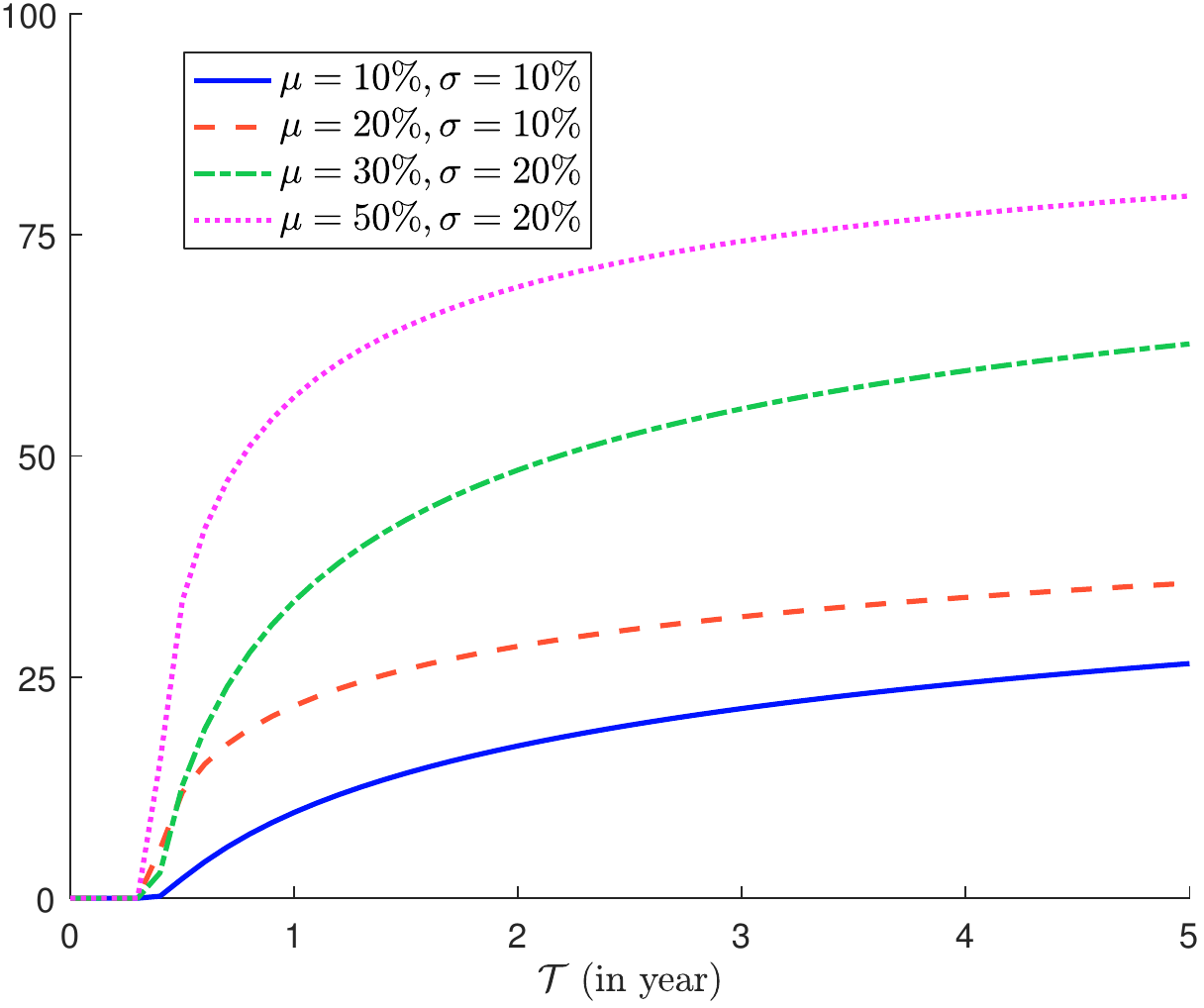}
\end{figure}

In Figure \ref{fig:inflated8}, we report the $99\%$ value-at-risk of the
redemption rate. As explained before, the $\mathbb{Q}$-measure highly depends
on the redemption frequency $p$. Again, we observe that the sensitivity of the
value-at-risk is particularly important when $p$ is small\footnote{Because of
the impact of $p$ on the confidence level $\alpha_{\mathbf{G}}$ --- see Figure
\ref{fig:inflated5} on page \pageref{fig:inflated5}.}. The ratio between the
$99\%$ conditional value-at-risk and the $99\%$ value-at-risk is given in
Figure \ref{fig:inflated9}. When the redemption frequency $p$ is high, the ratio is
less than $1.5$ and we retrieve the typical figures that we observe for market
and credit risks\footnote{When $p$ tends to one, the ratio is respectively
equal to $1.15$, $1.09$, $1.06$ and $1.03$ for the four probability
distributions of the redemption severity.}. When the redemption frequency $p$ is small, the ratio may be
greater than $2.0$. These results shows that the sensitivity to redemption risk
is very high when the observed redemption frequency is low. The stress
scenarios $\mathbb{S}\left(\mathcal{T}\right)$ are given in Figure
\ref{fig:inflated10} when the redemption frequency $p$ is equal to $1\%$. By
definition, $\mathbb{S}\left(\mathcal{T}\right)$ increases with the return time
$\mathcal{T}$. From a theoretical point of view, the limit of the stress
scenario is $100\%$:
\begin{equation*}
\lim_{\mathcal{T}\rightarrow \infty }\mathbb{S}\left( \mathcal{T}\right)
=\lim_{\mathcal{T}\rightarrow \infty }\mathbf{G}^{-1}\left( 1-\dfrac{1}{p%
\mathcal{T}}\right) =1
\end{equation*}%
However, we observe that stress scenarios reach a plateau at five years,
meaning that stress scenarios beyond $5$ years have no interest. This is true
for small values of $p$, but it is even more the case for larger values of $p$
as shown in Figures \ref{fig:inflated11} and \ref{fig:inflated12} on page
\pageref{fig:inflated12}.

\begin{remark}
In order to better understand the use of the $\mathbb{C}$-measure as a
stress scenario, we compute the implied return time such that the stress
scenario is exactly equal to the conditional value-at-risk:%
\begin{equation*}
\mathcal{T}_{\mathbb{C}\left( \alpha \right) }=\left\{ \mathcal{T}:\mathbb{S}%
\left( \mathcal{T}\right) =\mathbb{C}\left( \alpha \right) \right\}
\end{equation*}%
Results are given in Table \ref{tab:inflated13}. We notice that the value is
between $0.77$ and $1.03$. On average, we can consider that the return time of
the $99\%$ conditional value-at-risk is about one year. This is $2.6$ times the
return time of the $99\%$ value-at-risk\footnote{We recall that the return time
of the $99\%$ value-at-risk is equal to $100$ market days or $\dfrac{100}{260}
\approx 0.38$ years.}.
\end{remark}

\begin{table}[tbph]
\centering
\caption{Implied return time $\mathcal{T}_{\mathbb{C}\left( 99\%\right) }$ in year}
\label{tab:inflated13}
\begin{tabular}{cc|cccc}
\hline
\multicolumn{2}{c}{$\mu$}    & $10\%$ & $20\%$ & $30\%$ & $50\%$ \\
\multicolumn{2}{c}{$\sigma$} & $10\%$ & $10\%$ & $20\%$ & $20\%$ \\
\hline
      & ${\TsV}1\%$ & $1.03$ & $0.86$ & $0.87$ & $0.77$ \\
      & ${\TsV}2\%$ & $1.00$ & $0.94$ & $0.89$ & $0.85$ \\
      & ${\TsV}3\%$ & $0.99$ & $0.95$ & $0.90$ & $0.86$ \\
$p$   & ${\TsV}5\%$ & $0.99$ & $0.97$ & $0.90$ & $0.87$ \\
      &      $10\%$ & $0.99$ & $0.98$ & $0.90$ & $0.88$ \\
      &      $50\%$ & $0.98$ & $0.99$ & $0.91$ & $0.89$ \\
      &      $99\%$ & $0.98$ & $0.99$ & $0.91$ & $0.89$ \\
\hline
\end{tabular}
\end{table}

\subsubsection{Extension to other probability distributions}

The choice of the beta distribution is natural since the support is $\left[
0,1\right] $, but we can consider other continuous probability distributions
for modeling $\redemption^{\star }$. For example, the Kumaraswamy distribution
is another good candidate, but it is close to the beta distribution. When the
support of the probability distribution is $\left[ 0,\infty \right) $, we apply
the truncation formula\footnote{For the probability density function, we have:
\begin{equation*}
g_{\left[ 0,1\right] }\left( x\right) =\frac{g\left( x\right) }{\mathbf{G}%
\left( 1\right) }
\end{equation*}%
}:%
\begin{equation*}
\mathbf{G}_{\left[ 0,1\right] }\left( x\right) =\frac{\mathbf{G}\left(
x\right) }{\mathbf{G}\left( 1\right) }
\end{equation*}%
For instance, we can use the gamma or log-logistic distribution. However, our
experience shows that some continuous probability distributions are not adapted
such as the log-gamma and log-normal distributions, because the logarithm
transform performs a bad scale for random variables in $\left[ 0,1\right] $.
Finally, we can also use the logit transformation, which is very popular for
modeling the probability of default (PD) or the loss given default (LGD) in
credit risk. Following \citet[page 910]{Roncalli-2020}, we assume that
$\redemption^{\star }$ is a logit transformation of a random variable $X\in
\left( -\infty ,\infty \right) $, meaning that\footnote{We also have:
\begin{equation*}
\redemption^{\star }=\limfunc{logit}\nolimits^{-1}\left( X\right) =\frac{1}{1+e^{-X}}
\end{equation*}%
}:
\begin{equation*}
X=\func{logit}\left( \redemption^{\star }\right) =\ln \left( \frac{%
\redemption^{\star }}{1-\redemption^{\star }}\right)
\end{equation*}%
For instance, in the case of the logit-normal distribution, we have:%
\begin{equation*}
\func{logit}\left( \redemption^{\star }\right) \sim \mathcal{N}\left(
a,b^{2}\right)
\end{equation*}%
We deduce that:%
\begin{eqnarray*}
\mathbf{G}\left( x\right)  &=&\Pr \left( \redemption^{\star }\leq x\right) \\
&=&\Pr \left( X\leq \func{logit}\left( x\right) \right)  \\
&=&\Phi \left( \frac{\func{logit}\left( x\right) -a}{b}\right)
\end{eqnarray*}%
and:%
\begin{equation*}
g\left( x\right) =\frac{1}{bx\left( 1-x\right) }\phi \left( \frac{\func{logit%
}\left( x\right) -a}{b}\right)
\end{equation*}
A summary of these alternative approaches\footnote{$\gamma \left( \alpha
,x\right) $ is the lower incomplete gamma function.} is given in Table
\ref{tab:inflated-cdf}. In the sequel, we continue to use the beta distribution,
because it is easy to calibrate and it is the most popular approach when
modeling a random variable in $\left[0,1\right]$. However, we cannot claim that
it is the best fitting model. Such a debate has already taken place in
operational risk with the log-normal distribution and the modeling of the
severity distribution of operational risk losses \citep{Roncalli-2020}.
Nevertheless, we think that this debate is too early in the case of liability
stress testing, and can wait when we will have more comprehensive redemption
databases.

\begin{table}[tbph]
\centering
\caption{List of continuous probability distributions}
\label{tab:inflated-cdf}
\begin{tabular}{lcccc}
\hline
Distribution & Symbol & $\mathbf{G}\left( x\right) $ & $g\left( x\right) $ & Support \\ \hline
Beta &
$\mathcal{B}\left( a ,b \right) $ &
$\mathfrak{B}\left(x;a ,b \right) $ &
$\dfrac{x^{a -1}\left( 1-x\right) ^{b -1}}{\mathfrak{B}\left( a ,b \right) }$ &
$\left[0,1\right]$ \\
Gamma &
$\mathcal{G}\left( a ,b \right) $ &
$\dfrac{\gamma \left(a ,b x\right)}{\Gamma \left( a \right) }$ &
$\dfrac{b^{a }x^{a -1}e^{-b x}}{\Gamma \left( a \right) }$ &
$\left[0,\infty\right)$ \\
Kumaraswamy &
$\mathcal{K}\left( a,b\right) $ &
$1-\left( 1-x^{a}\right) ^{b}$ &
$abx^{a-1}\left( 1-x^{a}\right) ^{b-1}$ &
$\left[0,1\right]$ \\
Log-logistic &
$\mathcal{LL}\left( a ,b \right) $ &
$\dfrac{x^{b}}{a ^{b }+x^{b }}$ &
$\dfrac{b \left( x/a \right)^{b -1}}{a \left( 1+\left( x/a \right) ^{b }\right) ^{2}}$ &
$\left[0,\infty\right)$ \\
Logit-normal &
$\mathcal{LN}\left( a,b^{2}\right) $ &
$\Phi \left( \dfrac{\func{logit}\left( x\right) -a}{b}\right) $ &
$\dfrac{1}{bx\left( 1-x\right) }\phi \left( \dfrac{\func{logit}\left( x\right)
-a}{b}\right) $ &
$\left[0,1\right]$ \\
\hline
\end{tabular}%
\end{table}

\subsection{Parametric stress scenarios}

As explained previously, the zero-inflated beta model is appealing for producing
stress scenarios. For that, we proceed in two steps. We first calibrate the parameters of the model,
and then we compute the stress scenarios for a given return time.

\subsubsection{Estimation of the zero-inflated beta model}

Let $\Omega =\left\{ \redemption_{1},\ldots ,\redemption_{n}\right\} $ be the
sample of redemption rates for a given matrix cell. Three parameters have to be
estimated: the redemption frequency $p$ and the parameters $a$ and $b$ that
control the shape of the beta distribution. We note $n_{0}$ as the number of
observations that are equal to zero and $n_{1}=n-n_{0}$ as the
number of observations that are strictly
positive\footnote{We have $n_{0}=\sum_{i=1}^{n}\mathds{1}\left\{ \redemption_{i}=0\right\}= n - n_1 $ and
$n_{1}=\sum_{i=1}^{n}\mathds{1}\left\{ \redemption_{i}>0\right\} = \sum_{i=1}^{n}\mathcal{E}_{i} $.}. In
Appendix \ref{appendix:zero-inflated-ml} on page \pageref%
{appendix:zero-inflated-ml}, we show that the maximum likelihood estimates are:
\begin{equation*}
\hat{p}=\frac{n_{1}}{n_{0}+n_{1}}
\end{equation*}%
and:%
\begin{equation*}
\left\{ \hat{a},\hat{b}\right\} =\underset{a,b}{\arg \max }-n_1\ln \mathfrak{B}%
\left( a,b\right) +\sum_{\redemption_{i}>0}\left( a-1\right) \ln \redemption%
_{i}+\sum_{\redemption_{i}>0}\left( b-1\right) \ln \left( 1-\redemption%
_{i}\right)
\end{equation*}%
The estimates $\hat{a}$ and $\hat{b}$ can be found by numerical
optimization.\smallskip

This is the traditional approach for estimating a zero-inflated model. However,
it is not convenient since the parameters $\left(
\hat{p},\hat{a},\hat{b}\right) $ should be modified by risk managers and
business experts before computing redemption shocks. Indeed, the calibration
process of parametric stress scenarios follows the same process when one builds
historical stress scenarios, and estimated values $\left(
\hat{p},\hat{a},\hat{b}\right) $ cannot be directly used because they do not
necessarily respect some risk coherency principles and their robustness varies
across matrix cells.\smallskip

A second approach consists in using the method of moments. In this case, the
estimator of $p$ has the same expression:
\begin{equation}
\hat{p}=\frac{n_{1}}{n_{0}+n_{1}}  \label{eq:zero-inflated-p-mm}
\end{equation}%
For the parameters of the beta distribution, we first calculate the empirical
mean $\hat{\mu}$ and the standard deviation $\hat{\sigma}$ of the positive
redemption rates $\redemption^{\star }$, and then we use the following
relationships \citep[page 193]{Roncalli-2020}:
\begin{equation}
\hat{a}=\frac{\hat{\mu}^{2}\left( 1-\hat{\mu}\right) }{\hat{\sigma}^{2}}-%
\hat{\mu}  \label{eq:zero-inflated-mu-mm}
\end{equation}%
and:%
\begin{equation}
\hat{b}=\frac{\hat{\mu}\left( 1-\hat{\mu}\right) ^{2}}{\sigma ^{2}}-\left( 1-%
\hat{\mu}\right)   \label{eq:zero-inflated-sigma-mm}
\end{equation}%
The differences between the two methods are the following:
\begin{itemize}
\item In the case of the method of maximum likelihood, $a$ and $b$ are
    explicit parameters. Once the parameters $p$, $a$ and $b$ are estimated,
    we can calculate the mean $\mu $ and standard deviation $\sigma $ for the
    severity distribution. In this approach, $\mu $ and $\sigma $ are
    implicit, because they are deduced from $a$ and $b$.

\item In the case of the method of moments, $a$ and $b$ are implicit
    parameters. Indeed, they are calculated after having estimated the mean
    $\mu $ and standard deviation $\sigma $ for the severity distribution. In
    this approach, $\mu $ and $\sigma $ are explicit and define the severity
    distribution.
\end{itemize}
The first approach is known as the $p-a-b$ parameterization, whereas the second
approach corresponds to the $p-\mu -\sigma $ parameterization. By construction,
this last approach is more convenient in a liquidity stress testing framework,
because the parameters $\mu $ and $\sigma $ are intuitive and self-explanatory
measures, which is not the case of $a$ and $b$. Therefore, they can be
manipulated by risk managers and business experts.\smallskip

\begin{table}[p]
\centering
\caption{Estimated value of $p$ in \%}
\label{tab:inflated14-p}
\begin{tabular}{lcccccccc}
\hline
                        &          (1) &          (2) &          (3) &          (4) &          (5) &          (6) &          (7) &          (8) \\ \hline
Auto-consumption        &       $21.63$ &      $19.41$ &      $30.00$ &      $25.46$ &      $50.60$ & ${\TsV}6.39$ &  ${\TsX}   $ &      $22.16$  \\
Central bank            &  ${\TsV}0.16$ & ${\TsV}0.34$ &  ${\TsX}   $ & ${\TsV}1.47$ &  ${\TsX}   $ &  ${\TsX}   $ &  ${\TsX}   $ & ${\TsV}0.47$  \\
Corporate               &       $15.04$ & ${\TsV}6.19$ & ${\TsV}6.25$ & ${\TsV}2.87$ &      $39.81$ & ${\TsV}0.21$ &  ${\TsX}   $ &      $14.54$  \\
Corporate pension fund  &  ${\TsV}8.11$ & ${\TsV}3.38$ & ${\TsV}3.98$ & ${\TsV}3.37$ & ${\TsV}7.57$ & ${\TsV}0.00$ &  ${\TsX}   $ & ${\TsV}4.12$  \\
Employee savings plan   &  ${\TsV}2.67$ & ${\TsV}2.83$ & ${\TsV}2.97$ & ${\TsV}2.71$ & ${\TsV}2.29$ &  ${\TsX}   $ & ${\TsV}2.75$ & ${\TsV}2.69$  \\
Institutional           &       $19.36$ & ${\TsV}6.28$ & ${\TsV}1.96$ & ${\TsV}6.51$ &      $32.83$ & ${\TsV}1.04$ &  ${\TsX}   $ & ${\TsV}8.23$  \\
Insurance               &       $12.19$ & ${\TsV}6.72$ & ${\TsV}3.45$ & ${\TsV}7.22$ &      $27.92$ & ${\TsV}1.04$ &  ${\TsX}   $ & ${\TsV}9.71$  \\
Other                   &  ${\TsV}9.67$ & ${\TsV}3.87$ & ${\TsV}3.68$ &      $19.35$ &      $21.52$ & ${\TsV}2.22$ &  ${\TsX}   $ & ${\TsV}8.82$  \\
Retail                  &       $44.59$ &      $45.04$ &      $58.76$ &      $70.50$ &      $45.75$ &      $17.51$ &      $27.32$ &      $45.61$  \\
Sovereign               &       $16.30$ & ${\TsV}3.18$ & ${\TsV}1.05$ &      $10.07$ &      $18.23$ & ${\TsV}0.06$ &  ${\TsX}   $ &      $10.14$  \\
Third-party distributor &       $33.77$ &      $37.36$ &      $45.97$ &      $45.94$ &      $65.94$ &      $32.86$ & ${\TsV}6.52$ &      $40.61$  \\ \hline
Total                   &       $34.66$ &      $27.10$ &      $24.19$ &      $38.34$ &      $37.57$ &      $11.14$ &      $24.79$ &      $31.11$  \\
\hline
\end{tabular}
\end{table}

\begin{table}[p]
\centering
\caption{Estimated value of $\mu$ in \% (method of moments)}
\label{tab:inflated14-mu-mm}
\begin{tabular}{lcccccccc}
\hline
                        &         (1) &         (2) &          (3) &         (4) &          (5) &         (6) &         (7) &         (8) \\ \hline
Auto-consumption        &       $1.24$ &      $1.88$ &      $2.15$ &      $1.19$ & ${\TsV}3.11$ &      $2.81$ & ${\TsV}   $ &      $1.70$  \\
Central bank            &  ${\TsV}   $ & ${\TsV}   $ & ${\TsV}   $ & ${\TsV}   $ &  ${\TsX}   $ & ${\TsV}   $ & ${\TsV}   $ & ${\TsV}   $  \\
Corporate               &       $0.55$ &      $2.50$ & ${\TsV}   $ & ${\TsV}   $ & ${\TsV}3.82$ & ${\TsV}   $ & ${\TsV}   $ &      $3.73$  \\
Corporate pension fund  &  ${\TsV}   $ &      $1.54$ & ${\TsV}   $ &      $2.84$ & ${\TsV}7.26$ & ${\TsV}   $ & ${\TsV}   $ &      $3.21$  \\
Employee savings plan   &       $1.29$ & ${\TsV}   $ & ${\TsV}   $ &      $2.08$ &  ${\TsX}   $ & ${\TsV}   $ & ${\TsV}   $ &      $2.10$  \\
Institutional           &       $0.67$ &      $2.62$ & ${\TsV}   $ &      $2.80$ & ${\TsV}4.46$ & ${\TsV}   $ & ${\TsV}   $ &      $3.23$  \\
Insurance               &       $1.36$ &      $2.20$ & ${\TsV}   $ &      $2.19$ & ${\TsV}3.21$ & ${\TsV}   $ & ${\TsV}   $ &      $2.66$  \\
Other                   &       $0.87$ &      $2.60$ & ${\TsV}   $ &      $1.10$ & ${\TsV}3.51$ &      $0.99$ & ${\TsV}   $ &      $2.65$  \\
Retail                  &       $0.34$ &      $0.31$ &      $0.44$ &      $0.23$ & ${\TsV}1.98$ &      $0.43$ &      $0.15$ &      $0.33$  \\
Sovereign               &       $0.06$ & ${\TsV}   $ & ${\TsV}   $ &      $1.84$ &      $10.48$ & ${\TsV}   $ & ${\TsV}   $ &      $4.46$  \\
Third-party distributor &       $0.35$ &      $0.64$ &      $1.45$ &      $0.42$ & ${\TsV}1.40$ &      $0.86$ &      $1.21$ &      $0.56$  \\ \hline
Total                   &       $0.40$ &      $0.73$ &      $1.64$ &      $0.48$ & ${\TsV}2.82$ &      $0.98$ &      $0.18$ &      $0.72$  \\
\hline
\end{tabular}
\end{table}

\begin{table}[p]
\centering
\caption{Estimated value of $\sigma$ in \% (method of moments)}
\label{tab:inflated14-sigma-mm}
\begin{tabular}{lcccccccc}
\hline
                        &          (1) &          (2) &          (3) &          (4) &          (5) &          (6) &         (7) &          (8) \\ \hline
Auto-consumption        &       $7.38$ & ${\TsV}6.86$ &      $9.73$ & ${\TsV}5.98$ & ${\TsV}8.80$ &      $9.10$ & ${\TsV}   $ & ${\TsV}7.09$  \\
Central bank            &  ${\TsV}   $ &  ${\TsX}   $ & ${\TsV}   $ &  ${\TsX}   $ &  ${\TsX}   $ & ${\TsV}   $ & ${\TsV}   $ &  ${\TsX}   $  \\
Corporate               &       $5.55$ & ${\TsV}9.57$ & ${\TsV}   $ &  ${\TsX}   $ & ${\TsV}7.49$ & ${\TsV}   $ & ${\TsV}   $ & ${\TsV}8.70$  \\
Corporate pension fund  &  ${\TsV}   $ &      $10.36$ & ${\TsV}   $ &      $13.51$ &      $13.14$ & ${\TsV}   $ & ${\TsV}   $ &      $12.09$  \\
Employee savings plan   &       $3.26$ &  ${\TsX}   $ & ${\TsV}   $ & ${\TsV}8.40$ &  ${\TsX}   $ & ${\TsV}   $ & ${\TsV}   $ & ${\TsV}8.61$  \\
Institutional           &       $5.46$ & ${\TsV}9.99$ & ${\TsV}   $ & ${\TsV}9.23$ &      $11.46$ & ${\TsV}   $ & ${\TsV}   $ &      $10.86$  \\
Insurance               &       $8.66$ &      $10.56$ & ${\TsV}   $ &      $10.11$ & ${\TsV}8.13$ & ${\TsV}   $ & ${\TsV}   $ & ${\TsV}9.35$  \\
Other                   &       $3.61$ & ${\TsV}9.36$ & ${\TsV}   $ & ${\TsV}7.27$ &      $11.88$ &      $6.70$ & ${\TsV}   $ &      $10.68$  \\
Retail                  &       $2.80$ & ${\TsV}2.58$ &      $3.32$ & ${\TsV}2.10$ & ${\TsV}7.52$ &      $3.22$ &      $2.64$ & ${\TsV}2.88$  \\
Sovereign               &       $0.25$ &  ${\TsX}   $ & ${\TsV}   $ & ${\TsV}9.90$ &      $21.63$ & ${\TsV}   $ & ${\TsV}   $ &      $14.94$  \\
Third-party distributor &       $2.68$ & ${\TsV}3.48$ &      $7.63$ & ${\TsV}2.58$ & ${\TsV}4.71$ &      $5.84$ &      $6.98$ & ${\TsV}3.37$  \\ \hline
Total                   &       $3.31$ & ${\TsV}4.35$ &      $8.93$ & ${\TsV}3.50$ & ${\TsV}8.66$ &      $6.08$ &      $3.03$ & ${\TsV}4.55$  \\
\hline
\end{tabular}
\medskip

\begin{flushleft}
\begin{footnotesize}
(1) = balanced, (2) = bond, (3) = enhanced treasury, (4) = equity, (5) = money
market, (6) = other, (7) = structured, (8) = total
\end{footnotesize}
\end{flushleft}
\end{table}

We have estimated the parameters $p$, $a$, $b$, $\mu $ and $\sigma $ with the
two methods. Table \ref{tab:inflated14-p} shows the redemption frequency. On
average, $\hat{p}$ is equal to $31\%$, but we observe large differences between
the matrix cells. For instance, $\hat{p}$ is less than $5\%$ for central banks,
corporate pension funds and employee savings plans, whereas the largest values
of $\hat{p}$ are observed for retail investors and third-party distributors.
The values of $\hat{\mu}$ and $\hat{\sigma}$ are reported in Tables
\ref{tab:inflated14-mu-mm} and \ref{tab:inflated14-sigma-mm}. The average
redemption severity is $0.72\%$, whereas the redemption volatility is $4.55\%$.
Again, we observe some large differences between the matrix cells.

\begin{remark}
In Tables \ref{tab:inflated14-a-mm} and \ref{tab:inflated14-b-mm} on page
\pageref{tab:inflated14-a-mm}, we have also reported the implicit values of
$\hat{a}$ and $\hat{b}$ that are deduced from $\hat{\mu}$ and $\hat{\sigma}$.
Moreover, we have reported the estimated values by the method of maximum
likelihood on pages \pageref{tab:inflated14-a-ml} and
\pageref{tab:inflated14-mu-ml}.
\end{remark}

\subsubsection{Stress scenarios based on the $p-\mu-\sigma$ parameterization}

Using the previous estimates $\left( \hat{p},\hat{\mu},\hat{\sigma}\right) $,
risk managers and business experts can define the triplet $\left( p,\mu,\sigma
\right) $ for the different matrix cells. For that, they must assess the
confidence in estimated values with respect to the number of observations. For
the frequency parameter, we use the value of $n$, which has been already
reported in Table \ref{tab:historical5} on page \pageref{tab:historical5}. For the
severity parameters $\hat{\mu}$ and $\hat{\sigma}$, we use the value of
$n_{1}$, which is much smaller than $n$. Using the data given in Table
\ref{tab:data3-4} on page \pageref{tab:data3-4}, we have built the confidence
measure in Table \ref{tab:inflated15}. We confirm that the confidence measure
in $\hat{\mu}$ and $\hat{\sigma}$ is lower than the confidence measure in
$\hat{p}$. In particular, there are many matrix cells, where the number $n_1$
of observations is lower than $200$. This explains why Tables
\ref{tab:inflated14-mu-mm} and \ref{tab:inflated14-sigma-mm} contain a lot of
missing values. Therefore, except for a few matrix cells, the
estimated values $\hat{\mu}$ and $\hat{\sigma}$ must be challenged by risk
managers and business experts. Again, they can use risk coherency
principles\footnote{They are defined on page \pageref{marker:coherency}.}
$\mathcal{C}_{\mathrm{investor}}$ and $\mathcal{C}_{\mathrm{fund}}$ to build
their own figures of $p$, $\mu$ and $\sigma$.\smallskip

\begin{table}[tbph]
\centering
\caption{Confidence in estimated values $\hat{\mu}$ and $\hat{\sigma}$ with respect to the number $n_1$ of observations}
\label{tab:inflated15}
\begin{tabular}{lccccccc}
\hline
                        &   (1) &   (2) &   (3) &   (4) &   (5) &   (6) &   (7)  \\ \hline
Auto-consumption        &      \bII &     \bII &     \bII &    \bIII &     \bII &      \bI &    \bZZZ   \\
Central bank            &     \bZZZ &    \bZZZ &    \bZZZ &    \bZZZ &    \bZZZ &    \bZZZ &    \bZZZ   \\
Corporate               &       \bI &      \bI &      \bZ &      \bZ &     \bII &    \bZZZ &    \bZZZ   \\
Corporate pension fund  &       \bZ &      \bI &     \bZZ &      \bI &      \bI &    \bZZZ &    \bZZZ   \\
Employee savings plan   &       \bI &      \bZ &     \bZZ &      \bI &      \bZ &    \bZZZ &      \bZ   \\
Institutional           &      \bII &     \bII &      \bZ &     \bII &     \bII &      \bZ &    \bZZZ   \\
Insurance               &       \bI &      \bI &      \bZ &     \bII &     \bII &      \bZ &    \bZZZ   \\
Other                   &       \bI &      \bI &      \bZ &      \bI &     \bII &      \bI &    \bZZZ   \\
Retail                  &     \bIII &    \bIII &     \bII &    \bIII &     \bII &     \bII &    \bIII   \\
Sovereign               &       \bI &      \bZ &    \bZZZ &      \bI &      \bI &    \bZZZ &    \bZZZ   \\
Third-party distributor &     \bIII &    \bIII &     \bII &    \bIII &     \bII &     \bII &      \bI   \\
\hline
\end{tabular}
\begin{flushleft}
\begin{footnotesize}
\bZZZ\ $0-10$, \bZZ\ $11-50$, \bZ\ $51-200$, \bI\ $201-1\,000$, \bII\ $1\,001-10\,000$, \bIII\  $+10\,000$
\end{footnotesize}
\end{flushleft}
\vspace*{-10pt}
\end{table}

\begin{figure}[tbph]
\centering
\caption{Parametric stress scenarios $\mathbb{S}\left( \mathcal{T};p,\mu ,\sigma \right) $ in \%}
\label{fig:inflated16}
\includegraphics[width = \figurewidth, height = \figureheight]{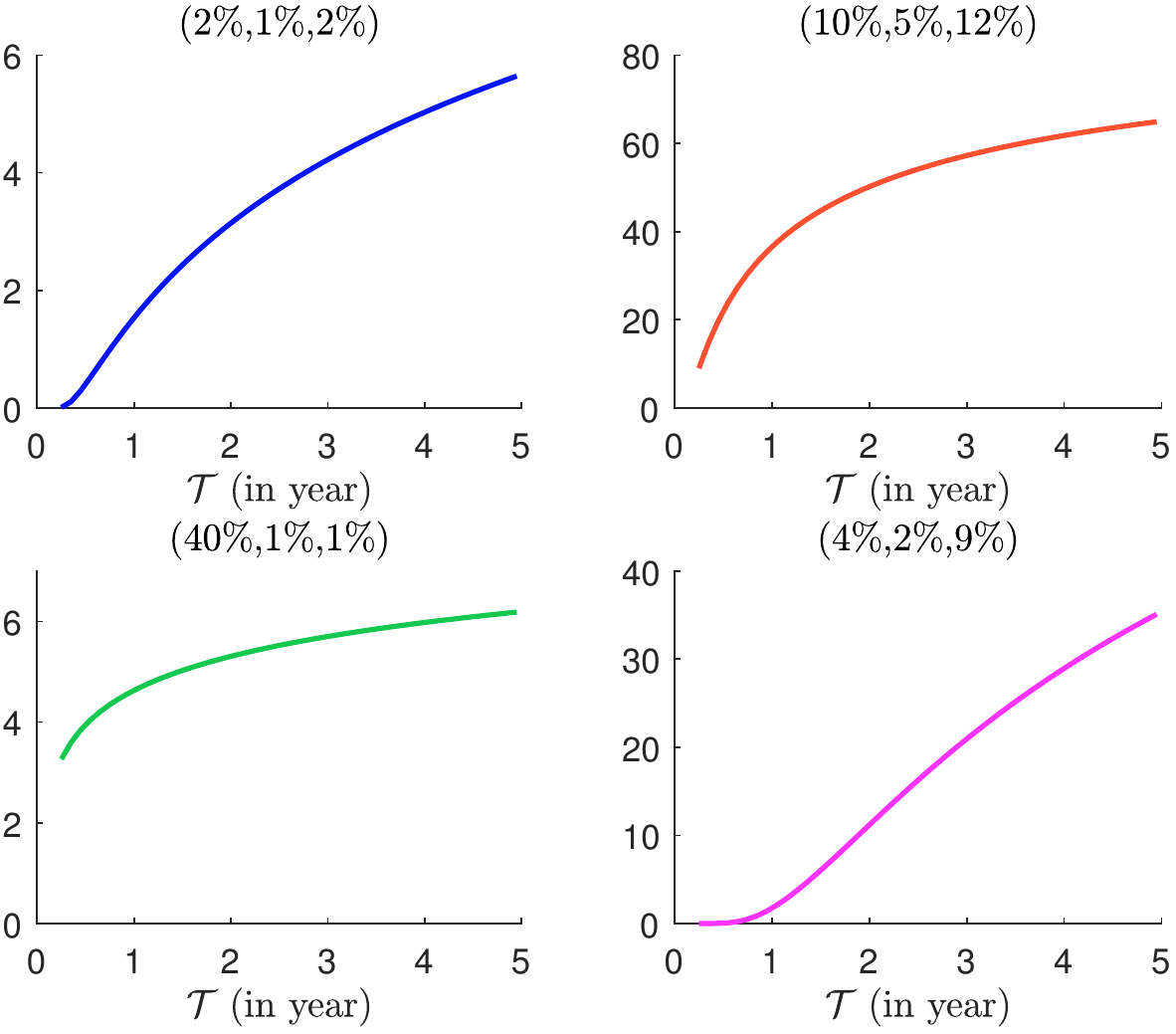}
\end{figure}

Once the triplet $\left( p,\mu ,\sigma \right) $ is defined for each matrix
cell, we compute stress scenarios using the following formula:%
\begin{equation*}
\mathbb{S}\left( \mathcal{T};p,\mu ,\sigma \right) =\mathcal{B}^{-1}\left( 1-%
\frac{1}{p\mathcal{T}};\frac{\mu ^{2}\left( 1-\mu \right) }{\sigma ^{2}}-\mu
,\frac{\mu \left( 1-\mu \right) ^{2}}{\sigma ^{2}}-\left( 1-\mu \right)
\right)
\end{equation*}%
where $\mathcal{B}^{-1}\left( \alpha ;a,b\right) $ is the $\alpha $-quantile of
the beta distribution with parameters $a$ and $b$. The parametric stress
scenario $\mathbb{S}\left( \mathcal{T};p,\mu ,\sigma \right) $ depends on the
return time $\mathcal{T}$ and the three parameters of the zero-inflated model.
An example is provided in Figure \ref{fig:inflated16}. For each plot, we
indicate the triplet $\left( p,\mu ,\sigma \right) $. For instance, the first
plot corresponds to the triplet ($2\%,1\%,2\%$), meaning that the daily
redemption frequency is $2\%$, the expected redemption severity is $1\%$ and
the redemption volatility is $2\%$. In particular, these plots illustrate the
high impact of $\sigma $, which is the key parameter when computing parametric
stress scenarios. The reason is that the parameters $p$ and $\mu$ determine
the mean $\mathbb{E}\left[\redemption\right]$, whereas the uncertainty around this number
is mainly driven by the parameter $\sigma$. The redemption volatility controls then the shape of the
probability distribution of the redemption rate (both the skewness and the kurtosis),
implying that $\sigma$ has a major impact on the stress scenario $\mathbb{S}\left( \mathcal{T}\right)$
when $\mathcal{T}$ is large.

\section{Behavioral modeling}

In this section, we go beyond the zero-inflated model by
considering the behavior of each investor. In particular, we show
that the redemption rate depends on the liability structure of the mutual fund.
Moreover, the behavior of investors may be correlated, in particular in a stress period.
In this case, the modeling of spillover effects  is important to define stress scenarios.

\subsection{The individual-based model}

The individual-based model and the zero-inflated model are highly connected.
Indeed, the zero-inflated model can be seen as a special case of the
individual-based model when we summarize the behavior of all unitholders by
the behavior of a single client.

\subsubsection{Definition}

Let $\limfunc{TNA}\left( t\right) $ be the assets under management of an
investment fund composed of $n$ clients:
\begin{equation*}
\limfunc{TNA}\left( t\right) =\sum_{i=1}^{n}\limfunc{TNA}\nolimits_{i}\left(
t\right)
\end{equation*}%
where $\limfunc{TNA}\nolimits_{i}\left( t\right) $ is the net asset
of the individual $i$. The redemption rate of the fund is equal to
the redemption flows divided by the total net assets:
\begin{eqnarray*}
\redemption &=&\frac{\sum_{i=1}^{n}\limfunc{TNA}\nolimits_{i}\cdot %
\redemption_{i}}{\limfunc{TNA}} \\
&=&\sum_{i=1}^{n}\omega _{i}\cdot \redemption_{i}
\end{eqnarray*}%
where $\omega _{i}$ represents the weight of the client $i$ in the fund:
\begin{equation*}
\omega _{i}=\frac{\limfunc{TNA}\nolimits_{i}}{\limfunc{TNA}}
\end{equation*}%
Since we have $\redemption_{i}=\mathcal{E}_{i}\cdot \redemption_{i}^{\star }$%
, we obtain:
\begin{equation*}
\redemption=\sum_{i=1}^{n}\omega _{i}\cdot \mathcal{E}_{i}\cdot \redemption%
_{i}^{\star }
\end{equation*}%
Generally, we assume that the clients are homogenous, meaning that
$\mathcal{E}_{i}$ and $\redemption_{i}^{\star }$ are \textit{iid} random variables. If
we denote by $\tilde{p}$ and $\mathbf{\tilde{G}}$  the individual redemption
probability and the individual severity distribution. The individual-based
model is then characterized by the 4-tuple $\left( n,\omega ,\tilde{p},
\mathbf{\tilde{G}}\right) $, where $n$ is the number of clients and $\omega
=\left( \omega _{1},\ldots ,\omega _{n}\right) $ is the vector of weights.
Like the zero-inflated model, we consider a $\tilde{\mu}-\tilde{\sigma}$
parameterization of $\mathbf{\tilde{G}}$, meaning that the model is denoted
by $\mathcal{IM}\left( n,\omega ,\tilde{p},\tilde{\mu},\tilde{\sigma}\right)
$.

\begin{remark}
When the individual severity distribution $\mathbf{\tilde{G}}$ is no specified, we
assume that it is a beta distribution $\mathcal{B}\left(
\tilde{a},\tilde{b}\right) $, whose parameters $\tilde{a}$ and $\tilde{b}$
are calibrated with respect to the severity mean $\tilde{\mu}$ and the
severity volatility $\tilde{\sigma} $ using the method of moments. In a
similar way, we assume that the vector of weights is equally-weighted when it
is not specified. In this case, the individual-based model is denoted by
$\mathcal{IM}\left( n,\tilde{p},\tilde{\mu},\tilde{\sigma}\right) $.
\end{remark}

\begin{figure}[tbph]
\centering
\caption{Histogram of the redemption rate in \%
($\tilde{p} = 50\%, \tilde{\mu} = 50\%, \tilde{\sigma} = 10\%$)}
\label{fig:individual1}
\includegraphics[width = \figurewidth, height = \figureheight]{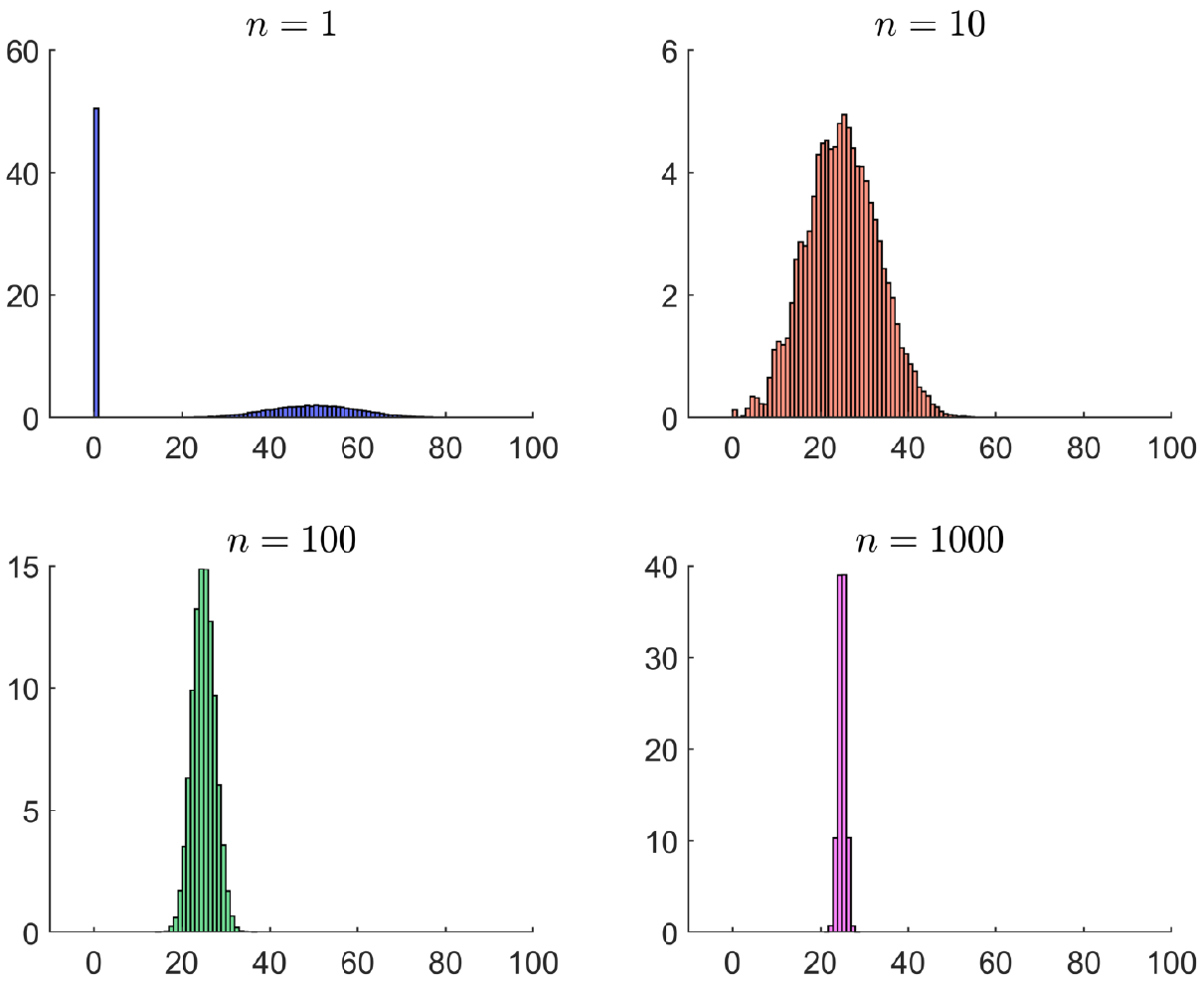}
\end{figure}

In Figure \ref{fig:individual1}, we report the histogram of the redemption
rate $\redemption$ in the case $\tilde{p}=50\%$, $\tilde{\mu} =50\%$ and
$\tilde{\sigma} =10\%$. In the case $n = 1$, we obtain a singular
distribution. Indeed, there is a probability of $50\%$ that there is no
redemption. The singularity decreases with respect to the number $n$ of
investors, because the probability to have a redemption increases. This is
normal since the redemption frequency of a mutual fund depends on the number
of unitholders. This explains that the redemption frequency is larger for a
retail fund than for an institutional fund.

\subsubsection{Statistical analysis}

\paragraph{The skewness effect}
The singularity of the distribution function $\mathbf{F}$ at the point
$\redemption=0$ is entirely explained by the two parameters $\tilde{p}$ and $n
$ as shown in Appendix \ref{appendix:individual2} on page
\pageref{appendix:individual2}, because we have:
\begin{equation*}
\Pr \left\{ \redemption=0\right\} =\left( 1-\tilde{p}\right) ^{n}
\end{equation*}%
The fact that the probability distribution is not continuous has an impact
on the skewness and the kurtosis. In Table \ref{tab:individual2}, we have reported the
probability $\Pr \left\{ \redemption=0\right\} $. If there is a few
investors in the fund, the probability to observe no redemption in the
fund is high. For instance, if $\tilde{p}=5\%$ and $n=10$, we have $\Pr
\left\{ \redemption=0\right\} =59.87\%$. If $\tilde{p}=1\%$ and $n=10$, we
have $\Pr \left\{ \redemption=0\right\} =90.44\%$. How to interpret these
results? Since $\tilde{p}$ is the individual redemption probability,
$1/\tilde{p}$ is the return time of a redemption at the investor level. For
example, $\tilde{p}=5\%$ (resp. $\tilde{p}=1\%$) means that investors redeem
every 20 days (resp. 100 days) on average. At the fund level, the return
time to observe a redemption is equal to $\left( 1-\Pr \left\{ \redemption
=0\right\} \right) ^{-1}$. For instance, in the case $\tilde{p}=5\%$ and
$n=10$, we observe a redemption two days per week in the fund on average%
\footnote{The exact value is equal to $1/\left( 1-59.87\%\right) =2.4919$.}. This
analysis may help to distinguish active and passive investors. In the case
of passive investors when the redemption event occurs once a year or less,
$\tilde{p}$ is less than 40 bps. In the case of active investors that redeem
once a month, $\tilde{p}$ is greater than $5\%$. Therefore, the skewness
effect depends if the fund has active investors or not, and if the fund is
granular or not.

\begin{table}[tbph]
\centering
\caption{Probability to observe no redemption $\Pr \left\{ \redemption=0\right\} $ in \%}
\label{tab:individual2}
\begin{tabular}{cccccccccc}
\hline
$p$          & \multicolumn{9}{c}{Number $n$ of investors} \\
(in \%)      &       $1$ &       $2$ &          $5$ &         $10$ &         $50$ &        $100$ &       $1000$ &      $10000$ \\ \hline
${\TsV}0.01$ &   $99.99$ &   $99.98$ &      $99.95$ &      $99.90$ &      $99.50$ &      $99.00$ &      $90.48$ &      $36.79$ \\
${\TsV}0.02$ &   $99.98$ &   $99.96$ &      $99.90$ &      $99.80$ &      $99.00$ &      $98.02$ &      $81.87$ &      $13.53$ \\
${\TsV}0.05$ &   $99.95$ &   $99.90$ &      $99.75$ &      $99.50$ &      $97.53$ &      $95.12$ &      $60.65$ & ${\TsV}0.67$ \\
${\TsV}0.10$ &   $99.90$ &   $99.80$ &      $99.50$ &      $99.00$ &      $95.12$ &      $90.48$ &      $36.77$ & ${\TsV}0.00$ \\
${\TsV}0.20$ &   $99.80$ &   $99.60$ &      $99.00$ &      $98.02$ &      $90.47$ &      $81.86$ &      $13.51$ & ${\TsV}0.00$ \\
${\TsV}0.50$ &   $99.50$ &   $99.00$ &      $97.52$ &      $95.11$ &      $77.83$ &      $60.58$ & ${\TsV}0.67$ & ${\TsV}0.00$ \\
${\TsV}1.00$ &   $99.00$ &   $98.01$ &      $95.10$ &      $90.44$ &      $60.50$ &      $36.60$ & ${\TsV}0.00$ & ${\TsV}0.00$ \\
${\TsV}2.00$ &   $98.00$ &   $96.04$ &      $90.39$ &      $81.71$ &      $36.42$ &      $13.26$ & ${\TsV}0.00$ & ${\TsV}0.00$ \\
${\TsV}5.00$ &   $95.00$ &   $90.25$ &      $77.38$ &      $59.87$ & ${\TsV}7.69$ & ${\TsV}0.59$ & ${\TsV}0.00$ & ${\TsV}0.00$ \\
     $10.00$ &   $90.00$ &   $81.00$ &      $59.05$ &      $34.87$ & ${\TsV}0.52$ & ${\TsV}0.00$ & ${\TsV}0.00$ & ${\TsV}0.00$ \\
     $25.00$ &   $75.00$ &   $56.25$ &      $23.73$ & ${\TsV}5.63$ & ${\TsV}0.00$ & ${\TsV}0.00$ & ${\TsV}0.00$ & ${\TsV}0.00$ \\
     $50.00$ &   $50.00$ &   $25.00$ & ${\TsV}3.13$ & ${\TsV}0.10$ & ${\TsV}0.00$ & ${\TsV}0.00$ & ${\TsV}0.00$ & ${\TsV}0.00$ \\
\hline
\end{tabular}
\end{table}

\paragraph{The mean effect}

\begin{figure}[tbph]
\centering
\caption{Mean of the redemption rate $\redemption$ in \%}
\label{fig:individual3}
\includegraphics[width = \figurewidth, height = \figureheight]{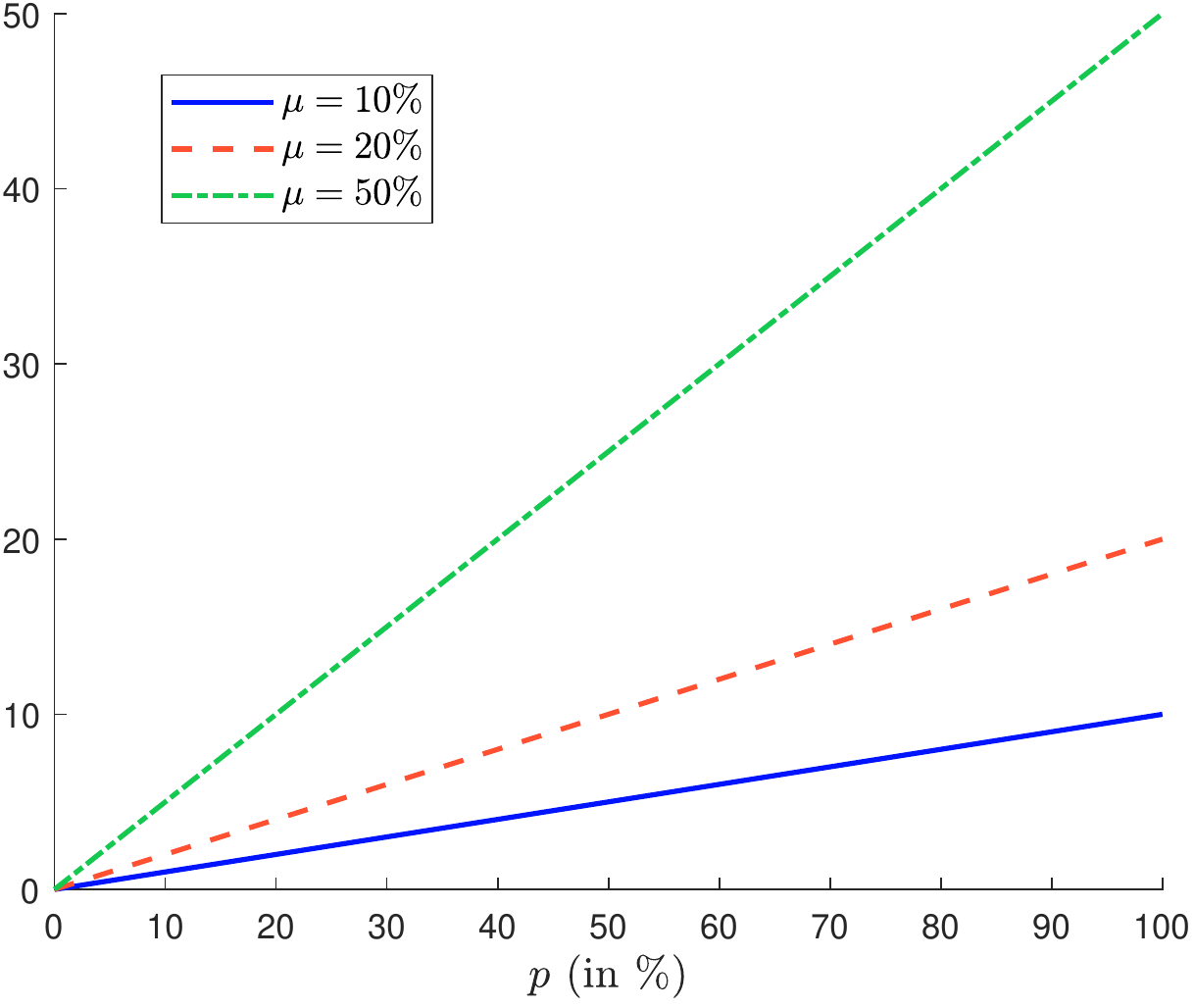}
\end{figure}

The mean shape is easy to understand since it is the product of the
redemption probability and the mean of the redemption severity:
\begin{equation*}
\mathbb{E}\left[ \redemption\right] =\tilde{p}\tilde{\mu}
\end{equation*}%
Curiously, it depends neither on the number of investors in the fund, nor on
the liability structure (see Figure \ref{fig:individual3}). Since $\tilde{\mu}\in
\left[ 0,1\right] $, we notice that $\mathbb{E}\left[ \redemption\right] \leq \tilde{p}$,
meaning that we must observe very low values of the redemption mean.
And we verify this property if we consider the results\footnote{Another way to
compute the empirical mean of $\redemption$ is to calculate
the product of the aggregate redemption frequency $p$
(Table \ref{tab:inflated14-p} on page \pageref{tab:inflated14-p})
and the aggregate severity mean (Table \ref{tab:inflated14-mu-mm} on page
\pageref{tab:inflated14-mu-mm}).} given in Table \ref{tab:historical4a} on
page \pageref{tab:historical4a}. If we consider all investor and fund categories, the mean is equal
to $22$ bps. The largest value is observed for the sovereign/money market
category and is equal to $1.91\%$.

\paragraph{The volatility effect}

\begin{figure}[tbph]
\centering
\caption{Volatility of the redemption rate $\redemption$ in \% ($n = 10$)}
\label{fig:individual5}
\includegraphics[width = \figurewidth, height = \figureheight]{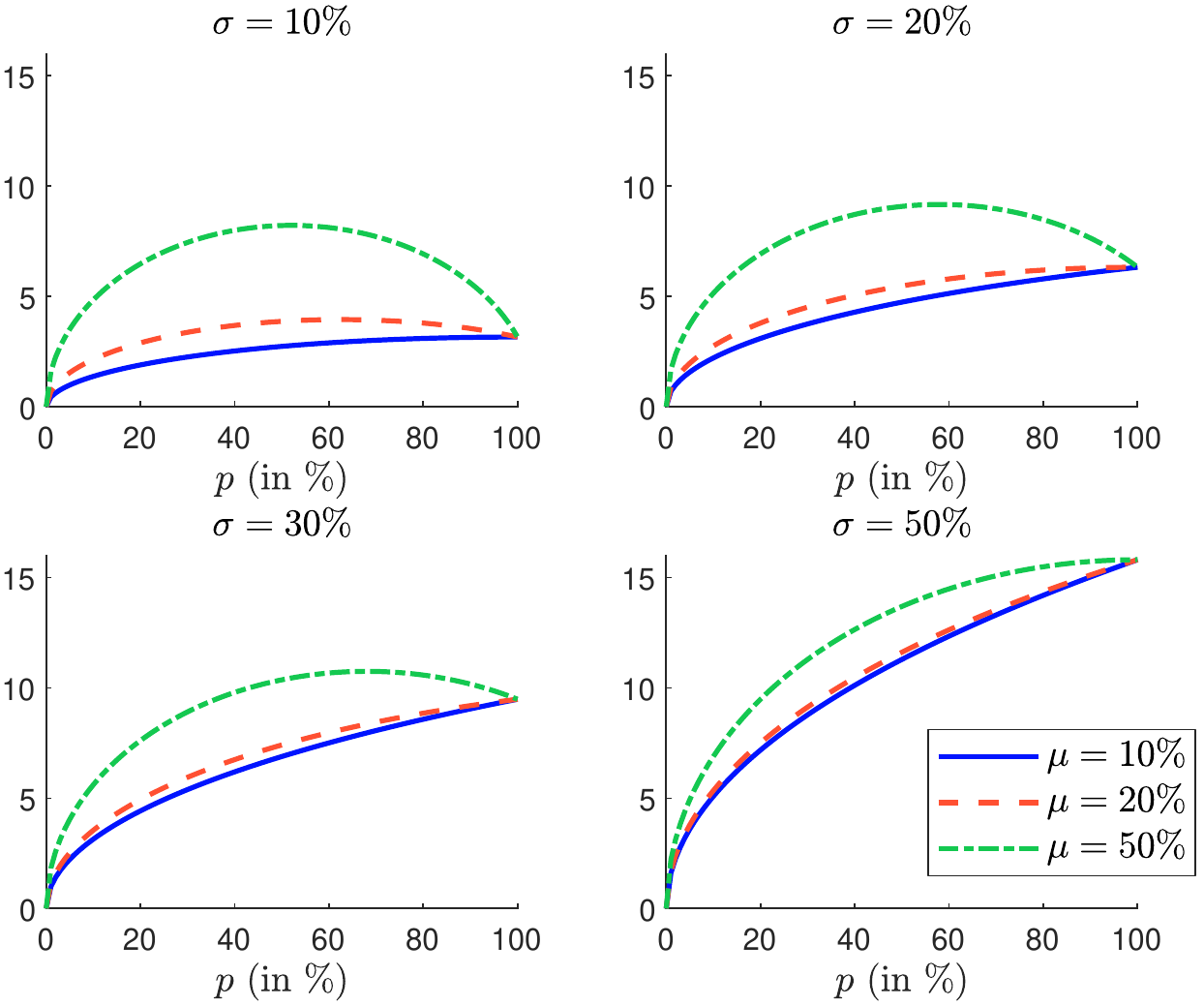}
\end{figure}

\begin{figure}[tbph]
\centering
\caption{Volatility of the redemption rate $\redemption$ in \% ($p = 10\%$, $\mu = 50\%$, $\sigma = 30\%$)}
\label{fig:individual6}
\includegraphics[width = \figurewidth, height = \figureheight]{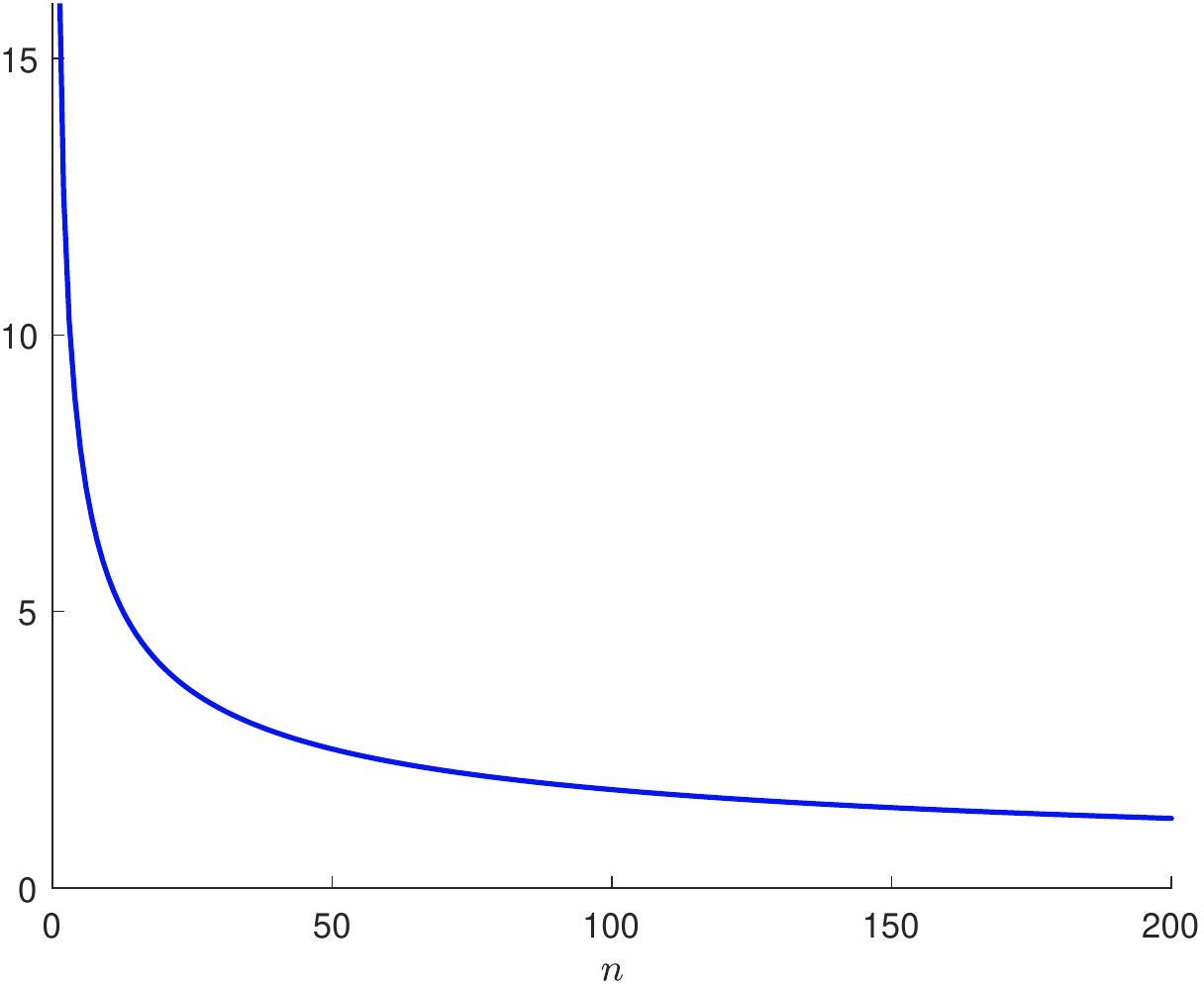}
\end{figure}

By assuming that the liability weights are equal ($\omega _{i}=1/n$),
the volatility of the redemption rate is equal to:
\begin{equation*}
\sigma ^{2}\left( \redemption\right) =\frac{\tilde{p}\left( \tilde{\sigma}%
^{2}+\left( 1-\tilde{p}\right) \tilde{\mu}^{2}\right) }{n}
\end{equation*}%
Globally, we observe that $\sigma ^{2}\left( \redemption\right) $ is an
increasing function of $\tilde{p}$, $\tilde{\mu}$ and $\tilde{\sigma}$ as
shown in Figure \ref{fig:individual5}. When the redemption probability increases,
we observe a convexity shape because we have:
\begin{equation*}
\sigma ^{2}\left( \redemption\right) =\frac{\tilde{p}\left( \tilde{\sigma}%
^{2}+\tilde{\mu}^{2}\right) }{n}-\frac{\tilde{p}^{2}\tilde{\mu}^{2}}{n}
\end{equation*}%
However, this is not realistic since $\tilde{p}\leq 20\%$ in practice.
Another interesting property is that $\sigma ^{2}\left( \redemption\right) $
tends to zero when the number of investors in the fund increases (Figure \ref{fig:individual6}).
If we compute the volatility of the redemption rate, we obtain the figures given
in Table \ref{tab:individual7} on page \pageref{tab:individual7}.
We observe that $\sigma \left( \redemption\right)
\gg \mathbb{E}\left[ \redemption\right] $, implying that $\redemption$ is a
high-skewed random variable. This challenges the use of the $\mathbb{SD}\left( c\right) $
measure presented on page \pageref{eq:SD-measure}.

\paragraph{Correspondence between zero-inflated and individual-based models}

We notice that the zero-inflated model $\mathcal{ZI}\left( p,\mu ,\sigma
\right) $ is a special case of the individual-based model by considering
only one unitholder. Therefore, it is obvious that the zero-inflated model
can not be seen as an explanatory model. It is a reduced-form model or a
parametric model that can fit the data, but the interpretation of the $p-\mu
-\sigma $ parameterization is not obvious, because $\mathcal{ZI}\left( p,\mu
,\sigma \right) $ is an aggregate population model.\smallskip

In this paragraph, we would like to find the relationships between the
parameters of the zero-inflated model and those of the individual-based
model, such that the two models are statistically equivalent:%
\begin{equation*}
\mathcal{ZI}\left( p,\mu ,\sigma \right) \equiv \mathcal{IM}\left( n,\omega ,%
\tilde{p},\tilde{\mu},\tilde{\sigma}\right)
\end{equation*}%
There are different approaches. A first one is to minimize the
Kolmogorov-Smirnov statistics between $\mathcal{ZI}\left( p,\mu ,\sigma
\right) $ and $\mathcal{IM}\left( n,\omega ,\tilde{p},\tilde{\mu},\tilde{%
\sigma}\right) $. Another approach consists in matching their moments. We
consider the second approach because we obtain analytical formulas, whereas
the solution of the first approach can only be numerical. In Appendix \ref{appendix:matching-zi-im} on
page \pageref{appendix:matching-zi-im}, we show that:%
\begin{equation*}
p=1-\left( 1-\tilde{p}\right) ^{n}
\end{equation*}%
and%
\begin{equation*}
\mu =\frac{\tilde{p}}{1-\left( 1-\tilde{p}\right) ^{n}}\tilde{\mu}
\end{equation*}%
whereas the parameter $\sigma $ satisfies the following relationship:%
\begin{eqnarray*}
\sigma ^{2} &=&\left( \frac{\tilde{p}\mathcal{H}\left( \omega \right) }{%
1-\left( 1-\tilde{p}\right) ^{n}}\right) \tilde{\sigma}^{2}+ \\
&&\left( \frac{\tilde{p}\left( \left( 1-\tilde{p}\right) -\left( 1-\tilde{p}%
\right) ^{n}\right) \mathcal{H}\left( \omega \right) -\tilde{p}^{2}\left( 1-%
\tilde{p}\right) ^{n}\left( 1-\mathcal{H}\left( \omega \right) \right) }{%
\left( 1-\left( 1-\tilde{p}\right) ^{n}\right) ^{2}}\right) \tilde{\mu}^{2}
\end{eqnarray*}%
where $\mathcal{H}\left( \omega \right) =\sum_{i=1}^{n}\omega _{i}^{2}$ is
the Herfindahl index. It is interesting to notice that $p$ is a function of $%
n$ and $\tilde{p}$, $\mu $ is a function of $n$, $\tilde{p}$ and $\tilde{\mu}
$, but $\sigma $ does not only depends on the parameters $n$, $\tilde{p}$, $%
\tilde{\mu}$ and $\tilde{\sigma}$:%
\begin{equation*}
\left\{
\begin{array}{l}
p=\varphi _{1}\left( n,\tilde{p}\right)  \\
\mu =\varphi _{2}\left( n,\tilde{p},\tilde{\mu}\right)  \\
\sigma =\varphi _{3}\left( n,\tilde{p},\tilde{\mu},\tilde{\sigma},\mathcal{H}%
\left( \omega \right) \right)
\end{array}%
\right.
\end{equation*}%
Indeed, the aggregate severity volatility also depends on the Herfindahl
index of the fund liability structure.

\begin{remark}
The previous relationships can be inverted in order to define the parameters
of the individual-based model with respect to the parameters of the
zero-inflated model:%
\begin{equation*}
\left\{
\begin{array}{l}
\tilde{p}=\varphi _{1}^{\prime }\left( p;n\right)  \\
\tilde{\mu}=\varphi _{2}^{\prime }\left( p,\mu ;n\right)  \\
\tilde{\sigma}=\varphi _{3}^{\prime }\left( p,\mu ,\sigma ;n,\mathcal{H}%
\left( \omega \right) \right)
\end{array}%
\right.
\end{equation*}%
However, we notice that there are two degrees of freedom -- $n$ and $\mathcal{%
H}\left( \omega \right) $ -- that must be fixed.
\end{remark}

In Tables \ref{tab:individual8a} and \ref{tab:individual8b}, we report
some examples of calibration when $n$ is equal to $10$ and $\omega_i$ is equal to $10\%$.
For instance, if the parameters of the individual-based model are
$\tilde{p} = 1.00\%$, $\tilde{\mu} = 50\%$ and $\tilde{\sigma} = 10\%$,
we obtain $p = 9.56\%$, $\mu = 5.23\%$ and $\sigma = 1.48\%$ for the zero-inflated model.
If we know the weights of the investors in the investment fund, we can therefore calibrate the
zero-inflated model from the individual-based model (Table \ref{tab:individual8a}),
but also the individual-based model from the zero-inflated model (Table \ref{tab:individual8b}).

\begin{table}[tbph]
\centering
\caption{Calibration of the zero-inflated model from the individual-based model}
\label{tab:individual8a}
\begin{tabular}{c|ccc|ccc}
\hline
& & & & & & \\[-2ex]
Parameter & \multicolumn{3}{c|}{$\mathcal{IM}\left(n,\tilde{p},\tilde{\mu},\tilde{\sigma}\right)$}
          & \multicolumn{3}{c}{$\mathcal{ZI}\left( p,\mu ,\sigma \right)$} \\
set & $\tilde{p}$ & $\tilde{\mu}$ & $\tilde{\sigma}$ & $p$ & $\mu$ & $\sigma$ \\[0.75ex] \hline
\#1 & $0.20\%$ & $50.00\%$ & $10.00\%$ & $1.98\%$ & $5.05\%$ & $1.11\%$ \\
\#2 & $1.00\%$ & $50.00\%$ & $10.00\%$ & $9.56\%$ & $5.23\%$ & $1.48\%$ \\
\#3 & $1.00\%$ & $30.00\%$ & $20.00\%$ & $9.56\%$ & $3.14\%$ & $2.14\%$ \\
\hline
\end{tabular}
\end{table}

\begin{table}[tbph]
\centering
\caption{Calibration of the individual-based model from the zero-inflated model}
\label{tab:individual8b}
\begin{tabular}{c|ccc|ccc}
\hline
& & & & & & \\[-2ex]
Parameter & \multicolumn{3}{c|}{$\mathcal{ZI}\left( p,\mu ,\sigma \right)$}
          & \multicolumn{3}{c}{$\mathcal{IM}\left(n,\tilde{p},\tilde{\mu},\tilde{\sigma}\right)$} \\
set & $p$ & $\mu$ & $\sigma$ & $\tilde{p}$ & $\tilde{\mu}$ & $\tilde{\sigma}$ \\[0.75ex] \hline
\#1 & ${\TsV}5.00\%$ & $2.00\%$ & ${\TsV}5.00\%$ & $0.51\%$ & $19.55\%$ & $49.34\%$ \\
\#2 &      $10.00\%$ & $2.00\%$ & ${\TsV}5.00\%$ & $1.05\%$ & $19.08\%$ & $48.67\%$ \\
\#3 &      $10.00\%$ & $5.00\%$ &      $10.00\%$ & $1.05\%$ & $47.71\%$ & $97.14\%$ \\
\hline
\end{tabular}
\end{table}

\subsubsection{On the importance of the liability structure}
\label{section:mandates}

We notice that the variance of the redemption rate depends on the Herfindahl
index:
\begin{equation*}
\mathcal{H}\left( \omega \right) =\sum_{i=1}^{n}\omega _{i}^{2}
\end{equation*}%
This implies that the liability structure $\omega $ is an important parameter
to understand the probability distribution of the redemption rate.

\paragraph{The arithmetics of the Herfindahl index}

We know that the Herfindahl index is bounded:
\begin{equation*}
\frac{1}{n}\leq \mathcal{H}\left( \omega \right) \leq 1
\end{equation*}%
$\mathcal{H}\left( \omega \right) $ is equal to one when one
investor holds 100\% of the investment fund ($\exists i:\omega
_{i}=1$), whereas the lower bound is reached for an equally-weighted
liability structure ($\omega _{i}=n^{-1}$). Therefore,
$\mathcal{H}\left( \omega \right) $ is a measure of concentration. A
related statistic is the \textquotedblleft \textit{effective number
of unitholders}\textquotedblright :
\begin{equation*}
\mathcal{N}\left( \omega \right) =\frac{1}{\mathcal{H}\left( \omega \right) }
\end{equation*}%
$\mathcal{N}\left( \omega \right) $ indicates how many equivalent investors
hold the investment fund. For instance, we consider two funds with the
following liability structures $\omega ^{\left( 1\right) }=\left(
25\%,25\%,25\%,25\%\right) $ and $\omega ^{\left( 2\right) }=\left(
42\%,17\%,15\%,13\%,9\%,3\%,1\%\right) $. Since we have $\mathcal{N}\left(
\omega ^{\left( 1\right) }\right) =4$ and $\mathcal{N}\left( \omega ^{\left(
2\right) }\right) =3.94$, we may consider that the first fund is not more
concentrated than the second fund even if the second fund has 7
unitholders.\smallskip

\begin{figure}[tbph]
\centering
\caption{Effective number of unitholders with a geometric liability structure $\omega
_{i}\propto q^{i}$}
\label{fig:herfindahl2}
\includegraphics[width = \figurewidth, height = \figureheight]{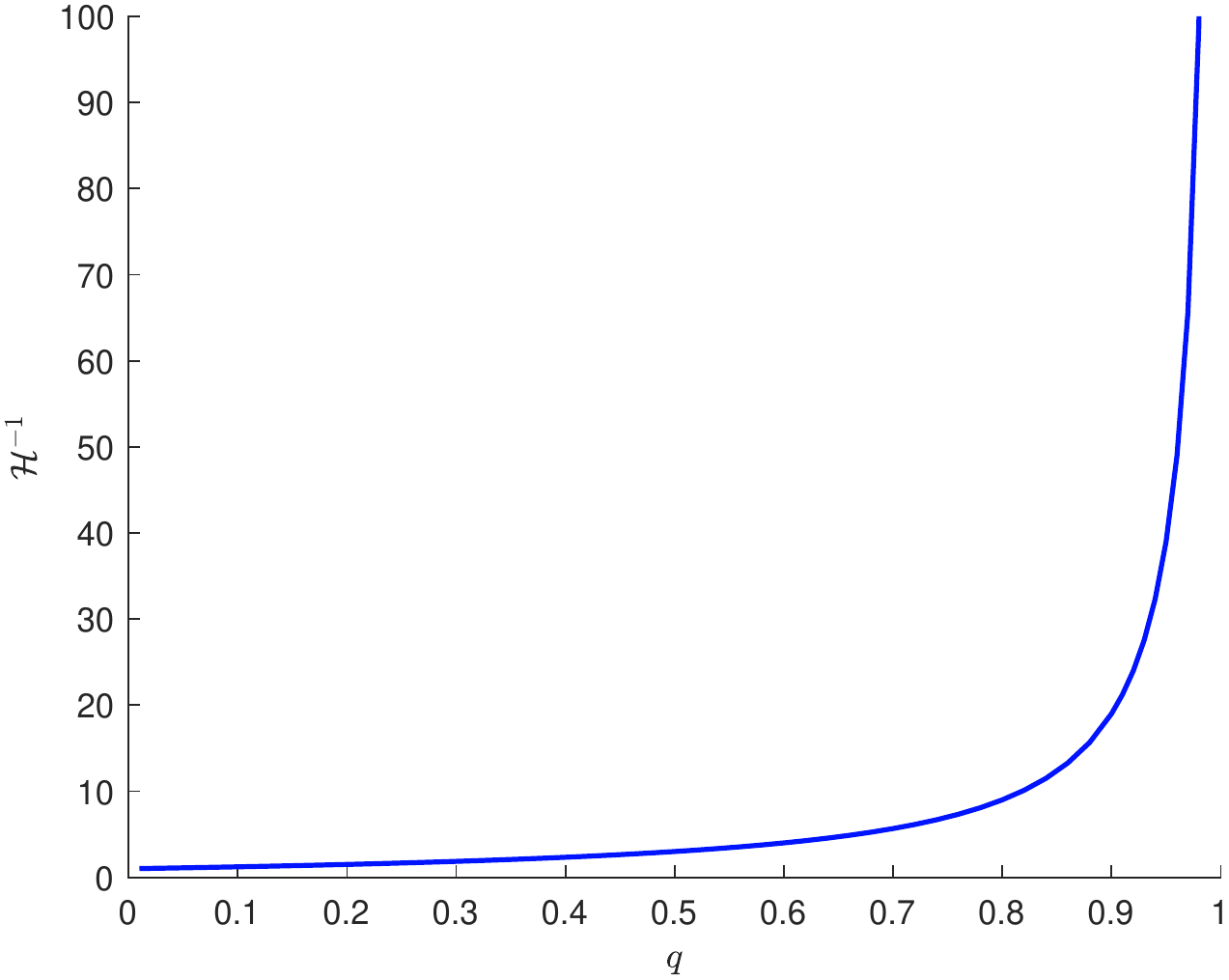}
\end{figure}

We assume that the liability weights follow a geometric series with $\omega
_{i}\propto q^{i}$ and $0<q<1$. We have\footnote{Because we have:
\begin{equation*}
\mathcal{H}\left( \omega \right)
= \frac{\left( 1-q\right) ^{2}}{q^{2}} \sum_{i=1}^{\infty }q^{2i}
= \frac{\left( 1-q\right) ^{2}}{q^{2}}\frac{q^{2}}{\left( 1-q^{2}\right) }
= \frac{\left( 1-q\right) ^{2}}{1-q^{2}}
\end{equation*}%
}:%
\begin{equation*}
\mathcal{N}\left( \omega \right) =\frac{1-q^{2}}{\left( 1-q\right) ^{2}}
\end{equation*}%
As shown in Figure \ref{fig:herfindahl2}, we have an infinite number of
unitholders, but a finite number of effective unitholders. For example, if
$q\leq 0.98$, then $\mathcal{N}\left( \omega \right) <100$.

\paragraph{Approximation of the probability distribution $\mathbf{\tilde{F}}\left( x\mid \omega \right)$}

We recall that the unconditional probability distribution of the redemption
rate is given by $\mathbf{F}\left( x\right) =\Pr \left\{ \redemption\leq
x\right\} $. Since the redemption rate depends on the liability structure
$\omega $ in the individual-based model $\mathcal{IM}\left( n,\omega
,\tilde{p},\tilde{\mu},\tilde{\sigma}\right) $, we note
$\mathbf{\tilde{F}}\left( x\mid \omega \right) $ the associated probability
distribution:
\begin{equation*}
\mathbf{\tilde{F}}\left( x\mid \omega \right) =\Pr \left\{
\sum_{i=1}^{n}\omega _{i}\cdot \mathcal{E}_{i}\cdot \redemption_{i}^{\star
}\leq x\right\}
\end{equation*}%
We now consider the model $\mathcal{IM}\left( \mathcal{N}\left( \omega
\right) ,\tilde{p},\tilde{\mu},\tilde{\sigma}\right) $ and define
$\mathbf{\tilde{F}}\left( x\mid \mathcal{H}\right) $ as follows:
\begin{eqnarray*}
\mathbf{\tilde{F}}\left( x\mid \mathcal{H}\right)  &=&\Pr \left\{ \frac{1}{%
\mathcal{N}\left( \omega \right) }\sum_{i=1}^{\mathcal{N}\left( \omega
\right) }\mathcal{E}_{i}\cdot \redemption_{i}^{\star }\leq x\right\}  \\
&=&\Pr \left\{ \mathcal{H}\left( \omega \right) \sum_{i=1}^{\mathcal{H}%
\left( \omega \right) ^{-1}}\mathcal{E}_{i}\cdot \redemption_{i}^{\star
}\leq x\right\}
\end{eqnarray*}%
The issue is to know under which conditions we can approximate
$\mathbf{\tilde{F}}\left( x\mid \omega \right) $ by $\mathbf{\tilde{F}}\left(
x\mid \mathcal{H}\right) $.\smallskip

\begin{figure}[tbph]
\centering
\caption{Comparison of $\mathbf{\tilde{F}}\left( x\mid \omega \right) $ and $\mathbf{\tilde{F}}\left( x\mid
\mathcal{H}\right) $ ($q = 0.95$ and $\mathcal{H}\left( \omega \right)^{-1} = 38$)}
\label{fig:herfindahl4}
\includegraphics[width = \figurewidth, height = \figureheight]{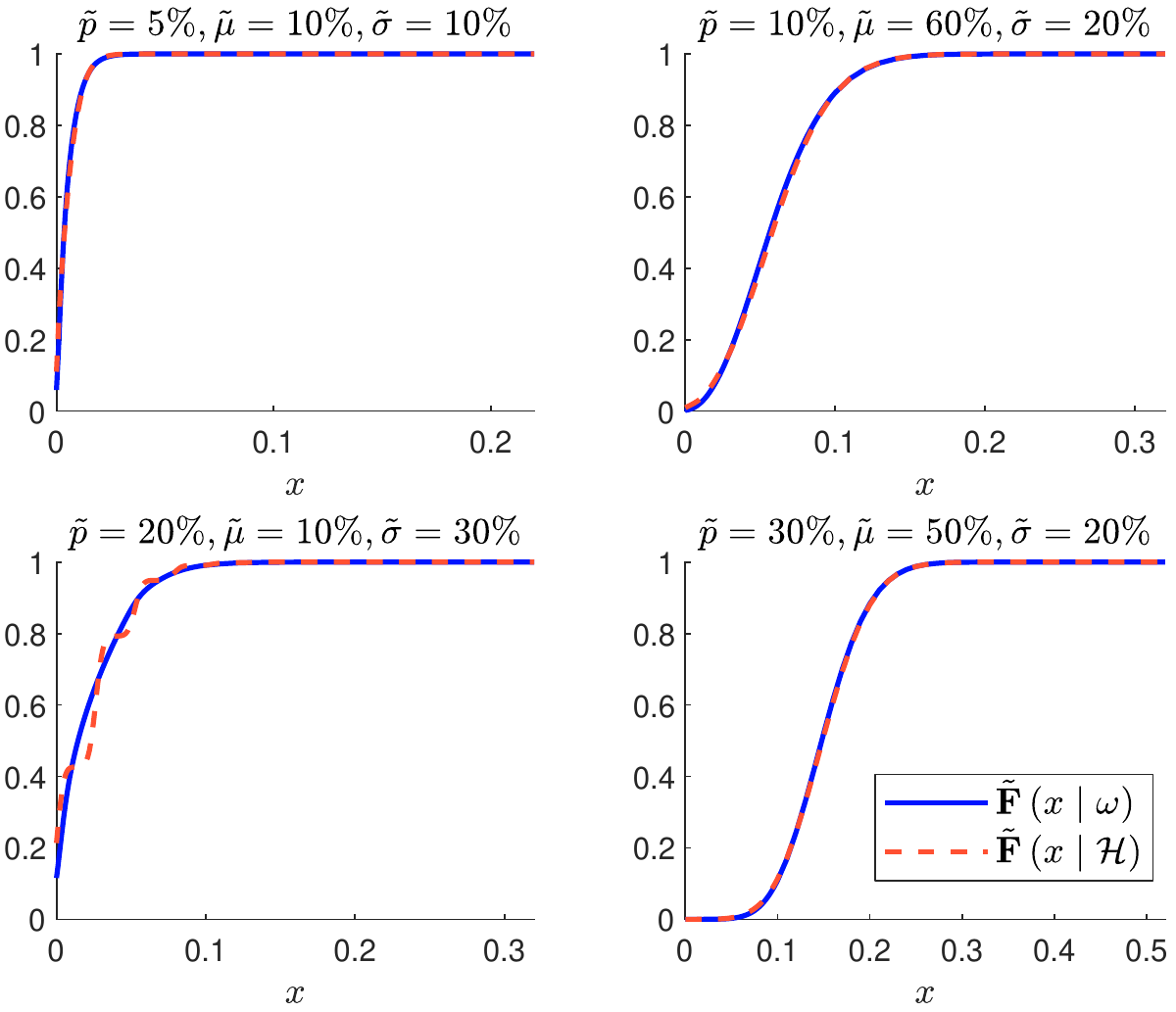}
\end{figure}

Let us consider some Monte Carlo experimentations. We assume that the
liability weights are geometric distributed: $\omega _{i}\propto q^{i}$. In
Figure \ref{fig:herfindahl4}, we compare the two probability distributions
$\mathbf{\tilde{F}}\left( x\mid \omega \right) $ and
$\mathbf{\tilde{F}}\left( x\mid \mathcal{H}\right) $ for several sets of
parameters\footnote{We recall that $\mathbf{\tilde{G}}$ is the beta
distribution by default.} $\left( \tilde{p},\tilde{\mu},\tilde{\sigma}\right)
$. The weights $\omega _{i}$ for $q=0.95$ are given in Figure
\ref{fig:herfindahl1} on page \pageref{fig:herfindahl1}. We notice that the
approximation of $\mathbf{\tilde{F}}\left( x\mid \omega \right) $ by
$\mathbf{\tilde{F}}\left( x\mid \mathcal{H}\right) $ is good and satisfies
the Kolmogorov-Smirnov test at the $99\%$ confidence level. This is not the
case if we assume that $q=0.90$ or $q=0.50$ (see Figures
\ref{fig:herfindahl5} and \ref{fig:herfindahl6} on page
\pageref{fig:herfindahl5}).\smallskip

\begin{figure}[tbph]
\centering
\caption{The case $\mathcal{H}\left( \omega \right)^{-1} = 3$}
\label{fig:herfindahl7}
\includegraphics[width = \figurewidth, height = \figureheight]{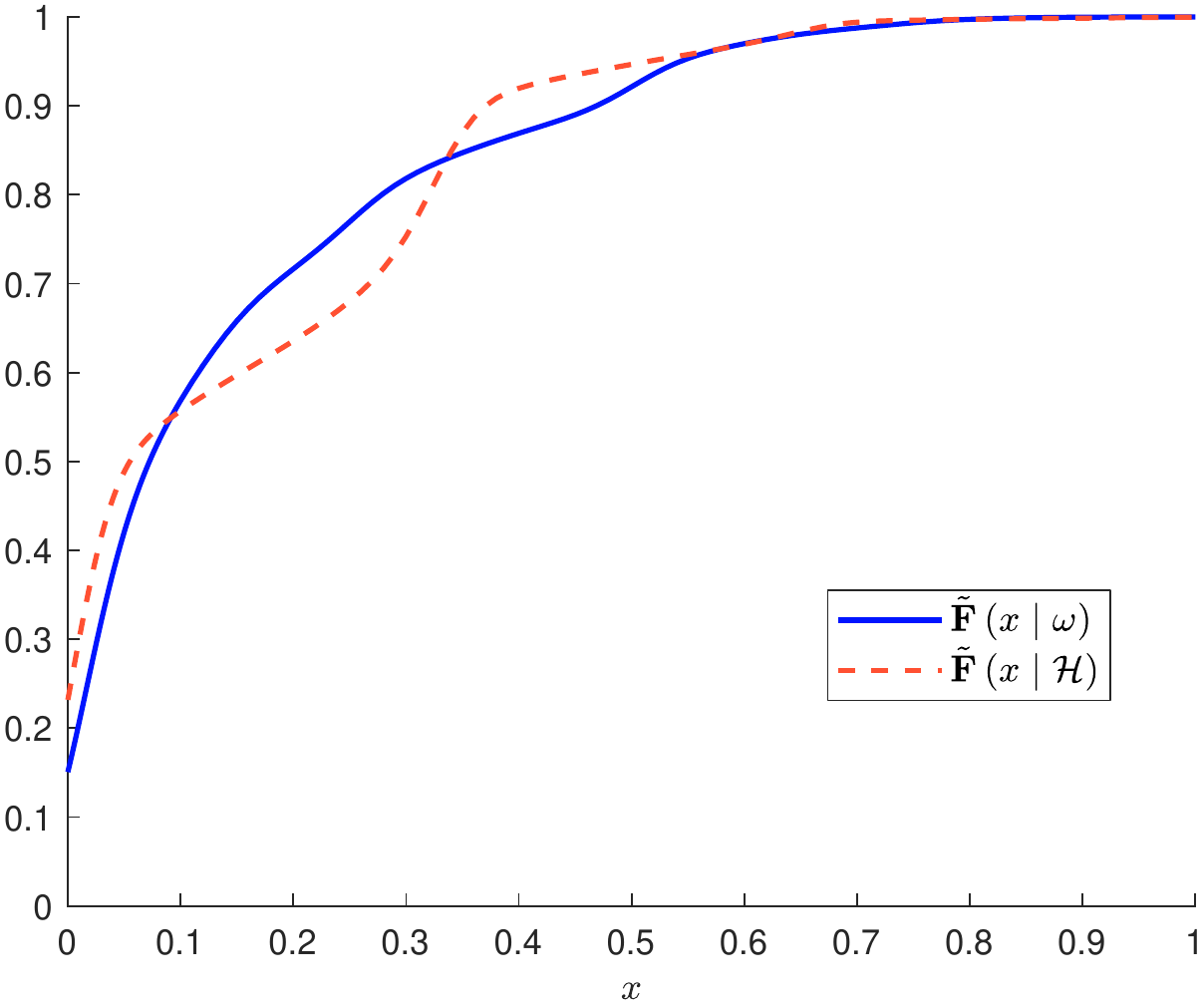}
\end{figure}

\begin{figure}[tbph]
\centering
\caption{The case $\mathcal{H}\left( \omega \right)^{-1} = 18$}
\label{fig:herfindahl8}
\includegraphics[width = \figurewidth, height = \figureheight]{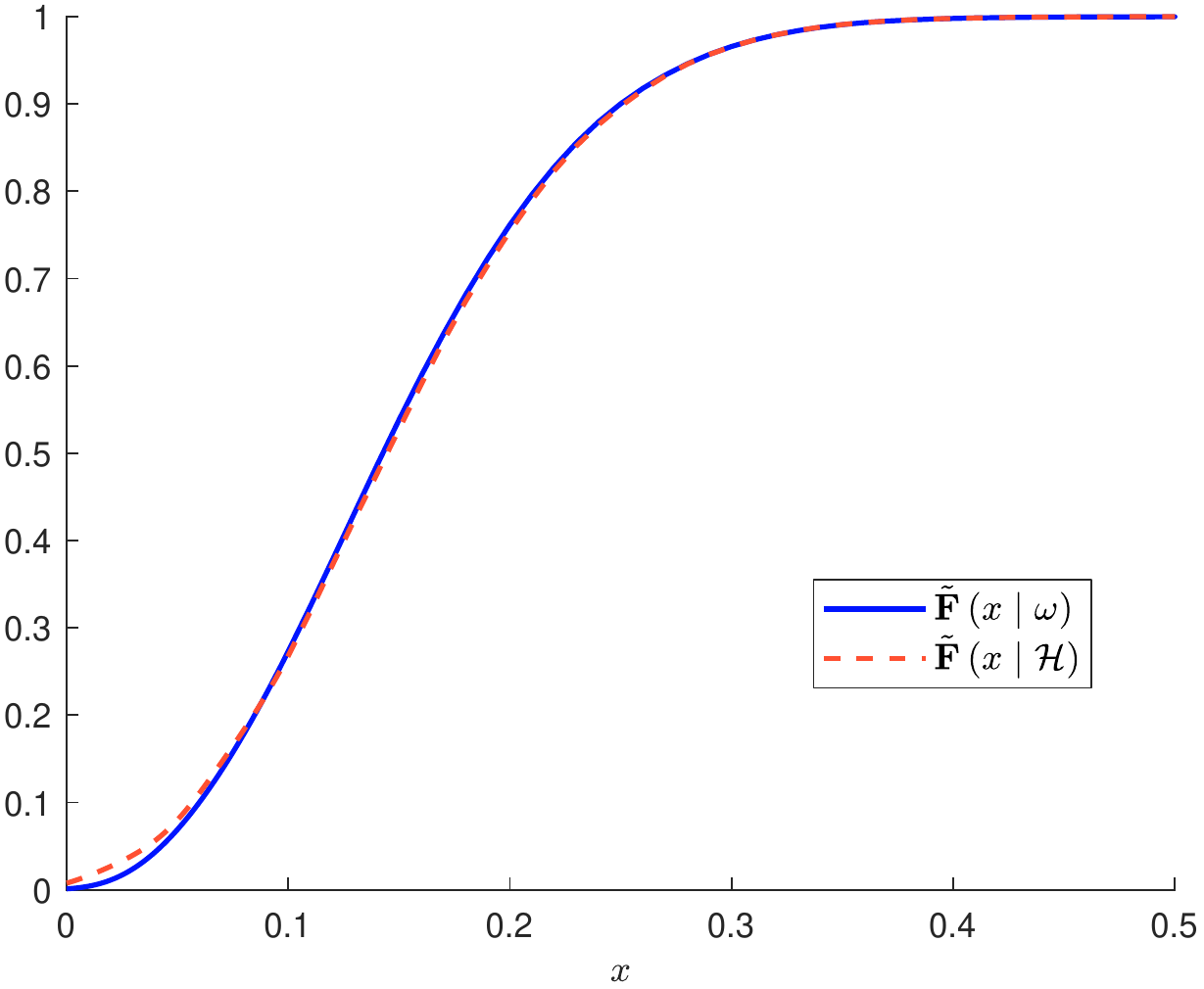}
\end{figure}

To better understand these results, we assume that $\tilde{p}=0.3$,
$\tilde{\mu}=0.5$ and $\tilde{\sigma}=0.4$. When $q$ is equal to $0.50$, the
effective number of unitholders is low and is equal to $3$. In this case,
the probability distribution $\mathbf{\tilde{F}}\left( x\mid \mathcal{H}%
\right) $ is far from the probability distribution $\mathbf{\tilde{F}}\left(
x\mid \omega \right) $ as shown in Figure \ref{fig:herfindahl7}. In fact,
this case corresponds to an investment fund which is highly concentrated. The
risk is then to observe redemptions from the largest unitholders. In
particular, we notice that $\mathbf{\tilde{F}}\left( x\mid \mathcal{H}\right)
$ presents some steps. The reason is that the redemption rate can be
explained by the redemption of one unitholder, the redemption of two
unitholders or the redemption of three unitholders. If we assume that $q$ is
equal to $0.90$, the effective number of unitholders is larger and becomes
$38$. In this case, the probability distribution $\mathbf{\tilde{F}}\left(
x\mid \mathcal{H}\right) $ is close to the probability distribution
$\mathbf{\tilde{F}}\left( x\mid \omega \right) $, because the step effects
disappear (see Figure \ref{fig:herfindahl8}). To summarize, the approximation
of $\mathbf{\tilde{F}}\left( x\mid \omega \right) $ by
$\mathbf{\tilde{F}}\left( x\mid \mathcal{H}\right) $ cannot be good when the
effective number of unitholders (or $\mathcal{H}\left( \omega \right) ^{-1}$)
is low.

\begin{remark}
In many cases, we don't know the comprehensive liability structure $\omega $,
but only the largest weights. In Appendix \ref{appendix:herfindahl} on page
\pageref{appendix:herfindahl}, we derive an upper bound
$\mathcal{H}_{m}^{+}\left( \omega \right) $ of the Herfindahl index
$\mathcal{H}\left( \omega \right) $ from the $m$ largest weights. Therefore,
we can deduce a lower bound of the effective number of unitholders:
\begin{equation*}
\mathcal{N}\left( \omega \right) >\mathcal{N}_{m}^{-}\left( \omega \right) =%
\frac{1}{\mathcal{H}_{m}^{+}\left( \omega \right) }
\end{equation*}
An example is provided in Table \ref{tab:herfindahl10} when we assume that
$\omega _{i}\propto q^{i}$. When the fund is highly concentrated, we obtain a
good approximation of $\mathcal{N}\left( \omega \right) $ with the $10^{th}$
or $20^{th}$ largest unitholders. Otherwise, $\mathcal{N}\left( \omega
\right) $ is underestimated. However, this is not a real issue because we can
think that generated stress scenarios will be overestimated. Indeed, using a
lower value $\mathcal{N}\left( \omega \right) $ increases $\sigma \left(
\redemption\right) $ as shown in Figure \ref{fig:individual6} on page
\pageref{fig:individual6}, implying that the redemption risk is generally
overestimated.
\end{remark}

\begin{table}[tbph]
\centering
\caption{Lower bound $\mathcal{N}_{m}^{-}\left( \omega \right)$ of the effective number of unitholders}
\label{tab:herfindahl10}
\begin{tabular}{ccccccc}
\hline
$m$       & $q = 0.50$ & $q = 0.90$ & $q = 0.95$ & $q = 0.97$ & $q = 0.99$ & $q = 0.995$ \\ \hline
${\TsV}5$ & $3$ & $14$ & $24$ & $37$ & $104$ & $204$ \\
     $10$ & $3$ & $17$ & $28$ & $42$ & $109$ & $209$ \\
     $20$ & $3$ & $19$ & $34$ & $50$ & $119$ & $219$ \\
     $50$ & $3$ & $19$ & $39$ & $63$ & $145$ & $248$ \\
 $\infty$ & $3$ & $19$ & $39$ & $66$ & $199$ & $399$ \\
\hline
\end{tabular}
\end{table}

\paragraph{Stress scenarios based on the largest unitholders}

The previous results show that the main risk in a concentrated fund comes
from the behavior of the largest unitholders. It justifies the use of stress
scenarios based on the order statistics $\omega _{k:n}$:
\begin{equation*}
\min \omega _{i}=\omega _{1:n}\leq \cdots \leq \omega _{k:n}\leq \omega
_{k+1:n}\leq \cdots \leq \omega _{n:n}=\max \omega _{i}
\end{equation*}%
Then, we can define the stress scenario that corresponds to the full
redemption of the $m$ largest unitholders:
\begin{equation*}
\mathbb{S}\left( m\right) =\sum_{k=1}^{m}\omega _{n-k+1:n}
\end{equation*}%
An example is given in Table \ref{tab:herfindahl11} when the liability structure is defined by
$\omega _{i}\propto q^{i}$. Of course, these stress scenarios
$\mathbb{S}\left( m\right) $ make sense only if the fund presents some
liability concentration. Otherwise, they are not informative.

\begin{table}[tbph]
\centering
\caption{Stress scenarios $\mathbb{S}\left( m\right) $ when $\omega _{i}\propto q^{i}$}
\label{tab:herfindahl11}
\begin{tabular}{ccccccc}
\hline
$m$       & $q = 0.50$ & $q = 0.90$ & $q = 0.95$ & $q = 0.97$ & $q = 0.99$ & $q = 0.995$ \\ \hline
${\TsV}1$ & $50.0\%$ & $10.0\%$ & ${\TsV}5.0\%$ & ${\TsV}3.0\%$ & $1.0\%$ & $0.5\%$ \\
${\TsV}2$ & $75.0\%$ & $19.0\%$ & ${\TsV}9.8\%$ & ${\TsV}5.9\%$ & $2.0\%$ & $1.0\%$ \\
${\TsV}5$ & $96.9\%$ & $41.0\%$ &      $22.6\%$ &      $14.1\%$ & $4.9\%$ & $2.5\%$ \\
     $10$ & $99.9\%$ & $65.1\%$ &      $40.1\%$ &      $26.3\%$ & $9.6\%$ & $4.9\%$ \\
\hline
\end{tabular}
\end{table}

\subsubsection{Calibration of stress scenarios}

\paragraph{Using collective and mutual funds}

The calibration of the individual-based model is much more complicated than
the calibration of the zero-inflated model. The reason is that it depends on
the liability structure of the funds, which are not necessarily the same for
the different funds. Let us consider the case of a single fund $f$. We can
estimate the parameters $\tilde{p}$, $\tilde{\mu}$ and $\hat{\sigma}$ using
the quadratic criterion:
\begin{eqnarray}
\left\{ \tilde{p}^{\star },\tilde{\mu}^{\star },\tilde{\sigma}^{\star
}\right\}  &=&\arg \min \varpi _{\tilde{p}}\left( \hat{p}_{\left( f\right)
}-1+(1-\tilde{p})^{\mathcal{H}_{\left( f\right) }^{-1}}\right) ^{2}+\varpi _{%
\tilde{\mu}}\left( \hat{p}_{\left( f\right) }\hat{\mu}_{\left( f\right) }-%
\tilde{p}\tilde{\mu}\right) ^{2}+  \notag \\
&&\varpi _{\tilde{\sigma}}\left( \hat{p}_{\left( f\right) }\left( \hat{\sigma%
}_{\left( f\right) }^{2}+\left( 1-\hat{p}_{\left( f\right) }\right) \hat{\mu}%
_{\left( f\right) }^{2}\right) -\tilde{p}\left( \tilde{\sigma}^{2}+\left( 1-\tilde{p}\right)
\tilde{\mu}^{2}\right) \mathcal{H}_{\left( f\right) }\right) ^{2}
\label{eq:im-calib1}
\end{eqnarray}%
where $\hat{p}_{\left( f\right) }$, $\hat{\mu}_{\left( f\right) }$ and
$\hat{\sigma}_{\left( f\right) }$ are the empirical estimates of the
parameters $p$, $\mu $ and $\sigma $, and $\mathcal{H}_{\left( f\right) }$ is
the Herfindahl index associated with the fund. In practice, the liability
structure changes every day, meaning that the Herfindahl index is
time-varying. Therefore, we can use the average of Herfindahl indices. The
weights $\varpi _{\tilde{p}}$, $\varpi _{\tilde{\mu}}$ and $\varpi
_{\tilde{\sigma}}$ indicate the relative importance of each moment condition.
If we consider several funds, the previous criterion becomes:
\begin{eqnarray}
\left\{ \tilde{p}^{\star },\tilde{\mu}^{\star },\tilde{\sigma}^{\star
}\right\}  &=&\arg \min \varpi _{\tilde{p}}\sum_{f}\varpi _{\left( f\right)
}\left( \hat{p}_{\left( f\right) }-1+(1-\tilde{p})^{\mathcal{H}_{\left(
f\right) }^{-1}}\right) ^{2}+  \notag \\
&&\varpi _{\tilde{\mu}}\sum_{f}\varpi _{\left( f\right) }\left( \hat{p}%
_{\left( f\right) }\hat{\mu}_{\left( f\right) }-\tilde{p}\tilde{\mu}\right)
^{2}+ \notag \\
&&\varpi _{\tilde{\sigma}}\sum_{f}\varpi _{\left( f\right) }\left( \hat{p}%
_{\left( f\right) }\left( \hat{\sigma}_{\left( f\right) }^{2}+\left( 1-\hat{p%
}_{\left( f\right) }\right) \hat{\mu}_{\left( f\right) }^{2}\right) -\tilde{p}\left( \tilde{%
\sigma}^{2}+\left( 1-\tilde{p}\right) \tilde{\mu}^{2}\right) \mathcal{H}%
_{\left( f\right) }\right) ^{2}  \notag \\
&& \label{eq:im-calib2}
\end{eqnarray}%
where $\varpi _{\left( f\right) }$ is the relative weight of the fund
$f$.\smallskip

In practice, the estimates $\tilde{p}^{\star }$, $\tilde{\mu}^{\star }$ and
$\tilde{\sigma}^{\star }$ are very sensitive to the Herfindahl index because
of the first and third moment conditions. To illustrate this point, we
consider the institutional category and we assume that there is only one
fund. On page \pageref{tab:inflated14-p}, we found that $\hat{p}_{\left(
f\right) }=8.23\%$, $\hat{\mu}_{\left( f\right) }=3.23\%$ and
$\hat{\sigma}_{\left( f\right) }=10.86\%$. If $\mathcal{H}_{\left( f\right)
}=5$, we obtain $\tilde{p}^{\star }=1.70\%$, $\tilde{\mu}^{\star }=15.61\%$
and $\tilde{\sigma}^{\star }=53.31\%$. If $\mathcal{H}_{\left( f\right)
}=20$, we obtain $\tilde{p}^{\star }=0.43\%$, $\tilde{\mu}^{\star }=62.04\%$
and $\tilde{\sigma}^{\star }=212.48\%$. In the case of the retail category,
we found that $\hat{p}_{\left( f\right) }=45.61\%$, $\hat{\mu}_{\left(
f\right) }=0.33\%$ and $\hat{\sigma}_{\left( f\right) }=2.88\%$. If
$\mathcal{H}_{\left( f\right) }=1\,000$, we obtain $\tilde{p}^{\star }=0.06\%$,
$\tilde{\mu}^{\star }=247\%$ and $\tilde{\sigma}^{\star }=2\,489\%$. If
$\mathcal{H}_{\left( f\right) }=10\,000$, we obtain $\tilde{p}^{\star
}=0.01\%$, $\tilde{\mu}^{\star }=2\,472\%$ and $\tilde{\sigma}^{\star
}=24\,891\%$. These results are not realistic since $\tilde{\mu}^{\star }>1$
and $\tilde{\sigma}^{\star }>1$.

\paragraph{Using mandates and dedicated funds}

Collective investment and mutual funds are pooled investment vehicles,
meaning that they are held by several investors. We now consider another type
of funds with a single unitholder. They correspond to mandates and funds that
are dedicated to a unique investor. In this case, the Herfindahl index is
equal to one, and the solution of Problem (\ref{eq:im-calib1}) corresponds to
the parameter set of the zero-inflated model:
\begin{equation*}
\left\{ \tilde{p}^{\star }=\hat{p}_{\left( f\right) },\tilde{\mu}^{\star }=%
\hat{\mu}_{\left( f\right) },\tilde{\sigma}^{\star }=\hat{\sigma}_{\left(
f\right) }\right\}
\end{equation*}

\begin{table}[h!]
\centering
\caption{Estimated value of $\tilde{p}$ in \%}
\label{tab:individual10-p}
\begin{tabular}{lcccccccc}
\hline
                        &          (1) &          (2) &          (3) &          (4) &          (5) &          (6) &          (7) &          (8) \\ \hline
Central bank            &  ${\TsV}0.13$ &      $0.21$ & ${\TsV}   $ & ${\TsV}0.73$ & ${\TsV}2.99$ & ${\TsV}   $ & ${\TsV}   $ & ${\TsV}0.49$   \\
Corporate               &  ${\TsV}0.49$ &      $1.14$ & ${\TsV}   $ & ${\TsV}0.13$ &  ${\TsX}   $ &      $0.57$ & ${\TsV}   $ & ${\TsV}0.71$   \\
Corporate pension fund  &  ${\TsV}2.16$ &      $1.40$ & ${\TsV}   $ & ${\TsV}1.60$ & ${\TsV}3.06$ &      $0.41$ &      $0.47$ & ${\TsV}1.57$   \\
Institutional           &  ${\TsV}1.47$ &      $1.35$ &      $0.41$ & ${\TsV}2.13$ & ${\TsV}1.65$ &      $0.40$ &      $0.00$ & ${\TsV}1.46$   \\
Insurance               &  ${\TsV}2.09$ &      $2.12$ & ${\TsV}   $ & ${\TsV}1.52$ & ${\TsV}0.59$ &      $0.13$ & ${\TsV}   $ & ${\TsV}1.93$   \\
Sovereign               &  ${\TsV}0.23$ &      $0.44$ & ${\TsV}   $ & ${\TsV}0.35$ & ${\TsV}0.16$ &      $0.03$ & ${\TsV}   $ & ${\TsV}0.32$   \\
Third-party distributor &       $12.71$ &      $8.07$ &      $3.46$ &      $25.40$ &      $11.68$ &      $7.17$ & ${\TsV}   $ &      $14.22$   \\ \hline
Total                   &  ${\TsV}3.95$ &      $2.63$ &      $1.73$ & ${\TsV}5.82$ & ${\TsV}2.92$ &      $0.68$ &      $7.46$ & ${\TsV}3.34$   \\
\hline
\end{tabular}
\vspace*{1cm}

\centering
\caption{Estimated value of $\tilde{\mu}$ in \%}
\label{tab:individual10-mu}
\begin{tabular}{lcccccccc}
\hline
                        &         (1) &         (2) &          (3) &         (4) &          (5) &         (6) &         (7) &         (8) \\ \hline
Central bank            &  ${\TsV}   $ & ${\TsV}   $ &      $   $ & ${\TsV}   $ & ${\TsV}   $ & ${\TsV}   $ &      $   $ & ${\TsV}   $  \\
Corporate               &  ${\TsV}   $ & ${\TsV}   $ &      $   $ & ${\TsV}   $ & ${\TsV}   $ & ${\TsV}   $ &      $   $ & ${\TsV}   $  \\
Corporate pension fund  &       $4.39$ &      $2.94$ &      $   $ & ${\TsV}   $ & ${\TsV}   $ & ${\TsV}   $ &      $   $ &      $4.11$  \\
Institutional           &       $3.88$ &      $4.05$ &      $   $ &      $3.29$ & ${\TsV}   $ & ${\TsV}   $ &      $   $ &      $4.00$  \\
Insurance               &  ${\TsV}   $ &      $3.46$ &      $   $ & ${\TsV}   $ & ${\TsV}   $ & ${\TsV}   $ &      $   $ &      $4.23$  \\
Sovereign               &  ${\TsV}   $ & ${\TsV}   $ &      $   $ & ${\TsV}   $ & ${\TsV}   $ & ${\TsV}   $ &      $   $ & ${\TsV}   $  \\
Third-party distributor &       $0.77$ &      $1.52$ &      $   $ &      $0.44$ & ${\TsV}   $ & ${\TsV}   $ &      $   $ &      $0.77$  \\ \hline
Total                   &       $1.89$ &      $2.47$ &      $   $ &      $1.48$ &      $2.64$ &      $3.91$ &      $   $ &      $2.13$  \\
\hline
\end{tabular}
\vspace*{1cm}

\centering
\caption{Estimated value of $\tilde{\sigma}$ in \%}
\label{tab:individual10-sigma}
\begin{tabular}{lcccccccc}
\hline
                        &          (1) &          (2) &          (3) &          (4) &          (5) &          (6) &         (7) &          (8) \\ \hline
Central bank            &   ${\TsX}   $ &  ${\TsX}   $ &      $   $ &  ${\TsX}   $ &  ${\TsX}   $ &  ${\TsX}   $ &      $   $ &  ${\TsX}   $  \\
Corporate               &   ${\TsX}   $ &  ${\TsX}   $ &      $   $ &  ${\TsX}   $ &  ${\TsX}   $ &  ${\TsX}   $ &      $   $ &  ${\TsX}   $  \\
Corporate pension fund  &       $15.37$ &      $10.65$ &      $   $ &  ${\TsX}   $ &  ${\TsX}   $ &  ${\TsX}   $ &      $   $ &      $14.53$  \\
Institutional           &       $16.42$ &      $13.88$ &      $   $ &      $12.30$ &  ${\TsX}   $ &  ${\TsX}   $ &      $   $ &      $14.64$  \\
Insurance               &   ${\TsX}   $ &      $13.01$ &      $   $ &  ${\TsX}   $ &  ${\TsX}   $ &  ${\TsX}   $ &      $   $ &      $14.08$  \\
Sovereign               &   ${\TsX}   $ &  ${\TsX}   $ &      $   $ &  ${\TsX}   $ &  ${\TsX}   $ &  ${\TsX}   $ &      $   $ &  ${\TsX}   $  \\
Third-party distributor &  ${\TsV}5.28$ & ${\TsV}5.84$ &      $   $ & ${\TsV}3.29$ &  ${\TsX}   $ &  ${\TsX}   $ &      $   $ & ${\TsV}4.65$  \\ \hline
Total                   &  ${\TsV}9.61$ &      $10.35$ &      $   $ & ${\TsV}8.29$ &      $10.39$ &      $15.06$ &      $   $ &      $10.27$  \\
\hline
\end{tabular}
\medskip

\begin{flushleft}
\begin{footnotesize}
(1) = balanced, (2) = bond, (3) = enhanced treasury, (4) = equity, (5) = money
market, (6) = other, (7) = structured, (8) = total
\end{footnotesize}
\end{flushleft}
\vspace*{-25pt}
\end{table}

In our database, we can separate the observations between collective and
mutual funds on one side and mandates and dedicated funds on the other side.
In Tables \ref{tab:individual10-p}, \ref{tab:individual10-mu} and
\ref{tab:individual10-sigma}, we have estimated the parameters $\tilde{p}$,
$\tilde{\mu}$ and $\tilde{\sigma}$\ by only considering mandates and
dedicated funds. These results highly differ than those obtained for
collective and mutual funds (Tables \ref{tab:inflated14-p},
\ref{tab:inflated14-mu-mm} and \ref{tab:inflated14-sigma-mm} on page
\pageref{tab:inflated14-p}). First, we can calibrate a smaller number of
cells. Indeed, we recall that the estimates are not calculated if the number
of observations is less than 200. Second, the magnitude of the estimates is
very different. If we consider all fund and investor categories, we obtain
$\tilde{p}=3.34\%$, $\tilde{\mu}=2.13\%$ and $\tilde{\sigma}=10.27\%$,
whereas we have previously found $p=31.11\%$, $\mu =0.72\%$ and $\sigma
=4.55\%$. As expected, we verify that $\tilde{p}\ll p$ and $\tilde{\sigma}\gg
\tilde{\sigma}$ because of the following reasons:
\begin{itemize}
\item the redemption probability is larger in a collective fund than in a
dedicated fund because they are several investors;

\item the volatility of the redemption severity is smaller in a collective
fund than in a dedicated fund because the behavior of the different
investors is averaged, implying that the dispersion of redemption is reduced.
\end{itemize}
Curiously, we do not observe that $\tilde{\mu}\approx \mu $. One explanation
may be that investors in mandates are not the same as investors in
collective funds. Indeed, we may consider that they are more sophisticated
and bigger when they are able to put in place a mandate or a dedicated fund.
For instance, they can be more active.\smallskip

The results obtained with data from mandates and dedicated funds are more
realistic than those obtained with data from collective and mutual funds. The
drawback is that they are based on a smaller number of observations and there
are many cells where we don't have enough observations for computing the
estimates. Therefore, these estimates must be completed by expert judgements.

\paragraph{Computing the stress scenarios}

Once we have estimated the parameters $\tilde{p}$, $\tilde{\mu}$ and
$\tilde{\sigma}$, we can compute the stress scenarios using the Monte
Carlo method. Nevertheless, we face an issue here, because the stress scenario is
not unique to an investor category. Indeed, it depends on the liability
structure of the fund. While the individual-based model is more realistic and
relevant than the zero-inflated model, then it appears to be limited from a
practical point of view. Nevertheless, it is useful to understand the
importance of the liability structure on the redemption rate.

\subsection{Correlation risk}

\subsubsection{Specification of the model}

We now consider an extension of the previous model by introducing
correlations between the investors. We obtain the same expression of the
redemption rate:
\begin{equation*}
\redemption=\sum_{i=1}^{n}\omega _{i}\cdot \mathcal{E}_{i}\cdot \redemption_{i}^{\star }
\end{equation*}%
However, the random variables $\left( \mathcal{E}_{1},\ldots
,\mathcal{E}_{n},\redemption_{1}^{\star },\ldots ,\redemption_{n}^{\star
}\right) $ are not necessarily independent. We discuss three different
correlation patterns:
\begin{enumerate}
\item We can assume that $\mathcal{E}_{i}$ and $\mathcal{E}_{j}$ are
    correlated. This is the simplest and most understandable case. Indeed, we
    generally observe long periods with low redemption frequencies followed
    by short periods with high redemption frequencies, in particular when
    there is a crisis or a panic.

\item We can assume that the redemption severities $\redemption_{i}^{\star
    }$ and $\redemption_{j}^{\star }$ are correlated. For example, it would
    mean that high (resp. low) redemptions are observed at the same time.
    Nevertheless, this severity correlation is different from the previous
    frequency correlation. Indeed, the severities are independent from the
    number of redemptions, implying that the severity correlation only
    concerns the unitholders that have already decided to redeem.

\item We can assume that $\mathcal{E}_{i}$ and $\redemption_{i}^{\star }$
    are correlated. We notice that we can write the redemption rate for a
    given category as follows:
\begin{equation*}
\redemption=\sum_{i=1}^{n}\omega _{i}\cdot \redemption_{i}
\end{equation*}%
where $\redemption_{i}=\mathcal{E}_{i}\cdot \redemption_{i}^{\star }$ is
the individual redemption rate for the $i^{\mathrm{th}}$ investor. The
breakdown between the binary variable $\mathcal{E}_{i}$ and the
continuous variable $\redemption_{i}^{\star }$ helps us to model the
\textquotedblleft \textit{clumping-at-zero}\textquotedblright\ pattern. But
there is no reason that the value taken by the redemption severity $%
\redemption_{i}^{\star }$ depends whether $\mathcal{E}_{i}$ takes the value
$0$ or $1$, because $\redemption_{i}^{\star }$ is observed only if
$\mathcal{E}_{i}=1$.
\end{enumerate}
Finally, only the first two correlation patterns are relevant from a
financial point of view, because the third correlation model has no
statistical meaning. Nevertheless, it is obvious that the first correlation model is
more appropriate because the second correlation model confuses low-severity
and high-severity regimes. During a liquidity crisis, both the redemption
frequency and the redemption severity increase \citep{Coval-2007,
Duarte-2013, Kacperczyk-2013, Roncalli-2015a, Schmidt-2016}. The first effect
may be obtained by stressing the parameter $\tilde{p}$ or by considering a
high-frequency regime deduced from the first correlation model, but the
second effect can only be obtained by stressing the parameter $\tilde{\mu}$
and cannot be explained by the second correlation model. Therefore, we only
consider the first correlation pattern by modeling the random vector $\left(
\mathcal{E}_{1},\ldots ,\mathcal{E}_{n}\right) $ with a copula decomposition.
We continue to assume that $\mathcal{E}_{i}\sim \mathcal{B}\left(
\tilde{p}\right) $ are identically distributed, but the dependence function
of $\left( \mathcal{E}_{1},\ldots ,\mathcal{E}_{n}\right) $ is given by the
copula function $\mathbf{C}\left( u_{1},\ldots ,u_{n}\right) $. The
individual-based model is then a special case of this copula-based model when
the copula function is the product copula $\mathbf{C}^{\bot }$.\smallskip

In what follows, we consider the Clayton copula\footnote{We use the notations
of \citet[Chapter 11]{Roncalli-2020}.}:
\begin{equation*}
\mathbf{C}_{\left( \theta _{c}\right) }\left( u_{1},\ldots ,u_{n}\right)
=\left( u_{1}^{-\theta _{c}}+\cdots +u_{n}^{-\theta _{c}}-n+1\right)
^{-1/\theta _{c}}
\end{equation*}%
or the Normal copula:%
\begin{equation*}
\mathbf{C}_{\left( \theta _{c}\right) }\left( u_{1},\ldots ,u_{n}\right)
=\Phi \left( \Phi ^{-1}\left( u_{1}\right) +\cdots +\Phi ^{-1}\left(
u_{n}\right) ;\mathcal{C}_n\left(\theta _{c}\right)\right)
\end{equation*}%
The Clayton parameter satisfies $\theta _{c}\geq 0$ whereas the Normal
parameter $\theta _{c}$ lies in the range $\left[ -1,1\right] $. These two
copula families are very interesting since they are positively ordered with
respect to the concordance stochastic ordering. For the Clayton copula, we
have:%
\begin{equation*}
\mathbf{C}_{\left( 0\right) }=\mathbf{C}^{\bot }\prec \mathbf{C}_{\left(
\theta _{c}\right) }\prec \mathbf{C}^{+}=\mathbf{C}_{\left( \infty \right) }
\end{equation*}%
meaning that the product copula is reached when $\theta _{c}=0$ and the
upper Fr\'echet bound corresponds to the limiting case $\theta _{c}\rightarrow
\infty $. For the Normal copula, we restrict our analysis to $\theta _{c}\in
\left[ 0,1\right] $ because there is no sense to obtain negative
correlations. Therefore, we have:%
\begin{equation*}
\mathbf{C}_{\left( 0\right) }=\mathbf{C}^{\bot }\prec \mathbf{C}_{\left(
\theta _{c}\right) }\prec \mathbf{C}^{+}=\mathbf{C}_{\left( 1\right) }
\end{equation*}
\smallskip

The Normal parameter $\theta _{c}$ is easy to interpret because it
corresponds to the Pearson linear correlation between two Gaussian random
variables. The interpretation of the Clayton copula $\theta _{c}$ is more
tricky. Nevertheless, we can compute the associated Kendall's tau and
Spearman's rho rank correlations\footnote{For the Clayton copula, we calculate an approximation
of the Spearman's rho:
\begin{equation*}
\varrho \approx \dfrac{6}{\pi }\arcsin \left( \frac{1}{2}\sin \left( \dfrac{%
\pi \theta _{c}}{2\theta _{c}+4}\right) \right) \approx \sin \left( \dfrac{%
\pi \theta _{c}}{2\theta _{c}+4}\right)
\end{equation*}%
}. Their expressions are given in Table \ref{tab:copula-rho1}. Therefore, we
can deduce the Pearson correlation $\rho $.

\begin{table}[tbph]
\centering
\caption{Relationship between the copula parameter $\theta _{c}$, the Kendall's tau
$\tau $, the Spearman's rho $\varrho $ and the Pearson correlation $\rho $}
\label{tab:copula-rho1}
\begin{tabular}{lccc}
\hline
& & &  \\[-2ex]
& $\tau $ & $\varrho $ & $\rho $ \\[0.75ex] \hline
& & &  \\[-2ex]
Clayton & $\dfrac{\theta _{c}}{\theta _{c}+2}$ &
$\sin \left( \dfrac{\pi \theta _{c}}{2\theta _{c}+4}\right) $ &
$\sin \left( \dfrac{\pi \theta _{c}}{2\theta _{c}+4}\right) $ \\[0.75ex]
& & &  \\[-2ex]
Normal & $\dfrac{2}{\pi }\arcsin \left( \theta _{c}\right) $ &
$\dfrac{6}{\pi }\arcsin \left( \dfrac{\theta _{c}}{2}\right) $ &
$\theta _{c}$ \\[0.75ex]
\hline
\end{tabular}
\end{table}

The previous formulas can be used to map the copula
parameter space into the Kendall, Spearman or Pearson correlation space. Some
numeric values are given in Table \ref{tab:copula-rho2}. For example, the
Clayton copula $\theta_c = 2$ corresponds to a Kendall's tau of 50\%, a
Spearman's rho of $69\%$ and a Pearson correlation of $70.7\%$. The following
analyses will present the results with respect to the Pearson correlation,
which is the most readable and known parameter.

\begin{table}[tbph]
\centering
\caption{Mapping of the copula parameter space}
\label{tab:copula-rho2}
\begin{tabular}{cccc|cccc}
\hline
\multicolumn{4}{c|}{Clayton copula} & \multicolumn{4}{c}{Normal copula} \\
$\theta_c$ & $\tau $ & $\varrho $ & $\rho $ & $\theta_c$ & $\tau $ & $\varrho $ & $\rho $ \\ \hline
${\TsV}0.00$ & ${\TsV}0.00\%$ & ${\TsV}0.00\%$ & ${\TsV}0.00\%$ & $0.00$ & ${\TsV}0.00\%$ & ${\TsV}0.00\%$ & ${\TsV}0.00\%$ \\
${\TsV}1.00$ &      $33.33\%$ &      $48.26\%$ &      $50.00\%$ & $0.20$ &      $12.82\%$ &      $19.13\%$ &      $20.00\%$ \\
${\TsV}2.00$ &      $50.00\%$ &      $69.02\%$ &      $70.71\%$ & $0.50$ &      $33.33\%$ &      $48.26\%$ &      $50.00\%$ \\
${\TsV}5.00$ &      $71.43\%$ &      $89.25\%$ &      $90.10\%$ & $0.75$ &      $53.99\%$ &      $73.41\%$ &      $75.00\%$ \\
     $10.00$ &      $83.33\%$ &      $96.26\%$ &      $96.59\%$ & $0.90$ &      $71.29\%$ &      $89.15\%$ &      $90.00\%$ \\
     $50.00$ &      $96.15\%$ &      $99.80\%$ &      $99.82\%$ & $0.99$ &      $90.99\%$ &      $98.90\%$ &      $99.00\%$ \\
\hline
\end{tabular}
\end{table}

\begin{remark}
We denote the copula-based model by $\mathcal{CM}\left( n,\omega
,\tilde{p},\tilde{\mu},\tilde{\sigma},\rho \right) $ (or $\mathcal{CM}\left(
n,\tilde{p},\tilde{\mu},\tilde{\sigma},\rho \right) $ when the vector of
weights are equally-weighted). We have the following equivalence:
\begin{equation*}
\mathcal{IM}\left( n,\omega ,\tilde{p},\tilde{\mu},\tilde{\sigma}\right) =
\mathcal{CM}\left( n,\omega ,\tilde{p},\tilde{\mu},\tilde{\sigma},0\right)
\end{equation*}
\end{remark}

\subsubsection{Statistical analysis}

\paragraph{The skewness effect}

\begin{figure}[tbph]
\centering
\caption{Probability to observe no redemption $\Pr \left\{ \redemption=0\right\} $ in \%
with respect to the frequency correlation $\rho$ ($\tilde{p} = 10\%$)}
\label{fig:copula4a}
\includegraphics[width = \figurewidth, height = \figureheight]{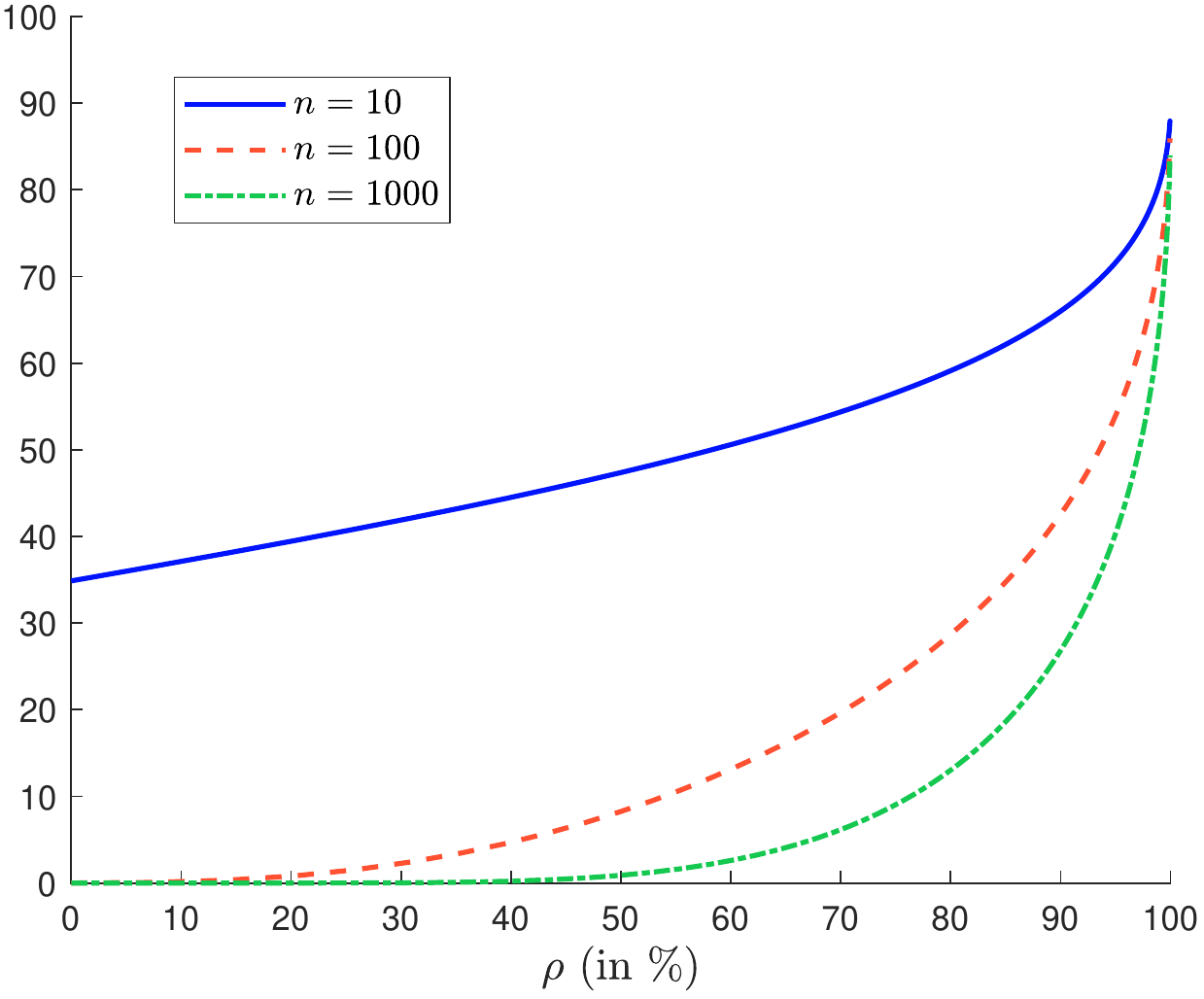}
\end{figure}

In Appendix \ref{appendix:correlation2} on page
\pageref{appendix:correlation2}, we show that:
\begin{equation*}
\Pr \left\{ \redemption=0\right\} =\mathbf{C}_{\left( \theta _{c}\right)
}\left( 1-\tilde{p},\ldots ,1-\tilde{p}\right)
\end{equation*}%
Since $\mathbf{C}^{\bot }\mathbf{\prec C}_{\left( \theta _{c}\right) }%
\mathbf{\prec C}^{+}$, we obtain the following bounds\footnote{%
Because we have:%
\begin{equation*}
\mathbf{C}^{\bot }\left( 1-\tilde{p},\ldots ,1-\tilde{p}\right)
=\prod_{i=1}^{n}\left( 1-\tilde{p}\right) =\left( 1-\tilde{p}\right) ^{n}
\end{equation*}%
and:%
\begin{equation*}
\mathbf{C}^{+}\left( 1-\tilde{p},\ldots ,1-\tilde{p}\right) =\min \left( 1-%
\tilde{p},\ldots ,1-\tilde{p}\right) =1-\tilde{p}
\end{equation*}%
}:%
\begin{equation*}
\left( 1-\tilde{p}\right) ^{n}\leq \Pr \left\{ \redemption=0\right\} \leq 1-%
\tilde{p}
\end{equation*}
We notice that the probability to observe zero redemptions converges to zero
only when the number $n$ of unitholders tends to $\infty $ and the copula is
the product copula. By assuming that the redemption frequency $\tilde{p}$ is
equal to 10\%, we obtain the results given in Figure \ref{fig:copula3} on
page \pageref{fig:copula3} and we verify the previous statistical property.
In Figure \ref{fig:copula4a}, we show the relationship between the Pearson
correlation $\rho$ and the probability $\Pr \left\{ \redemption=0\right\}$.
As expected, it is an increasing function. We notice that the introduction of
the correlation is very important to understand the empirical results we have
calculated in Table \ref{tab:inflated14-p} on page \pageref{tab:inflated14-p}
and some unrealistic values we have obtained in Table \ref{tab:individual2}
on page \pageref{tab:individual2}. For instance, the fact that $\Pr \left\{
\redemption=0\right\}$ is equal to $54.39\%$ for the retail category can only
be explained by a significant frequency correlation since the number $n$ of
unitholders is high for this category.\smallskip

\begin{figure}[tbph]
\centering
\caption{Redemption frequencies in \%
with respect to the frequency correlation $\rho$ ($\tilde{p} = 20\%, n = 1\,000$)}
\label{fig:copula4b}
\includegraphics[width = \figurewidth, height = \figureheight]{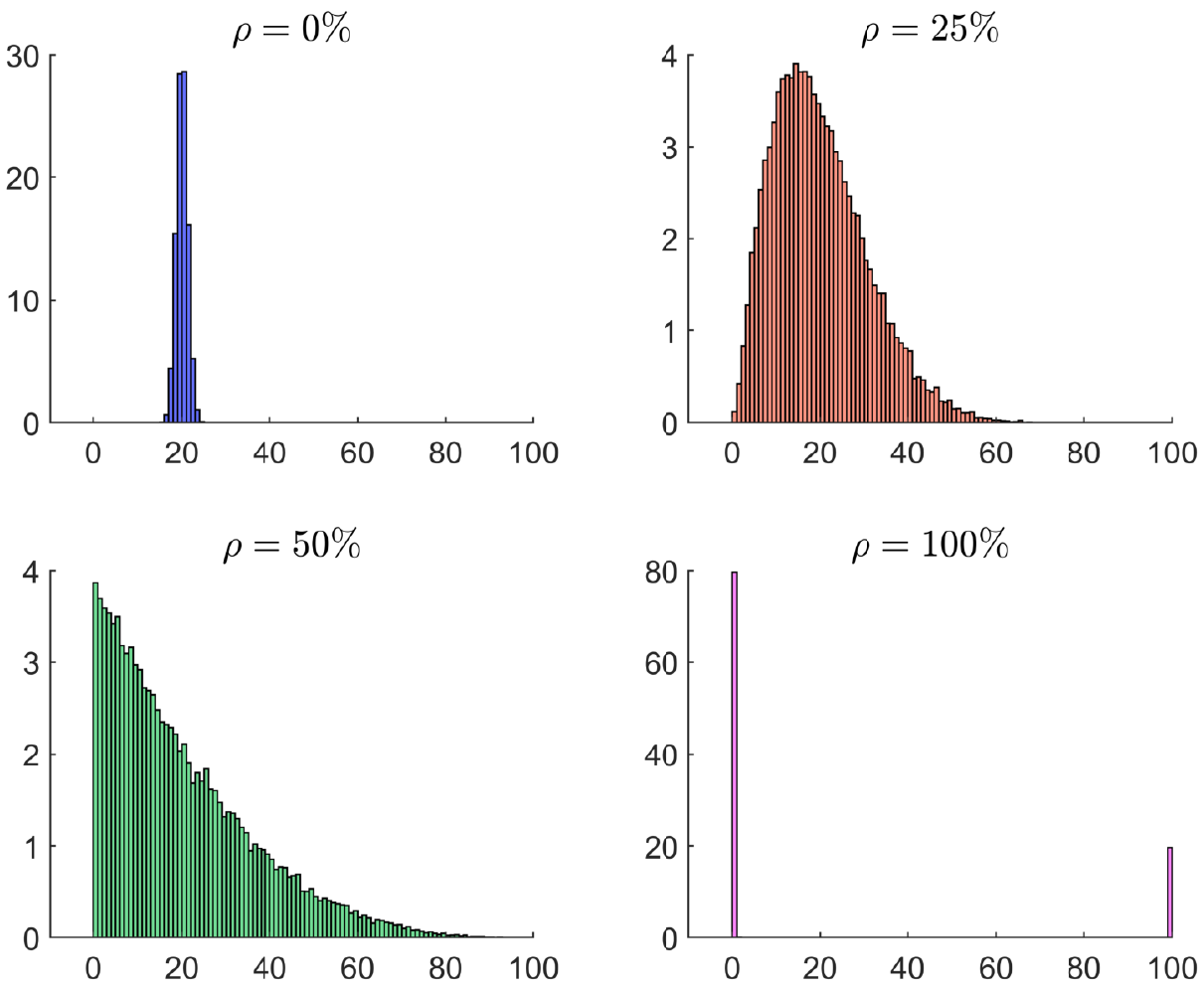}
\end{figure}

By construction, the frequency correlation modifies the probability
distribution of the redemption frequency $\frequency$, which is defined as
the proportion of unitholders that redeem:
\begin{equation*}
\frequency=\frac{1}{n}\sum_{i=1}^{n}\mathds{1}\left\{
\mathcal{E}_{i}=1\right\}
\end{equation*}%
$\frequency$ is a random variable, whose range is between $0$ and $1$.
$\frequency$ depends on the frequency parameter $\tilde{p}$, the number $n$
of unitholders and the copula function $\mathbf{C}_{\left( \theta _{c}\right)
}$ (or the Pearson correlation $\rho $). When $\mathbf{C}_{\left( \theta
_{c}\right) }$ is the product copula $\mathbf{C}^{\bot }$, the redemption
events are independent and we obtain:
\begin{equation*}
\frequency\sim \frac{\mathcal{B}\left( n,\tilde{p}\right) }{n}
\end{equation*}%
because the sum of independent Bernoulli random variables is a binomial
random variable. Therefore, we obtain the following approximation when $n$ is
sufficiently large:
\begin{eqnarray*}
\frac{\mathcal{B}\left( n,\tilde{p}\right) }{n} &\approx &\frac{\mathcal{N}%
\left( n\tilde{p},n\tilde{p}\left( 1-\tilde{p}\right) \right) }{n} \\
&=&\mathcal{N}\left( \tilde{p},\frac{\tilde{p}\left( 1-\tilde{p}\right) }{n}%
\right)
\end{eqnarray*}%
When the copula $\mathbf{C}_{\left( \theta _{c}\right) }$ corresponds to the
upper Fr\'echet bound $\mathbf{C}^{+}$, the redemption frequency follows the
Bernoulli distribution and does not depend on the number of unitholders:
\begin{equation*}
\frequency\sim \mathcal{B}\left( \tilde{p}\right)
\end{equation*}%
We have represented these two extreme cases in Figure \ref{fig:copula4b} when
$\tilde{p}=20\%$ and $n=1\,000$. We have also reported the probability
distribution of $\frequency$ when the Pearson correlation of the copula
function is equal to $25\%$ and $50\%$. We notice that the skewness risk
increases with the frequency correlation. Therefore, the parameter $\rho$
will have a high impact on the stress testing results. In particular, when
the frequency correlation is high, the risk is to observe a large proportion
of redemptions even if the number of unitholders is large. In this case, the
diversification effect across unitholders is limited. An illustration is
provided in Figure \ref{fig:copula4c} on page \pageref{fig:copula4c} that
shows the probability to observe $100\%$ of redemptions\footnote{It
corresponds to the statistic $\Pr \left\{ \frequency=1\right\} $.} when $n$
is set to $20$.

\paragraph{The mean effect}

In Appendix \ref{appendix:correlation4} on page
\pageref{appendix:correlation4}, we show that the frequency correlation has
no impact on the average redemption rate since we obtain the same expression
as previously:
\begin{equation*}
\mathbb{E}\left[ \redemption\right] =\tilde{p}\tilde{\mu}
\end{equation*}%
Therefore, the redemption frequency changes the shape of the probability
distribution of $\redemption$, but not its mean.

\paragraph{The volatility effect}

The volatility of the redemption rate is equal to:
\begin{equation*}
\sigma ^{2}\left( \redemption\right) =\left( \tilde{p}\tilde{\sigma}%
^{2}+\left( \tilde{p}-\mathbf{\breve{C}}_{\left( \theta _{c}\right) }\left(
\tilde{p},\tilde{p}\right) \right) \tilde{\mu}^{2}\right) \mathcal{H}\left(
\omega \right) +\left( \mathbf{\breve{C}}_{\left( \theta _{c}\right) }\left(
\tilde{p},\tilde{p}\right) -\tilde{p}^{2}\right) \tilde{\mu}^{2}
\end{equation*}%
where $\mathbf{\breve{C}}_{\left( \theta _{c}\right) }$ is the survival
copula associated to $\mathbf{C}_{\left( \theta _{c}\right) }$. Since we have
$\mathbf{C}^{\top }\mathbf{\prec C}_{\left( \theta _{c}\right) }\prec
\mathbf{C}^{+}$, we obtain the following inequalities:
\begin{equation*}
\tilde{p}\left( \tilde{\sigma}^{2}+\left( 1-\tilde{p}\right) \tilde{\mu}
^{2}\right) \mathcal{H}\left( \omega \right) \leq \sigma ^{2}\left(
\redemption\right) \leq \tilde{p}\tilde{\sigma}^{2}\mathcal{H}\left( \omega
\right) +\tilde{p}\left( 1-\tilde{p}\right) \tilde{\mu}^{2}
\end{equation*}%
If we consider the equally-weighted case and assume that $n$ tends to
infinity, we obtain:
\begin{equation*}
0\leq \sigma ^{2}\left( \redemption\right) =\left( \mathbf{\breve{C}}
_{\left( \theta _{c}\right) }\left( \tilde{p},\tilde{p}\right) -\tilde{p}
^{2}\right) \tilde{\mu}^{2}\leq \tilde{p}\left( 1-\tilde{p}\right) \tilde{\mu}^{2}
\end{equation*}%
This implies that the volatility risk is not equal to zero for an
infinitely fine-grained liability structure if the frequency correlation is different from
zero.\smallskip

\begin{figure}[tbph]
\centering
\caption{Volatility of the redemption rate $\redemption$ in \% with respect to the number $n$ of unitholders
($\tilde{p} =10\%, \tilde{\mu} = 50\%, \tilde{\sigma} = 30\%$)}
\label{fig:copula5}
\includegraphics[width = \figurewidth, height = \figureheight]{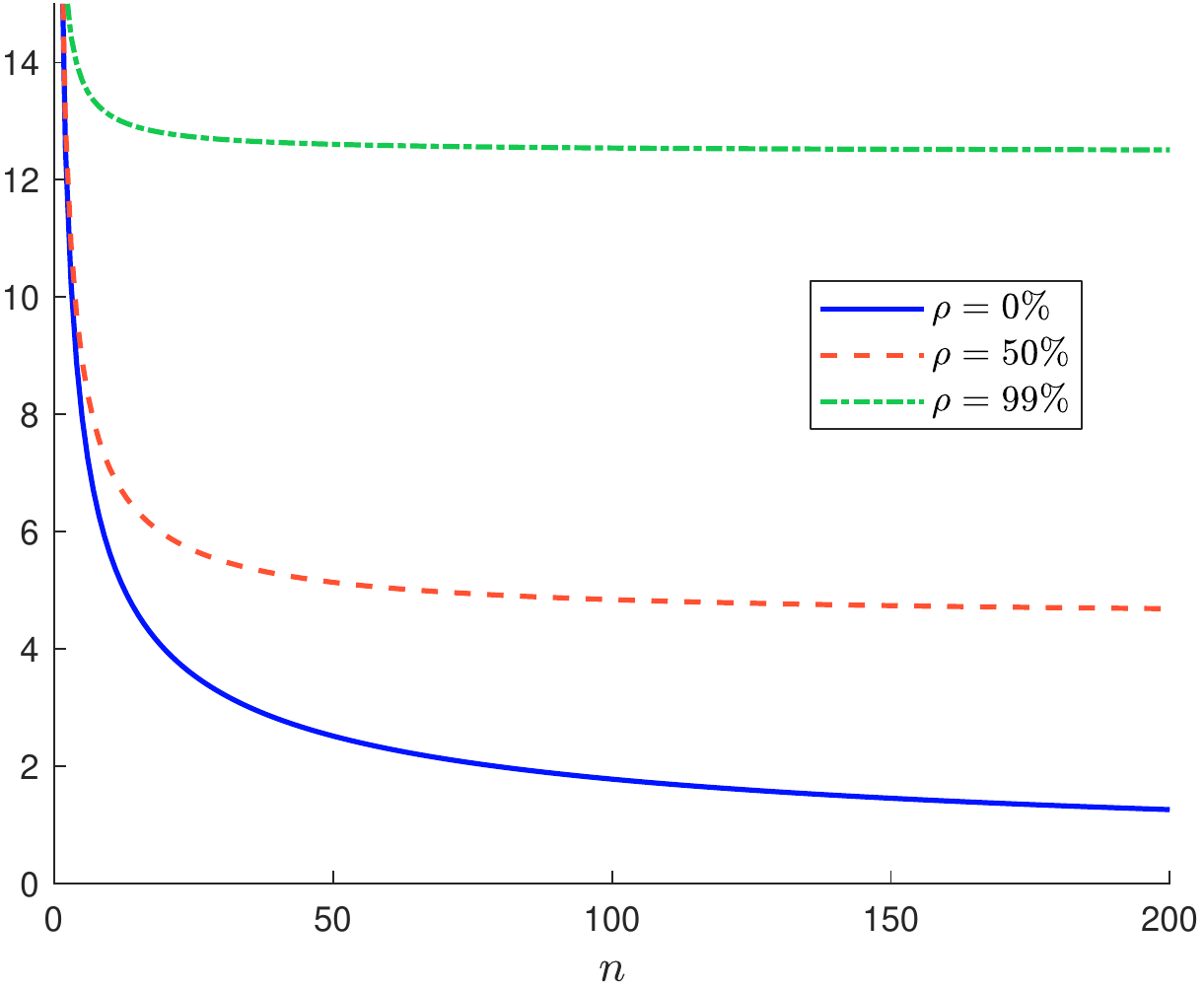}
\end{figure}

\begin{figure}[tbph]
\centering
\caption{Volatility of the redemption rate $\redemption$
in \% with respect to the frequency correlation
($\tilde{p} =10\%, \tilde{\mu} = 50\%, \tilde{\sigma} = 10\%$)}
\label{fig:copula6}
\includegraphics[width = \figurewidth, height = \figureheight]{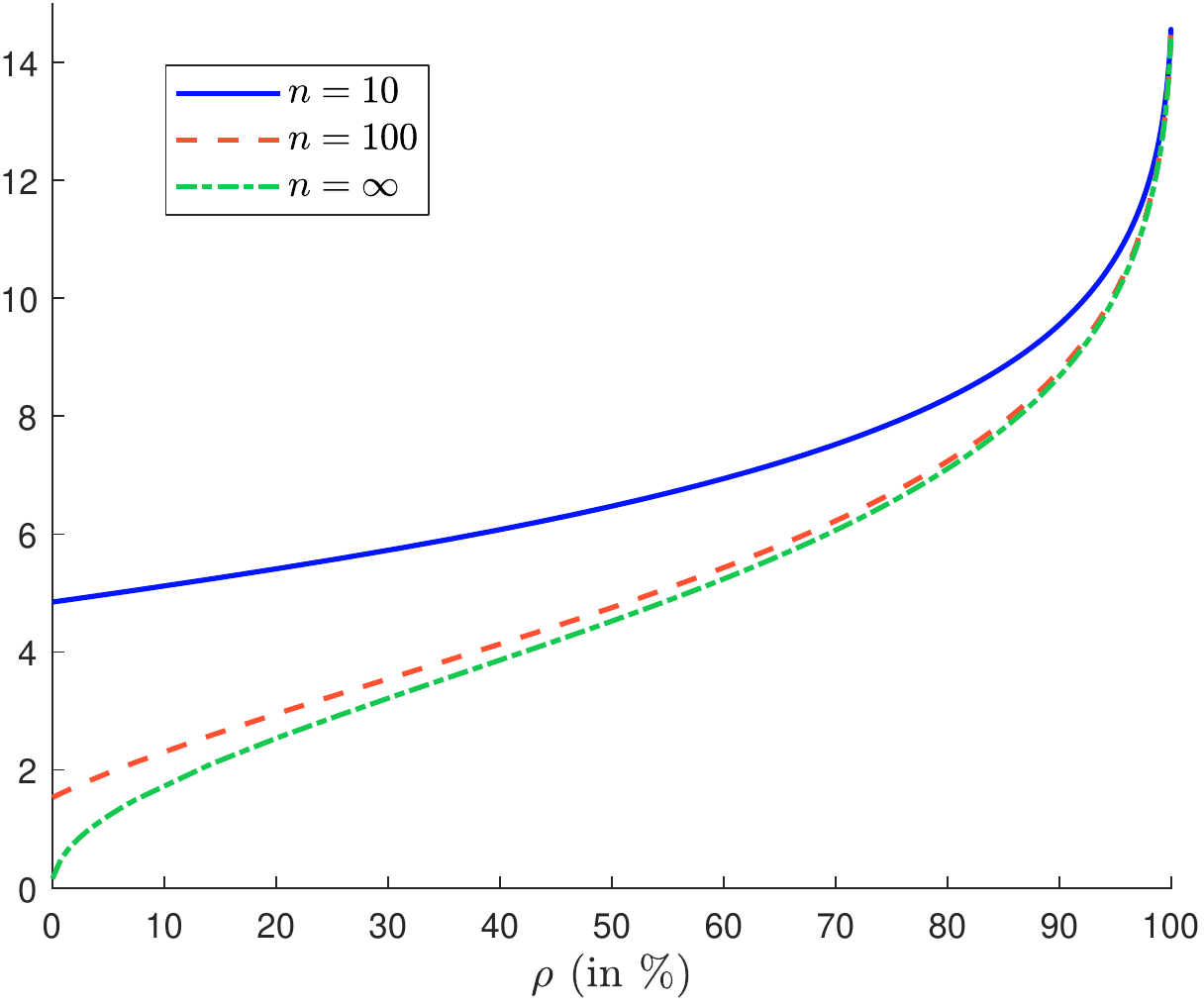}
\end{figure}

The impact of the frequency correlation is illustrated in Figures
\ref{fig:copula5} and \ref{fig:copula6}. We notice that the decrease of the
volatility risk highly depends on the correlation parameter $\rho $. These
figures confirm that the volatility risk is minimum when the frequency
correlation is equal to zero. The consequence is that the frequency
correlation is a key parameter when building stress testing scenarios. This
is perfectly normal since $\rho $ can been seen as a parameter that controls
spillover effects and the magnitude of redemption contagion. All these
results corroborate the previous intuition that the individual-based model
without redemption correlation may be not appropriate for
building a robust stress testing program.

\paragraph{The shape effect}

\begin{figure}[tbph]
\centering
\caption{Histogram of the redemption rate in \% with respect to the number $n$ of unitholders
($\tilde{p} =50\%, \tilde{\mu} = 50\%, \tilde{\sigma} = 10\%, \rho = 50\%$)}
\label{fig:copula7a}
\includegraphics[width = \figurewidth, height = \figureheight]{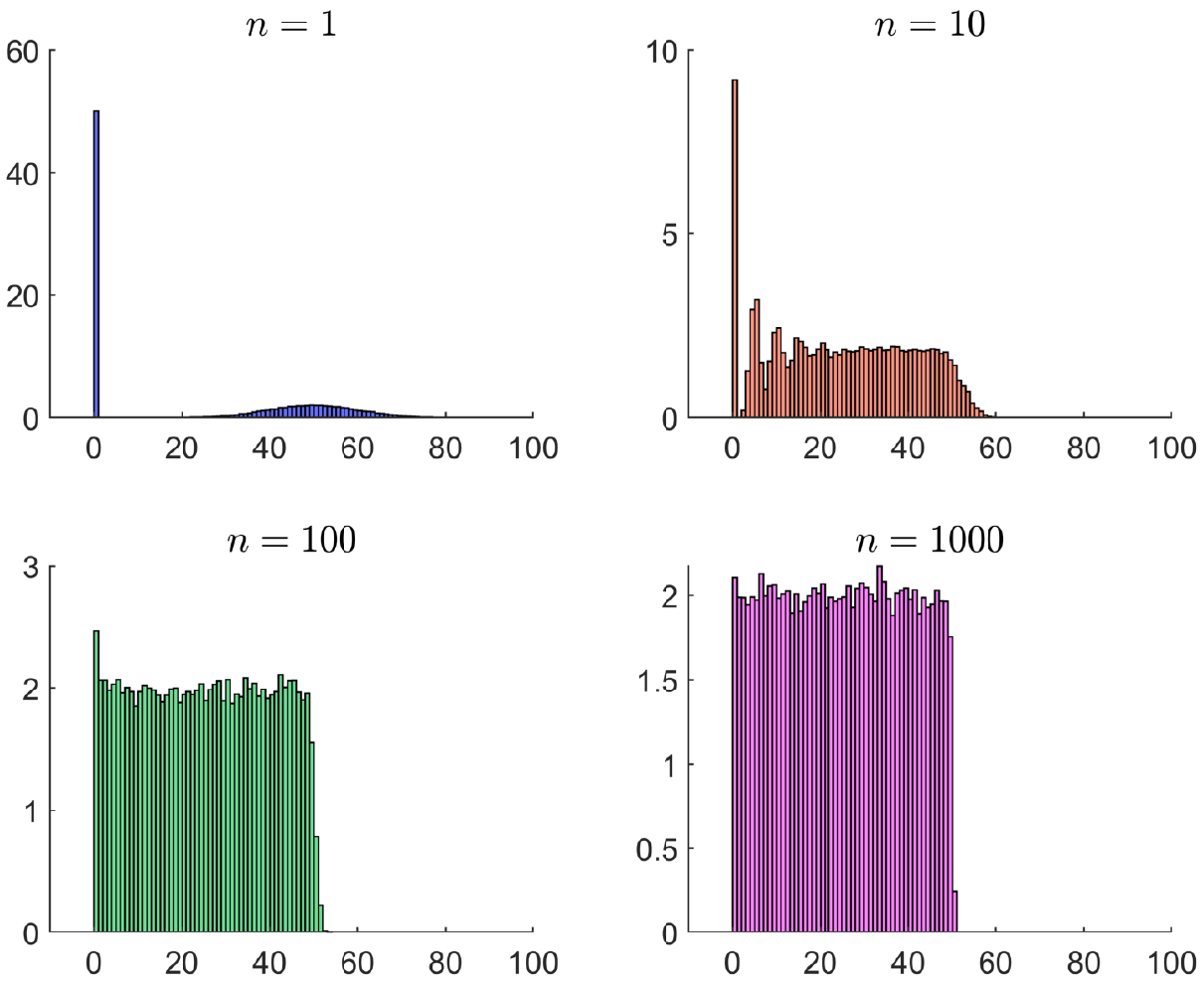}
\end{figure}

\begin{figure}[tbph]
\centering
\caption{Histogram of the redemption rate in \% with respect to the frequency correlation
($\tilde{p} =50\%, \tilde{\mu} = 50\%, \tilde{\sigma} = 10\%, n = 10$)}
\label{fig:copula8}
\includegraphics[width = \figurewidth, height = \figureheight]{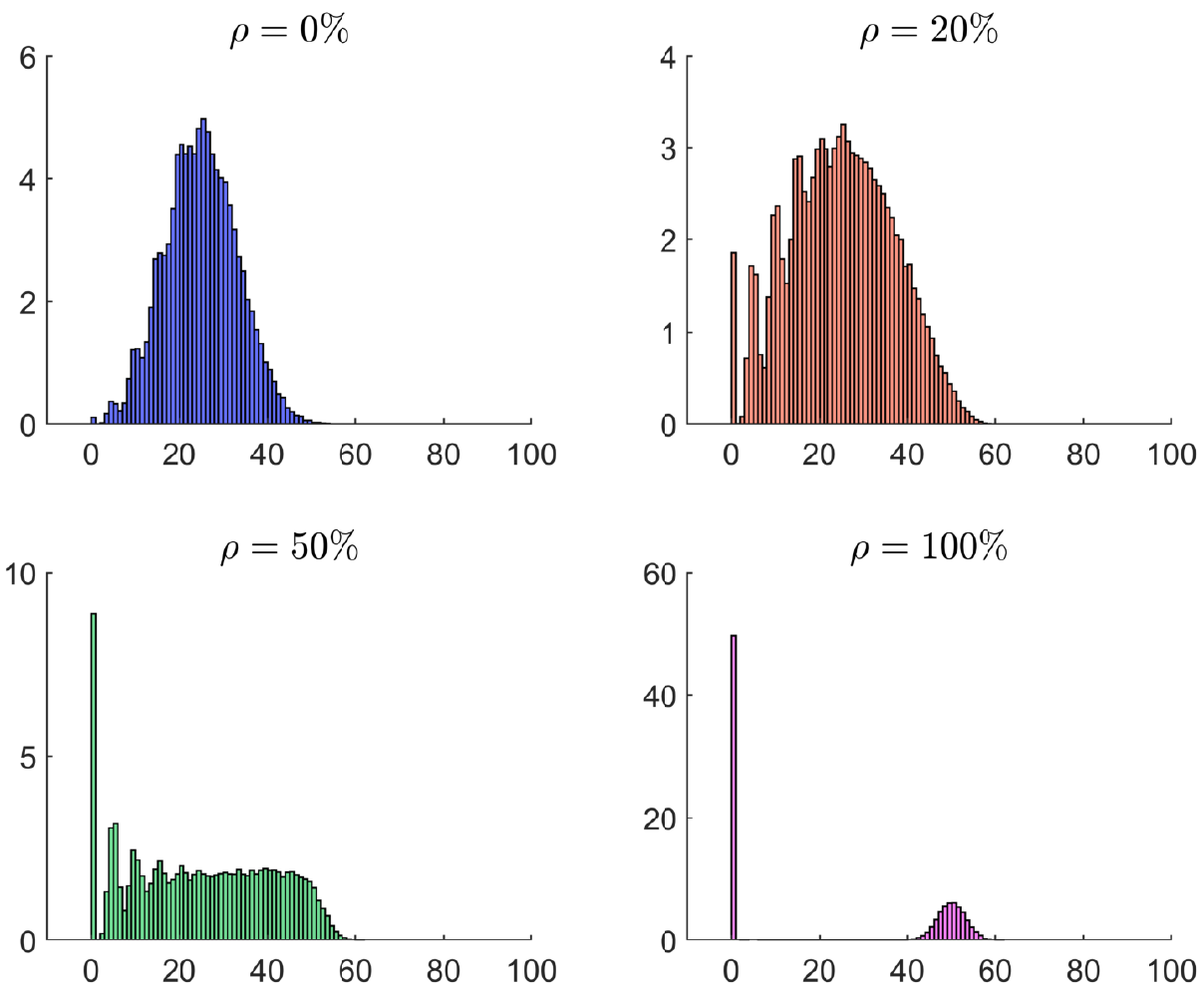}
\end{figure}

The impact of the frequency correlation on the skewness and the
volatility can then change dramatically the shape of the probability
distribution of the redemption rate. In Figure \ref{fig:individual1}
on page \pageref{fig:individual1}, we have already studied the
histogram of the redemption rate in the case $\tilde{p}=50\%$,
$\tilde{\mu} =50\%$ and $\tilde{\sigma}=10\%$. Let us reproduce the
same exercise by assuming that the frequency correlation is equal to
$50\%$. The results are given in Figure \ref{fig:copula7a}. The
shape of the probability distributions is completely different
except in the case of a single unitholder\footnote{Other
illustrations are provided in Appendix
\ref{appendix:additional-results} on page \pageref{fig:copula7b}.
Figures \ref{fig:copula7b}, \ref{fig:copula7c} and
\ref{fig:copula7d} correspond to the cases $\rho = 25\%$, $\rho =
75\%$ and $\rho = 90\%$.}. To better illustrate the impact of the
frequency correlation, we report in Figure \ref{fig:copula8} the
histogram of the redemption rate by fixing $n=10$. In the case of a
perfect correlation of $100\%$ and an equally-weighted liability
structure, we obtain two different cases:
\begin{enumerate}
\item there is zero redemption with a probability $1-\tilde{p}$;

\item there are $n$ redemptions with a probability $\tilde{p}$, and
the redemption severity $\redemption^{\star }$ is the average
of the individual redemption severities:
\begin{equation*}
\redemption^{\star }=\frac{1}{n}\sum_{i=1}^{n}\redemption^{\star }_i
\end{equation*}
\end{enumerate}
It follows that the probability distribution of the redemption rate
is equal to:
\begin{equation*}
\mathbf{F}\left( x\right) =\mathds{1}\left\{ x\geq 0\right\} \cdot \left( 1-%
\tilde{p}\right) +\mathds{1}\left\{ x>0\right\} \cdot \tilde{p}\cdot \mathbf{%
\bar{G}}\left( x\right)
\end{equation*}%
We retrieve the zero-inflated model $\mathcal{ZI}\left(
\tilde{p},\tilde{\mu},n^{-1/2}\tilde{\sigma}\right) $ or the
individual-based model with a single unitholder $\mathcal{IM}\left(
1,\tilde{p},\tilde{\mu},n^{-1/2}\tilde{\sigma}\right) $. The only
difference is the severity distribution $\mathbf{\bar{G}}$, whose
variance is divided by a factor $n$. Spillover and contagion risks
come then from the herd behavior of unitholders. Instead of having
$n$ different investors, we have a unique investor in the fund,
because the decision to redeem by one investor induces the decision
to redeem by all the other remaining investors.

\subsubsection{Evidence of the correlation risk}

\paragraph{Correlation risk within the same investor category}

In order to illustrate that redemption frequencies are correlated, we build
the time series of the frequency rate $\frequency_{t}$ for a given
category\footnote{We can use an equally-weighted scheme $\omega _{i,t}=1/n$.}:
\begin{equation*}
\frequency_{t}=\sum_{i=1}^{n}\omega _{i,t}\cdot \mathds{1}\left\{ \mathcal{E}%
_{i,t}=1\right\} =\sum_{i=1}^{n}\omega _{i,t}\mathcal{E}_{i,t}
\end{equation*}%
where $\mathcal{E}_{i,t}$ is the redemption indicator for the investor $i$
at time $t$. Using the sample $\left( \frequency_{1},\ldots ,\frequency_{T}\right) $,
we compute the empirical mean $\overline{\frequency}$ and the
standard deviation $\hat{\sigma}\left( \frequency\right) $. Then, the copula
parameter $\theta _{c}$ can be calibrated by solving the following nonlinear
equation\footnote{See Equation (\ref{eq:appendix-pearson1}) on page
\pageref{eq:appendix-pearson1}.}:
\begin{equation*}
\mathbf{C}_{\left( \theta _{c}\right) }\left( \overline{\frequency},%
\overline{\frequency}\right) =\frac{\hat{\sigma}^{2}\left( \frequency\right)
-\overline{\frequency}\left( \mathcal{H}\left( \omega \right) -\overline{%
\frequency}\right) }{1-\mathcal{H}\left( \omega \right) }
\end{equation*}%
The copula parameter $\theta _{c}$ can be transformed into the Kendall,
Spearman or Pearson correlation using the standard formulas given in Table
\ref{tab:copula-rho1} on page \pageref{tab:copula-rho1}.
For instance, if $\mathbf{C}_{\left( \theta _{c}\right) }$ is
the Clayton copula, the Pearson correlation is equal to:
\begin{equation*}
\rho =\sin \left( \frac{\pi \theta _{c}}{2\theta _{c}+4}\right)
\end{equation*}
An example is provided in Table \ref{tab:correl2b} when the fund liability structure is equally-weighted
and has 20 unitholders. For instance, if the empirical mean $\overline{\frequency}$
and the standard deviation $\hat{\sigma}\left( \frequency\right)$ are equal to
$25\%$ and $20\%$, the calibrated Pearson correlation is equal to $44.5\%$.

\begin{table}[tbph]
\centering
\caption{Calibrated Pearson correlation (Clayton copula, $\mathcal{H}\left( \omega \right)=1/20$)}
\label{tab:correl2b}
\begin{tabular}{c|ccccc}
\hline
& & & & & \\[-2ex]
\multirow{2}{*}{$\hat{\sigma}\left( \frequency\right)$}
         & \multicolumn{5}{c}{$\overline{\frequency}$} \\[0.75ex]
         & ${\TsV}10.0\%$ & ${\TsV}20.0\%$ &      $25.0\%$ & $30.0\%$ & $40.0\%$ \\
& & &  \\[-2ex]  \hline
$10.0\%$ & ${\TsV}39.1\%$ &  ${\TsX}5.1\%$ & ${\TsV}1.1\%$ &          &          \\
$20.0\%$ & ${\TsV}93.9\%$ & ${\TsV}58.7\%$ &      $44.5\%$ & $34.7\%$ & $23.5\%$ \\
$30.0\%$ &      $100.0\%$ & ${\TsV}91.5\%$ &      $82.3\%$ & $72.8\%$ & $57.7\%$ \\
$40.0\%$ &                &      $100.0\%$ &      $98.7\%$ & $95.6\%$ & $87.4\%$ \\
\hline
\end{tabular}
\end{table}

\begin{remark}
At first sight, calibrating the frequency correlation seems to be an easy task.
However, it is very sensitive to the different parameters $\overline{\frequency}$,
$\hat{\sigma}\left( \frequency\right)$ and $\mathcal{H}\left( \omega \right)$. Moreover,
it depends on the copula specification. For instance, we obtain the results given in Table
\ref{tab:correl2c} on page \pageref{tab:correl2c} when the dependence function
is the Normal copula. We observe that the Pearson correlations
calibrated with the Clayton copula are different from those calibrated
with the Normal copula.
\end{remark}

\begin{remark}
Another way to illustrate the frequency correlation is to split a given
investor category into two subsamples $\mathcal{S}_{1}$ and $\mathcal{S}_{2}$
and calculate the time series of the redemption frequency for the two
subsamples $\mathcal{S}_{k}$ ($k=1,2$):
\begin{equation*}
\frequency_{k,t}=\frac{1}{\sum_{i\in \mathcal{S}_{k}}\omega _{i,t}}%
\sum_{i\in \mathcal{S}_{k}}\omega _{i,t}\mathcal{E}_{i,t}
\end{equation*}%
Then, we can calculate the Pearson correlation
$\rho \left( \frequency_{1},\frequency_{2}\right) $ and
calibrate the associated copula parameter $\theta _{c}$ using Equation
(\ref{eq:appendix-pearson3}) on page \pageref{eq:appendix-pearson3}.
\end{remark}

\paragraph{Correlation risk between investor categories}

\begin{table}[tbph]
\centering
\caption{Intra-class Spearman correlation}
\label{tab:correl2}
\begin{tabular}{llcccc}
\hline
\multirow{2}{*}{Category \#1} & \multirow{2}{*}{Category \#2} & \multirow{2}{*}{Balanced} &
\multirow{2}{*}{Bond} & \multirow{2}{*}{Equity} & Money \\
                         &                         &               &          &          & Market \\ \hline
Retail                   & Third-party distributor &      $53.0\%$ & $52.9\%$ & $52.1\%$ &  ${\TsXIII}$$3.3\%$ \\
Retail                   & Institutional           &      $10.4\%$ & $23.2\%$ & $22.0\%$ &    ${\TsV}$$-6.5\%$ \\
Retail                   & Insurance               & ${\TsV}3.0\%$ & $18.8\%$ & $31.6\%$ &           $-12.3\%$ \\
Third-party distributor  & Institutional           &      $13.5\%$ & $48.0\%$ & $54.1\%$ &   ${\TsVIII}24.0\%$ \\
Third-party distributor  & Insurance               &      $23.1\%$ & $21.5\%$ & $22.8\%$ &   ${\TsVIII}39.2\%$ \\
Institutional            & Insurance               & ${\TsV}2.5\%$ & $16.2\%$ & $16.4\%$ &   ${\TsVIII}29.8\%$ \\ \hline
\multicolumn{2}{c}{Average}                        &      $17.6\%$ & $30.1\%$ & $33.2\%$ &   ${\TsVIII}12.9\%$ \\
\hline
\end{tabular}

\vspace*{-30pt}
\end{table}

The correlation risk is present within a given investor category, but
it may also concern two different investor categories. In order to
distinguish them, we use the classical statistical jargon of inter-class and
intra-class correlations. In Table \ref{tab:correl2}, we report the intra-class Spearman
correlation\footnote{The correlations of retail/insurance and institutional/insurance
for balanced funds and the correlations of retail/third-party distributor and
retail/insurance for money market funds are not significant at the
confidence level of $95\%$.} for four
investor categories (retail, third-party distributor, institutional and
insurance) and four fund categories (balanced, bond, equity and money
market). We observe a high inter-class correlation between retail investors and third-party distributors except for money market funds. We notice that equity
and bond funds present very similar frequency correlations. On average, it
is equal to $30\%$. For balanced and money market funds, we obtain lower
figures less than $20\%$. These results are coherent with the academic
research, since redemption runs and contagions in bond and equity funds have
been extensively studied and illustrated
\citep{Lakonishok-1992, Wermers-1999, Sias-2004, Wylie-2005, Coval-2007, Shleifer-2011, Cai-2019}.

\begin{figure}[tbph]
\centering
\caption{Dependogram of redemption frequencies between retail investors and third-party distributors}
\label{fig:correl4}
\includegraphics[width = \figurewidth, height = \figureheight]{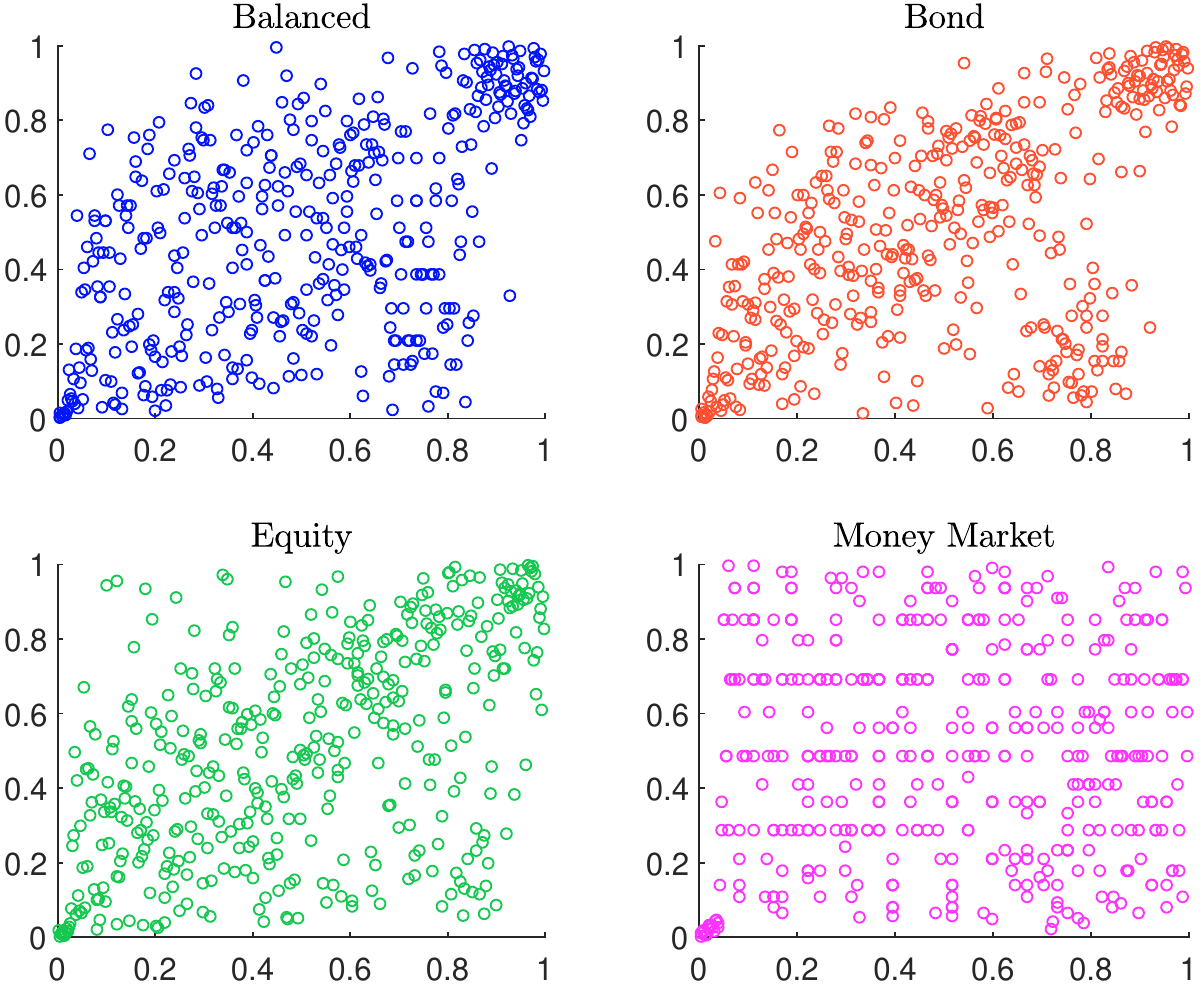}
\end{figure}

\begin{remark}
Another way to illustrate the intra-class correlation is to report
the dependogram (or empirical copula) of redemption frequencies.
An example is provided in Figure \ref{fig:correl4} for retail investors
and third-party distributors. We observe that these dependogram
does not correspond to the product copula\,\footnote{Examples of dependogram with the Normal copula and different
correlations are provided in Figure \ref{fig:correl5} on page \pageref{fig:correl5}.}.
\end{remark}

\subsubsection{Computing the stress scenarios}

\begin{figure}[tbph]
\centering
\caption{Conditional value-at-risk $\mathbb{C}\left(99\%\right)$ with respect
to the frequency correlation}
\label{fig:copula11}
\includegraphics[width = \figurewidth, height = \figureheight]{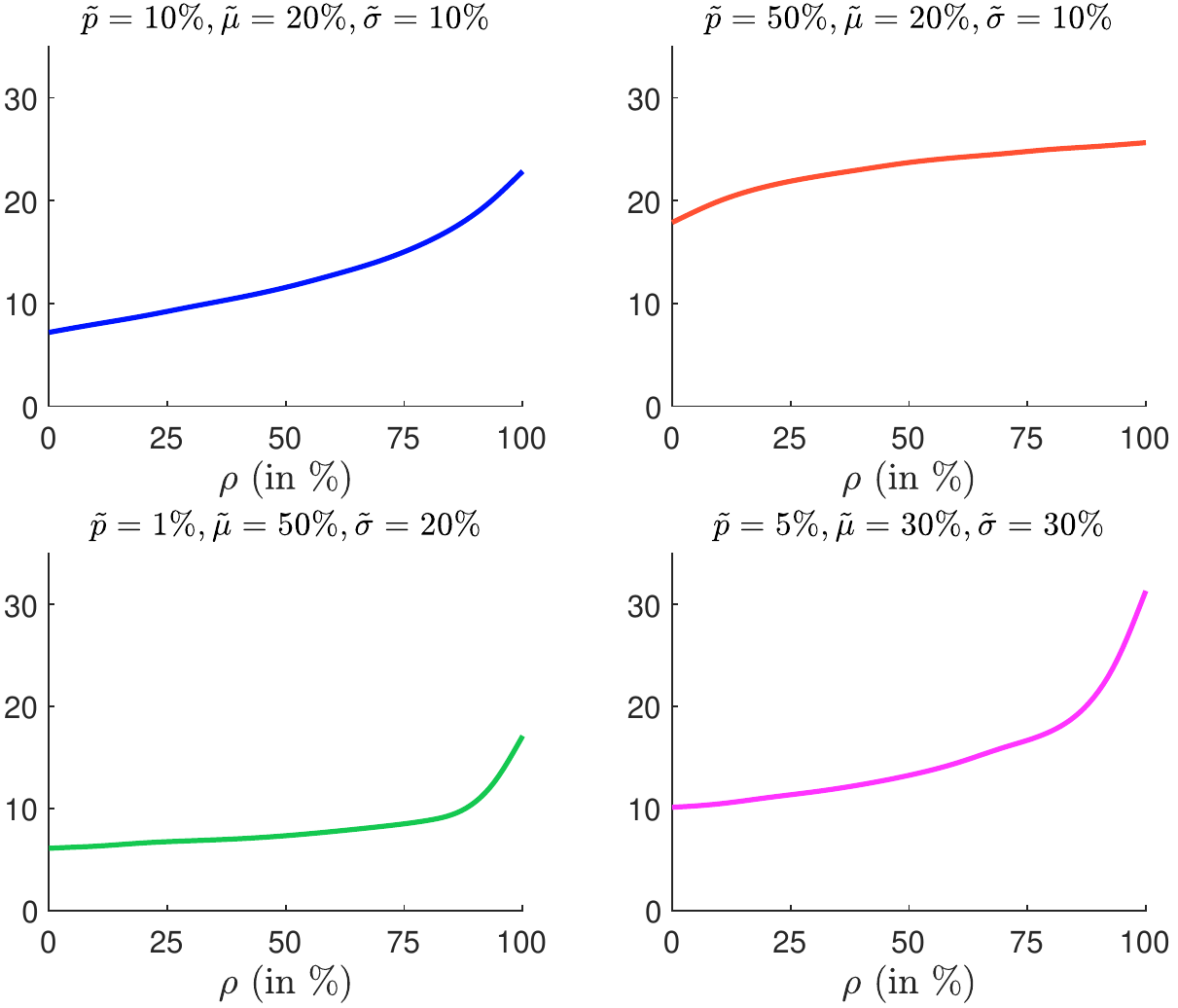}
\end{figure}

The parameters of the copula-based model is made up by the parameters of the
individual-based model ($\tilde{p}$, $\tilde{\mu}$ and $\tilde{\sigma}$) and
the copula parameter $\theta _{c}$ (or the associated frequency
correlation). Once these parameters are estimated for a given investor/fund
category, we transform the $\tilde{\mu}-\tilde{\sigma}$ parameterization
into the $a-b$ parameterization of the beta distribution\ and compute the
risk measures $\mathbb{M}$, $\mathbb{Q}\left( \alpha \right) $,
$\mathbb{C}\left( \alpha \right) $ and $\mathbb{S}\left( \mathcal{T}\right) $ by using
the following Monte Carlo algorithm:
\begin{enumerate}
\item we set $k\longleftarrow 1$;

\item we generate\footnote{%
Clayton and Normal copulas are easy to simulate using the method of
transformation \citep[page 803]{Roncalli-2020}.} $\left( u_{1},\ldots
,u_{n}\right) \sim \mathbf{C}_{\left( \theta _{c}\right) }$;

\item we compute the redemption events $\left( \mathcal{E}_{1},\ldots ,
\mathcal{E}_{n}\right) $ such that:%
\begin{equation*}
\mathcal{E}_{i}=\mathds{1}\left\{ u_{i}\geq 1-\tilde{p}\right\}
\end{equation*}

\item we simulate the redemption severities $\left( \redemption_{1}^{\star},
\ldots ,\redemption_{n}^{\star }\right) $ from the beta
distribution\footnote{Generally, the generation of beta random numbers is present in mathematical
programming languages (Matlab, Python). Otherwise, we can use the method of
rejection sampling \citep[pages 886-887]{Roncalli-2020}.} $\mathcal{B}\left(
a,b\right) $;

\item we compute the redemption rate for the $k^{\mathrm{th}}$ simulation
iteration:
\begin{equation*}
\redemption_{\left( k\right) }=\sum_{i=1}^{n}\omega _{i}\mathcal{E}_{i}%
\redemption_{i}^{\star }
\end{equation*}

\item if $k$ is equal to $n_{S}$, we return the simulated sample
$\left(\redemption_{\left( 1\right) },\ldots ,\redemption_{\left( n_S\right) }\right)$,
otherwise we set $k\longleftarrow k+1$ and go back to step 2.
\end{enumerate}
Figure \ref{fig:copula11} shows the relationship between the correlation
frequency\footnote{It corresponds to the Pearson correlation of the Clayton copula.}
and $\mathbb{C}\left( 99\%\right) $ for different parameter sets when the
liability structure has 20 equally-weighted unitholders. The impact of the
correlation risk is not negligible in some cases. This is particularly true
when the frequency correlation is close to $100\%$, but its impact is also significant
when the frequency correlation is larger than $20\%$. On average,
we observe that the risk measure $\mathbb{C}\left( 99\%\right) $
increases by $15\%$, $20\%$ and $35\%$ when the frequency correlation is respectively
equal to $20\%$, $30\%$ and $50\%$ compared the independent case.

\begin{remark}
The algorithm to simulate the copula-based model
$\mathcal{CM}\left(n,\omega ,\tilde{p},\tilde{\mu},\tilde{\sigma},\rho \right) $
can be used to simulate the individual-based model
$\mathcal{IM}\left( n,\omega ,\tilde{p},\tilde{\mu},\tilde{\sigma}\right) $
by setting $\mathbf{C}_{\left( \theta_{c}\right) }=\mathbf{C}^{\bot }$.
This is equivalent to replace step 2 and
simulate $n$ independent uniform random numbers $\left( u_{1},\ldots,u_{n}\right) $.
\end{remark}

\subsection{Time aggregation risk}

In the case of daily redemptions, the correlation risk only concerns the
cross-correlation between investors for a given market day. When we consider
fire sales or liquidity crisis, the one-day study period is not adapted and
must be extended to a weekly or monthly basis. In this case, we may face
time aggregation risk, meaning that redemption flows for the subsequent market
days may depend on the current redemption flows.

\subsubsection{Analysis of non-daily redemptions}

We recall that the total net assets at time $t+1$ can be decomposed
as follows:
\begin{equation*}
\limfunc{TNA}\left( t+1\right) =\left( 1+R\left( t+1\right) \right) \cdot
\limfunc{TNA}\left( t\right) +\mathcal{F}^{+}\left( t+1\right) -\mathcal{F}%
^{-}\left( t+1\right)
\end{equation*}%
By assuming that $\mathcal{F}^{+}\left( t+1\right) =0$, we obtain:%
\begin{equation*}
\limfunc{TNA}\left( t+1\right) \approx \left( 1+R\left( t+1\right) -%
\redemption\left( t+1\right) \right) \cdot \limfunc{TNA}\left( t\right)
\end{equation*}%
This formula is valid on a daily basis. If we consider a period of $n_{h}$
market days (e.g. a weekly period), we have:%
\begin{equation*}
\limfunc{TNA}\left( t+n_{h}\right) \approx \limfunc{TNA}\left( t\right)
\prod_{h=1}^{n_{h}}\left( 1+R\left( t+h\right) -\redemption\left( t+h\right)
\right)
\end{equation*}%
Therefore, it is not obvious to decompose the difference $\limfunc{TNA}%
\left( t+n_{h}\right) -\limfunc{TNA}\left( t\right) $ into a
\textquotedblleft \textit{performance}\textquotedblright\ effect and a
\textquotedblleft \textit{redemption}\textquotedblright\ effect since the
two effects are related. Indeed, the mathematical definition of the $n_{h}$%
-day redemption rate is:%
\begin{equation*}
\redemption\left( t;t+n_{h}\right) =\frac{\sum_{h=1}^{n_{h}}\mathcal{F}%
^{-}\left( t+h\right) }{\limfunc{TNA}\left( t\right) }
\end{equation*}%
whereas the fund return over the period $\left[ t,t+n_{h}\right] $ is given
by the compound formula:%
\begin{equation*}
R\left( t;t+h\right) =\prod_{h=1}^{n_{h}}\left( 1+R\left( t+h\right) \right)
-1
\end{equation*}%
Because of the cross-products \citep{Brinson-1991}, we cannot separate the
two effects:%
\begin{equation*}
\limfunc{TNA}\left( t+n_{h}\right) \neq \left( 1+R\left( t;t+n_{h}\right) -%
\redemption\left( t;t+n_{h}\right) \right) \cdot \limfunc{TNA}\left(
t\right)
\end{equation*}

\subsubsection{The autocorrelation risk}

\begin{figure}[tbph]
\centering
\caption{Histogram of the weekly redemption rate in \% with respect to the autocorrelation $\rho _{\mathrm{time}}$
($\tilde{p} = 50\%, \tilde{\mu} = 50\%, \tilde{\sigma} = 10\%, \rho = 0\%, n = 10$)}
\label{fig:autocor1a}
\includegraphics[width = \figurewidth, height = \figureheight]{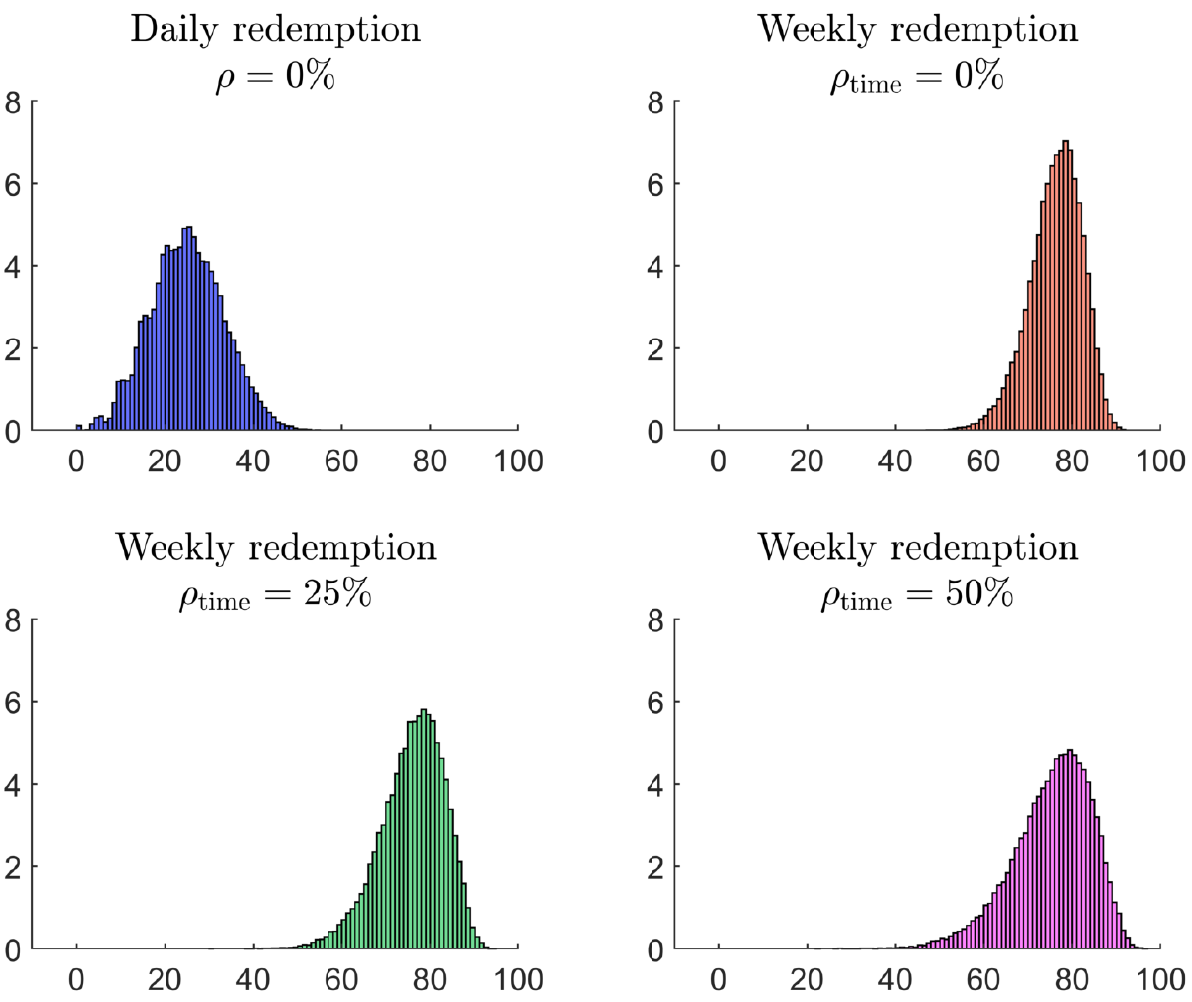}
\end{figure}

In the case where the performance effect is negligible --- $R\left(
t+h\right) \ll \redemption\left( t+h\right) $, we have:
\begin{equation}
\redemption\left( t,t+n_{h}\right) \approx
1-\prod_{h=1}^{n_{h}}\left( 1-\redemption\left( t+h\right) \right)
\label{eq:time-aggregation1}
\end{equation}%
We can then calculate the probability distribution of $\redemption\left(
t,t+n_{h}\right) $ by the Monte Carlo method. A first solution is to
consider that the redemption rates are time-independent. A second solution
is to consider that redemption rates are auto-correlated:
\begin{equation}
\redemption\left( t\right) =\rho _{\mathrm{time}}\redemption\left(
t-1\right) +\varepsilon \left( t\right)   \label{eq:time-aggregation2}
\end{equation}%
where $\rho _{\mathrm{time}}$ is the autocorrelation parameter and
$\varepsilon \left( t\right) $ is a random variable such that
$\redemption\left( t\right) \in \left[ 0,1\right] $. Such modeling is complex because of
the specification of $\varepsilon \left( t\right) $. However, this approach
can be approximated by considering a time-series copula representation:
\begin{equation}
\left( \redemption\left( t+1\right) ,\ldots ,\redemption\left(
t+n_{n}\right) \right) \sim \mathbf{C}\left( \mathbf{\tilde{F}}\left(
x\right) ,\ldots ,\mathbf{\tilde{F}}\left( x\right) ;\Sigma _{\mathrm{time}%
}\left( n_{h}\right) \right)   \label{eq:time-aggregation3}
\end{equation}%
where $\mathbf{\tilde{F}}$ is the probability distribution of
$\redemption\left( t\right) $ defined by the individual-based (or copula-based) model,
$\mathbf{C}$ is the Normal copula, whose parameters are given by the Toeplitz
correlation matrix\footnote{ For instance, in the case of a weekly
period, the Toeplitz correlation matrix is equal to:
\begin{equation*}
\Sigma _{\mathrm{time}}\left( 5\right) =\left(
\begin{array}{ccccc}
1 & \rho_{\mathrm{time}} & \rho_{\mathrm{time}}^{2} & \rho_{\mathrm{time}}^{3} & \rho_{\mathrm{time}}^{4} \\
\rho_{\mathrm{time}} & 1 & \rho_{\mathrm{time}} & \rho_{\mathrm{time}}^{2} & \rho_{\mathrm{time}}^{3} \\
\rho_{\mathrm{time}}^{2} & \rho_{\mathrm{time}} & 1 & \rho_{\mathrm{time}} & \rho_{\mathrm{time}}^{2} \\
\rho_{\mathrm{time}}^{3} & \rho_{\mathrm{time}}^{2} & \rho_{\mathrm{time}} & 1 & \rho_{\mathrm{time}} \\
\rho_{\mathrm{time}}^{4} & \rho_{\mathrm{time}}^{3} & \rho_{\mathrm{time}}^{2} & \rho_{\mathrm{time}} & 1
\end{array}%
\right)
\end{equation*}}
$\Sigma _{\mathrm{time}}\left( n_{h}\right) $ such that $\Sigma
_{\mathrm{time}}\left( n_{h}\right) _{i,j}= \rho _{\mathrm{time}}^{\left\vert
i-j\right\vert }$. To calculate the probability distribution of
$\redemption\left( t,t+n_{h}\right) $, we first simulate the individual-based
(or copula-based) model in order to estimate the probability distribution
$\mathbf{\tilde{F}}\left( x\right) $ of daily redemptions. Then, we generate
the sample of the time-series $\left( \redemption\left( t+1\right) ,\ldots
,\redemption\left(t+n_{n}\right) \right) $ by using the method of the
empirical quantile function \citep[pages 806-809]{Roncalli-2020}. Finally, we
calculate the redemption rate $\redemption\left( t,t+n_{h}\right) $ using
Equation (\ref{eq:time-aggregation1}). An example is provided in Figure
\ref{fig:autocor1a} when the correlation between investors is equal to
zero\footnote{The same example with a correlation of $50\%$ between investors
is given in Figure \ref{fig:autocor1c} on page \pageref{fig:autocor1c}.}. We
have also measured the impact of the autocorrelation value $\rho
_{\mathrm{time}}$ on the value-at-risk and the conditional value-at-risk.
Results are given in Tables \ref{tab:autocor3a} and \ref{tab:autocor3b} for
six different individual-based models $\mathcal{IM}\left(
n,\tilde{p},\tilde{\mu},\tilde{\sigma}\right)$. When the value of the risk
measure is small, we notice that the impact of $\rho _{\mathrm{time}}$ is high. For
instance, when $n = 500$, $\tilde{p} = 1\%$, $\tilde{\mu} = 25\%$ and
$\tilde{\sigma} = 10\%$, the value-at-risk $\mathbb{Q}\left(99\%\right)$ is
equal to $1.9\%$ in the independent case. This figure increases respectively
by $+9\%$ and $+19\%$ when $\rho _{\mathrm{time}}$ is equal to $25\%$ and
$50\%$. We also notice that the impact on the conditional value-at-risk is
close to that on the value-at-risk.

\begin{table}[tbph]
\centering
\caption{Impact of the autocorrelation $\rho_{\mathrm{time}}$ on the value-at-risk $\mathbb{Q}\left(99\%\right)$}
\label{tab:autocor3a}
\begin{tabular}{cccc|ccccc}
\multirow{2}{*}{$n$} & \multirow{2}{*}{$\tilde{p}$} & \multirow{2}{*}{$\tilde{\mu}$} &
\multirow{2}{*}{$\tilde{\sigma}$} & \multicolumn{5}{c}{$\rho_{\mathrm{time}}$} \\
& & & & $0\%$ & $25\%$ & $50\%$ & $75\%$ & $100\%$ \\ \hline
    $10\,000$ & ${\TsV}0.1\%$ & $25\%$ & $10\%$ & ${\TsV}0.2\%$ & ${\TsV}$$+6\%$ &        $+14\%$ &        $+24\%$ &        $+36\%$ \\
${\TsX}\,500$ & ${\TsV}1.0\%$ & $25\%$ & $10\%$ & ${\TsV}1.9\%$ & ${\TsV}$$+9\%$ &        $+19\%$ &        $+33\%$ &        $+50\%$ \\
${\TsXV}\,50$ & ${\TsV}2.0\%$ & $50\%$ & $10\%$ &      $10.5\%$ &        $+12\%$ &        $+29\%$ &        $+49\%$ &        $+79\%$ \\
${\TsX}\,100$ & ${\TsV}5.0\%$ & $50\%$ & $30\%$ &      $18.2\%$ & ${\TsV}$$+8\%$ &        $+18\%$ &        $+29\%$ &        $+45\%$ \\
${\TsXV}\,10$ &      $20.0\%$ & $50\%$ & $30\%$ &      $65.8\%$ & ${\TsV}$$+6\%$ &        $+13\%$ &        $+21\%$ &        $+28\%$ \\
${\TsXV}\,10$ &      $50.0\%$ & $50\%$ & $30\%$ &      $90.1\%$ & ${\TsV}$$+2\%$ & ${\TsV}$$+4\%$ & ${\TsV}$$+6\%$ & ${\TsV}$$+8\%$ \\ \hline
\end{tabular}
\end{table}

\begin{table}[tbph]
\centering
\caption{Impact of the autocorrelation $\rho_{\mathrm{time}}$ on the conditional value-at-risk $\mathbb{C}\left(99\%\right)$}
\label{tab:autocor3b}
\begin{tabular}{cccc|ccccc}
\multirow{2}{*}{$n$} & \multirow{2}{*}{$\tilde{p}$} & \multirow{2}{*}{$\tilde{\mu}$} &
\multirow{2}{*}{$\tilde{\sigma}$} & \multicolumn{5}{c}{$\rho_{\mathrm{time}}$} \\
& & & & $0\%$ & $25\%$ & $50\%$ & $75\%$ & $100\%$ \\ \hline
    $10\,000$ & ${\TsV}0.1\%$ & $25\%$ & $10\%$ & ${\TsV}0.2\%$ & ${\TsV}$$+6\%$ &        $+16\%$ &        $+27\%$ &        $+41\%$ \\
${\TsX}\,500$ & ${\TsV}1.0\%$ & $25\%$ & $10\%$ & ${\TsV}2.0\%$ & ${\TsV}$$+9\%$ &        $+21\%$ &        $+37\%$ &        $+56\%$ \\
${\TsXV}\,50$ & ${\TsV}2.0\%$ & $50\%$ & $10\%$ &      $11.4\%$ &        $+13\%$ &        $+32\%$ &        $+54\%$ &        $+84\%$ \\
${\TsX}\,100$ & ${\TsV}5.0\%$ & $50\%$ & $30\%$ &      $19.2\%$ & ${\TsV}$$+9\%$ &        $+20\%$ &        $+32\%$ &        $+50\%$ \\
${\TsXV}\,10$ &      $20.0\%$ & $50\%$ & $30\%$ &      $68.8\%$ & ${\TsV}$$+6\%$ &        $+13\%$ &        $+21\%$ &        $+28\%$ \\
${\TsXV}\,10$ &      $50.0\%$ & $50\%$ & $30\%$ &      $91.3\%$ & ${\TsV}$$+2\%$ & ${\TsV}$$+4\%$ & ${\TsV}$$+6\%$ & ${\TsV}$$+7\%$ \\ \hline
\end{tabular}
\end{table}

\begin{remark}
The compound approach defined by Equation (\ref{eq:time-aggregation1}) certainly overestimates
stress scenarios. Indeed, we implicitly assume that the redemptions rates $\redemption\left( t+h\right)$
are identically distributed, meaning that there is no time effect on the individual redemption behaviour.
However, we can think that an investor that redeems at time $t+1$ will not redeem at time $t+2$ and $t+3$.
In practice, we observe that redemptions of a given investor are mutually exclusive during a short period of time.
This property is not verified by Equation (\ref{eq:time-aggregation1}).
At time $t+h$, we notice $\mathcal{IS}\left( t+h\right) $ the set of
investors that have redeemed some units before $t+h$. We have
$\mathcal{IS}\left( t+1\right) =\left\{ 1,\ldots ,n\right\} $.
The  mutually exclusive property implies that\footnote{For instance,
if the investor has done a redemption at time $t+1$, the
probability that he will perform a new redemption at time $t+2$ is very
small, meaning that:%
\begin{equation*}
\mathcal{E}_{i}\left( t+1\right) =1\Rightarrow \mathcal{E}_{i}\left(
t+2\right) =\ldots =\mathcal{E}_{i}\left( t+n_{h}\right) =0
\end{equation*}%
}:%
\begin{equation*}
i\in \mathcal{IS}\left( t+h\right) \Rightarrow \mathcal{E}_{i}\left(
t+1\right) =\ldots =\mathcal{E}_{i}\left( t+n_{h}\right) =0
\end{equation*}%
It follows that:%
\begin{equation*}
\redemption\left( t+h\right) =\sum\limits_{i\notin \mathcal{IS}\left(
t+h\right) }\omega _{i}\left( t+h\right) \cdot \mathcal{E}_{i}\left(
t+h\right) \cdot \redemption_{i}^{\star }\left( t+h\right)
\end{equation*}%
and:%
\begin{equation*}
\omega _{i}\left( t+h+1\right) =\frac{\omega _{i}\left( t+h\right) }{%
\sum\limits_{i\notin \mathcal{IS}\left( t+h\right) }\omega _{i}\left(
t+h\right) }
\end{equation*}%
Because $\omega _{i}\left( t+h-1\right) \neq \omega _{i}\left( t+h\right)$
and $\mathcal{IS}\left( t+h-1\right) \neq \mathcal{IS}\left( t+h\right) $,
it is obvious that $\redemption\left( t+h-1\right) \neq \redemption\left(t+h\right) $.
Therefore, the redemption decisions taken in the recent past
(e.g. two or three days ago) have an impact on the future redemptions for
the next days. This is a limit of the compound approach. The solution would
be to develop a comprehensive individual-based model, whose random variables
are replaced by stochastic processes. Nevertheless, the complexity of such
model is not worth it with respect to the large uncertainty of stress
testing exercises.
\end{remark}

\subsubsection{The sell-herding behavior risk}

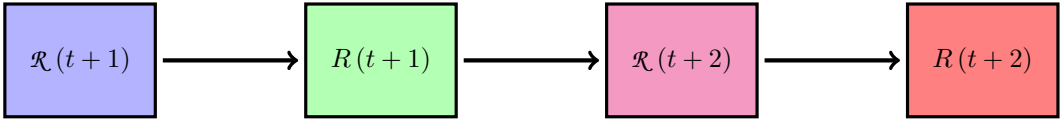
\begin{figure}
\centering
\caption{Spillover between fund redemptions and fund returns}
\label{fig:spillover}

\begin{tikzpicture}

\fill[color=blue!30] (0,0) rectangle (2.0,1.5);
\draw[very thick, black] (0,0) rectangle (2.0,1.5);
\draw[ultra thick, ->, black] (2.10,0.75) -- (3.90,0.75);

\fill[color=green!30] (4.0,0) rectangle (6.0,1.5);
\draw[very thick, black] (4.0,0) rectangle (6.0,1.5);
\draw[ultra thick, ->, black] (6.10,0.75) -- (7.90,0.75);

\fill[color=magenta!50] (8.0,0) rectangle (10.0,1.5);
\draw[very thick, black] (8.0,0) rectangle (10.0,1.5);
\draw[ultra thick, ->, black] (10.10,0.75) -- (11.90,0.75);

\fill[color=red!50] (12.0,0) rectangle (14.0,1.5);
\draw[very thick, black] (12.0,0) rectangle (14.0,1.5);

\draw (1.00,0.75) node{$\redemption\left( t+1\right)$};
\draw (5.00,0.75) node{$R\left( t+1\right)$};
\draw (9.00,0.75) node{$\redemption\left( t+2\right)$};
\draw (13.00,0.75) node{$R\left( t+2\right)$};

\end{tikzpicture}
\end{figure}

Herding risk is related to momentum trading. According to
\citet{Grinblatt-1995}, herding behavior corresponds to the situation where
investors buy and sell the same securities at the same time. Herding risk
happens during good and bad times, and is highly documented in economic
research \citep{Wermers-1999, ONeal-2004, Ivkovic-2009, Ferreira-2012,
Lou-2012, Cashman-2014, Chen-2017, Goldstein-2017, Choi-2019, Dotz-2019}.
However, we generally notice that sell herding may have more impact on asset
prices than buy herding. Therefore, the sell-herding behavior risk may be
associated to a price destabilizing or spillover effect. In the case of
redemption risk, the spillover mechanism corresponds to two related effects:

\begin{itemize}
\item A first spillover effect is that the unconditional probability of
    redemption is not equal to the conditional probability of the
    redemption given the returns of the fund during the recent past period:
\begin{equation*}
\Pr \left\{ \redemption\left( t+h\right) \leq x\right\} \neq \Pr \left\{ %
\redemption\left( t+h\right) \leq x\mid \left( R\left( t+1\right) ,\ldots
,R\left( t+h-1\right) \right) \right\}
\end{equation*}

\item A second spillover effect is that the unconditional probability of
    return is not equal to the conditional probability of the return given
    the redemptions of the fund during the recent past period:
\begin{equation*}
\Pr \left\{ R\left( t+h\right) \leq x\right\} \neq \Pr \left\{ R\left(
t+h\right) \leq x\mid \left( \redemption\left( t+1\right) ,\ldots ,%
\redemption\left( t+h-1\right) \right) \right\}
\end{equation*}
\end{itemize}
This implies that the transmission of a negative shock on the redemption rate
$\redemption\left( t+1\right) $ may also impact the redemption rates $\left\{
\redemption\left( t+2\right) ,\redemption\left( t+3\right) ,\ldots \right\} $
because of the feedback loop on the fund performance. An illustration is
provided in Figure \ref{fig:spillover}. A large negative redemption
$\redemption\left( t+1\right) $ may induce a negative abnormal performance
$R\left( t+1\right) $, and this negative performance may encourage the
remaining investors of the fund to redeem, because negative returns
accelerate redemption flows. This type of behavior is generally observed in
the case of fire sales and less liquid markets.\smallskip

As explained in the introduction, an integrated model that combines liability
risk and asset risk is too ambitious and too complex. Moreover, this means
modeling the policy reaction function of other investors and asset managers.
Nevertheless, if we want to take into account sell herding, spillover
or fire sales, we must build an econometric model. For example, the simplest
way is to consider the linear dynamic model:
\begin{equation*}
\left\{
\begin{array}{l}
R\left( t\right) =\phi _{1}\redemption\left( t\right) +u_{1}\left( t\right)
\\
\redemption\left( t+1\right) =\overline{\redemption}+\phi _{2}R\left(
t\right) +u_{2}\left( t+1\right)
\end{array}%
\right.
\end{equation*}%
We obtain an AR(1) process:%
\begin{equation*}
\redemption\left( t\right) =\overline{\redemption}+\phi \redemption\left(
t-1\right) +u\left( t\right)
\end{equation*}%
where $\phi =\phi _{1}\phi _{2}$ and $u\left( t\right) =u_{2}\left( t\right)
+\phi _{2}u_{1}\left( t-1\right) $ is a white noise process. It follows that:
\begin{equation*}
\mathbb{E}\left[ \redemption\left( t+h\right) \right] =\frac{1}{1-\phi
_{1}\phi _{2}}\overline{\redemption}
\end{equation*}%
Therefore, spillover scenarios can be estimated by applying a scaling
factor to the initial shock\footnote{The previous analysis can be extended to
more sophisticated process, e.g. $\limfunc{VAR}\left( p\right) $ processes.}.

\subsubsection{Empirical results}
\label{section:autocorrelation}

In order to illustrate the time dependency between redemptions, we build the
time series of the redemption rate $\redemption_{\left( j,k\right) }\left(
t\right) $, the redemption frequency $\frequency_{\left( j,k\right) }\left(
t\right) $ and the redemption severities $\redemption_{\left( j,k\right)
}^{\star }\left( t\right) $ for each classification matrix cell $\left(
j,k\right) $, which is defined by a fund category $\mathcal{FC}_{\left(
j\right) }$ and an investor category $\mathcal{IC}_{\left( k\right) }$. For
that, we calculate $\redemption_{\left( f,k\right) }\left( t\right) $ the
redemption rate of the fund $f$ for the investor category $\mathcal{IC}%
_{\left( k\right) }$ at time $t$. Then, we estimate the daily redemption
rate $\redemption_{\left( j,k\right) }\left( t\right) $ as the average of
the redemption rates of all funds that belong to the fund category
$\mathcal{FC}_{\left( j\right) }$:
\begin{equation}
\redemption_{\left( j,k\right) }\left( t\right) =\frac{1}{\left\vert
\mathcal{S}_{\left( j,k\right) }\left( t\right) \right\vert }\sum_{f\in
\mathcal{S}_{\left( j,k\right) }\left( t\right) }\redemption_{\left(
f,k\right) }\left( t\right)   \label{eq:time-series-R}
\end{equation}%
where\footnote{%
We only consider funds which have unitholders that belong to the
investor category $\mathcal{IC}_{\left( k\right) }$. This is
equivalent to impose that the assets under management held by the
investor category $\mathcal{IC}_{\left( k\right) }$ are strictly
positive: $\limfunc{TNA}_{\left( f,k\right) }>0$.}
$\mathcal{S}_{\left( j,k\right) }\left( t\right) =\left\{ f:f\in
\mathcal{FC}_{\left( j\right) },\limfunc{TNA}_{\left( f,k\right)
}\left( t\right) >0\right\} $. We also estimate the daily redemption
frequency as follows:
\begin{equation}
\frequency_{\left( j,k\right) }\left( t\right) =\frac{1}{\left\vert \mathcal{%
S}_{\left( j,k\right) }\left( t\right) \right\vert }\sum_{f\in \mathcal{S}%
_{\left( j,k\right) }\left( t\right) }\mathbf{1}\left\{ \redemption_{\left(
f,k\right) }\left( t\right) >0\right\}   \label{eq:time-series-p}
\end{equation}%
whereas the daily redemption severity is given by the following formula:
\begin{equation}
\redemption_{\left( j,k\right) }^{\star }\left( t\right) =\frac{1}{%
\left\vert \mathcal{S}_{\left( j,k\right) }^{\star }\left( t\right)
\right\vert }\sum_{f\in \mathcal{S}_{\left( j,k\right) }^{\star }\left(
t\right) }\redemption_{\left( f,k\right) }\left( t\right)
\label{eq:time-series-mu}
\end{equation}%
where $\mathcal{S}_{\left( j,k\right) }^{\star }\left( t\right) =\left\{
f:f\in \mathcal{FC}_{\left( j\right) },\limfunc{TNA}_{\left( f,k\right) }>0,
\redemption_{\left( f,k\right) }\left( t\right) >0\right\} $.\smallskip

\begin{table}[tbph]
\centering
\caption{Autocorrelation of the redemption rate in \%}
\label{tab:time5-R}
\begin{tabular}{lcccc}
\hline
                        & Balanced & Bond & Equity & Money market  \\ \hline
Institutional           &     ${\TsIII}$$25.9^{**}$ &         $-2.2$ &         $-1.5$ & $24.2^{**}$ \\
Insurance               &           $-1.5{\TsVIII}$ & ${\TsVIII}9.9$ & ${\TsVIII}5.4$ & $17.8^{**}$ \\
Retail                  &   ${\TsVIII}1.9{\TsVIII}$ &         $-2.1$ & ${\TsVIII}9.8$ & $11.7^{**}$ \\
Third-party distributor &   ${\TsVIII}2.7{\TsVIII}$ & ${\TsVIII}7.4$ & ${\TsVIII}5.5$ & $23.2^{**}$ \\ \hline
\end{tabular}
\end{table}

The computation of $\redemption_{\left( j,k\right) }\left( t\right) $,
$\frequency_{\left( j,k\right) }\left( t\right) $ and $\redemption_{\left(
j,k\right) }^{\star }\left( t\right) $ does make sense only if there is
enough observations $\left\vert \mathcal{S}_{\left( j,k\right) }\left(
t\right) \right\vert $ and $\left\vert \mathcal{S}_{\left( j,k\right)
}^{\star }\left( t\right) \right\vert $ at time $t$. This is why we focus on
the most representative investor categories (retail, third-party distributor,
institutional and insurance) and fund categories (balanced, bond, equity and
money market). In Table \ref{tab:time5-R}, we report the maximum between the
autocorrelation $\rho \left( \redemption\left( t\right) , \redemption\left(
t-1\right) \right) $ of order one and the autocorrelation $\rho \left(
\redemption\left( t\right) ,\redemption\left( t-2\right) \right) $ of order
two. Moreover, we indicate with the symbol $^{\ast \ast }$ the matrix cells
where the $p$-value of the autocorrelation is lower than $5\%$. Except for
money market funds and the institutional/balanced matrix cell, redemptions
are not significantly autocorrelated. If we consider redemption frequencies
and severities, we observe more autocorrelation (see Tables \ref{tab:time5-p}
and \ref{tab:time5-mu} on page \pageref{tab:time5-p}). However, for bond and
equity funds, the results show that the autocorrelation is significant and
high for the redemption frequency, but low for the redemption severity.

\section{Factor-based liquidity stress testing}

The last section of this article is dedicated to the factors that may
explain a redemption stress. First, we investigate whether it is due to a
redemption frequency shock or a redemption severity shock.
Second, we study how market risk may explain extreme redemption rates, and we
focus on three factors: stock returns, bond returns and volatility levels.

\subsection{Where does the stress come from?}

We may wonder whether the time variation of redemption rates is explained by
the time variation of redemption frequencies or redemption severities. Using
the time series built in Section \ref{section:autocorrelation} on page
\pageref{section:autocorrelation}, we consider three linear regression
models:
\begin{equation*}
\left\{
\begin{array}{l}
\redemption\left( t\right) =\beta _{0}+\beta _{1}\frequency\left( t\right)
+u\left( t\right)  \\
\redemption\left( t\right) =\beta _{0}+\beta _{1}\redemption^{\star }\left(
t\right) +u\left( t\right)  \\
\redemption\left( t\right) =\beta _{0}+\beta _{1}\frequency\left( t\right)
+\beta _{2}\redemption^{\star }\left( t\right) +u\left( t\right)
\end{array}%
\right.
\end{equation*}%
In the first model, we explain the redemption rate using the redemption
frequency. In the second model, the explanatory variable is the
redemption severity. Finally, the third model combines the two
previous models. For each classification matrix cell $\left(
j,k\right) $, we have reported the centered coefficient
of determination $\mathfrak{R}_{c}^{2}$ in Tables \ref{tab:time7-p},
\ref{tab:time7-mu} and \ref{tab:time7-p-mu}.\smallskip

\begin{table}[h]
\centering
\caption{Coefficient of determination $\mathfrak{R}_{c}^{2}$ in \% ---
$\redemption\left(t\right) = \beta_0 + \beta_1 \frequency\left(t\right) + u\left(t\right)$}
\label{tab:time7-p}
\begin{tabular}{lcccc}
\hline
                        & Balanced & Bond & Equity & Money market  \\ \hline
Institutional           & ${\TsV}2.4$ & $36.2$ & $53.4$ & $17.2$ \\
Insurance               & ${\TsV}0.9$ & $11.6$ & $10.8$ & $17.8$ \\
Retail                  &      $37.2$ & $34.5$ & $14.7$ & $18.4$ \\
Third-party distributor &      $11.5$ & $31.6$ & $17.7$ & $11.5$ \\ \hline
\end{tabular}
\vspace*{35pt}

\centering
\caption{Coefficient of determination $\mathfrak{R}_{c}^{2}$ in \% ---
$\redemption\left(t\right) = \beta_0 + \beta_1 \redemption^{\star}\left(t\right) + u\left(t\right)$}
\label{tab:time7-mu}
\begin{tabular}{lcccc}
\hline
                        & Balanced & Bond & Equity & Money market  \\ \hline
Institutional           & $87.2$ & $74.8$ & $44.5$ & $87.5$ \\
Insurance               & $99.2$ & $84.0$ & $83.3$ & $90.1$ \\
Retail                  & $77.6$ & $88.4$ & $98.1$ & $80.8$ \\
Third-party distributor & $93.1$ & $91.5$ & $92.1$ & $95.0$ \\ \hline
\end{tabular}
\vspace*{35pt}

\centering
\caption{Coefficient of determination $\mathfrak{R}_{c}^{2}$ in \% ---
$\redemption\left(t\right) = \beta_0 + \beta_1 \frequency\left(t\right)
+ \beta_2 \redemption^{\star}\left(t\right) + u\left(t\right)$}
\label{tab:time7-p-mu}
\begin{tabular}{lcccc}
\hline
                        & Balanced & Bond & Equity & Money market  \\ \hline
Institutional           & $88.2$ & $84.7$ & $81.7$ & $93.3$ \\
Insurance               & $99.3$ & $86.2$ & $86.4$ & $94.9$ \\
Retail                  & $92.5$ & $95.4$ & $99.3$ & $92.3$ \\
Third-party distributor & $97.0$ & $96.3$ & $95.7$ & $97.3$ \\ \hline
\end{tabular}
\end{table}

\begin{figure}[tbph]
\centering
\caption{Relationship between $\redemption\left(t\right)$, $\frequency\left(t\right)$
and $\redemption^{\star}\left(t\right)$ (retail/equity)}
\label{fig:time8-retail-equity}
\includegraphics[width = \figurewidth, height = \figureheight]{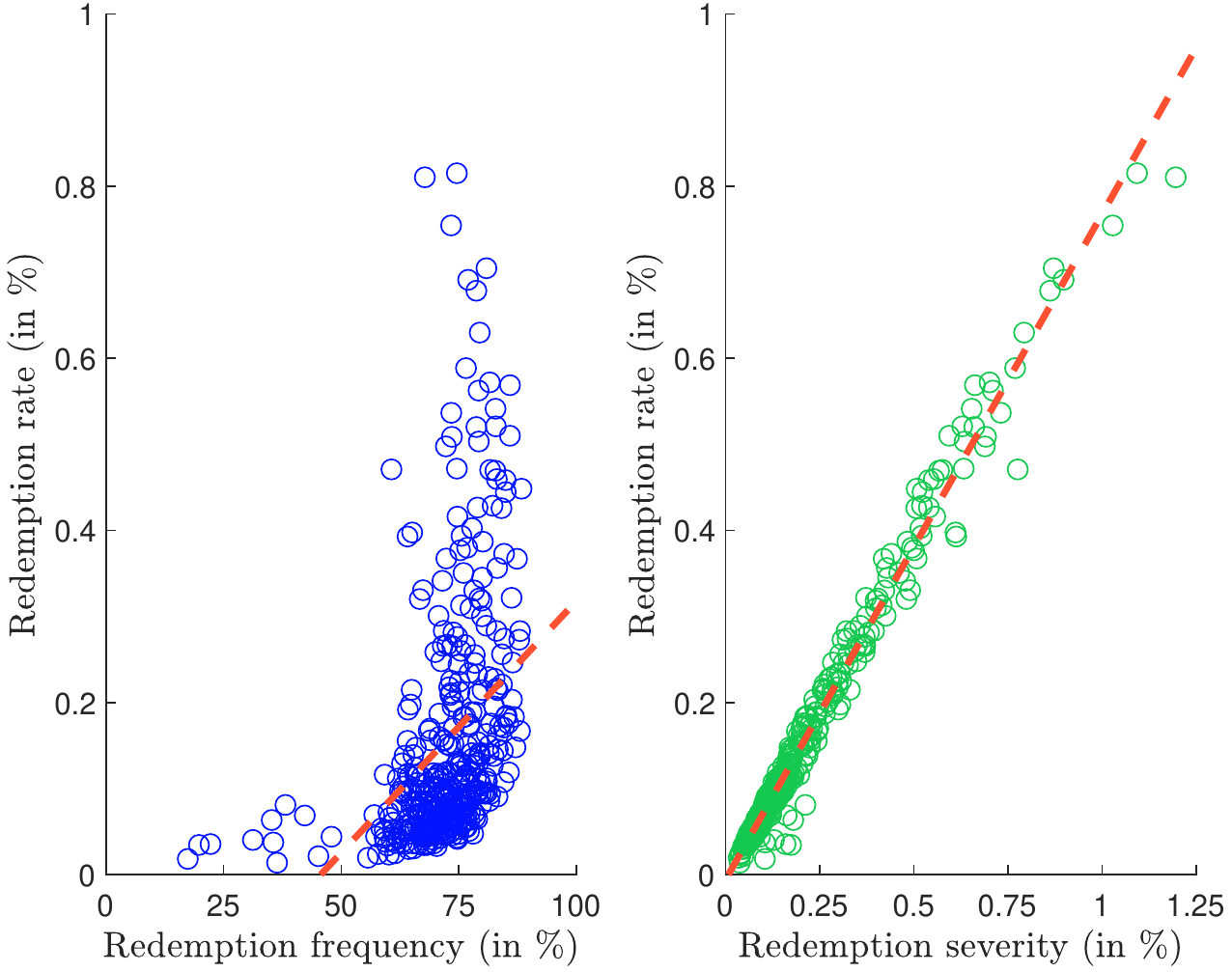}
\end{figure}

\begin{figure}[tbph]
\centering
\caption{Relationship between $\redemption\left(t\right)$, $\frequency\left(t\right)$
and $\redemption^{\star}\left(t\right)$ (institutional/equity)}
\label{fig:time8-institutional-equity}
\includegraphics[width = \figurewidth, height = \figureheight]{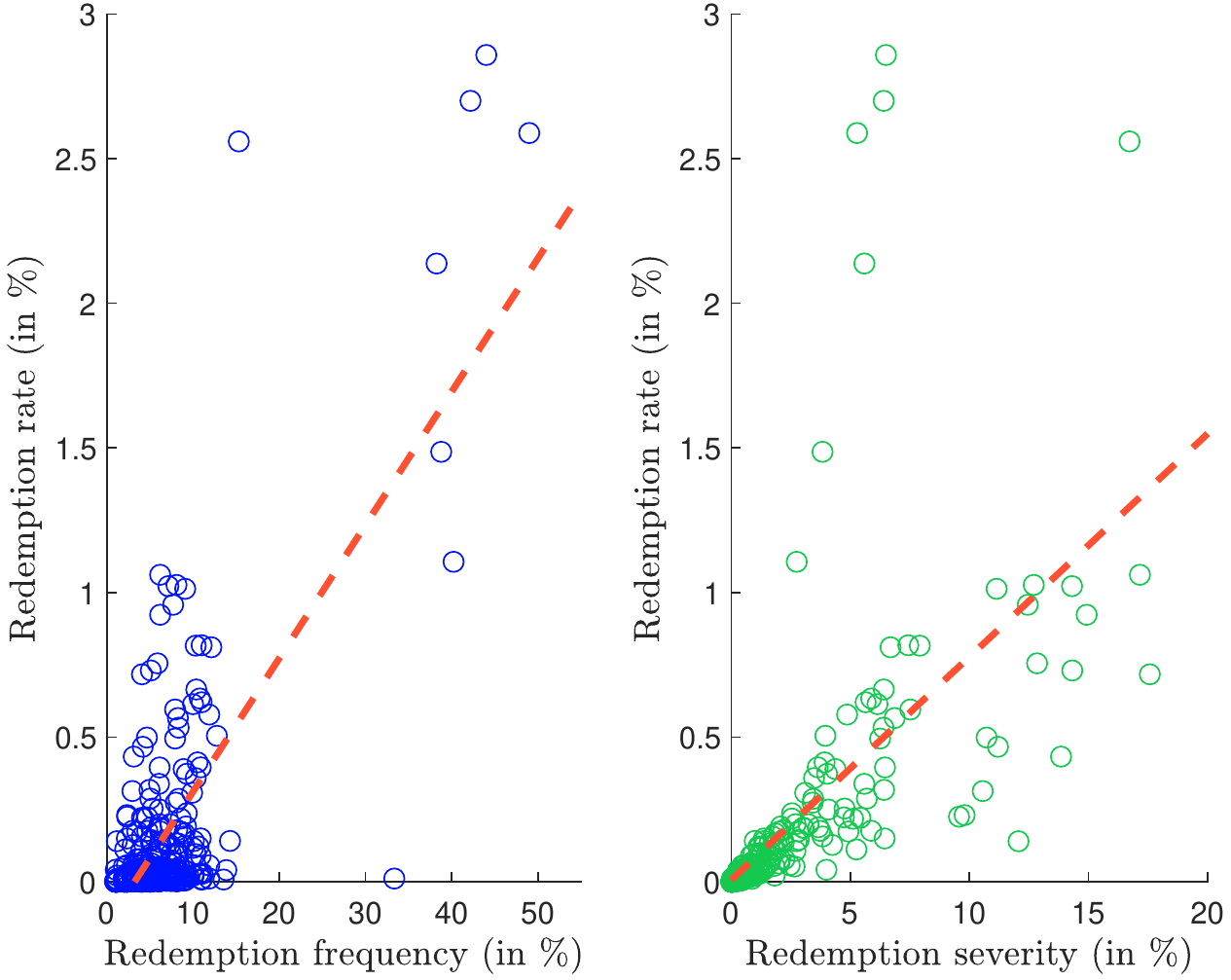}
\end{figure}

If we consider the first linear regression model, we notice that
$\mathfrak{R}_{c}^{2}$ is greater than $50\%$ only for the institutional/equity
category. $\mathfrak{R}_{c}^{2}$ takes a value around $35\%$ for the
retail/balanced, retail/bond and institutional/bond categories, otherwise it is less
than $20\%$. Results for the second linear regression are better. This
indicates that the redemption severity is a better explanatory variable than
the redemption frequency. The only exception is the institutional/equity
category. The combination of the two variables allows us to improve the
explanatory power of the model, but we also notice that the redemption
severity is the primary factor. The matrix cell with the highest
$\mathfrak{R}_{c}^{2}$ is retail/equity, whereas the matrix cell with the lowest
$\mathfrak{R}_{c}^{2}$ is institutional/equity. The scatter plot between
$\redemption\left( t\right) $, $\frequency\left( t\right) $ and
$\redemption^{\star }\left( t\right) $ for these two extreme cases are
reported in Figures \ref{fig:time8-retail-equity} and
\ref{fig:time8-institutional-equity}. For the retail/equity category, we
verify that the redemption severity explains the redemption rate. For the
institutional/equity category, the redemption severity is not able to explain
the high values of the redemption rate.\smallskip

The previous results are very interesting since the redemption severity is
the primary factor for explaining the redemption shocks. Therefore, a high
variation of the redemption rate is generally due to an increase of the
redemption severity. Nevertheless, there are some exceptions where stress
scenarios are also explained by an increase in the redemption frequency.

\begin{remark}
We have used the coefficient $\mathfrak{R}_{c}^{2}$  to show the power explanation
of the two variables $\frequency\left(t\right)$ and $\redemption^{\star}\left(t\right)$
without considering the effect of the constant.
For some matrix cells, we notice that the constant may be important (see Tables \ref{tab:time7b-p},
\ref{tab:time7b-mu} and \ref{tab:time7b-p-mu} on page \pageref{tab:time7b-p}).
\end{remark}

\clearpage

\subsection{What market risk factors matter in stress testing?}

\subsubsection{The flow-performance relationship}

Numerous academic research papers suggest that investor flows depend on
past performance. According to \citet{Sirri-1998} and
\citet{Huang-2007}, there is an asymmetry concerning the
flow-performance relationship: equity mutual funds with good
performance gain a lot of money inflows, while equity mutual funds
with poor performance suffer smaller outflows. However, this
asymmetry concerns relative performance. Indeed, according to
\citet{Ivkovic-2009}, \textquotedblleft \textsl{inflows are related
only to relative performance}\textquotedblright\ while
\textquotedblleft \textsl{outflows are related only to absolute fund
performance}\textquotedblright . Therefore, these authors suggest
that investors sell the asset class when this one has a bad
performance. In the case of corporate bonds, \citet{Goldstein-2017}
find that relative performance also matters in terms of explaining outflows.
In order to better understand these results, we consider the
following analytical model\footnote{See \citet{Arora-2019}.}:
\begin{equation*}
\left\{
\begin{array}{l}
R_{f}\left( t\right) =\alpha _{f}\left( t\right) +\beta _{f}\left(
t\right)
R_{\mathrm{mkt}}\left( t\right) +\varepsilon \left( t\right)  \\
\redemption_{f}\left( t\right) =\gamma _{f}+\delta _{f}\alpha
_{f}\left( t-1\right) +\varphi _{f}R_{f}\left( t-1\right) +\eta
\left( t\right)
\end{array}%
\right.
\end{equation*}%
where $R_{f}\left( t\right) $ is the return of the fund $f$,
$R_{\mathrm{mkt}}\left( t\right) $ is the return of the market risk
factor and $\redemption_{f}\left( t\right) $ is the redemption rate
of the fund $f$. $\varepsilon \left( t\right) $ and $\eta \left(
t\right) $ are two independent white noise processes. Using the
first equation, we can estimate the relative performance of the
fund, which is measured by its alpha component $\alpha _{f}\left(
t\right) $. The second equation states that the redemption rate
$\redemption_{f}\left( t\right) $ of the fund depends on the past
relative performance $\alpha _{f}\left( t-1\right) $ and the past
absolute performance $R_{f}\left( t-1\right) $. Then, we can test
two assumptions: $\mathcal{H}_{1}:\delta _{f}<0$ and
$\mathcal{H}_{2}:\varphi _{f}<0$. Accepting $\mathcal{H}_{1}$
implies that outflows depend on the relative performance, while
accepting $\mathcal{H}_{2}$ implies that outflows depend on the
absolute performance. In both cases, the value of the coefficient is
negative, because we expect that a negative performance will
increase the redemption rate. The previous framework can be extended
to take into account a more sophisticated model for determining the
relative performance\footnote{For instance, we can use the
three-factor Fama-French model or the four-factor Carhart model.}
$\alpha _{f}\left( t\right) $ or to consider lagged variables
\citep{Bellando-2011, Ferreira-2012, Lou-2012, Cashman-2014, Barber-2016, Fricke-2017}. More
generally, we have:
\begin{equation}
\redemption_{f}\left( t\right) =\gamma _{f}+\sum_{h=1}^{p}\left(
\phi _{f}^{\left( h\right) }\redemption_{f}\left( t-h\right) +\delta
_{f}^{\left( h\right) }\alpha _{f}\left( t-h\right) +\varphi
_{f}^{\left( h\right) }R_{f}\left( t-h\right) \right) +\eta \left(
t\right)   \label{eq:factor1}
\end{equation}%
Even if this type of flow-performance relationship is interesting to
understand the investor behavior, it is however not adapted in the
case of a stress testing program for two reasons. The first
reason is that Equation (\ref{eq:factor1}) is calibrated using low
frequency data, e.g. quarterly or monthly data. Therefore, the goal
of Equation (\ref{eq:factor1}) is to describe long-term behavior of
investors, whereas stress testing of liabilities concerns short-term
periods. The second reason is the inadequacy of this approach with
macro stress testing approaches developed by regulators and
institutional bodies.

\subsubsection{The macro stress testing approach}

If we consider stress testing programs developed in the banking
sector \citep[pages 893-922]{Roncalli-2020}, we distinguish
historical, probabilistic and macroeconomic approaches. While the
first two methods have been developed in the previous sections, we
focus on the third method, which is the approach used by the
regulators \citep{FRB-2017, EBA-2020a, EBA-2020b, ECB-2019,
Ong-2014}. The macroeconomic approach consists in defining stress
scenarios by a set of risk factors corresponding to some exogenous
shocks. In this article, we focus on three market risk factors:
\begin{itemize}
\item the performance of the bond market;
\item the performance of the stock market;
\item market volatility.
\end{itemize}
Therefore, we assume that there is a linear relationship between
the redemption rate and these factors:
\begin{equation}
\redemption\left( t\right) =\beta _{0}+\beta _{1}\mathcal{F}_{\mathrm{bond}%
}\left( t\right) +\beta _{2}\mathcal{F}_{\mathrm{stock}}\left( t\right)
+\beta _{3}\mathcal{F}_{\mathrm{vol}}\left( t\right) +u\left( t\right)
\label{eq:factor2}
\end{equation}%
where $\mathcal{F}_{\mathrm{bond}}\left( t\right) $ and
$\mathcal{F}_{\mathrm{stock}}\left( t\right) $ are the $h$-day total returns of the FTSE
World Broad Investment-Grade Bond index and the MSCI\ World index,
$\mathcal{F}_{\mathrm{vol}}\left( t\right) $ is the difference of the VIX index
between $t-h$ and $t$, and $h$ is the time horizon.\smallskip

In Table \ref{tab:factor1a}, we report the coefficient of determination
$\mathfrak{R}_{c}^{2}$ for the one-day time horizon. These figures are disappointing since
the impact of the market risk factors are very low\footnote{Nevertheless,
we verify that $\beta _{2}$ is negative for equity funds, even
though the relationship between redemption rate and stock returns is not
convincing as shown in Figure \ref{fig:factor2} on page \pageref{fig:factor2}.}.
For instance, the highest $\mathfrak{R}$-squared is reached for the third-party distributor/money
market category, but it is equal to $4.4\%$. If we consider a longer time
horizon, results do not improve and we always have
$\mathfrak{R}_{c}^{2}\ll 5\%$ (see Tables \ref{tab:factor1b} and \ref{tab:factor1c}
on page \pageref{tab:factor1b}).

\begin{table}[tbph]
\centering
\caption{Coefficient of determination $\mathfrak{R}_{c}^{2}$ in \% --- Equation (\ref{eq:factor2}), one-day time horizon}
\label{tab:factor1a}
\begin{tabular}{lcccc}
\hline
                        & Balanced & Bond & Equity & Money market  \\ \hline
Institutional           & $0.3$ & $0.8$ & $1.6$ & $1.9$ \\
Insurance               & $0.1$ & $0.1$ & $0.6$ & $0.8$ \\
Retail                  & $0.5$ & $3.1$ & $1.4$ & $0.6$ \\
Third-party distributor & $0.7$ & $1.5$ & $1.3$ & $4.4$ \\ \hline
\end{tabular}
\end{table}

\begin{remark}
The previous results suggest that redemption rates do not depend on market
risk factors on a short-term basis. However, fund managers generally have
the feeling that redemption rates increase when there is a stress on market
returns. Nevertheless, we know that returns are more or less independent from
one day to another. Therefore, we consider another approach using market
sentiment. For that, we compute the average redemption rate when the VIX
index is above 30, and calculate its relative variation with respect to the
entire period. Results are given in Table \ref{tab:factor3}. We observe an impact in
particular for bond/equity funds and institutional/third-party distributor
investors.
\end{remark}

\begin{table}[tbph]
\centering
\caption{Relative variation of the redemption rate $\redemption\left( t\right) $ when
$\limfunc{VIX}\geq 30$}
\label{tab:factor3}
\begin{tabular}{lcccc}
\hline
                        & Balanced & Bond & Equity & Money market  \\ \hline
Institutional           &         $+$$17.3\%$ &      $+$$54.7\%$ & $+$$74.3\%$ &      $+$$64.7\%$ \\
Insurance               &         $-$$63.4\%$ & ${\TsV}-$$1.1\%$ & $-$$14.2\%$ &      $+$$75.7\%$ \\
Retail                  &  ${\TsVII}+$$6.1\%$ &      $+$$21.5\%$ & $+$$13.8\%$ & ${\TsV}-$$4.5\%$ \\
Third-party distributor &         $+$$37.6\%$ &      $+$$43.6\%$ & $+$$49.5\%$ &      $+$$22.7\%$ \\ \hline
\end{tabular}
\end{table}

\clearpage

\section{Conclusion}

Liquidity stress testing is a recent topic in asset management, which has given
rise to numerous publications from regulators \citep{AMF-2017, BaFin-2017, ESMA-2019,
FSB-2017, IOSCO-2015, IOSCO-2018}, investment management associations \citep{AFG-2015,
EFAMA-2020} and affiliated researchers from central banks and international bodies
\citep{Arora-2019, Baranova-2017, Bouveret-2017, Fricke-2017, Gourdel-2018}.
On the academic side, few studies specifically concern liquidity stress testing
in asset management\footnote{Because data on liabilities are not publicly available.
However, we can cite \citet{Christoffersen-2017} and \citet{Darolles-2018}, who
specifically study asset management flows with respect to the liability structure
of the investment fund.}. Therefore, we observe a gap between general concepts and
specific measurement models. As such, the purpose of our study is to propose several
analytical approaches in order to implement LST practical programs.\smallskip

Besides the historical approach that considers non-parametric risk measures, we have
developed a frequency-severity model that is useful when building parametric risk measures
of liquidity stress testing. This statistical approach can be seen as a reduced-form
model based on three parameters: the redemption frequency, the expected redemption
severity and the redemption uncertainty. Like the historical approach, the frequency-severity
model requires some expert judgements to correct some data biases. Nevertheless,
both historical and analytical approaches are simple enough to verify properties
of risk ordering coherency between fund and investor categories.\smallskip

We have also developed an individual-based behavioral model, which is an extension
of the frequency-severity model. We have shown that redemption risk depends
on the fund liability structure, and is related to the Herfindahl index of assets
under management held by unitholders. Even if this model is hard to implement
because it requires knowing the comprehensive liability structure, it allows us to
justify liquidity stress testing based on the largest fund holders. Moreover, this model
shows the importance of cross-correlation between unitholders of a same investor
category, but also of several investor categories. Nevertheless, the individual-based
behavioral model is flexible enough that it can easily take into account dependencies
between investors by incorporating a copula model. Again, the issue with this extended
individual-based behavioral model lies in the knowledge of the liability structure.\smallskip

The production of stress scenarios can be obtained by considering a risk measure applied
to the redemption rate. For the historical approach, we can use a value-at-risk or a
conditional value-at-risk figure, which is estimated with non-parametric statistical methods.
For the frequency-severity and individual-based behavioral models, the estimation of the
VaR or CVaR is based on analytical formulas. Moreover, these models may produce parametric
stress scenarios for a given return time. Another issue concerns the choice of data
between gross or net redemption rates for calibrating these stress scenarios. For some categories,
net redemption rates may be used to proxy gross redemption rates, because they are very close
in stress periods. However, we also demonstrate that it is better to use gross redemption rates
for some investor or fund categories (e.g. retail investors or money market funds).\smallskip

The design of macro stress testing programs is more complicated than
expected. Since the flow-performance relationship is extensively
documented by academic research, it is valid at low frequencies,
typically on a quarterly or annual basis. In this case, we may
observe inflows towards the best fund managers. However, this
relationship mainly concerns relative performance, whereas macro
stress testing programs deal with absolute performance. Indeed,
relative performance is a key parameter when we want to
analyze the idiosyncratic liability liquidity risk at the fund
level. Nevertheless, the liquidity risk in asset management primarily
involves systemic periods of liquidity shortage that impact a given
asset class. Our empirical results are mixed since
drawing a relationship between redemption rates and market risk
factors in stress periods is not obvious because there are lead/lag
effects and liquidity stress periods never look the same. For
instance, the redemption stress scenario on money market funds
during the covid-19 crisis and the first quarter of 2020 is very
different from the redemption stress scenario during the Lehman
Brothers' bankruptcy in September and October 2008. Indeed, we
observe a significant lag of one/two months in the case of the
covid-19 crisis. In a similar way, the liquidity stress transmission
to equity funds has not been immediate and has been delayed by
several weeks.\smallskip

The current interest in liquidity stress testing is related to the Financial Stability Board's tasks
on systemic risk \citep{FSB-2010, FSB-2015} and shadow banking supervision \citep{FSB-2017, FSB-2020}.
As explained by \citet{Blanque-2019b}, \textquotedblleft \textsl{regulation of asset managers has been lagging
behind that of banks since the global financial crisis}\textquotedblright. The implementation
of the liquidity coverage ratio (LCR) and the net stable funding ratio (NSFR), the use of liquidity and
high-quality liquid assets (HQLA) buffers and the definition of regulatory monitoring tools date back to 2010
for the banking industry\footnote{The LCR became a minimum requirement for BCBS member countries in
January 2015.} \citep{BCBS-2010, BCBS-2013}. The regulatory framework on liquidity stress testing
proposed by \citet{ESMA-2019} is an important step for the development
of liquidity measurement in the asset management industry. In this paper, we
develop an analytical framework and give some answers. However, it is still early days and
much remains to be done.

\clearpage

\bibliographystyle{apalike}

\clearpage

\appendix

\section*{Appendix}

\section{Mathematical results}

\subsection{Granularity and the $\mathbb{X}$-statistic}
\label{appendix:x-statistic}

We consider $n$ funds whose redemption rate is equal to $p$. The
assets under management of each fund are set to $\$1$. The maximum
redemption rate of $n$ funds is equal to the mathematical
expectation of $n$ Bernoulli random variables:
\begin{eqnarray*}
p\left( \max \right)  &=&\mathbb{E}\left[ \max \left( \mathcal{B}_{1}\left(
p\right) ,\ldots ,\mathcal{B}_{n}\left( p\right) \right) \right]  \\
&=&1-\left( 1-p\right) ^{n}
\end{eqnarray*}%
whereas the redemption rate of the sum of $n$ funds is equal to the expected
frequency of a Binomial random variable:%
\begin{equation*}
p\left( \limfunc{sum}\right) =\frac{\mathbb{E}\left[ \mathcal{B}\left(
n,p\right) \right] }{n}=p
\end{equation*}
In Table \ref{tab:xstatistic1}, we report the value taken by the ratio
$p\left( \max \right) /p\left( \limfunc{sum}\right) $. For example, this ratio
is equal to $3.71$ if $p=5\%$ and $n=4$. To understand this ratio, we can
consider a large fund whose redemption probability is $p$. This fund is split
into $n$ funds of the same size. The ratio indicates the multiplication factor
to obtain the maximum of the redemption rates among the $n$ funds.

\begin{table}[tbph]
\centering
\caption{Value of the ratio $p\left( \max \right) /p\left( \limfunc{sum}\right) $}
\label{tab:xstatistic1}
\begin{tabular}{cccccccc}
\hline
\multirow{2}{*}{$n$} & \multicolumn{7}{c}{Probability $p$} \\ \cline{2-8}
            &       $1$ bp &     $10$ bps &        $1\%$ &        $5\%$ &       $10\%$ &       $20\%$ &       $50\%$  \\
\hline
  ${\TsX}1$ & ${\TsV}1.00$ & ${\TsV}1.00$ & ${\TsV}1.00$ & ${\TsV}1.00$ & ${\TsV}1.00$ & ${\TsV}1.00$ & ${\TsV}1.00$  \\
  ${\TsX}2$ & ${\TsV}2.00$ & ${\TsV}2.00$ & ${\TsV}1.99$ & ${\TsV}1.95$ & ${\TsV}1.90$ & ${\TsV}1.80$ & ${\TsV}1.50$  \\
  ${\TsX}3$ & ${\TsV}3.00$ & ${\TsV}3.00$ & ${\TsV}2.97$ & ${\TsV}2.85$ & ${\TsV}2.71$ & ${\TsV}2.44$ & ${\TsV}1.75$  \\
  ${\TsX}4$ & ${\TsV}4.00$ & ${\TsV}3.99$ & ${\TsV}3.94$ & ${\TsV}3.71$ & ${\TsV}3.44$ & ${\TsV}2.95$ & ${\TsV}1.88$  \\
  ${\TsX}5$ & ${\TsV}5.00$ & ${\TsV}4.99$ & ${\TsV}4.90$ & ${\TsV}4.52$ & ${\TsV}4.10$ & ${\TsV}3.36$ & ${\TsV}1.94$  \\
 ${\TsV}10$ &      $10.00$ & ${\TsV}9.96$ & ${\TsV}9.56$ & ${\TsV}8.03$ & ${\TsV}6.51$ & ${\TsV}4.46$ & ${\TsV}2.00$  \\
 ${\TsV}50$ &      $49.88$ &      $48.79$ &      $39.50$ &      $18.46$ & ${\TsV}9.95$ & ${\TsV}5.00$ & ${\TsV}2.00$  \\
      $100$ &      $99.51$ &      $95.21$ &      $63.40$ &      $19.88$ &      $10.00$ & ${\TsV}5.00$ & ${\TsV}2.00$  \\
\hline
\end{tabular}
\end{table}

\subsection{Statistical moments of zero-inflated probability distribution}

\subsubsection{General formulas}
\label{appendix:zero-inflated-moments}

A zero-inflated random variable $Z$ can be written as the product of a
Bernoulli random variables $X\sim \mathcal{B}\left( p\right) $ and a positive
random
variable $Y$:%
\begin{equation*}
Z=X Y
\end{equation*}%
Let $\mu _{m}^{\prime }\left( Z\right) $ for the $m$-th moment of $Z
$. Using the previous relationship, we deduce that:%
\begin{eqnarray}
\mu _{m}^{\prime }\left( Z\right)  &=&\mathbb{E}\left[ Z^{m}\right]   \notag
\\
&=&\mathbb{E}\left[ X^{m} Y^{m}\right]   \notag \\
&=&\mathbb{E}\left[ X^{m}\right]  \mathbb{E}\left[ Y^{m}\right]   \notag
\\
&=&p\mu _{m}^{\prime }\left( Y\right)   \label{eq:inflated-zero-moment1}
\end{eqnarray}%
because $X$ and $Y$ are independent by definition, and $X^{m}=X$, implying that
$X^{m}$ follows a Bernoulli distribution $\mathcal{B}\left( p\right) $. From
Equation (\ref{eq:inflated-zero-moment1}), we can compute the $m$-th
centered moment $\mu _{m}\left( Z\right) $. For that, we recall that:%
\begin{eqnarray*}
\mu _{1} &=&\mu _{1}^{\prime } \\
\mu _{2} &=&\mu _{2}^{\prime }-\mu _{1}^{2} \\
\mu _{3} &=&\mu _{3}^{\prime }-3\mu _{2}^{\prime }\mu _{1}+2\mu _{1}^{3} \\
\mu _{4} &=&\mu _{4}^{\prime }-4\mu _{3}^{\prime }\mu _{1}+6\mu _{2}^{\prime
}\mu _{1}^{2}-3\mu _{1}^{4}
\end{eqnarray*}%
We deduce the expression of the second moment:%
\begin{equation*}
\mu _{2}^{\prime }=\mu _{2}+\mu _{1}^{2}
\end{equation*}%
For the third moment, we have:%
\begin{eqnarray*}
\mu _{3}^{\prime } &=&\mu _{3}+3\mu _{2}^{\prime }\mu _{1}-2\mu _{1}^{3} \\
&=&\mu _{3}+3\left( \mu _{2}+\mu _{1}^{2}\right) \mu _{1}-2\mu _{1}^{3} \\
&=&\mu _{3}+3\mu _{2}\mu _{1}+\mu _{1}^{3} \\
&=&\gamma _{1}\mu _{2}^{3/2}+3\mu _{2}\mu _{1}+\mu _{1}^{3}
\end{eqnarray*}%
where $\gamma _{1}$ is the skewness coefficient. For the fourth moment, it
follows that:%
\begin{eqnarray*}
\mu _{4}^{\prime } &=&\mu _{4}+4\mu _{3}^{\prime }\mu _{1}-6\mu _{2}^{\prime
}\mu _{1}^{2}+3\mu _{1}^{4} \\
&=&\mu _{4}+4\left( \gamma _{1}\mu _{2}^{3/2}+3\mu _{2}\mu _{1}+\mu
_{1}^{3}\right) \mu _{1}-6\left( \mu _{2}+\mu _{1}^{2}\right) \mu
_{1}^{2}+3\mu _{1}^{4} \\
&=&\mu _{4}+4\gamma _{1}\mu _{2}^{3/2}\mu _{1}+12\mu _{2}\mu _{1}^{2}+4\mu
_{1}^{4}-6\mu _{2}\mu _{1}^{2}-6\mu _{1}^{4}+3\mu _{1}^{4} \\
&=&\mu _{4}+4\gamma _{1}\mu _{2}^{3/2}\mu _{1}+6\mu _{2}\mu _{1}^{2}+\mu
_{1}^{4} \\
&=&\left( \gamma _{2}+3\right) \mu _{2}^{2}+4\gamma _{1}\mu _{2}^{3/2}\mu
_{1}+6\mu _{2}\mu _{1}^{2}+\mu _{1}^{4}
\end{eqnarray*}%
where $\gamma _{2}$ is the excess kurtosis coefficient. We can then compute the
moments of $Z$. For the mean, we have:
\begin{eqnarray}
\mu _{1}\left( Z\right)  &=&\mu _{1}^{\prime }\left( Z\right)   \notag \\
&=&p\mu _{1}\left( Y\right)
\end{eqnarray}%
We deduce that the variance of $Z$ is equal to:%
\begin{eqnarray}
\mu _{2}\left( Z\right)  &=&\mu _{2}^{\prime }\left( Z\right) -\mu
_{1}^{2}\left( Z\right)   \notag \\
&=&p\mu _{2}^{\prime }\left( Y\right) -p^{2}\mu _{1}^{2}\left( Y\right)
\notag \\
&=&p\mu _{2}\left( Y\right) +p\left( 1-p\right) \mu _{1}^{2}\left( Y\right)
\end{eqnarray}%
For the third moment, we have:%
\begin{eqnarray*}
\mu _{3}\left( Z\right)  &=&\mu _{3}^{\prime }\left( Z\right) -3\mu
_{2}^{\prime }\left( Z\right) \mu _{1}\left( Z\right) +2\mu _{1}^{3}\left(
Z\right)  \\
&=&p\mu _{3}^{\prime }\left( Y\right) -3p^{2}\mu _{2}^{\prime }\left(
Y\right) \mu _{1}\left( Y\right) +2p^{3}\mu _{1}^{3}\left( Y\right)  \\
&=&p\left( \gamma _{1}\left( Y\right) \mu _{2}^{3/2}\left( Y\right) +3\mu
_{2}\left( Y\right) \mu _{1}\left( Y\right) +\mu _{1}^{3}\left( Y\right)
\right) - \\
&&3p^{2}\left( \mu _{2}\left( Y\right) +\mu _{1}^{2}\left( Y\right) \right)
\mu _{1}\left( Y\right) +2p^{3}\mu _{1}^{3}\left( Y\right)  \\
&=&p\gamma _{1}\left( Y\right) \mu _{2}^{3/2}\left( Y\right) +3p\left(
1-p\right) \mu _{2}\left( Y\right) \mu _{1}\left( Y\right) +p\left(
1-p\right) \left( 1-2p\right) \mu _{1}^{3}\left( Y\right)
\end{eqnarray*}%
It follows that the skewness coefficient is equal to:%
\begin{eqnarray}
\gamma _{1}\left( Z\right)  &=&\frac{\mu _{3}\left( Z\right) }{\mu
_{2}^{3/2}\left( Z\right) }  \notag \\
&=&\frac{\vartheta _{1}\left( Z\right) }{\left( p\mu _{2}\left( Y\right)
+p\left( 1-p\right) \mu _{1}^{2}\left( Y\right) \right) ^{3/2}}
\end{eqnarray}%
where:%
\begin{equation*}
\vartheta _{1}\left( Z\right) =p\gamma _{1}\left( Y\right) \mu
_{2}^{3/2}\left( Y\right) +3p\left( 1-p\right) \mu _{2}\left( Y\right) \mu
_{1}\left( Y\right) +p\left( 1-p\right) \left( 1-2p\right) \mu
_{1}^{3}\left( Y\right)
\end{equation*}%
For the fourth moment, we have:%
\begin{eqnarray}
\mu _{4}\left( Z\right)  &=&\mu _{4}^{\prime }\left( Z\right) -4\mu
_{3}^{\prime }\left( Z\right) \mu _{1}\left( Z\right) +6\mu _{2}^{\prime
}\left( Z\right) \mu _{1}^{2}\left( Z\right) -3\mu _{1}^{4}\left( Z\right)
\notag \\
&=&p\mu _{4}^{\prime }\left( Y\right) -4p^{2}\mu _{3}^{\prime }\left(
Y\right) \mu _{1}\left( Y\right) +6p^{3}\mu _{2}^{\prime }\left( Y\right)
\mu _{1}^{2}\left( Y\right) -3p^{4}\mu _{1}^{4}\left( Y\right)   \notag \\
&=&p\left( \gamma _{2}\left( Y\right) +3\right) \mu _{2}^{2}\left( Y\right)
+4p\gamma _{1}\left( Y\right) \mu _{2}^{3/2}\left( Y\right) \mu _{1}\left(
Y\right) +6p\mu _{2}\left( Y\right) \mu _{1}^{2}\left( Y\right) +p\mu
_{1}^{4}\left( Y\right) -  \notag \\
&&4p^{2}\gamma _{1}\left( Y\right) \mu _{2}^{3/2}\left( Y\right) \mu
_{1}\left( Y\right) -12p^{2}\mu _{2}\left( Y\right) \mu _{1}^{2}\left(
Y\right) -4p^{2}\mu _{1}^{4}\left( Y\right) +  \notag \\
&&6p^{3}\mu _{2}\left( Y\right) \mu _{1}^{2}\left( Y\right) +6p^{3}\mu
_{1}^{4}\left( Y\right) -3p^{4}\mu _{1}^{4}\left( Y\right)   \notag \\
&=&p\left( \gamma _{2}\left( Y\right) +3\right) \mu _{2}^{2}\left( Y\right)
+4p\left( 1-p\right) \gamma _{1}\left( Y\right) \mu _{2}^{3/2}\left(
Y\right) \mu _{1}\left( Y\right) +  \notag \\
&&6p\left( 1-p\right) ^{2}\mu _{2}\left( Y\right) \mu _{1}^{2}\left(
Y\right) +p\left( 1-p\right) \left( 1-3p+3p^{2}\right) \mu _{1}^{4}\left(
Y\right)
\end{eqnarray}%
We deduce that the excess kurtosis coefficient is equal to:%
\begin{eqnarray}
\gamma _{2}\left( Z\right)  &=&\frac{\mu _{4}\left( Z\right) }{\mu
_{2}^{2}\left( Z\right) }-3  \notag \\
&=&\frac{\vartheta _{2}\left( Z\right) }{\left( p\mu _{2}\left( Y\right)
+p\left( 1-p\right) \mu _{1}^{2}\left( Y\right) \right) ^{2}}
\end{eqnarray}%
where:%
\begin{eqnarray*}
\vartheta _{2}\left( Z\right)  &=&p\left( \gamma _{2}\left( Y\right)
+3\right) \mu _{2}^{2}\left( Y\right) +4p\left( 1-p\right) \gamma _{1}\left(
Y\right) \mu _{2}^{3/2}\left( Y\right) \mu _{1}\left( Y\right) + \\
&&6p\left( 1-p\right) ^{2}\mu _{2}\left( Y\right) \mu _{1}^{2}\left(
Y\right) +p\left( 1-p\right) \left( 1-3p+3p^{2}\right) \mu _{1}^{4}\left(
Y\right) - \\
&&3p^{2}\mu _{2}^{2}\left( Y\right) -6p^{2}\left( 1-p\right) \mu _{2}\left(
Y\right) \mu _{1}^{2}\left( Y\right) -3p^{2}\left( 1-p\right) ^{2}\mu
_{1}^{4}\left( Y\right)  \\
&=&\left( p\gamma _{2}\left( Y\right) +3p\left( 1-p\right) \right) \mu
_{2}^{2}\left( Y\right) +4p\left( 1-p\right) \gamma _{1}\left( Y\right) \mu
_{2}^{3/2}\left( Y\right) \mu _{1}\left( Y\right) + \\
&&6p\left( 1-p\right) \left( 1-2p\right) \mu _{2}\left( Y\right) \mu
_{1}^{2}\left( Y\right) +
p\left( 1-p\right) \left( 1-6p+6p^{2}\right) \mu _{1}^{4}\left( Y\right)
\end{eqnarray*}
\smallskip

We can deduce the following properties:
\begin{enumerate}
\item The skewness of $Z$ is equal to zero if and only if:
\begin{enumerate}
\item the skewness of $Y$ is equal to zero and the frequency probability
    $p$ is equal to one;
\item the frequency probability $p$ is equal to zero, meaning that $Z$ is
    always equal to zero.
\end{enumerate}
\item The excess kurtosis of $Z$ is equal to zero if and only if:
\begin{enumerate}
\item the kurtosis of $Y$ is equal to $3$ and the frequency probability $p$
    is equal to one;
\item the frequency probability $p$ is equal to zero, meaning that $Z$ is
    always equal to zero.
\end{enumerate}
\end{enumerate}
In other cases, the skewness and excess kurtosis coefficients of $Z $ are
different from zero even if the random variable $Y$ is not skewed and has not
fat tails.

\begin{remark}
The previous results seem to be contradictory with the properties given in
Equation (\ref{eq:zero-inflated-limits}) on page
\pageref{eq:zero-inflated-limits}. In fact, the limit case $p \rightarrow
0^{+}$ is not equal to $p = 0$, because there is a singularity at the point $p
= 0$.
\end{remark}

\subsubsection{Application to the beta distribution}
\label{appendix:zero-inflated-moments-beta}

We assume that $Y\sim \mathcal{B}\left( a,b\right) $. Since we have:%
\begin{equation*}
\mu _{1}\left( Y\right) =\frac{a}{a+b}
\end{equation*}%
we deduce that:%
\begin{equation*}
\mu _{1}\left( Z\right) =p\frac{a}{a+b}
\end{equation*}%
For the second moment, we have:%
\begin{equation*}
\mu _{2}\left( Y\right) =\frac{ab}{\left( a+b\right) ^{2}\left( a+b+1\right)
}
\end{equation*}%
and:%
\begin{eqnarray*}
\mu _{2}\left( Z\right)  &=&p\frac{ab}{\left( a+b\right) ^{2}\left(
a+b+1\right) }+p\left( 1-p\right) \left( \frac{a}{a+b}\right) ^{2} \\
&=&p\frac{ab+\left( 1-p\right) a^{2}\left( a+b+1\right) }{\left( a+b\right)
^{2}\left( a+b+1\right) }
\end{eqnarray*}%
This formula has been already found by \citet{Ospina-2010}. The skewness and
excess kurtosis coefficients of the beta distribution are equal to:
\begin{equation*}
\gamma _{1}\left( Y\right) =\frac{2\left( b-a\right) \sqrt{a+b+1}}{\left(
a+b+2\right) \sqrt{ab}}
\end{equation*}%
and:%
\begin{equation*}
\gamma _{2}\left( Y\right) =\frac{6\left( a-b\right) ^{2}\left( a+b+1\right)
}{ab\left( a+b+2\right) \left( a+b+3\right) }-\frac{6}{\left( a+b+3\right) }
\end{equation*}%
We plug these different expressions into the general formulas\footnote{The
formulas are not reported here because they don't have a lot of interest.} to
obtain $\gamma _{1}\left( Z\right) $ and $\gamma _{2}\left( Z\right) $.

\subsection{Maximum likelihood of the zero-inflated model}
\label{appendix:zero-inflated-ml}

We consider a sample $\left\{ x_{1},\ldots ,x_{n}\right\} $ of $n$
observations, and we assume that $X$ follows a zero-inflated model, whose
frequency and probability distributions are $p$ and $\mathbf{G}\left( x;\theta
\right) $. The log-likelihood of the $i^{\mathrm{th}}$ observation is equal to:
\begin{eqnarray*}
\ell _{i}\left( p,\theta \right)  &=&\ln \Pr \left\{ X=x_{i}\right\}  \\
&=&\ln f\left( x_{i}\right)  \\
&=&\mathds{1}\left\{ x_{i}=0\right\} \cdot \ln \left( 1-p\right) +\mathds{1}%
\left\{ x_{i}>0\right\} \cdot \ln \left( pg\left( x_{i};\theta \right)
\right)  \\
&=&\mathds{1}\left\{ x_{i}=0\right\} \cdot \ln \left( 1-p\right) +\mathds{1}%
\left\{ x_{i}>0\right\} \cdot \ln p+\mathds{1}\left\{ x_{i}>0\right\} \cdot
\ln g\left( x_{i};\theta \right)
\end{eqnarray*}%
We deduce that the log-likelihood function is equal to:%
\begin{eqnarray*}
\ell \left( p,\theta \right)  &=&\sum_{i=1}^{n}\ell _{i}\left( p,\theta
\right)  \\
&=&n_{0}\ln \left( 1-p\right) +\left( n-n_{0}\right) \ln p+\sum_{x_{i}>0}\ln
g\left( x_{i}\right)
\end{eqnarray*}%
where $n_{0}$ is the number of observations $x_{i}$ that are equal to zero. The
maximum likelihood estimator $\left( \hat{p},\hat{\theta}\right) $ is
defined as follows:%
\begin{equation*}
\left\{ \hat{p},\hat{\theta}\right\} =\arg \max \ell \left( p,\theta \right)
\end{equation*}%
and satisfies the first-order conditions:%
\begin{equation*}
\left\{
\begin{array}{c}
\partial _{p}\ell \left( \hat{p};\hat{\theta}\right) =0 \\
\partial _{\theta }\ell \left( \hat{p};\hat{\theta}\right) =\mathbf{0}%
\end{array}%
\right.
\end{equation*}%
We deduce that:%
\begin{eqnarray}
\partial _{p}\ell \left( \hat{p};\hat{\theta}\right)  &=&0\Leftrightarrow -%
\frac{n_{0}}{1-\hat{p}}+\frac{n-n_{0}}{\hat{p}}=0  \notag \\
&\Leftrightarrow &\hat{p}=\frac{n-n_{0}}{n}  \label{eq:appendix-ml1}
\end{eqnarray}%
The concentrated log-likelihood function becomes:%
\begin{equation*}
\ell \left( \hat{p},\theta \right) =n_{0}\ln n_{0}+\left( n-n_{0}\right) \ln
\left( n-n_{0}\right) -n\ln n+\sum_{x_{i}>0}\ln g\left( x_{i}\right)
\end{equation*}%
Therefore, the ML estimator $\hat{\theta}$ corresponds to the ML estimator of
$\theta $ when considering only the observations $x_{i}$ that are
strictly positive:%
\begin{eqnarray}
\hat{\theta} &=&\arg \max \ell \left( \hat{p},\theta \right)   \notag \\
&=&\arg \max \sum_{x_{i}>0}\ln g\left( x_{i}\right)   \label{eq:appendix-ml2}
\end{eqnarray}

\begin{remark}
In the case of the zero-inflated beta model, we have $\theta =\left(
a,b\right) $ and:%
\begin{equation}
\left\{ \hat{a},\hat{b}\right\} =\arg \max \sum_{x_{i}>0}\left( \left(
a-1\right) \ln x_{i}+\left( b-1\right) \ln \left( 1-x_{i}\right) -\ln
\mathfrak{B}\left( a,b\right) \vphantom{\int_{0}^{1}}\right)   \label{eq:appendix-ml3}
\end{equation}
\end{remark}

\subsection{Statistical properties of the individual-based model}
\label{appendix:individual1}

We define the random variable $\tilde{Z}$ as the sum of products of two random
variables:
\begin{equation*}
\tilde{Z}=\sum_{i=1}^{n}\omega _{i}\tilde{X}_{i}\tilde{Y}_{i}
\end{equation*}%
where $\tilde{X}_{i}\sim \mathcal{B}\left( \tilde{p}\right) $ and
$\tilde{Y}_{i}$ are \textit{iid} random variables. Moreover, we assume that
$\omega _{i}>0$ and $\sum_{i=1}^{n}\omega _{i}=1$.

\subsubsection{Computation of $\Pr \left\{ \tilde{Z}=0\right\} $}
\label{appendix:individual2}

This case corresponds to the situation where no client redeems:
\begin{eqnarray}
\Pr \left\{ \tilde{Z}=0\right\} &=&\Pr \left\{ \sum_{i=1}^{n}\omega _{i}\tilde{X}_{i}\tilde{Y}_{i}=0\right\} \notag \\
&=&\Pr \left\{ \tilde{X}_{1}=0,\ldots ,\tilde{X}_{n}=0\right\} \notag \\
&=&\prod_{i=1}^{n}\Pr \left\{ \tilde{X}_{i}=0\right\} \notag \\
&=&(1-\tilde{p})^{n}
\end{eqnarray}

\subsubsection{Statistical moments}
\label{appendix:individual3}

\paragraph{First moment}

For the mean, we have:
\begin{eqnarray*}
\mathbb{E}\left[ \tilde{Z}\right] &=&\mathbb{E}\left[ \sum_{i=1}^{n}\omega
_{i}\tilde{X}_{i}\tilde{Y}_{i}\right] \\
&=&\sum_{i=1}^{n}\omega _{i}\mathbb{E}\left[ \tilde{X}_{i}\right] \mathbb{E}\left[ \tilde{Y}_{i}\right]
\end{eqnarray*}
We deduce that:
\begin{equation}
\mu _{1}\left( \tilde{Z}\right) =\tilde{p}\mu _{1}\left( \tilde{Y}\right) \label{eq:individual-moment1}
\end{equation}

\paragraph{Second moment}

Since we have $\mathbb{E}\left[ \tilde{X}_{i}^{2}\right] =\tilde{p}$ and
$\mathbb{E}\left[ \tilde{Y}_{i}^{2}\right] =\mu _{2}^{\prime }\left(
\tilde{Y}\right) $, it follows that:
\begin{eqnarray*}
\mathbb{E}\left[ \tilde{Z}^{2}\right]  &=&\mathbb{E}\left[ \left(
\sum_{i=1}^{n}\omega _{i}\tilde{X}_{i}\tilde{Y}_{i}\right) ^{2}\right]  \\
&=&\mathbb{E}\left[ \sum_{i=1}^{n}\omega _{i}^{2}\tilde{X}_{i}^{2}\tilde{Y}%
_{i}^{2}+2\sum_{j>i}\omega _{i}\omega _{j}\tilde{X}_{i}\tilde{X}_{j}\tilde{Y}%
_{i}\tilde{Y}_{j}\right]  \\
&=&\tilde{p}\mu _{2}^{\prime }\left( \tilde{Y}\right) \sum_{i=1}^{n}\omega
_{i}^{2}+2\tilde{p}^{2}\mu _{1}^{2}\left( \tilde{Y}\right) \sum_{j>i}\omega
_{i}\omega _{j}
\end{eqnarray*}%
We notice that:
\begin{eqnarray*}
1 &=&\sum_{i=1}^{n}\omega _{i} \\
&=&\left( \sum_{i=1}^{n}\omega _{i}\right) ^{2} \\
&=&\sum_{i=1}^{n}\omega _{i}^{2}+2\sum_{j>i}\omega _{i}\omega _{j}
\end{eqnarray*}%
We deduce that:%
\begin{eqnarray*}
\mu _{2}\left( \tilde{Z}\right)  &=&\mathbb{E}\left[ \tilde{Z}^{2}\right] -%
\mathbb{E}^{2}\left[ \tilde{Z}\right]  \\
&=&\tilde{p}\mu _{2}^{\prime }\left( \tilde{Y}\right) \sum_{i=1}^{n}\omega
_{i}^{2}+2\tilde{p}^{2}\mu _{1}^{2}\left( \tilde{Y}\right) \sum_{j>i}\omega
_{i}\omega _{j}-\tilde{p}^{2}\mu _{1}^{2}\left( \tilde{Y}\right)  \\
&=&\tilde{p}\mu _{2}^{\prime }\left( \tilde{Y}\right) \sum_{i=1}^{n}\omega
_{i}^{2}+2\tilde{p}^{2}\mu _{1}^{2}\left( \tilde{Y}\right) \sum_{j>i}\omega
_{i}\omega _{j}- \\
&&\tilde{p}^{2}\mu _{1}^{2}\left( \tilde{Y}\right) \left(
\sum_{i=1}^{n}\omega _{i}^{2}+2\sum_{j>i}\omega _{i}\omega _{j}\right)
\end{eqnarray*}%
Therefore, the variance of $\tilde{Z}$ is equal to:
\begin{eqnarray}
\mu _{2}\left( \tilde{Z}\right)  &=&\left( \tilde{p}\mu _{2}^{\prime }\left(
\tilde{Y}\right) -\tilde{p}^{2}\mu _{1}^{2}\left( \tilde{Y}\right) \right)
\sum_{i=1}^{n}\omega _{i}^{2} \notag \\
&=&\tilde{p}\left( \mu _{2}\left( \tilde{Y}\right) +\left( 1-\tilde{p}%
\right) \mu _{1}^{2}\left( \tilde{Y}\right) \right) \sum_{i=1}^{n}\omega
_{i}^{2} \label{eq:individual-moment2}
\end{eqnarray}

\begin{remark}
In the equally-weighted case, we obtain:
\begin{equation*}
\mu _{2}\left( \tilde{Z}\right) =\frac{\tilde{p}\left( \mu _{2}\left( \tilde{%
Y}\right) +\left( 1-\tilde{p}\right) \mu _{1}^{2}\left( \tilde{Y}\right)
\right) }{n}
\end{equation*}
\end{remark}

\paragraph{Application to the beta severity distribution}

If we assume that $\tilde{Y}_{i}\sim \mathcal{B}\left(
\tilde{a},\tilde{b}\right) $, we have:
\begin{equation*}
\mu _{1}\left( \tilde{Y}\right) =\frac{\tilde{a}}{\tilde{a}+\tilde{b}}
\end{equation*}%
and:%
\begin{equation*}
\mu _{2}\left( \tilde{Y}\right) =\frac{\tilde{a}\tilde{b}}{\left( \tilde{a}+
\tilde{b}\right) ^{2}\left( \tilde{a}+\tilde{b}+1\right) }
\end{equation*}%
We deduce that:%
\begin{equation*}
\mu _{1}\left( \tilde{Z}\right) =\tilde{p}\frac{\tilde{a}}{\tilde{a}+\tilde{b}}
\end{equation*}%
and:%
\begin{eqnarray*}
\mu _{2}\left( \tilde{Z}\right)  &=&\frac{\tilde{p}}{n}\left( \frac{\tilde{a}%
\tilde{b}}{\left( \tilde{a}+\tilde{b}\right) ^{2}\left( \tilde{a}+\tilde{b}%
+1\right) }+\left( 1-\tilde{p}\right) \frac{\tilde{a}^{2}}{\left( \tilde{a}+%
\tilde{b}\right) ^{2}}\right)  \\
&=&\tilde{p}\frac{\tilde{a}}{n}\left( \frac{\tilde{b}+\left( 1-\tilde{p}%
\right) \tilde{a}\left( \tilde{a}+\tilde{b}+1\right) }{\left( \tilde{a}+%
\tilde{b}\right) ^{2}\left( \tilde{a}+\tilde{b}+1\right) }\right)
\end{eqnarray*}

\subsection{Moment matching between the zero-inflated model and the individual-based model}
\label{appendix:matching-zi-im}

In order to calibrate the probability $p$, we match the redemption
probability $\Pr \left\{ \redemption>0\right\} $. Using the results in
Appendix \ref{appendix:individual2} on page \pageref{appendix:individual2}, we obtain:
\begin{eqnarray*}
p &=&1-\Pr \left\{ \redemption=0\right\}  \\
&=&1-\left( 1-\tilde{p}\right) ^{n}
\end{eqnarray*}%
For the first moment, we have:%
\begin{equation*}
\mathbb{E}\left[ \redemption\right] =p\mu =\tilde{p}\tilde{\mu}
\end{equation*}%
We deduce that:%
\begin{equation*}
\mu =\frac{\tilde{p}}{1-\left( 1-\tilde{p}\right) ^{n}}\tilde{\mu}
\end{equation*}%
For the second moment, we have:%
\begin{equation*}
\sigma ^{2}\left( \redemption\right) =p\sigma ^{2}+p\left( 1-p\right) \mu
^{2}=\tilde{p}\left( \tilde{\sigma}^{2}+\left( 1-\tilde{p}\right) \tilde{\mu}%
^{2}\right) \sum_{i=1}^{n}\omega _{i}^{2}
\end{equation*}%
It follows that:%
\begin{eqnarray*}
\sigma ^{2} &=&\frac{\tilde{p}\left( \tilde{\sigma}^{2}+\left( 1-\tilde{p}%
\right) \tilde{\mu}^{2}\right) \sum_{i=1}^{n}\omega _{i}^{2}-p\left(
1-p\right) \mu ^{2}}{p} \\
&=&\frac{\tilde{p}\left( \tilde{\sigma}^{2}+\left( 1-\tilde{p}\right) \tilde{%
\mu}^{2}\right) \sum_{i=1}^{n}\omega _{i}^{2}}{1-\left( 1-\tilde{p}\right)
^{n}}-\frac{(1-\tilde{p})^{n}\tilde{p}^{2}}{\left( 1-\left( 1-\tilde{p}%
\right) ^{n}\right) ^{2}}\tilde{\mu}^{2} \\
&=&\left( \frac{\tilde{p}\mathcal{H}\left( \omega \right) }{1-\left( 1-%
\tilde{p}\right) ^{n}}\right) \tilde{\sigma}^{2}+ \\
&&\left( \frac{\tilde{p}\left( \left( 1-\tilde{p}\right) -\left( 1-\tilde{p}%
\right) ^{n}\right) \mathcal{H}\left( \omega \right) -\tilde{p}^{2}\left( 1-%
\tilde{p}\right) ^{n}\left( 1-\mathcal{H}\left( \omega \right) \right) }{%
\left( 1-\left( 1-\tilde{p}\right) ^{n}\right) ^{2}}\right) \tilde{\mu}^{2}
\end{eqnarray*}%
where $\mathcal{H}\left( \omega \right) =\sum_{i=1}^{n}\omega _{i}^{2}$ is
the Herfindahl index.

\begin{remark}
If we consider the equally-weighted case $\omega _{i}=n^{-1}$, we have $%
\mathcal{H}\left( \omega \right) =n^{-1}$ and:%
\begin{equation*}
\sigma ^{2}=\frac{1}{n}\left( \frac{\tilde{p}}{1-\left( 1-\tilde{p}\right)
^{n}}\right) \tilde{\sigma}^{2}+\frac{1}{n}\left( \frac{\tilde{p}\left(
\left( 1-\tilde{p}\right) -\left( 1-\tilde{p}\right) ^{n}\right) -\tilde{p}%
^{2}\left( 1-\tilde{p}\right) ^{n}\left( n-1\right) }{\left( 1-\left( 1-%
\tilde{p}\right) ^{n}\right) ^{2}}\right) \tilde{\mu}^{2}
\end{equation*}
\end{remark}

When $\tilde{p}\neq 0$, the limit cases are:%
\begin{equation*}
\lim_{n\rightarrow \infty }p=1
\end{equation*}%
and:%
\begin{equation*}
\lim_{n\rightarrow \infty }\mu =\tilde{p}\tilde{\mu}
\end{equation*}%
For the parameter $\sigma $, we obtain:%
\begin{equation*}
\lim_{n\rightarrow \infty }\sigma ^{2}=\tilde{p}\left( \tilde{\sigma}%
^{2}+\left( 1-\tilde{p}\right) \tilde{\mu}^{2}\right) \mathcal{H}\left(
\omega \right)
\end{equation*}%
For an infinitely fine-grained liability structure, we have:%
\begin{equation*}
\lim_{n\rightarrow \infty }\sigma ^{2}=0
\end{equation*}

\subsection{Upper bound of the Herfindahl index under partial information}
\label{appendix:herfindahl}

Let $\pi _{k}$ be a probability distribution, meaning that $\pi _{k}\geq 0$ and
$\sum_{k=1}^{n}\pi _{k}=1$. The Herfindahl index is equal to:
\begin{eqnarray*}
\mathcal{H} &=&\sum_{k=1}^{n}\pi _{k}^{2} \\
&=&\sum_{k=1}^{n}\pi _{k:n}^{2} \\
&=&\sum_{k=1}^{n}\pi _{n-k+1:n}^{2}
\end{eqnarray*}%
where:%
\begin{equation*}
0\leq \min \pi _{k}=\pi _{1:n}\leq \pi _{2:n}\leq \cdots \leq \pi _{k:n}\leq
\pi _{k+1:n}\leq \cdots \leq \pi _{n:n}=\max \pi _{k}
\end{equation*}%
We have:%
\begin{equation*}
\mathcal{H}=\sum_{k=1}^{m}\pi _{n-k+1:n}^{2}+\sum_{k=m+1}^{n}\pi
_{n-k+1:n}^{2}
\end{equation*}%
where $k=1:m$ denotes the largest contributions that are known, meaning that
we don't know the values taken by $\left\{ \pi _{1:n},\ldots ,\pi
_{n-m:n}\right\} $. Since we have $\pi _{n-k:n}\leq \pi _{n-k+1:n}$, we
deduce that:%
\begin{eqnarray*}
\sum_{k=m+1}^{n}\pi _{n-k+1:n}^{2} &\leq &\left( \frac{1-\sum_{k=1}^{m}\pi
_{n-k+1:n}}{\pi _{n-m+1:n}}\right) \pi _{n-m+1:n}^{2} \\
&=&\left( 1-\sum_{k=1}^{m}\pi _{n-k+1:n}\right) \pi _{n-m+1:n}
\end{eqnarray*}%
and\footnote{We verify that $\mathcal{H}_{n}^{+}=\mathcal{H}$.}:
\begin{equation}
\mathcal{H}\leq \mathcal{H}_{m}^{+}=\sum_{k=1}^{m}\pi _{n-k+1:n}^{2}+\left(
1-\sum_{k=1}^{m}\pi _{n-k+1:n}\right) \pi _{n-m+1:n} \label{eq:app-Herfindahl1}
\end{equation}
An example is given in Table \ref{tab:herfindahl1}. The Herfindahl index is
equal to $17.96\%$. Using the first three largest values, we obtain an estimate
of $20.50\%$.

\begin{table}[tbph]
\centering
\caption{Example of partial Herfindahl index computation}
\label{tab:herfindahl1}
\begin{tabular}{ccccccccc}
\hline
$m$                           &       1 &       2 &       3 &       4 &            5 &            6 &            7 &            8 \\ \hline
$\pi_m$ (in \%)               & $30.00$ & $20.00$ & $15.00$ & $10.00$ & ${\TsV}9.00$ & ${\TsV}7.00$ & ${\TsV}5.00$ & ${\TsV}4.00$ \\
$\mathcal{H}_{m}^{+}$ (in \%) & $30.00$ & $23.00$ & $20.50$ & $18.75$ &      $18.50$ &      $18.18$ &      $18.00$ &      $17.96$ \\
\hline
\end{tabular}
\end{table}

\subsection{Correlated redemptions with copula functions}
\label{appendix:correlation1}

We define the random variable $\tilde{Z}$ as previously:%
\begin{equation*}
\tilde{Z}=\sum_{i=1}^{n}\omega _{i}\tilde{X}_{i}\tilde{Y}_{i}
\end{equation*}%
where $\tilde{Y}_{i}$ are \textit{iid} random variables. We assume that
$\tilde{X}_{i}\sim \mathcal{B}\left( \tilde{p}\right) $ are identically
distributed, but not independent. We note $\mathbf{C}\left( u_{1},\ldots
,u_{n}\right) $ the copula function of the random vector
$\left( \tilde{X}_{1},\ldots ,\tilde{X}_{n}\right) $ and $\mathbf{B}\left( x\right) $ the
cumulative distribution function of the Bernoulli random variable
$\mathcal{B}\left( \tilde{p}\right) $. This means that
$\mathbf{B}\left( 0\right) =1-\tilde{p}$ and $\mathbf{B}\left( 1\right) =1$.\smallskip

In practice we use the Clayton copula:%
\begin{equation*}
\mathbf{C}_{\left(\theta_c\right)}\left( u_{1},\ldots ,u_{n}\right) =\left( u_{1}^{-\theta
_{c}}+\cdots +u_{n}^{-\theta _{c}}-n+1\right) ^{-1/\theta _{c}}
\end{equation*}%
or the Normal copula\footnote{The Normal copula depends on the correlation matrix $\Sigma $.
Here, we assume a uniform redemption correlation, implying that $\Sigma $
is the constant correlation matrix $\mathcal{C}_{n}\left( \theta _{c}\right) $
where $\theta _{c}$ is the pairwise correlation.}:
\begin{equation*}
\mathbf{C}_{\left( \theta _{c}\right) }\left( u_{1},\ldots ,u_{n}\right)
=\Phi \left( \Phi ^{-1}\left( u_{1}\right) +\cdots +\Phi ^{-1}\left(
u_{n}\right) ;\mathcal{C}_{n}\left( \theta _{c}\right) \right)
\end{equation*}
The Clayton parameter satisfies $\theta _{c}\geq 0$ whereas the Normal
parameter $\theta _{c}$ lies in the range $\left[ -1,1\right] $. Moreover,
we notice that the expressions of the bivariate copula functions are:%
\begin{equation*}
\mathbf{C}_{\left(\theta_c\right)}\left( u_{1},u_{2}\right) =
\mathbf{C}_{\left(\theta_c\right)}\left( u_{1},u_{2},1,\ldots
,1\right) =\left( u_{1}^{-\theta _{c}}+u_{2}^{-\theta _{c}}-1\right)
^{-1/\theta _{c}}
\end{equation*}%
and:%
\begin{equation*}
\mathbf{C}_{\left(\theta_c\right)}\left( u_{1},u_{2}\right) =
\mathbf{C}_{\left(\theta_c\right)}\left( u_{1},u_{2},1,\ldots
,1\right) =\Phi \left( \Phi ^{-1}\left( u_{1}\right) +\Phi ^{-1}\left(
u_{2}\right) ;\mathcal{C}_{2}\left( \theta _{c}\right)\right)
\end{equation*}

\subsubsection{Joint probability of two $\tilde{X}_{i}$'s}
\label{appendix:correlation2}

We consider the bivariate case. The probability mass function is described
by the following contingency table:%
\begin{equation}
\begin{tabular}{c|cc|c}
& $\tilde{X}_{2}=0$ & $\tilde{X}_{1}=1$ &  \\ \hline
$\tilde{X}_{1}=0$ & $\pi _{0,0}$ & $\pi _{0,1}$ & $\pi _{0}=1-\tilde{p}$ \\
$\tilde{X}_{1}=1$ & $\pi _{1,0}$ & $\pi _{1,1}$ & $\pi _{1}=\tilde{p}$ \\
\hline
& $\pi _{0}=1-\tilde{p}$ & $\pi _{1}=\tilde{p}$ & $1$%
\end{tabular}
\label{eq:corr1}
\end{equation}%
Since we have $\Pr \left\{ \tilde{X}_{1}\leq u_{1},\tilde{X}_{2}\leq
u_{2}\right\} =\mathbf{C}_{\left(\theta_c\right)}\left( \mathbf{B}\left( u_{1}\right) ,\mathbf{B}%
\left( u_{2}\right) \right) $, we deduce that:%
\begin{eqnarray*}
\mathbf{C}_{\left(\theta_c\right)}\left( \mathbf{B}\left( 0\right) ,\mathbf{B}\left( 0\right)
\right)  &=&\mathbf{C}_{\left(\theta_c\right)}\left( 1-\tilde{p},1-\tilde{p}\right)  \\
\mathbf{C}_{\left(\theta_c\right)}\left( \mathbf{B}\left( 0\right) ,\mathbf{B}\left( 1\right)
\right)  &=&\mathbf{C}_{\left(\theta_c\right)}\left( 1-\tilde{p},1\right) =1-\tilde{p} \\
\mathbf{C}_{\left(\theta_c\right)}\left( \mathbf{B}\left( 1\right) ,\mathbf{B}\left( 0\right)
\right)  &=&\mathbf{C}_{\left(\theta_c\right)}\left( 1,1-\tilde{p}\right) =1-\tilde{p} \\
\mathbf{C}_{\left(\theta_c\right)}\left( \mathbf{B}\left( 1\right) ,\mathbf{B}\left( 1\right)
\right)  &=&\mathbf{C}_{\left(\theta_c\right)}\left( 1,1\right) =1
\end{eqnarray*}%
and:%
\begin{equation}
\begin{tabular}{c|cc|c}
& $\tilde{X}_{2}=0$ & $\tilde{X}_{1}=1$ &  \\ \hline
$\tilde{X}_{1}=0$ & $\mathbf{C}_{\left(\theta_c\right)}\left( 1-\tilde{p},1-\tilde{p}\right) $ & $1-%
\tilde{p}-\mathbf{C}_{\left(\theta_c\right)}\left( 1-\tilde{p},1-\tilde{p}\right) $ & $1-\tilde{p}$
\\
$\tilde{X}_{1}=1$ & $1-\tilde{p}-\mathbf{C}_{\left(\theta_c\right)}\left( 1-\tilde{p},1-\tilde{p}%
\right) $ & $\mathbf{C}_{\left(\theta_c\right)}\left( 1-\tilde{p},1-\tilde{p}\right) +2\tilde{p}-1$
& $\tilde{p}$ \\ \hline
& $1-\tilde{p}$ & $\tilde{p}$ & $1$%
\end{tabular}
\label{eq:corr2}
\end{equation}%
In the case where $\mathbf{C}_{\left(\theta_c\right)}=\mathbf{C}^{\bot }$, $\tilde{X}_{1}$ and $%
\tilde{X}_{2}$ are independent, we retrieve the results obtained for the
individual-based model:%
\begin{equation}
\begin{tabular}{c|cc|c}
& $\tilde{X}_{2}=0$ & $\tilde{X}_{1}=1$ &  \\ \hline
$\tilde{X}_{1}=0$ & $\left( 1-\tilde{p}\right) ^{2}$ & $\left( 1-\tilde{p}%
\right) \tilde{p}$ & $1-\tilde{p}$ \\
$\tilde{X}_{1}=1$ & $\left( 1-\tilde{p}\right) \tilde{p}$ & $\tilde{p}^{2}$
& $\tilde{p}$ \\ \hline
& $1-\tilde{p}$ & $\tilde{p}$ & $1$%
\end{tabular}
\label{eq:corr3}
\end{equation}%
because $\mathbf{C}^{\bot }\left( u_{1},u_{2}\right) =u_{1}u_{2}$. In the
case where $\mathbf{C}_{\left(\theta_c\right)}=\mathbf{C}^{+}$, $\tilde{X}_{1}$ and $\tilde{X}_{2}$
are perfectly dependent and we obtain the following contingency table:%
\begin{equation}
\begin{tabular}{c|cc|c}
& $\tilde{X}_{2}=0$ & $\tilde{X}_{1}=1$ &  \\ \hline
$\tilde{X}_{1}=0$ & $1-\tilde{p}$ & $0$ & $1-\tilde{p}$ \\
$\tilde{X}_{1}=1$ & $0$ & $\tilde{p}$ & $\tilde{p}$ \\ \hline
& $1-\tilde{p}$ & $\tilde{p}$ & $1$%
\end{tabular}
\label{eq:corr4}
\end{equation}%
because $\mathbf{C}^{+}\left( u_{1},u_{2}\right) =\min \left(
u_{1},u_{2}\right) $. The contingency tables (\ref{eq:corr3}) and (\ref%
{eq:corr4}) represent the two extremes cases.\smallskip

\begin{remark}
If we use a radially symmetric copula \citep{Nelsen-2006} such that:%
\begin{equation*}
\mathbf{C}_{\left(\theta_c\right)}\left( u_{1},u_{2}\right) =u_{1}+u_{2}-1+\mathbf{C}_{\left(\theta_c\right)}\left(
1-u_{1},1-u_{2}\right)
\end{equation*}%
the contingency table (\ref{eq:corr2}) becomes:%
\begin{equation*}
\begin{tabular}{c|cc|c}
& $\tilde{X}_{2}=0$ & $\tilde{X}_{1}=1$ &  \\ \hline
$\tilde{X}_{1}=0$ & $1-2\tilde{p}+\mathbf{C}_{\left(\theta_c\right)}\left( \tilde{p},\tilde{p}%
\right) $ & $\tilde{p}-\mathbf{C}_{\left(\theta_c\right)}\left( \tilde{p},\tilde{p}\right) $ & $1-%
\tilde{p}$ \\
$\tilde{X}_{1}=1$ & $p-\mathbf{C}_{\left(\theta_c\right)}\left( \tilde{p},\tilde{p}\right) $ & $%
\mathbf{C}_{\left(\theta_c\right)}\left( \tilde{p},\tilde{p}\right) $ & $\tilde{p}$ \\ \hline
& $1-\tilde{p}$ & $\tilde{p}$ & $1$%
\end{tabular}%
\end{equation*}%
In the general case, we obtain a similar contingency table by replacing the
copula function $\mathbf{C}_{\left(\theta_c\right)}\left( u_{1},u_{2}\right) $ by its corresponding
survival function $\mathbf{\breve{C}}_{\left(\theta_c\right)}\left( u_{1},u_{2}\right) $ because we
have \citep{Nelsen-2006}:%
\begin{equation*}
\mathbf{\breve{C}}_{\left(\theta_c\right)}\left( u_{1},u_{2}\right) =u_{1}+u_{2}-1
+\mathbf{C}_{\left(\theta_c\right)}\left(
1-u_{1},1-u_{2}\right)
\end{equation*}
\end{remark}

\subsubsection{Computation of $\Pr \left\{ \tilde{Z}=0\right\} $}
\label{appendix:correlation3}

This case corresponds to the situation where no client redeems:
\begin{eqnarray}
\Pr \left\{ \tilde{Z}=0\right\}  &=&\Pr \left\{ \sum_{i=1}^{n}\omega _{i}%
\tilde{X}_{i}\tilde{Y}_{i}=0\right\}  \notag \\
&=&\Pr \left\{ \tilde{X}_{1}=0,\ldots ,\tilde{X}_{n}=0\right\}  \notag \\
&=&\mathbf{C}_{\left(\theta_c\right)}\left( 1-\tilde{p},\ldots ,1-\tilde{p}\right)
\end{eqnarray}
In the case where $\mathbf{C}_{\left(\theta_c\right)}=\mathbf{C}^{\bot }$, we retrieve the result $%
\Pr \left\{ \tilde{Z}=0\right\} =\left( 1-\tilde{p}\right) ^{n}$. In the
case where $\mathbf{C}_{\left(\theta_c\right)}=\mathbf{C}^{+}$, we obtain $\Pr \left\{ \tilde{Z}%
=0\right\} =1-\tilde{p}$.

\subsubsection{Statistical moments}
\label{appendix:correlation4}

\paragraph{First moment}

For the mean, we have:
\begin{eqnarray*}
\mathbb{E}\left[ \tilde{Z}\right] &=&\mathbb{E}\left[ \sum_{i=1}^{n}\omega
_{i}\tilde{X}_{i}\tilde{Y}_{i}\right] \\
&=&\sum_{i=1}^{n}\omega _{i}\mathbb{E}\left[ \tilde{X}_{i}\right] \mathbb{E}%
\left[ \tilde{Y}_{i}\right]
\end{eqnarray*}
We deduce that:
\begin{equation}
\mu _{1}\left( \tilde{Z}\right) =\tilde{p}\mu _{1}\left( \tilde{Y}\right)
\end{equation}

\paragraph{Second moment}

Using the contingency table (\ref{eq:corr2}), we have:
\begin{eqnarray*}
\mathbb{E}\left[ \tilde{X}_{1}\tilde{X}_{2}\right]  &=&\mathbf{C}_{\left(\theta_c\right)}\left( 1-%
\tilde{p},1-\tilde{p}\right) +2\tilde{p}-1 \\
&=&\mathbf{\breve{C}}_{\left(\theta_c\right)}\left( \tilde{p},\tilde{p}\right)
\end{eqnarray*}
It follows that:
\begin{eqnarray*}
\mathbb{E}\left[ \tilde{Z}^{2}\right]  &=&\mathbb{E}\left[ \left(
\sum_{i=1}^{n}\omega _{i}\tilde{X}_{i}\tilde{Y}_{i}\right) ^{2}\right]  \\
&=&\mathbb{E}\left[ \sum_{i=1}^{n}\omega _{i}^{2}\tilde{X}_{i}^{2}\tilde{Y}%
_{i}^{2}+2\sum_{j>i}\omega _{i}\omega _{j}\tilde{X}_{i}\tilde{X}_{j}\tilde{Y}%
_{i}\tilde{Y}_{j}\right]  \\
&=&\tilde{p}\mu _{2}^{\prime }\left( \tilde{Y}\right) \sum_{i=1}^{n}\omega
_{i}^{2}+2\mathbf{\breve{C}}_{\left(\theta_c\right)}\left( \tilde{p},\tilde{p}\right) \mu
_{1}^{2}\left( \tilde{Y}\right) \sum_{j>i}\omega _{i}\omega _{j}
\end{eqnarray*}%
and%
\begin{eqnarray}
\mu _{2}\left( \tilde{Z}\right)  &=&\mathbb{E}\left[ \tilde{Z}^{2}\right] -%
\mathbb{E}^{2}\left[ \tilde{Z}\right] \notag \\
&=&\tilde{p}\mu _{2}^{\prime }\left( \tilde{Y}\right) \sum_{i=1}^{n}\omega
_{i}^{2}+2\mathbf{\breve{C}}_{\left(\theta_c\right)}\left( \tilde{p},\tilde{p}\right) \mu
_{1}^{2}\left( \tilde{Y}\right) \sum_{j>i}\omega _{i}\omega _{j}-\tilde{p}%
^{2}\mu _{1}^{2}\left( \tilde{Y}\right) \notag \\
&=&\tilde{p}\left( \mu _{2}\left( \tilde{Y}\right) +\mu _{1}^{2}\left(
\tilde{Y}\right) \right) \mathcal{H}\left( \omega \right) +\mathbf{\breve{C}}_{\left(\theta_c\right)}%
\left( \tilde{p},\tilde{p}\right) \mu _{1}^{2}\left( \tilde{Y}\right) \left(
1-\mathcal{H}\left( \omega \right) \right) -\tilde{p}^{2}\mu _{1}^{2}\left(
\tilde{Y}\right) \notag \\
&=&\tilde{p}\mu _{2}\left( \tilde{Y}\right) \mathcal{H}\left( \omega \right)
+\left( \tilde{p}\mathcal{H}\left( \omega \right) +\mathbf{\breve{C}}_{\left(\theta_c\right)}\left(
\tilde{p},\tilde{p}\right) \left( 1-\mathcal{H}\left( \omega \right) \right)
-\tilde{p}^{2}\right) \mu _{1}^{2}\left( \tilde{Y}\right) \notag \\
&=&\left( \tilde{p}\mu _{2}\left( \tilde{Y}\right) +\left( \tilde{p}-\mathbf{%
\breve{C}}_{\left(\theta_c\right)}\left( \tilde{p},\tilde{p}\right) \right) \mu _{1}^{2}\left(
\tilde{Y}\right) \right) \mathcal{H}\left( \omega \right) +\left( \mathbf{%
\breve{C}}_{\left(\theta_c\right)}\left( \tilde{p},\tilde{p}\right) -\tilde{p}^{2}\right) \mu
_{1}^{2}\left( \tilde{Y}\right) \notag \\
\end{eqnarray}%
In the case where $\mathbf{C}_{\left(\theta_c\right)}=\mathbf{C}^{\bot }$, we have
$\mathbf{\breve{C}_{\left(\theta_c\right)}}\left( \tilde{p},\tilde{p}\right) =\tilde{p}^{2}$.
Therefore, we retrieve the result found in Equation (\ref{eq:individual-moment2})
on page \pageref{eq:individual-moment2}:
\begin{equation*}
\mu _{2}\left( \tilde{Z}\right) =\tilde{p}\left( \mu _{2}\left( \tilde{Y}%
\right) +\left( 1-\tilde{p}\right) \mu _{1}^{2}\left( \tilde{Y}\right)
\right) \mathcal{H}\left( \omega \right)
\end{equation*}%
In the case where $\mathbf{C}_{\left(\theta_c\right)}=\mathbf{C}^{+}$, we have
$\mathbf{\breve{C}}_{\left(\theta_c\right)}\left( \tilde{p},\tilde{p}\right) =\tilde{p}$
and we obtain:
\begin{equation*}
\mu _{2}\left( \tilde{Z}\right) =\tilde{p}\mu _{2}\left( \tilde{Y}\right)
\mathcal{H}\left( \omega \right) +\tilde{p}\left( 1-\tilde{p}\right) \mu
_{1}^{2}\left( \tilde{Y}\right)
\end{equation*}

\subsection{Statistical moments of the redemption frequency}
\label{appendix:redemption-frequency1}

We recall that $\tilde{X}_{i}\sim \mathcal{B}\left( \tilde{p}\right) $,
meaning that $\mathbb{E}\left[ \tilde{X}_{i}\right] =\mathbb{E}\left[ \tilde{X}_{i}^{2}\right] =\tilde{p}$. The weighted redemption frequency is defined
as follows:
\begin{equation*}
\frequency=\sum_{i=1}^{n}\omega _{i}\tilde{X}_{i}
\end{equation*}%
We have:%
\begin{eqnarray*}
\mathbb{E}\left[ \frequency\right]  &=&\mathbb{E}\left[ \sum_{i=1}^{n}\omega
_{i}\tilde{X}_{i}\right]  \\
&=&\tilde{p}
\end{eqnarray*}%
and:%
\begin{eqnarray*}
\mathbb{E}\left[ \frequency^{2}\right]  &=&\mathbb{E}\left[ \left(
\sum_{i=1}^{n}\omega _{i}\tilde{X}_{i}\right) ^{2}\right]  \\
&=&\mathbb{E}\left[ \sum_{i=1}^{n}\omega _{i}^{2}\tilde{X}%
_{i}^{2}+2\sum_{j>i}\omega _{i}\omega _{j}\tilde{X}_{i}\tilde{X}_{j}\right] \\
&=&\tilde{p}\mathcal{H}\left( \omega \right) +\mathbf{\breve{C}}_{\left(\theta_c\right)}\left(
\tilde{p},\tilde{p}\right) \left( 1-\mathcal{H}\left( \omega \right) \right)
\end{eqnarray*}%
We deduce that:%
\begin{equation*}
\mu _{2}\left( \frequency\right) =\tilde{p}\mathcal{H}\left( \omega \right) +%
\mathbf{\breve{C}}_{\left(\theta_c\right)}\left( \tilde{p},\tilde{p}\right) \left( 1-\mathcal{H}%
\left( \omega \right) \right) -\tilde{p}^{2}
\end{equation*}

\begin{remark}
We notice that the expected value and the volatility of the redemption
frequency are related in the following way:
\begin{equation}
\mu _{2}\left( \frequency\right) =\mathbb{E}\left[ \frequency\right] \left(
\mathcal{H}\left( \omega \right) -\mathbb{E}\left[ \frequency\right] \right)
+\mathbf{\breve{C}}\left( \mathbb{E}\left[ \frequency\right] ,\mathbb{E}%
\left[ \frequency\right] \right) \left( 1-\mathcal{H}\left( \omega \right)
\right)   \label{eq:appendix-pearson1}
\end{equation}
\end{remark}

\subsection{Pearson correlation between two redemption frequencies}
\label{appendix:redemption-frequency2}

We consider two redemption frequencies $\frequency_{1}$ and $\frequency_{2}$.
The redemption frequency $\frequency_{k}$ is associated to the liability
structure $\left( \omega _{k,1},\ldots ,\omega _{k,n_{k}}\right) $ and
corresponds to an investor category, whose redemption probability is
$\tilde{p}_{k}$ and frequency correlation is characterized by the copula function
$\mathbf{C}_{\left( \theta _{k}\right) }$ ($k=1,2$). We also assume that the
redemption correlation between the two investor categories is defined by the
copula function $\mathbf{C}_{\left( \theta _{12}\right) }$. It follows that
we have three copula functions:
\begin{itemize}
\item $\mathbf{C}_{\left( \theta _{1}\right) }$ is the copula function that
defines the frequency correlation between the investors of the first
category;

\item $\mathbf{C}_{\left( \theta _{2}\right) }$ is the copula function that
defines the frequency correlation between the investors of the second
category;

\item $\mathbf{C}_{\left( \theta _{12}\right) }$ is the copula function that
defines the frequency correlation between the investors of the first
category and those of the second category.
\end{itemize}
In the case where the two categories are the same, we have $%
\mathbf{C}_{\left( \theta _{1}\right) }=\mathbf{C}_{\left( \theta
_{2}\right) }=\mathbf{C}_{\left( \theta _{12}\right) }=\mathbf{C}_{\left(
\theta _{c}\right) }$.\smallskip

To compute the covariance between $\frequency_{1}$ and $\frequency_{2}$, we
calculate the mathematical expectation of the cross product:
\begin{eqnarray*}
\mathbb{E}\left[ \frequency_{1}\frequency_{2}\right]  &=&\mathbb{E}\left[
\left( \sum_{i=1}^{n_{1}}\omega _{1,i}\tilde{X}_{1,i}\right) \left(
\sum_{j=1}^{n_{2}}\omega _{2,j}\tilde{X}_{2,j}\right) \right]  \\
&=&\mathbb{E}\left[ \sum_{i=1}^{n_{1}}\sum_{j=1}^{n_{2}}\omega _{1,i}\omega
_{2,j}\tilde{X}_{1,i}\tilde{X}_{2,j}\right]  \\
&=&\mathbb{E}\left[ \tilde{X}_{1,i}\tilde{X}_{2,j}\right] \left(
\sum_{i=1}^{n_{1}}\sum_{j=1}^{n_{2}}\omega _{1,i}\omega _{2,j}\right)  \\
&=&\mathbf{\breve{C}}_{\left( \theta _{12}\right) }\left( \tilde{p}_{1},%
\tilde{p}_{2}\right)
\end{eqnarray*}%
because $\sum_{i=1}^{n_{1}}\sum_{j=1}^{n_{2}}\omega _{1,i}\omega _{2,j}=1$.
We deduce the expression of the Pearson correlation:
\begin{equation}
\rho \left( \frequency_{1},\frequency_{2}\right) =\frac{\mathbf{\breve{C}}%
_{\left( \theta _{12}\right) }\left( \tilde{p}_{1},\tilde{p}_{2}\right) -%
\tilde{p}_{1}\tilde{p}_{2}}{\sqrt{\mu _{2}\left( \frequency_{1}\right) \mu
_{2}\left( \frequency_{2}\right) }}  \label{eq:appendix-pearson2}
\end{equation}%
where:%
\begin{equation*}
\mu _{2}\left( \frequency_{k}\right) =\tilde{p}_{k}\left( \mathcal{H}\left(
\omega _{k}\right) -\tilde{p}_{k}\right) +\mathbf{\breve{C}}_{\left( \theta
_{k}\right) }\left( \tilde{p}_{k},\tilde{p}_{k}\right) \left( 1-\mathcal{H}%
\left( \omega _{k}\right) \right) \qquad k=1,2
\end{equation*}

\begin{remark}
The Pearson correlation $\rho \left( \frequency_{1},\frequency_{2}\right) $
is equal to zero if only if\,\footnote{We recall that
$\mathbf{C}_{\left( \theta _{k}\right) }$ is the Clayton or
the Normal copula. In the general case, this property does not hold.}
$\mathbf{C}_{\left( \theta _{k}\right) }$ is the product copula $\mathbf{C}^{\bot }$.
\end{remark}

\begin{remark}
In the case where the two investor categories are the same and the liability
structures are equally-weighted, we have
$\tilde{p}_{1}=\tilde{p}_{2}=\tilde{p}$ and
$\mathbf{C}_{\left( \theta _{1}\right) }=\mathbf{C}_{\left( \theta
_{2}\right) }=\mathbf{C}_{\left( \theta _{12}\right) }=\mathbf{C}_{\left(
\theta _{c}\right) }$, and we obtain:
\begin{equation}
\rho \left( \frequency_{1},\frequency_{2}\right) =\frac{\mathbf{\breve{C}}%
_{\left( \theta _{c}\right) }\left( \tilde{p},\tilde{p}\right) -\tilde{p}^{2}%
}{\sqrt{\mu _{2}\left( \frequency_{1}\right) \mu _{2}\left( \frequency%
_{2}\right) }}  \label{eq:appendix-pearson3}
\end{equation}%
where:%
\begin{equation*}
\mu _{2}\left( \frequency_{k}\right) =\mathbf{\breve{C}}_{\left( \theta
_{c}\right) }\left( \tilde{p},\tilde{p}\right) -\tilde{p}^{2}+\frac{\tilde{p}%
-\mathbf{\breve{C}}_{\left( \theta _{c}\right) }\left( \tilde{p},\tilde{p}%
\right) }{n_{k}}\qquad k=1,2
\end{equation*}%
The limiting case $n_k \rightarrow \infty $ is equal to
$\rho \left( \frequency_{1},\frequency_{2}\right) =1$. This is normal since
$\frequency_{1}$ and $\frequency_{2}$ converges to $\tilde{p}$ when the liability structure is
infinitely fine-grained.
\end{remark}

\section{Data}

\landscape

\begin{table}
\centering
\caption{Breakdown of the liability dataset by investor and fund categories}
\label{tab:data1-2}
\begin{tabular}{lrrrrrrrr}
\hline
Total number $n$         & \mr{Balanced} & \mr{Bond} & Enhanced & \mr{Equity} &    Money &  \mr{Other} & \mr{Structured} & \mr{Total} \\
of observations          &          &                & Treasury &             &   Market &             &                 &            \\ \hline
Auto-consumption         &  ${\TsV}22\,762$ & ${\TsV}46\,651$ & ${\TsV}3\,784$ & ${\TsV}46\,678$ & ${\TsV}6\,175$ & ${\TsV}34\,064$ &          $\,0$ & ${\TsV}\,160\,114$  \\
Central bank             &   ${\TsX}2\,791$ &  ${\TsX}7\,400$ &          $\,0$ &  ${\TsX}4\,730$ &  ${\TsX}\,602$ &           $\,0$ &          $\,0$ &  ${\TsX}\,15\,523$  \\
Corporate                &  ${\TsV}10\,780$ & ${\TsV}13\,457$ & ${\TsV}2\,305$ &  ${\TsX}6\,962$ & ${\TsV}7\,812$ &  ${\TsX}6\,164$ &          $\,0$ &  ${\TsX}\,47\,480$  \\
Corporate pension fund   &  ${\TsV}14\,827$ & ${\TsV}24\,429$ &  ${\TsX}\,427$ & ${\TsV}17\,975$ & ${\TsV}3\,029$ &  ${\TsX}5\,474$ &  ${\TsX}\,427$ &  ${\TsX}\,66\,588$  \\
Employee savings plan    &   ${\TsX}9\,894$ &  ${\TsX}4\,240$ & ${\TsV}1\,349$ & ${\TsV}19\,145$ & ${\TsV}3\,232$ &           $\,0$ & ${\TsV}5\,279$ &  ${\TsX}\,43\,139$  \\
Institutional            &  ${\TsV}50\,813$ & ${\TsV}95\,013$ & ${\TsV}3\,961$ & ${\TsV}76\,057$ & ${\TsV}9\,542$ & ${\TsV}31\,973$ &  ${\TsX}\,241$ & ${\TsV}\,267\,600$  \\
Insurance                &  ${\TsV}10\,577$ & ${\TsV}45\,494$ & ${\TsV}3\,303$ & ${\TsV}23\,145$ &      $12\,633$ &  ${\TsX}6\,528$ &          $\,0$ & ${\TsV}\,101\,680$  \\
Other                    &  ${\TsV}27\,938$ & ${\TsV}29\,817$ & ${\TsV}5\,816$ &  ${\TsX}4\,898$ & ${\TsV}9\,347$ & ${\TsV}18\,717$ &          $\,0$ &  ${\TsX}\,96\,533$  \\
Retail                   &       $140\,023$ & ${\TsV}86\,937$ & ${\TsV}7\,531$ & ${\TsV}99\,624$ &      $15\,418$ & ${\TsV}31\,370$ &      $83\,496$ & ${\TsV}\,464\,399$  \\
Sovereign                &   ${\TsX}7\,291$ & ${\TsV}12\,788$ &  ${\TsX}\,854$ & ${\TsV}14\,183$ & ${\TsV}3\,471$ &  ${\TsX}5\,308$ &          $\,0$ &  ${\TsX}\,43\,895$  \\
Third-party distributor  &  ${\TsV}63\,792$ & ${\TsV}86\,716$ & ${\TsV}5\,247$ &      $123\,004$ &      $11\,160$ & ${\TsV}15\,407$ & ${\TsV}5\,126$ & ${\TsV}\,310\,452$  \\ \hline
Total                    &       $361\,488$ &      $452\,942$ &      $34\,577$ &      $436\,401$ &      $82\,421$ &      $155\,005$ &      $94\,569$ &      $1\,617\,403$  \\ \hline
& \\
\hline
Total number $n_1$      & \mr{Balanced} & \mr{Bond} & Enhanced & \mr{Equity} &    Money &  \mr{Other} & \mr{Structured} & \mr{Total} \\
of redemptions          &          &                & Treasury &             &   Market &             &                 &            \\ \hline
Auto-consumption        &  ${\TsV}3\,744$ & ${\TsV}8\,796$ &      $1\,135$ & ${\TsV}11\,871$ & ${\TsV}3\,040$ &  ${\TsX}\,883$ &          $\,0$ & ${\TsV}29\,469$  \\
Central bank            &           $\,4$ &         $\,16$ &         $\,0$ &          $\,38$ &         $\,18$ &          $\,0$ &          $\,0$ &          $\,76$  \\
Corporate               &   ${\TsX}\,324$ &  ${\TsX}\,484$ & ${\TsV}\,144$ &         $\,159$ & ${\TsV}3\,110$ &         $\,20$ &          $\,0$ &  ${\TsX}4\,241$  \\
Corporate pension fund  &   ${\TsX}\,460$ &  ${\TsX}\,513$ &  ${\TsX\,}17$ &         $\,447$ &  ${\TsX}\,213$ &         $\,17$ &          $\,2$ &  ${\TsX}1\,669$  \\
Employee savings plan   &   ${\TsX}\,264$ &  ${\TsX}\,120$ &  ${\TsX\,}40$ &         $\,519$ &         $\,74$ &          $\,0$ &  ${\TsX}\,145$ &  ${\TsX}1\,162$  \\
Institutional           &  ${\TsV}1\,973$ & ${\TsV}3\,098$ &  ${\TsX\,}74$ &  ${\TsX}3\,422$ & ${\TsV}2\,754$ &  ${\TsX}\,229$ &          $\,0$ & ${\TsV}11\,550$  \\
Insurance               &   ${\TsX}\,568$ & ${\TsV}1\,562$ & ${\TsV}\,114$ &  ${\TsX}1\,596$ & ${\TsV}2\,409$ &         $\,61$ &          $\,0$ &  ${\TsX}6\,310$  \\
Other                   &  ${\TsV}1\,145$ &  ${\TsX}\,926$ & ${\TsV}\,219$ &         $\,805$ & ${\TsV}2\,009$ &  ${\TsX}\,278$ &          $\,0$ &  ${\TsX}5\,382$  \\
Retail                  &       $54\,095$ &      $36\,018$ &      $3\,932$ & ${\TsV}67\,862$ & ${\TsV}6\,882$ & ${\TsV}5\,030$ &      $22\,783$ &      $196\,602$  \\
Sovereign               &   ${\TsX}\,494$ &  ${\TsX}\,118$ &         $\,9$ &         $\,381$ &  ${\TsX}\,521$ &          $\,2$ &          $\,0$ &  ${\TsX}1\,525$  \\
Third-party distributor &       $19\,837$ &      $29\,140$ &      $2\,277$ & ${\TsV}54\,689$ & ${\TsV}7\,127$ & ${\TsV}4\,569$ &  ${\TsX}\,334$ &      $117\,973$  \\ \hline
Total                   &       $82\,908$ &      $80\,791$ &      $7\,961$ &      $141\,789$ &      $28\,157$ &      $11\,089$ &      $23\,264$ &      $375\,959$  \\ \hline
\end{tabular}
\smallskip

\begin{flushleft}
{\small \textit{Source}: Amundi Cube Database (2020) and authors' calculation.}
\end{flushleft}
\end{table}

\endlandscape

\landscape

\begin{table}
\centering
\caption{Breakdown of the liability dataset by investor and fund categories (without mandates and dedicated mutual funds)}
\label{tab:data3-4}
\begin{tabular}{lrrrrrrrr}
\hline
Total number $n$         & \mr{Balanced} & \mr{Bond} & Enhanced & \mr{Equity} &    Money &  \mr{Other} & \mr{Structured} & \mr{Total} \\
of observations          &          &                & Treasury &             &   Market &             &                 &            \\ \hline
Auto-consumption         &   ${\TsV}16\,147$ & ${\TsV}43\,189$ & ${\TsV}3\,783$ & ${\TsV}43\,737$ & ${\TsV}6\,008$ &      $13\,793$ &          $\,0$ & ${\TsV}\,126\,657$  \\
Central bank             &    ${\TsX}1\,281$ &         $\,580$ &          $\,0$ &         $\,476$ &          $\,0$ &          $\,0$ &          $\,0$ &         $\,2\,337$  \\
Corporate                &    ${\TsX}1\,862$ &  ${\TsX}6\,542$ & ${\TsV}2\,305$ &  ${\TsX}5\,468$ & ${\TsV}7\,812$ & ${\TsV}4\,235$ &          $\,0$ &  ${\TsX}\,28\,224$  \\
Corporate pension fund   &    ${\TsX}2\,344$ &  ${\TsX}8\,650$ &  ${\TsX}\,427$ &  ${\TsX}9\,031$ & ${\TsV}2\,670$ & ${\TsV}1\,277$ &          $\,0$ &  ${\TsX}\,24\,399$  \\
Employee savings plan    &    ${\TsX}9\,894$ &  ${\TsX}4\,240$ & ${\TsV}1\,349$ & ${\TsV}19\,145$ & ${\TsV}3\,232$ &          $\,0$ & ${\TsV}5\,279$ &  ${\TsX}\,43\,139$  \\
Institutional            &    ${\TsX}6\,858$ & ${\TsV}36\,792$ & ${\TsV}3\,716$ & ${\TsV}41\,104$ & ${\TsV}8\,329$ &      $16\,029$ &          $\,0$ & ${\TsV}\,112\,828$  \\
Insurance                &    ${\TsX}3\,436$ & ${\TsV}13\,011$ & ${\TsV}3\,303$ & ${\TsV}21\,832$ & ${\TsV}8\,543$ & ${\TsV}5\,750$ &          $\,0$ &  ${\TsX}\,55\,875$  \\
Other                    &    ${\TsX}7\,577$ & ${\TsV}12\,751$ & ${\TsV}5\,428$ &  ${\TsX}4\,155$ & ${\TsV}9\,333$ &      $11\,788$ &          $\,0$ &  ${\TsX}\,51\,032$  \\
Retail                   &        $115\,394$ & ${\TsV}77\,879$ & ${\TsV}6\,692$ & ${\TsV}95\,393$ &      $14\,798$ &      $27\,834$ &      $83\,118$ & ${\TsV}\,421\,108$  \\
Sovereign                &    ${\TsX}2\,969$ &  ${\TsX}2\,261$ &  ${\TsX}\,854$ &  ${\TsX}3\,405$ & ${\TsV}2\,853$ & ${\TsV}1\,746$ &          $\,0$ &  ${\TsX}\,14\,088$  \\
Third-party distributor  &   ${\TsV}55\,696$ & ${\TsV}75\,591$ & ${\TsV}4\,929$ &      $114\,171$ &      $10\,732$ &      $13\,483$ & ${\TsV}5\,126$ & ${\TsV}\,279\,728$  \\ \hline
Total                    &        $223\,458$ &      $281\,486$ &      $32\,786$ &      $357\,917$ &      $74\,310$ &      $95\,935$ &      $93\,523$ &      $1\,159\,415$  \\ \hline
& \\
\hline
Total number $n_1$      & \mr{Balanced} & \mr{Bond} & Enhanced & \mr{Equity} &    Money &  \mr{Other} & \mr{Structured} & \mr{Total} \\
of redemptions          &          &                & Treasury &             &   Market &             &                 &            \\ \hline
Auto-consumption        &  ${\TsV}3\,492$ & ${\TsV}8\,385$ &      $1\,135$ & ${\TsV}11\,137$ & ${\TsV}3\,040$ &  ${\TsX}\,881$ &          $\,0$ & ${\TsV}28\,070$  \\
Central bank            &           $\,2$ &          $\,2$ &         $\,0$ &           $\,7$ &          $\,0$ &          $\,0$ &          $\,0$ &          $\,11$  \\
Corporate               &   ${\TsX}\,280$ &  ${\TsX}\,405$ & ${\TsV}\,144$ &         $\,157$ & ${\TsV}3\,110$ &          $\,9$ &          $\,0$ &  ${\TsX}4\,105$  \\
Corporate pension fund  &   ${\TsX}\,190$ &  ${\TsX}\,292$ &  ${\TsX\,}17$ &         $\,304$ &  ${\TsX}\,202$ &          $\,0$ &          $\,0$ &  ${\TsX}1\,005$  \\
Employee savings plan   &   ${\TsX}\,264$ &  ${\TsX}\,120$ &  ${\TsX\,}40$ &         $\,519$ &         $\,74$ &          $\,0$ &  ${\TsX}\,145$ &  ${\TsX}1\,162$  \\
Institutional           &  ${\TsV}1\,328$ & ${\TsV}2\,312$ &  ${\TsX\,}73$ &  ${\TsX}2\,677$ & ${\TsV}2\,734$ &  ${\TsX}\,166$ &          $\,0$ &  ${\TsX}9\,290$  \\
Insurance               &   ${\TsX}\,419$ &  ${\TsX}\,874$ & ${\TsV}\,114$ &  ${\TsX}1\,576$ & ${\TsV}2\,385$ &         $\,60$ &          $\,0$ &  ${\TsX}5\,428$  \\
Other                   &   ${\TsX}\,733$ &  ${\TsX}\,493$ & ${\TsV}\,200$ &         $\,804$ & ${\TsV}2\,008$ &  ${\TsX}\,262$ &          $\,0$ &  ${\TsX}4\,500$  \\
Retail                  &       $51\,454$ &      $35\,079$ &      $3\,932$ & ${\TsV}67\,250$ & ${\TsV}6\,770$ & ${\TsV}4\,875$ &      $22\,707$ &      $192\,067$  \\
Sovereign               &   ${\TsX}\,484$ &         $\,72$ &         $\,9$ &         $\,343$ &  ${\TsX}\,520$ &          $\,1$ &          $\,0$ &  ${\TsX}1\,429$  \\
Third-party distributor &       $18\,808$ &      $28\,242$ &      $2\,266$ & ${\TsV}52\,445$ & ${\TsV}7\,077$ & ${\TsV}4\,431$ &  ${\TsX}\,334$ &      $113\,603$  \\ \hline
Total                   &       $77\,454$ &      $76\,276$ &      $7\,930$ &      $137\,219$ &      $27\,920$ &      $10\,685$ &      $23\,186$ &      $360\,670$  \\ \hline
\end{tabular}
\smallskip

\begin{flushleft}
{\small \textit{Source}: Amundi Cube Database (2020) and authors' calculation.}
\end{flushleft}
\end{table}

\endlandscape

\section{Additional results}
\label{appendix:additional-results}

\begin{figure}[tbph]
\centering
\caption{Third-party distributor}
\label{fig:rate3}
\includegraphics[width = \figurewidth, height = \figureheight]{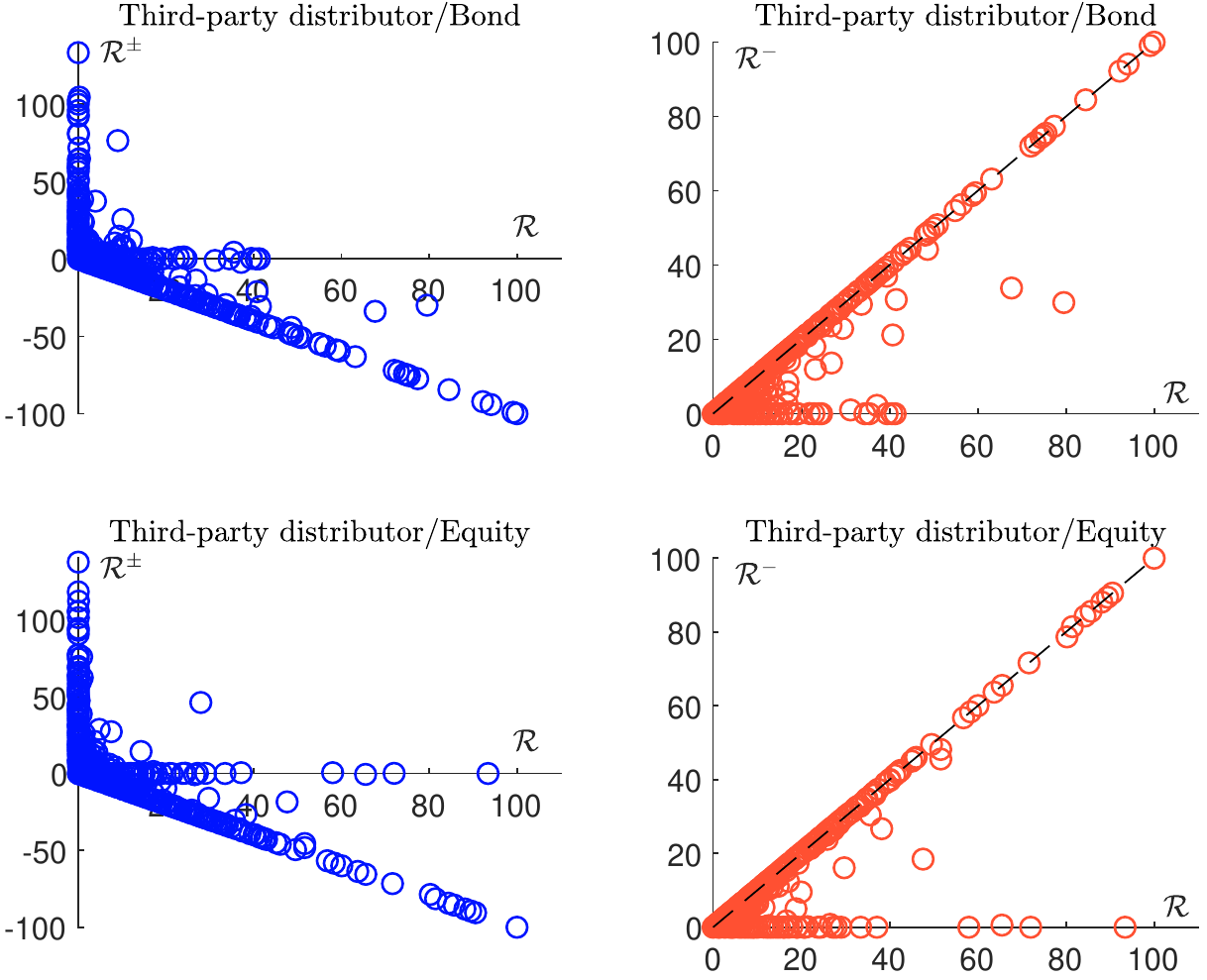}
\end{figure}

\begin{figure}[tbph]
\centering
\caption{Relationship between the stress scenario of the big fund and the stress scenario of $n$ equivalent small funds}
\label{fig:xstatistic2}
\includegraphics[width = \figurewidth, height = \figureheight]{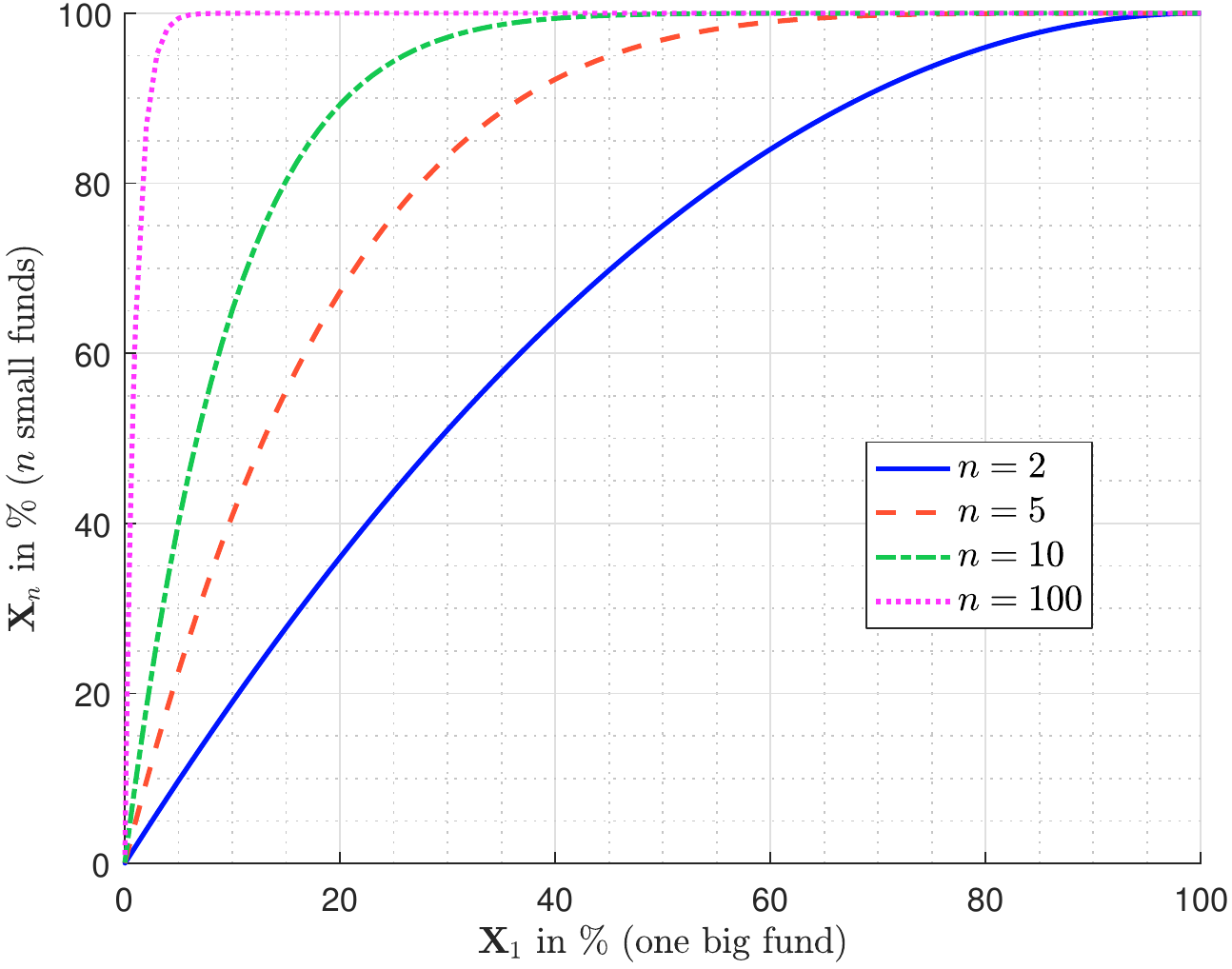}
\end{figure}

\begin{figure}[tbph]
\centering
\caption{Relationship between the confidence level $\alpha$ of $\mathbf{F}^{-1}\left(\alpha\right)$
and the confidence level $\alpha_{\mathbf{G}}$ of $\mathbf{G}^{-1}\left(\alpha_{\mathbf{G}}\right)$}
\label{fig:inflated4}
\includegraphics[width = \figurewidth, height = \figureheight]{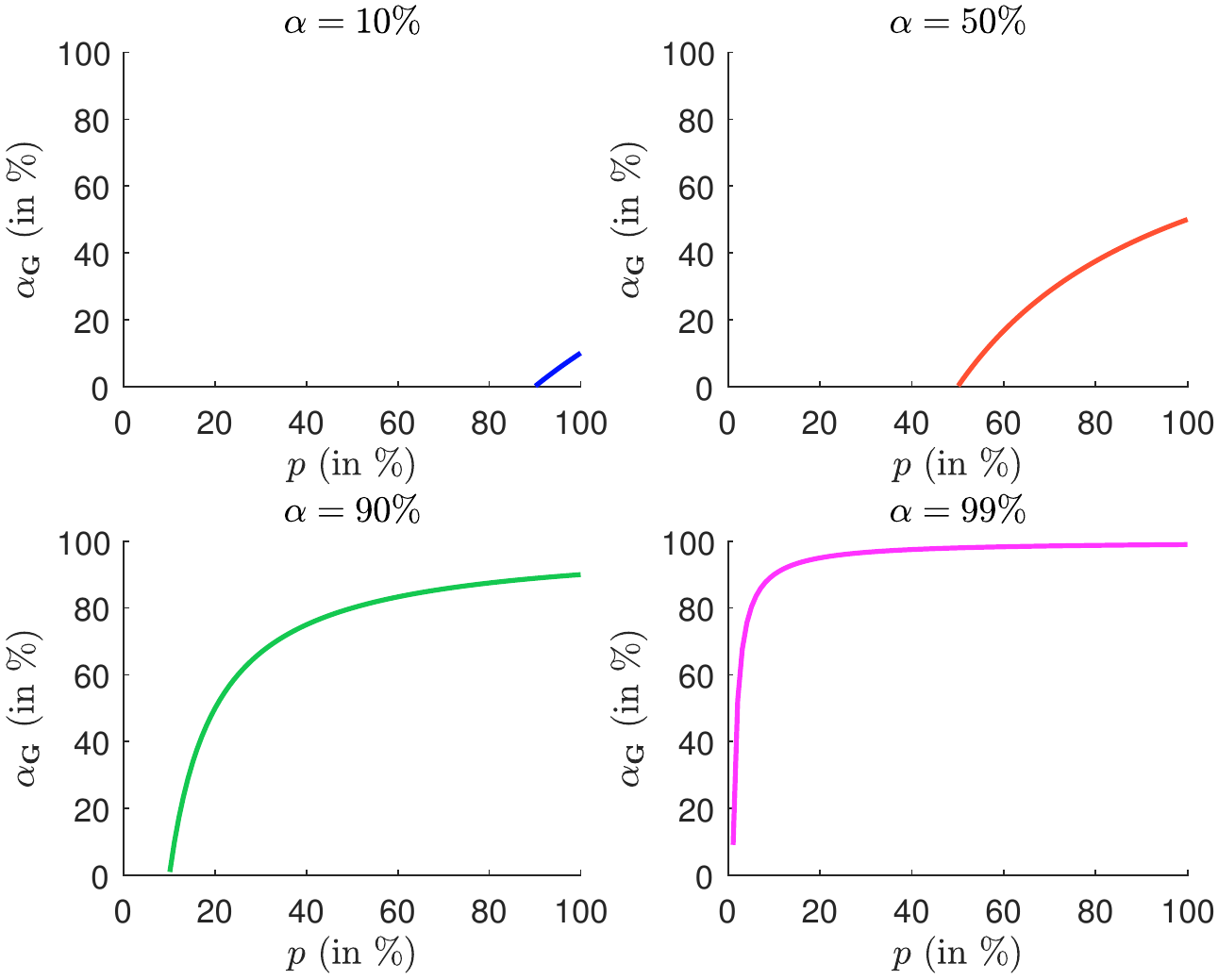}
\end{figure}

\begin{figure}[tbph]
\centering
\caption{Stress scenario $\mathbb{S}\left(\mathcal{T}\right)$ in \% ($p = 5\%$)}
\label{fig:inflated11}
\includegraphics[width = \figurewidth, height = \figureheight]{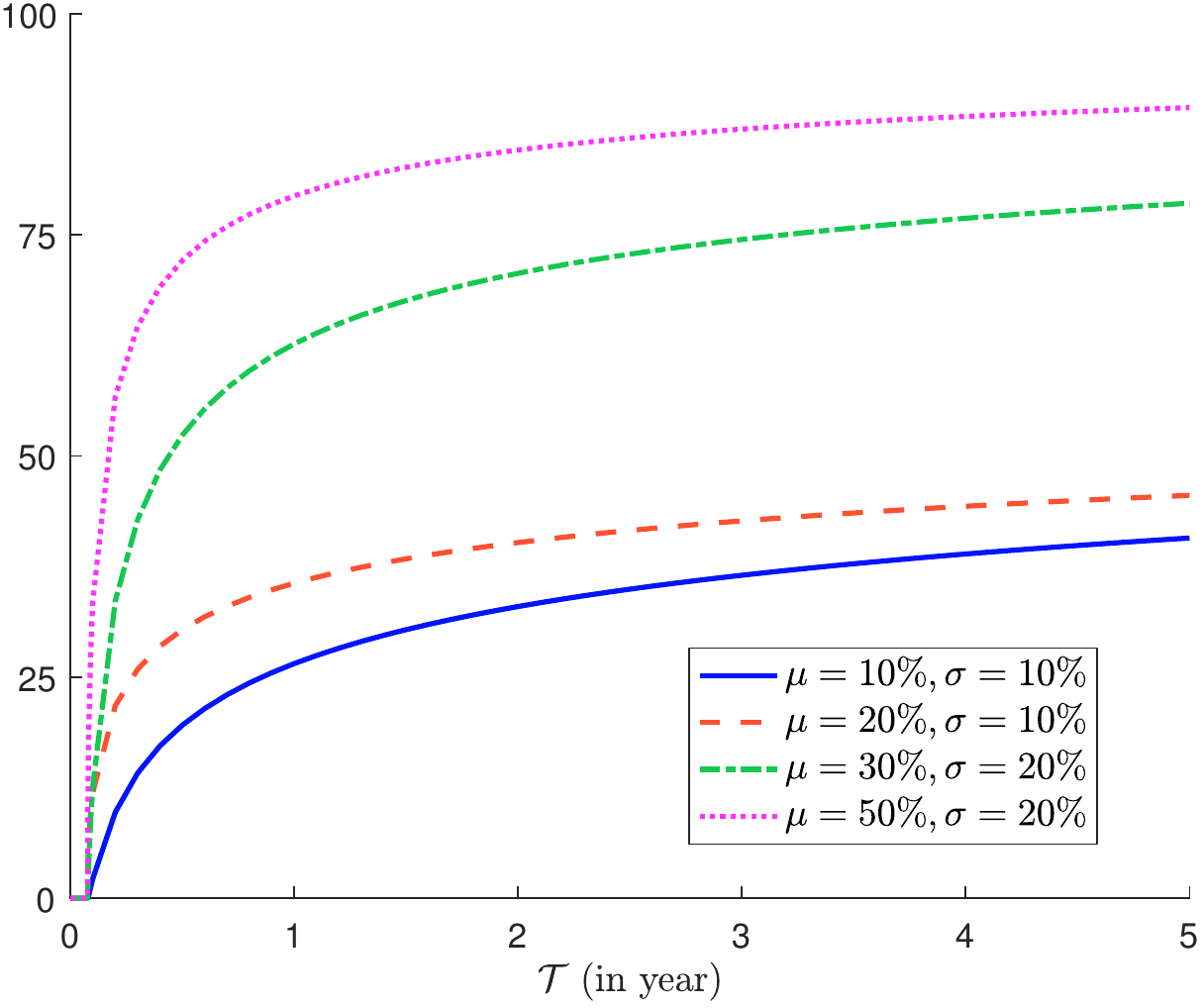}
\end{figure}

\begin{figure}[tbph]
\centering
\caption{Stress scenario $\mathbb{S}\left(\mathcal{T}\right)$ in \% ($p = 50\%$)}
\label{fig:inflated12}
\includegraphics[width = \figurewidth, height = \figureheight]{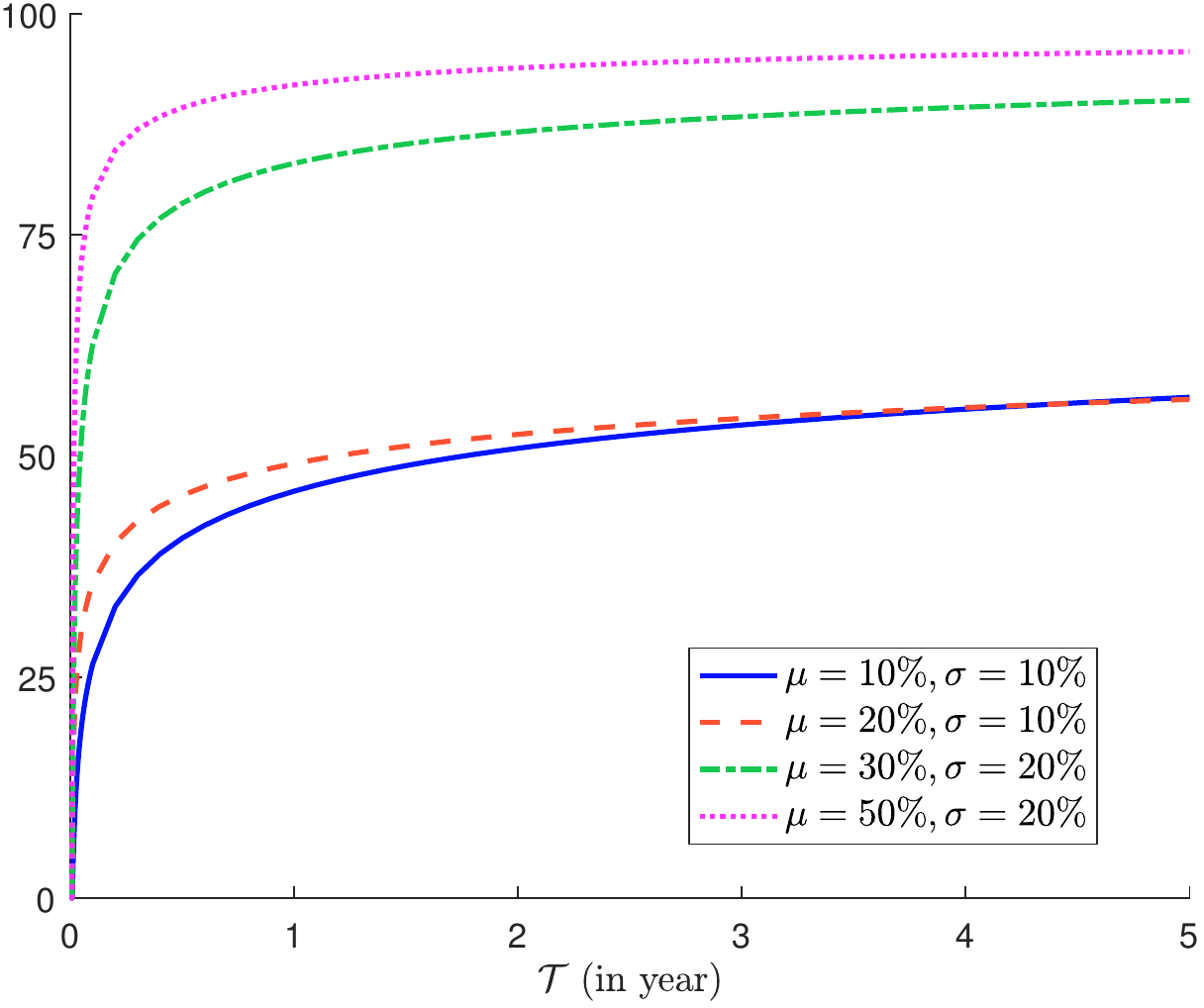}
\end{figure}

\begin{table}[tbph]
\centering
\caption{Estimated value of $a$ (method of moments)}
\label{tab:inflated14-a-mm}
\begin{tabular}{lcccccccc}
\hline
                        &         (1) &         (2) &         (3) &         (4) &         (5) &         (6) &         (7) &         (8)  \\ \hline
Auto-consumption        &       $0.02$ &      $0.05$ &      $0.03$ &      $0.03$ &      $0.09$ &      $0.06$ & ${\TsV}   $ &      $0.04$  \\
Central bank            &  ${\TsV}   $ & ${\TsV}   $ & ${\TsV}   $ & ${\TsV}   $ & ${\TsV}   $ & ${\TsV}   $ & ${\TsV}   $ & ${\TsV}   $  \\
Corporate               &       $0.00$ &      $0.04$ & ${\TsV}   $ & ${\TsV}   $ &      $0.21$ & ${\TsV}   $ & ${\TsV}   $ &      $0.14$  \\
Corporate pension fund  &  ${\TsV}   $ &      $0.01$ & ${\TsV}   $ &      $0.01$ &      $0.21$ & ${\TsV}   $ & ${\TsV}   $ &      $0.04$  \\
Employee savings plan   &       $0.14$ & ${\TsV}   $ & ${\TsV}   $ &      $0.04$ & ${\TsV}   $ & ${\TsV}   $ & ${\TsV}   $ &      $0.04$  \\
Institutional           &       $0.01$ &      $0.04$ & ${\TsV}   $ &      $0.06$ &      $0.10$ & ${\TsV}   $ & ${\TsV}   $ &      $0.05$  \\
Insurance               &       $0.01$ &      $0.02$ & ${\TsV}   $ &      $0.02$ &      $0.12$ & ${\TsV}   $ & ${\TsV}   $ &      $0.05$  \\
Other                   &       $0.05$ &      $0.05$ & ${\TsV}   $ &      $0.01$ &      $0.05$ &      $0.01$ & ${\TsV}   $ &      $0.03$  \\
Retail                  &       $0.01$ &      $0.01$ &      $0.01$ &      $0.01$ &      $0.05$ &      $0.01$ &      $0.00$ &      $0.01$  \\
Sovereign               &       $0.05$ & ${\TsV}   $ & ${\TsV}   $ &      $0.02$ &      $0.11$ & ${\TsV}   $ & ${\TsV}   $ &      $0.04$  \\
Third-party distributor &       $0.01$ &      $0.03$ &      $0.02$ &      $0.02$ &      $0.07$ &      $0.01$ &      $0.02$ &      $0.02$  \\ \hline
Total                   &       $0.01$ &      $0.02$ &      $0.02$ &      $0.01$ &      $0.07$ &      $0.02$ &      $0.00$ &      $0.02$  \\
\hline
\end{tabular}
\medskip

\begin{flushleft}
\begin{footnotesize}
(1) = balanced, (2) = bond, (3) = enhanced treasury, (4) = equity, (5) = money
market, (6) = other, (7) = structured, (8) = total
\end{footnotesize}
\end{flushleft}
\end{table}

\begin{table}[tbph]
\centering
\caption{Estimated value of $b$ (method of moments)}
\label{tab:inflated14-b-mm}
\begin{tabular}{lcccccccc}
\hline
                        &           (1) &         (2) &         (3) &         (4) &         (5) &         (6) &         (7) &         (8) \\ \hline
Auto-consumption        &  ${\TsV}1.23$ &      $2.86$ &      $1.20$ &      $2.25$ &      $2.81$ &      $2.23$ & ${\TsV}   $ &      $2.28$  \\
Central bank            &   ${\TsX}   $ & ${\TsV}   $ & ${\TsV}   $ & ${\TsV}   $ & ${\TsV}   $ & ${\TsV}   $ & ${\TsV}   $ & ${\TsV}   $  \\
Corporate               &  ${\TsV}0.78$ &      $1.62$ & ${\TsV}   $ & ${\TsV}   $ &      $5.34$ & ${\TsV}   $ & ${\TsV}   $ &      $3.61$  \\
Corporate pension fund  &   ${\TsX}   $ &      $0.41$ & ${\TsV}   $ &      $0.50$ &      $2.69$ & ${\TsV}   $ & ${\TsV}   $ &      $1.09$  \\
Employee savings plan   &       $10.89$ & ${\TsV}   $ & ${\TsV}   $ &      $1.84$ & ${\TsV}   $ & ${\TsV}   $ & ${\TsV}   $ &      $1.73$  \\
Institutional           &  ${\TsV}1.21$ &      $1.52$ & ${\TsV}   $ &      $2.13$ &      $2.15$ & ${\TsV}   $ & ${\TsV}   $ &      $1.60$  \\
Insurance               &  ${\TsV}0.78$ &      $0.91$ & ${\TsV}   $ &      $1.07$ &      $3.58$ & ${\TsV}   $ & ${\TsV}   $ &      $1.91$  \\
Other                   &  ${\TsV}5.58$ &      $1.84$ & ${\TsV}   $ &      $1.04$ &      $1.35$ &      $1.17$ & ${\TsV}   $ &      $1.23$  \\
Retail                  &  ${\TsV}3.29$ &      $3.56$ &      $2.98$ &      $4.11$ &      $2.39$ &      $3.07$ &      $1.11$ &      $3.00$  \\
Sovereign               &       $86.69$ & ${\TsV}   $ & ${\TsV}   $ &      $0.83$ &      $0.90$ & ${\TsV}   $ & ${\TsV}   $ &      $0.87$  \\
Third-party distributor &  ${\TsV}3.83$ &      $4.21$ &      $1.44$ &      $5.22$ &      $5.14$ &      $1.48$ &      $1.43$ &      $3.89$  \\ \hline
Total                   &  ${\TsV}2.68$ &      $2.80$ &      $1.01$ &      $2.89$ &      $2.58$ &      $1.60$ &      $0.97$ &      $2.43$  \\
\hline
\end{tabular}
\medskip

\begin{flushleft}
\begin{footnotesize}
(1) = balanced, (2) = bond, (3) = enhanced treasury, (4) = equity, (5) = money
market, (6) = other, (7) = structured, (8) = total
\end{footnotesize}
\end{flushleft}
\end{table}

\begin{table}[tbph]
\centering
\caption{Estimated value of $a$ (method of maximum likelihood)}
\label{tab:inflated14-a-ml}
\begin{tabular}{lcccccccc}
\hline
                        &         (1) &         (2) &         (3) &         (4) &         (5) &         (6) &         (7) &         (8) \\ \hline
Auto-consumption        &       $0.20$ &      $0.26$ &      $0.26$ &      $0.23$ &      $0.32$ &      $0.25$ & ${\TsV}   $ &      $0.24$  \\
Central bank            &  ${\TsV}   $ & ${\TsV}   $ & ${\TsV}   $ & ${\TsV}   $ & ${\TsV}   $ & ${\TsV}   $ & ${\TsV}   $ & ${\TsV}   $  \\
Corporate               &       $0.23$ &      $0.19$ & ${\TsV}   $ & ${\TsV}   $ &      $0.39$ & ${\TsV}   $ & ${\TsV}   $ &      $0.30$  \\
Corporate pension fund  &  ${\TsV}   $ &      $0.13$ & ${\TsV}   $ &      $0.13$ &      $0.37$ & ${\TsV}   $ & ${\TsV}   $ &      $0.16$  \\
Employee savings plan   &       $1.03$ & ${\TsV}   $ & ${\TsV}   $ &      $0.52$ & ${\TsV}   $ & ${\TsV}   $ & ${\TsV}   $ &      $0.57$  \\
Institutional           &       $0.22$ &      $0.19$ & ${\TsV}   $ &      $0.21$ &      $0.28$ & ${\TsV}   $ & ${\TsV}   $ &      $0.22$  \\
Insurance               &       $0.14$ &      $0.15$ & ${\TsV}   $ &      $0.17$ &      $0.28$ & ${\TsV}   $ & ${\TsV}   $ &      $0.19$  \\
Other                   &       $0.26$ &      $0.21$ & ${\TsV}   $ &      $0.27$ &      $0.25$ &      $0.28$ & ${\TsV}   $ &      $0.23$  \\
Retail                  &       $0.31$ &      $0.30$ &      $0.26$ &      $0.33$ &      $0.27$ &      $0.27$ &      $0.36$ &      $0.29$  \\
Sovereign               &       $0.68$ & ${\TsV}   $ & ${\TsV}   $ &      $0.17$ &      $0.31$ & ${\TsV}   $ & ${\TsV}   $ &      $0.19$  \\
Third-party distributor &       $0.40$ &      $0.28$ &      $0.24$ &      $0.30$ &      $0.34$ &      $0.27$ &      $0.26$ &      $0.30$  \\ \hline
Total                   &       $0.29$ &      $0.25$ &      $0.23$ &      $0.27$ &      $0.29$ &      $0.24$ &      $0.32$ &      $0.25$  \\
\hline
\end{tabular}
\medskip

\begin{flushleft}
\begin{footnotesize}
(1) = balanced, (2) = bond, (3) = enhanced treasury, (4) = equity, (5) = money
market, (6) = other, (7) = structured, (8) = total
\end{footnotesize}
\end{flushleft}
\end{table}

\begin{table}[tbph]
\centering
\caption{Estimated value of $b$ (method of maximum likelihood)}
\label{tab:inflated14-b-ml}
\begin{tabular}{lcccccccc}
\hline
                        &           (1) &          (2) &          (3) &           (4) &          (5) &          (6) &           (7) &          (8) \\ \hline
Auto-consumption        &          $6.53$ &      $11.36$ &      $17.89$ & ${\TsV}15.80$ & ${\TsV}7.50$ & ${\TsV}8.40$ &         $   $ &      $10.51$  \\
Central bank            &           $   $ &  ${\TsX}   $ &  ${\TsX}   $ &         $   $ &  ${\TsX}   $ &  ${\TsX}   $ &         $   $ &  ${\TsX}   $  \\
Corporate               &   ${\TsX}26.32$ & ${\TsV}6.03$ &  ${\TsX}   $ &         $   $ & ${\TsV}8.96$ &  ${\TsX}   $ &         $   $ & ${\TsV}6.55$  \\
Corporate pension fund  &           $   $ & ${\TsV}1.66$ &  ${\TsX}   $ &  ${\TsX}3.12$ & ${\TsV}4.14$ &  ${\TsX}   $ &         $   $ & ${\TsV}2.70$  \\
Employee savings plan   &   ${\TsX}74.62$ &  ${\TsX}   $ &  ${\TsX}   $ & ${\TsV}24.41$ &  ${\TsX}   $ &  ${\TsX}   $ &         $   $ &      $30.29$  \\
Institutional           &   ${\TsX}16.24$ & ${\TsV}4.94$ &  ${\TsX}   $ &  ${\TsX}5.65$ & ${\TsV}4.64$ &  ${\TsX}   $ &         $   $ & ${\TsV}5.04$  \\
Insurance               &          $3.56$ & ${\TsV}5.42$ &  ${\TsX}   $ &  ${\TsX}5.26$ & ${\TsV}7.14$ &  ${\TsX}   $ &         $   $ & ${\TsV}5.46$  \\
Other                   &   ${\TsX}28.00$ & ${\TsV}7.99$ &  ${\TsX}   $ & ${\TsV}31.90$ & ${\TsV}4.82$ &      $43.34$ &         $   $ & ${\TsV}6.72$  \\
Retail                  &   ${\TsX}56.99$ &      $82.20$ &      $51.08$ &      $116.86$ &      $10.51$ &      $48.15$ &      $309.26$ &      $65.57$  \\
Sovereign               &       $1225.65$ &  ${\TsX}   $ &  ${\TsX}   $ &  ${\TsX}4.84$ & ${\TsV}2.06$ &  ${\TsX}   $ &         $   $ & ${\TsV}2.92$  \\
Third-party distributor &  ${\TsV}111.38$ &      $39.19$ &      $15.23$ & ${\TsV}64.70$ &      $21.61$ &      $36.15$ & ${\TsV}12.11$ &      $47.77$  \\ \hline
Total                   &   ${\TsX}44.28$ &      $26.79$ &      $15.16$ & ${\TsV}44.67$ & ${\TsV}7.80$ &      $20.02$ &      $206.39$ &      $26.76$  \\
\hline
\end{tabular}
\medskip

\begin{flushleft}
\begin{footnotesize}
(1) = balanced, (2) = bond, (3) = enhanced treasury, (4) = equity, (5) = money
market, (6) = other, (7) = structured, (8) = total
\end{footnotesize}
\end{flushleft}
\end{table}

\begin{table}[tbph]
\centering
\caption{Estimated value of $\mu$ in \% (method of maximum likelihood)}
\label{tab:inflated14-mu-ml}
\begin{tabular}{lcccccccc}
\hline
                        &         (1) &         (2) &         (3) &         (4) &          (5) &         (6) &         (7) &         (8) \\ \hline
Auto-consumption        &       $2.92$ &      $2.23$ &      $1.45$ &      $1.43$ & ${\TsV}4.08$ &      $2.91$ & ${\TsV}   $ &      $2.20$  \\
Central bank            &  ${\TsV}   $ & ${\TsV}   $ & ${\TsV}   $ & ${\TsV}   $ &  ${\TsX}   $ & ${\TsV}   $ & ${\TsV}   $ & ${\TsV}   $  \\
Corporate               &       $0.87$ &      $3.10$ & ${\TsV}   $ & ${\TsV}   $ & ${\TsV}4.12$ & ${\TsV}   $ & ${\TsV}   $ &      $4.43$  \\
Corporate pension fund  &  ${\TsV}   $ &      $7.47$ & ${\TsV}   $ &      $4.03$ & ${\TsV}8.29$ & ${\TsV}   $ & ${\TsV}   $ &      $5.58$  \\
Employee savings plan   &       $1.36$ & ${\TsV}   $ & ${\TsV}   $ &      $2.07$ &  ${\TsX}   $ & ${\TsV}   $ & ${\TsV}   $ &      $1.85$  \\
Institutional           &       $1.32$ &      $3.78$ & ${\TsV}   $ &      $3.55$ & ${\TsV}5.77$ & ${\TsV}   $ & ${\TsV}   $ &      $4.11$  \\
Insurance               &       $3.78$ &      $2.62$ & ${\TsV}   $ &      $3.08$ & ${\TsV}3.80$ & ${\TsV}   $ & ${\TsV}   $ &      $3.44$  \\
Other                   &       $0.93$ &      $2.56$ & ${\TsV}   $ &      $0.83$ & ${\TsV}4.99$ &      $0.64$ & ${\TsV}   $ &      $3.26$  \\
Retail                  &       $0.54$ &      $0.36$ &      $0.51$ &      $0.28$ & ${\TsV}2.47$ &      $0.56$ &      $0.12$ &      $0.44$  \\
Sovereign               &       $0.06$ & ${\TsV}   $ & ${\TsV}   $ &      $3.47$ &      $13.01$ & ${\TsV}   $ & ${\TsV}   $ &      $6.08$  \\
Third-party distributor &       $0.36$ &      $0.72$ &      $1.55$ &      $0.46$ & ${\TsV}1.55$ &      $0.75$ &      $2.13$ &      $0.62$  \\ \hline
Total                   &       $0.66$ &      $0.92$ &      $1.46$ &      $0.59$ & ${\TsV}3.53$ &      $1.16$ &      $0.15$ &      $0.92$  \\
\hline
\end{tabular}
\medskip

\begin{flushleft}
\begin{footnotesize}
(1) = balanced, (2) = bond, (3) = enhanced treasury, (4) = equity, (5) = money
market, (6) = other, (7) = structured, (8) = total
\end{footnotesize}
\end{flushleft}
\end{table}

\begin{table}[tbph]
\centering
\caption{Estimated value of $\sigma$ in \% (method of maximum likelihood)}
\label{tab:inflated14-sigma-ml}
\begin{tabular}{lcccccccc}
\hline
                        &          (1) &          (2) &          (3) &          (4) &          (5) &          (6) &         (7) &          (8) \\ \hline
Auto-consumption        &       $6.05$ & ${\TsV}4.16$ &      $2.74$ &      $2.88$ & ${\TsV}6.66$ &      $5.41$ & ${\TsV}   $ & ${\TsV}4.28$  \\
Central bank            &  ${\TsV}   $ &  ${\TsX}   $ & ${\TsV}   $ & ${\TsV}   $ &  ${\TsX}   $ & ${\TsV}   $ & ${\TsV}   $ &  ${\TsX}   $  \\
Corporate               &       $1.77$ & ${\TsV}6.45$ & ${\TsV}   $ & ${\TsV}   $ & ${\TsV}6.18$ & ${\TsV}   $ & ${\TsV}   $ & ${\TsV}7.34$  \\
Corporate pension fund  &  ${\TsV}   $ &      $15.74$ & ${\TsV}   $ &      $9.53$ &      $11.74$ & ${\TsV}   $ & ${\TsV}   $ &      $11.68$  \\
Employee savings plan   &       $1.32$ &  ${\TsX}   $ & ${\TsV}   $ &      $2.80$ &  ${\TsX}   $ & ${\TsV}   $ & ${\TsV}   $ & ${\TsV}2.39$  \\
Institutional           &       $2.73$ & ${\TsV}7.70$ & ${\TsV}   $ &      $7.06$ & ${\TsV}9.59$ & ${\TsV}   $ & ${\TsV}   $ & ${\TsV}7.94$  \\
Insurance               &       $8.80$ & ${\TsV}6.23$ & ${\TsV}   $ &      $6.82$ & ${\TsV}6.58$ & ${\TsV}   $ & ${\TsV}   $ & ${\TsV}7.07$  \\
Other                   &       $1.78$ & ${\TsV}5.21$ & ${\TsV}   $ &      $1.58$ & ${\TsV}8.84$ &      $1.20$ & ${\TsV}   $ & ${\TsV}6.30$  \\
Retail                  &       $0.96$ & ${\TsV}0.65$ &      $0.99$ &      $0.48$ & ${\TsV}4.52$ &      $1.06$ &      $0.19$ & ${\TsV}0.81$  \\
Sovereign               &       $0.07$ &  ${\TsX}   $ & ${\TsV}   $ &      $7.46$ &      $18.33$ & ${\TsV}   $ & ${\TsV}   $ &      $11.79$  \\
Third-party distributor &       $0.56$ & ${\TsV}1.33$ &      $3.04$ &      $0.83$ & ${\TsV}2.58$ &      $1.41$ &      $3.95$ & ${\TsV}1.12$  \\ \hline
Total                   &       $1.20$ & ${\TsV}1.80$ &      $2.97$ &      $1.13$ & ${\TsV}6.13$ &      $2.33$ &      $0.27$ & ${\TsV}1.81$  \\
\hline
\end{tabular}
\medskip

\begin{flushleft}
\begin{footnotesize}
(1) = balanced, (2) = bond, (3) = enhanced treasury, (4) = equity, (5) = money
market, (6) = other, (7) = structured, (8) = total
\end{footnotesize}
\end{flushleft}
\end{table}

\begin{table}[tbph]
\centering
\caption{Volatility of the redemption rate in \%}
\label{tab:individual7}
\begin{tabular}{lcccccccc}
\hline
                        &          (1) &          (2) &          (3) &          (4) &          (5) &          (6) &         (7) &          (8) \\ \hline
Auto-consumption        &       $3.47$ &      $3.11$ &      $5.42$ &      $3.06$ & ${\TsV}6.45$ &      $2.40$ & ${\TsV}   $ &      $3.41$  \\
Central bank            &       $0.33$ &      $1.25$ & ${\TsV}   $ &      $2.29$ &  ${\TsX}   $ & ${\TsV}   $ & ${\TsV}   $ &      $1.23$  \\
Corporate               &       $2.16$ &      $2.45$ &      $3.43$ &      $3.07$ & ${\TsV}5.08$ &      $2.22$ & ${\TsV}   $ &      $3.57$  \\
Corporate pension fund  &       $3.02$ &      $1.92$ &      $1.03$ &      $2.53$ & ${\TsV}4.09$ &      $0.00$ & ${\TsV}   $ &      $2.53$  \\
Employee savings plan   &       $0.57$ &      $0.41$ &      $2.75$ &      $1.42$ & ${\TsV}0.59$ & ${\TsV}   $ &      $2.65$ &      $1.45$  \\
Institutional           &       $2.42$ &      $2.58$ &      $7.07$ &      $2.45$ & ${\TsV}6.89$ &      $1.87$ & ${\TsV}   $ &      $3.24$  \\
Insurance               &       $3.05$ &      $2.79$ &      $1.49$ &      $2.77$ & ${\TsV}4.53$ &      $1.89$ & ${\TsV}   $ &      $3.02$  \\
Other                   &       $1.15$ &      $1.90$ &      $4.79$ &      $3.22$ & ${\TsV}5.70$ &      $1.01$ & ${\TsV}   $ &      $3.26$  \\
Retail                  &       $1.88$ &      $1.74$ &      $2.55$ &      $1.76$ & ${\TsV}5.18$ &      $1.36$ &      $1.38$ &      $1.95$  \\
Sovereign               &       $0.10$ &      $0.45$ &      $1.66$ &      $3.19$ &      $10.07$ &      $2.39$ & ${\TsV}   $ &      $4.94$  \\
Third-party distributor &       $1.57$ &      $2.15$ &      $5.22$ &      $1.76$ & ${\TsV}3.88$ &      $3.37$ &      $1.80$ &      $2.17$  \\ \hline
Total                   &       $1.96$ &      $2.29$ &      $4.45$ &      $2.18$ & ${\TsV}5.48$ &      $2.05$ &      $1.51$ &      $2.56$  \\
\hline
\end{tabular}
\medskip

\begin{flushleft}
\begin{footnotesize}
(1) = balanced, (2) = bond, (3) = enhanced treasury, (4) = equity, (5) = money
market, (6) = other, (7) = structured, (8) = total
\end{footnotesize}
\end{flushleft}
\end{table}

\begin{figure}[tbph]
\centering
\caption{Liability weights in the case of the geometric liability structure $\omega
_{i}\propto q^{i}$}
\label{fig:herfindahl1}
\includegraphics[width = \figurewidth, height = \figureheight]{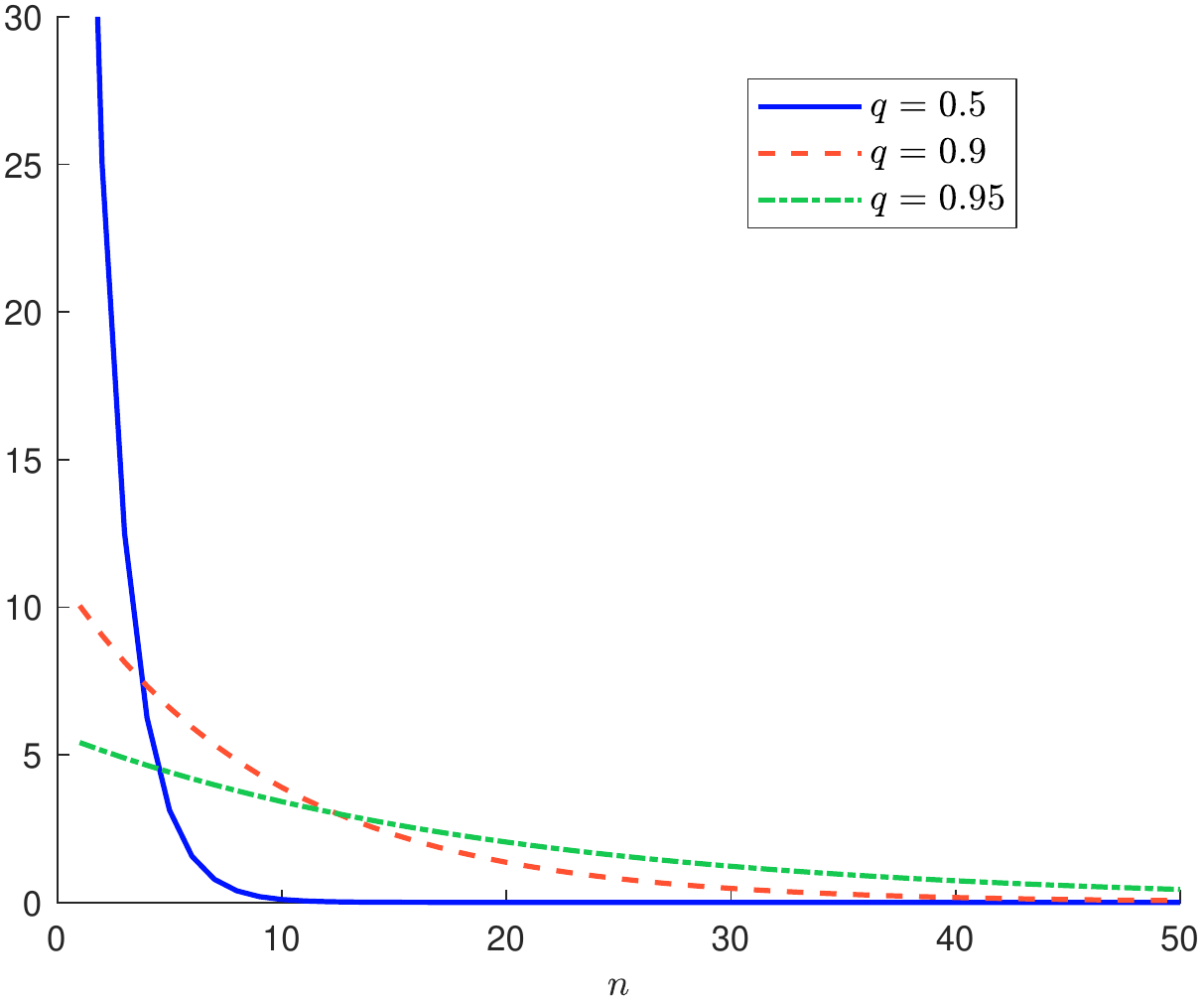}
\end{figure}

\begin{figure}[tbph]
\centering
\caption{Comparison of $\mathbf{\tilde{F}}\left( x\mid \omega \right) $ and $\mathbf{\tilde{F}}\left( x\mid
\mathcal{H}\right) $ ($q = 0.9$ and $\mathcal{H}\left( \omega \right)^{-1} = 18$))}
\label{fig:herfindahl5}
\includegraphics[width = \figurewidth, height = \figureheight]{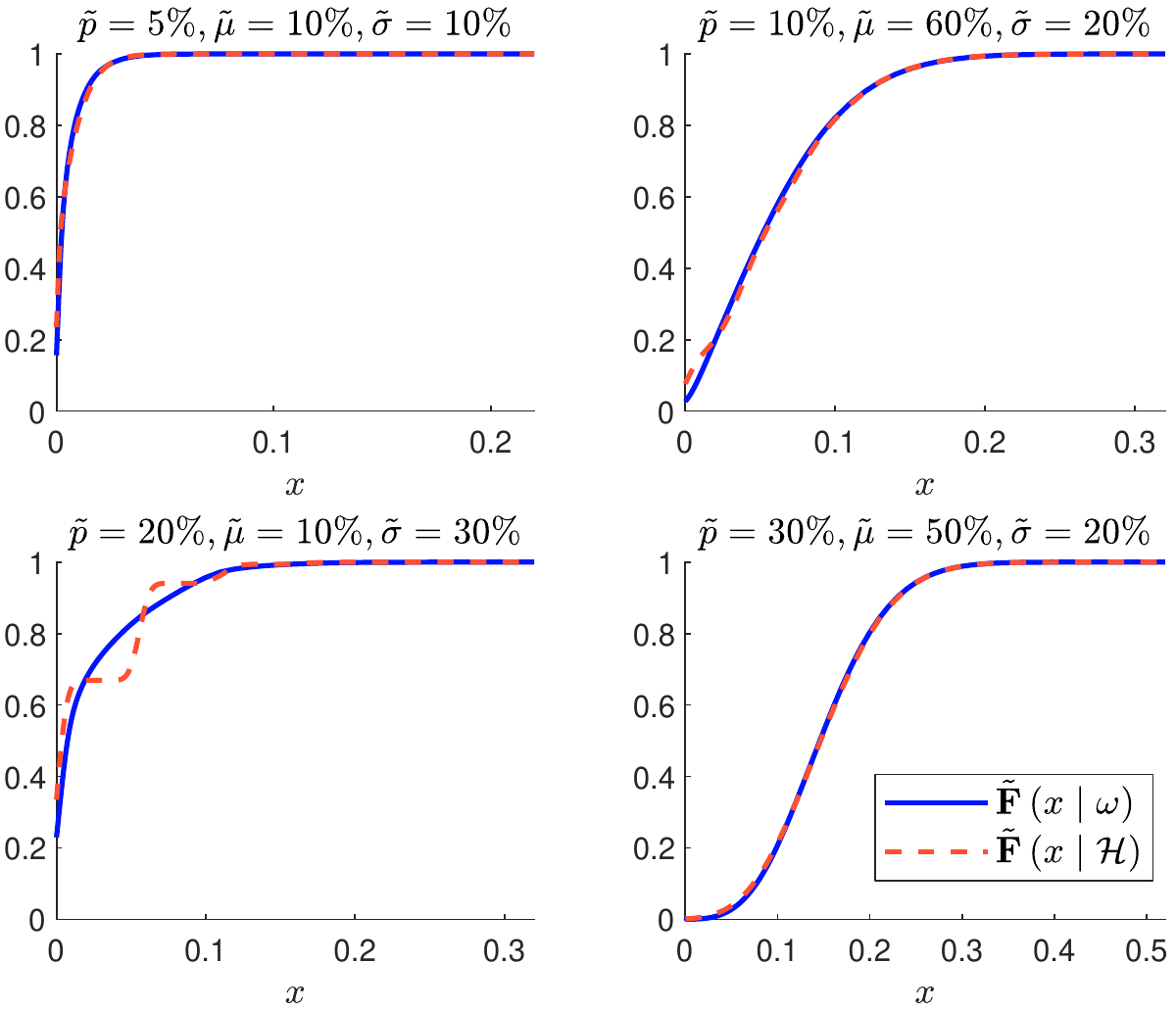}
\end{figure}

\begin{figure}[tbph]
\centering
\caption{Comparison of $\mathbf{\tilde{F}}\left( x\mid \omega \right) $ and $\mathbf{\tilde{F}}\left( x\mid
\mathcal{H}\right) $ ($q = 0.5$ and $\mathcal{H}\left( \omega \right)^{-1} = 3$))}
\label{fig:herfindahl6}
\includegraphics[width = \figurewidth, height = \figureheight]{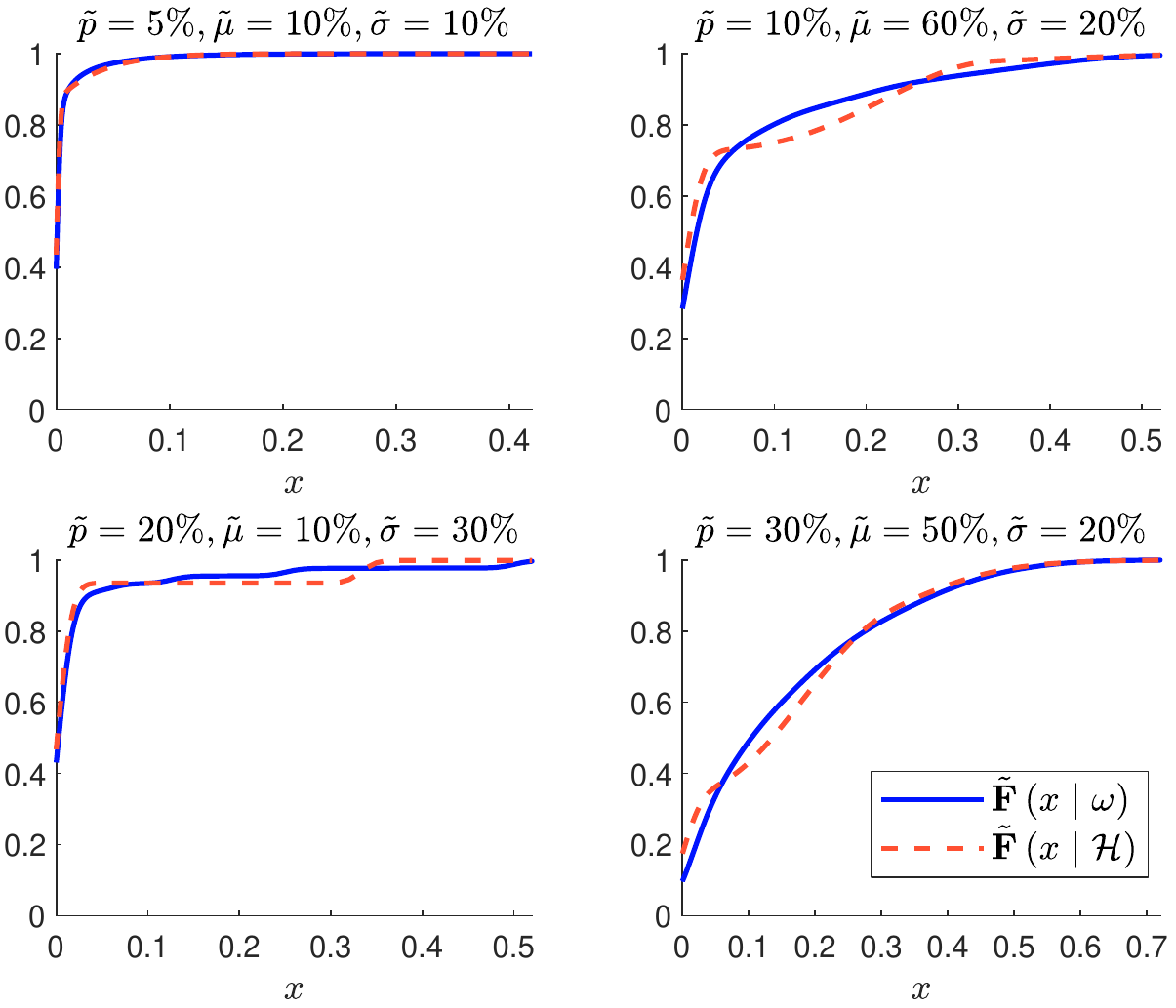}
\end{figure}

\begin{figure}[tbph]
\centering
\caption{Probability to observe no redemption $\Pr \left\{ \redemption=0\right\} $ in \%
with respect to the number $n$ of unitholders ($\tilde{p} = 10\%$)}
\label{fig:copula3}
\includegraphics[width = \figurewidth, height = \figureheight]{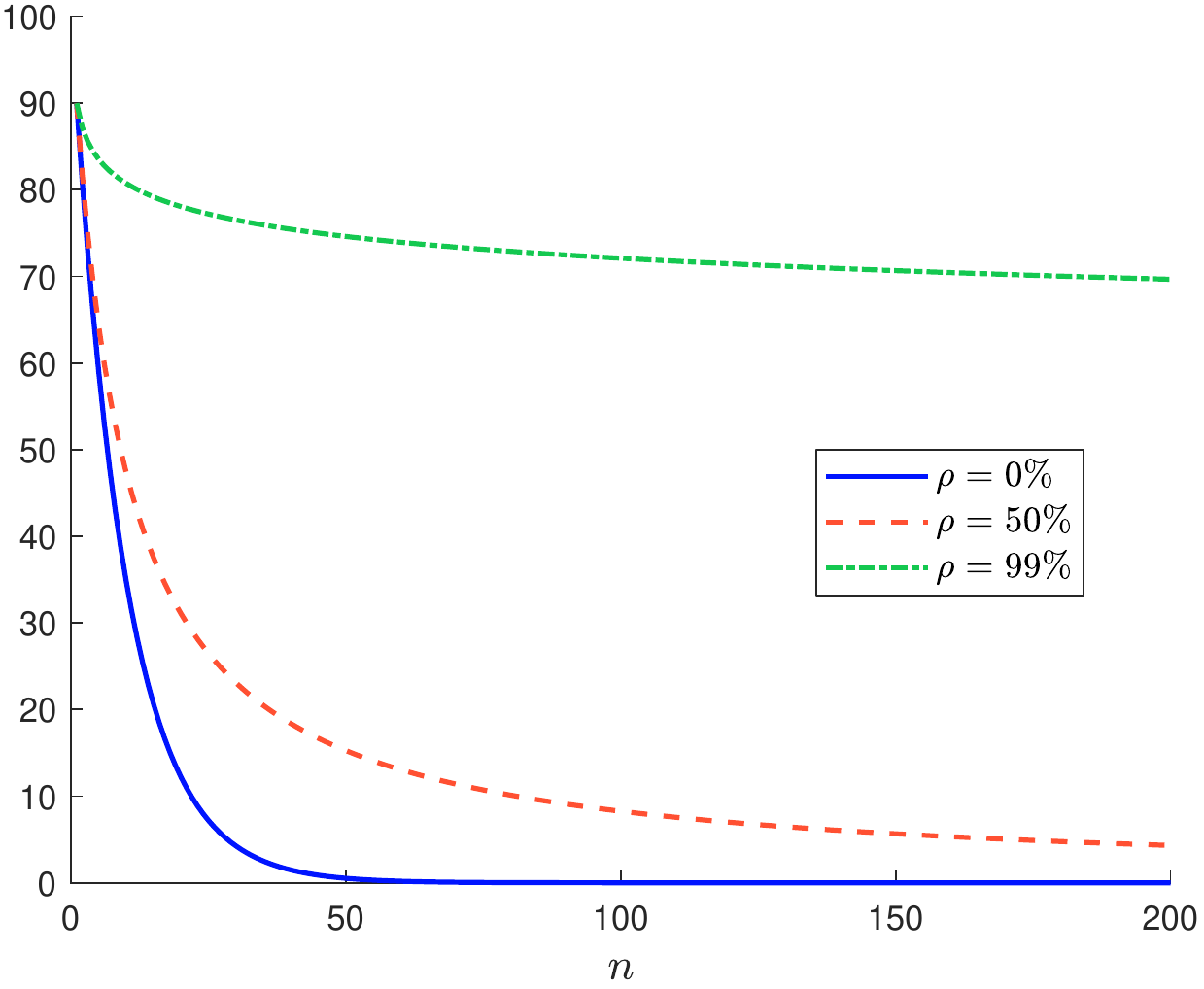}
\end{figure}

\begin{figure}[tbph]
\centering
\caption{Probability to observe 100\% of redemptions $\Pr \left\{ \frequency=1\right\} $ in \%
($n = 20$)}
\label{fig:copula4c}
\includegraphics[width = \figurewidth, height = \figureheight]{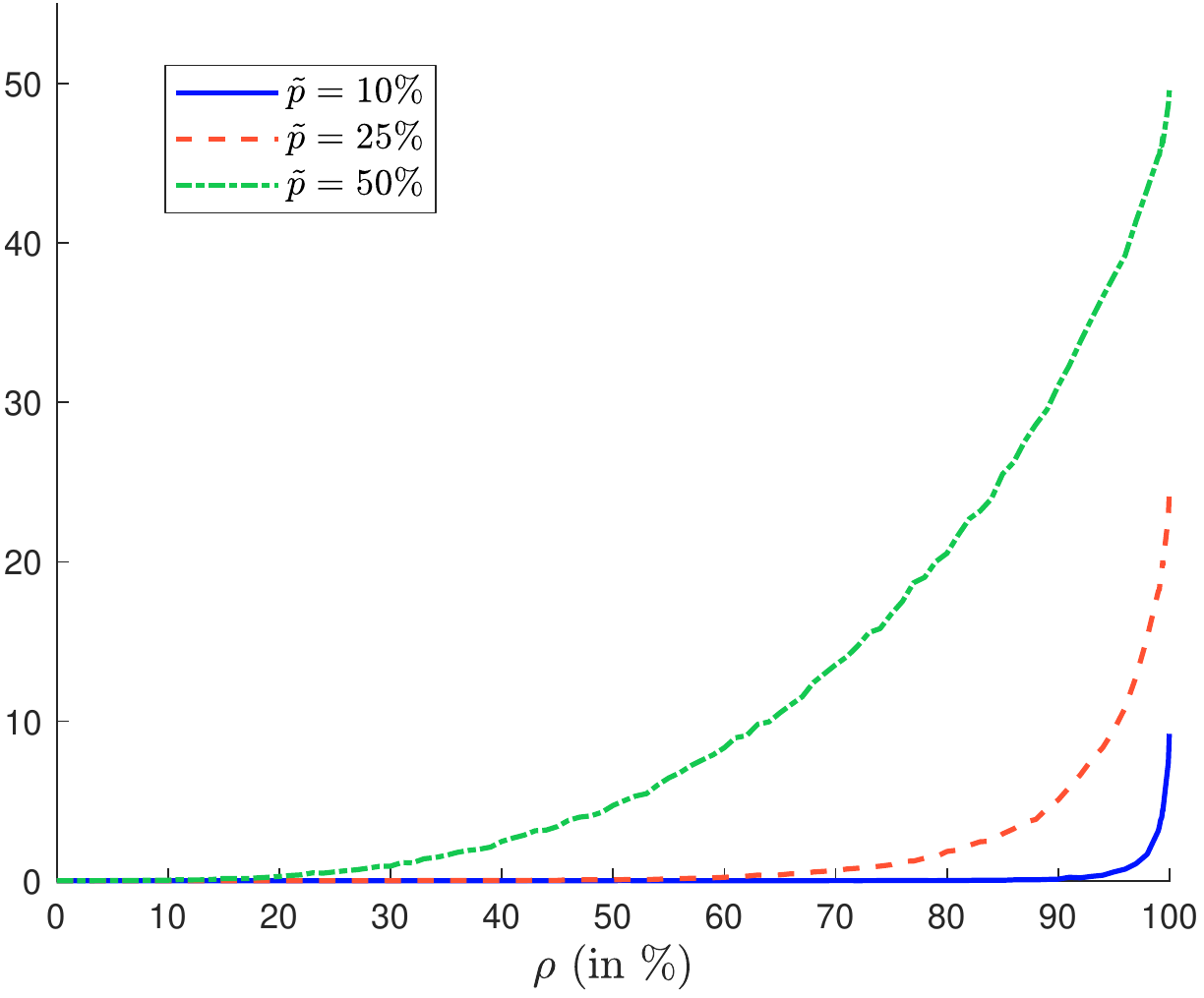}
\end{figure}

\begin{figure}[tbph]
\centering
\caption{Histogram of the redemption rate in \% with respect to the number $n$ of unitholders
($\tilde{p} = 50\%, \tilde{\mu} = 50\%, \tilde{\sigma} = 10\%, \rho = 25\%$)}
\label{fig:copula7b}
\includegraphics[width = \figurewidth, height = \figureheight]{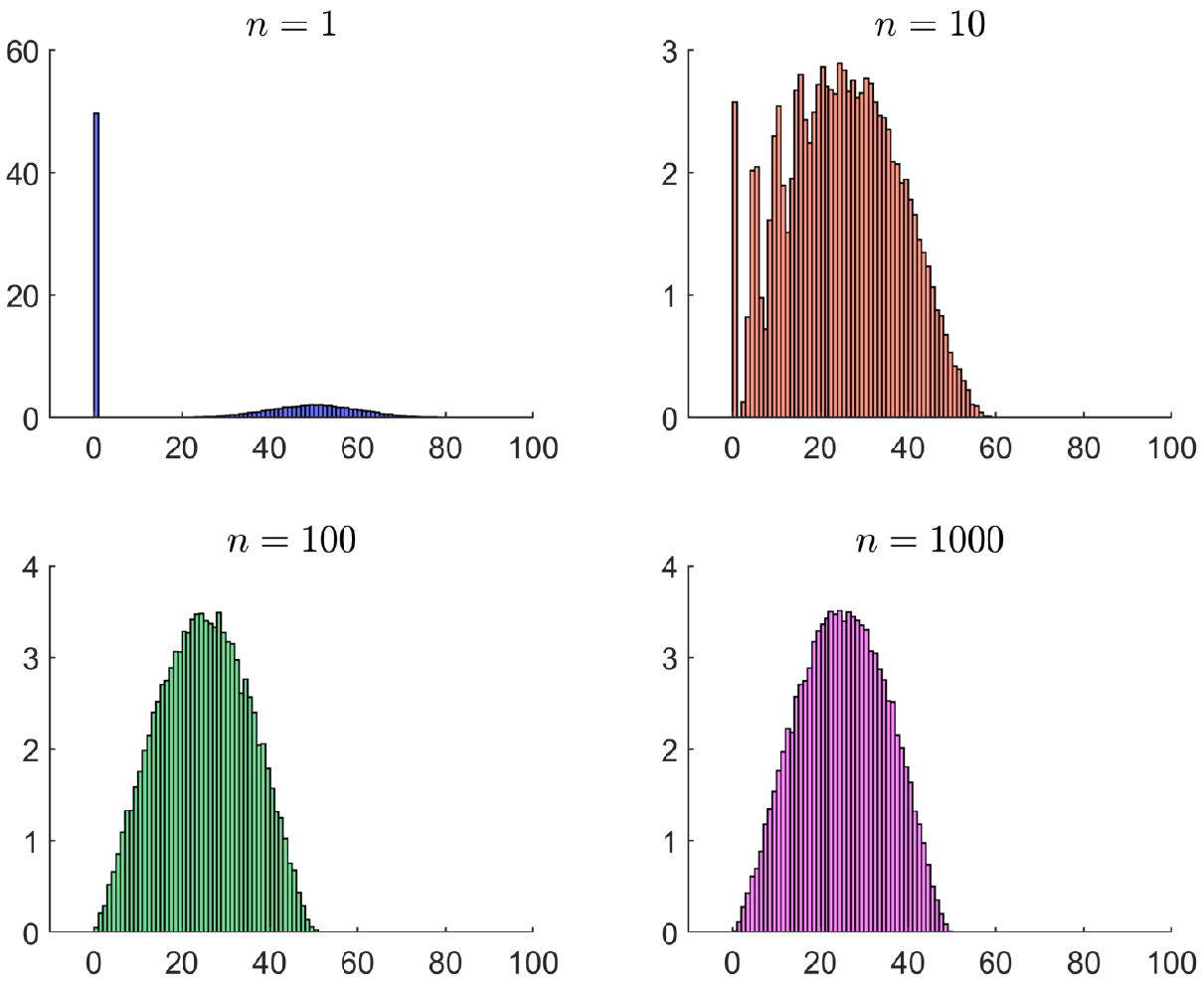}
\end{figure}

\begin{figure}[tbph]
\centering
\caption{Histogram of the redemption rate in \% with respect to the number $n$ of unitholders
($\tilde{p} = 50\%, \tilde{\mu} = 50\%, \tilde{\sigma} = 10\%, \rho = 75\%$)}
\label{fig:copula7c}
\includegraphics[width = \figurewidth, height = \figureheight]{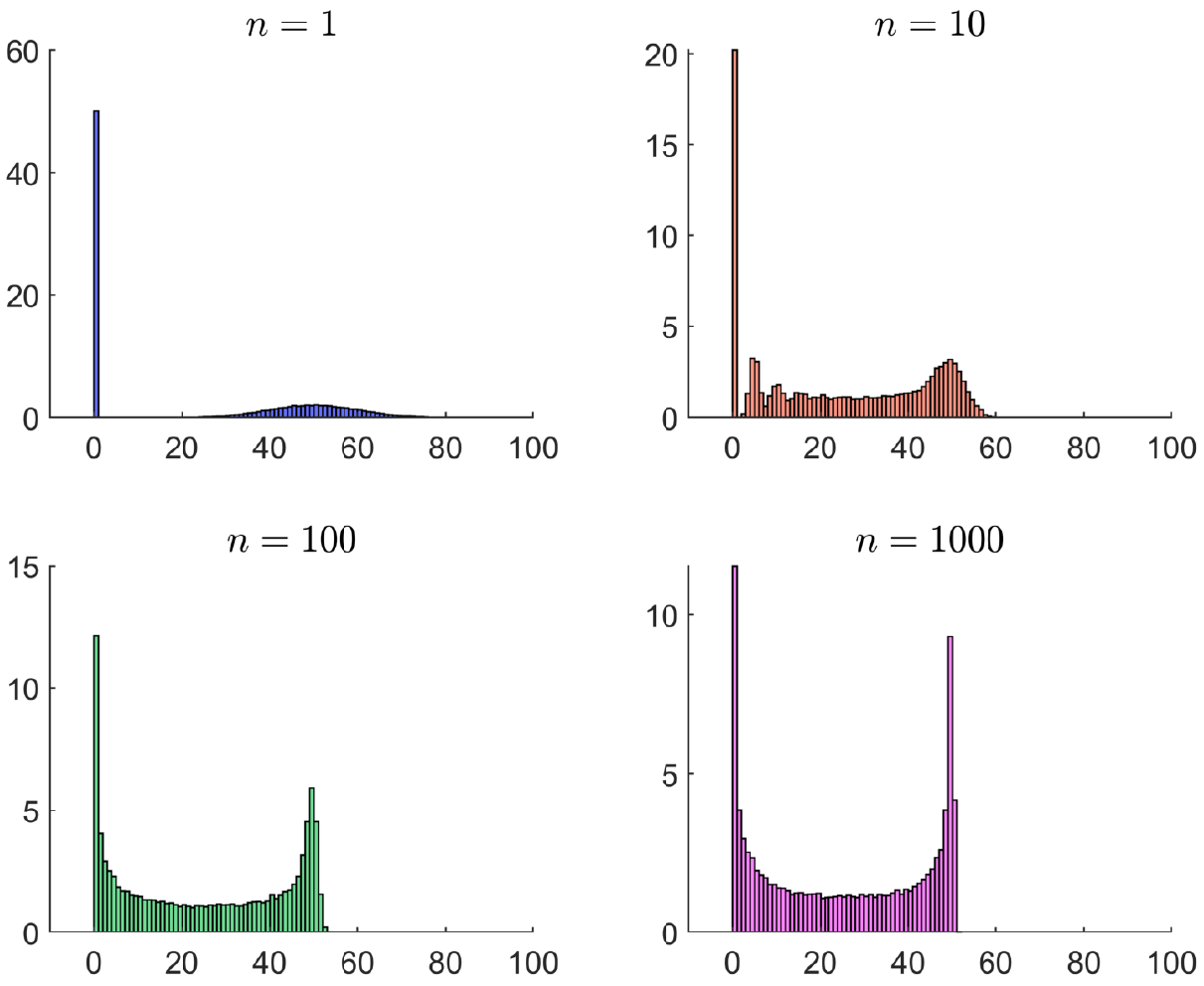}
\end{figure}

\clearpage

\begin{figure}[tbph]
\centering
\caption{Histogram of the redemption rate in \% with respect to the number $n$ of unitholders
($\tilde{p} = 50\%, \tilde{\mu} = 50\%, \tilde{\sigma} = 10\%, \rho = 90\%$)}
\label{fig:copula7d}
\includegraphics[width = \figurewidth, height = \figureheight]{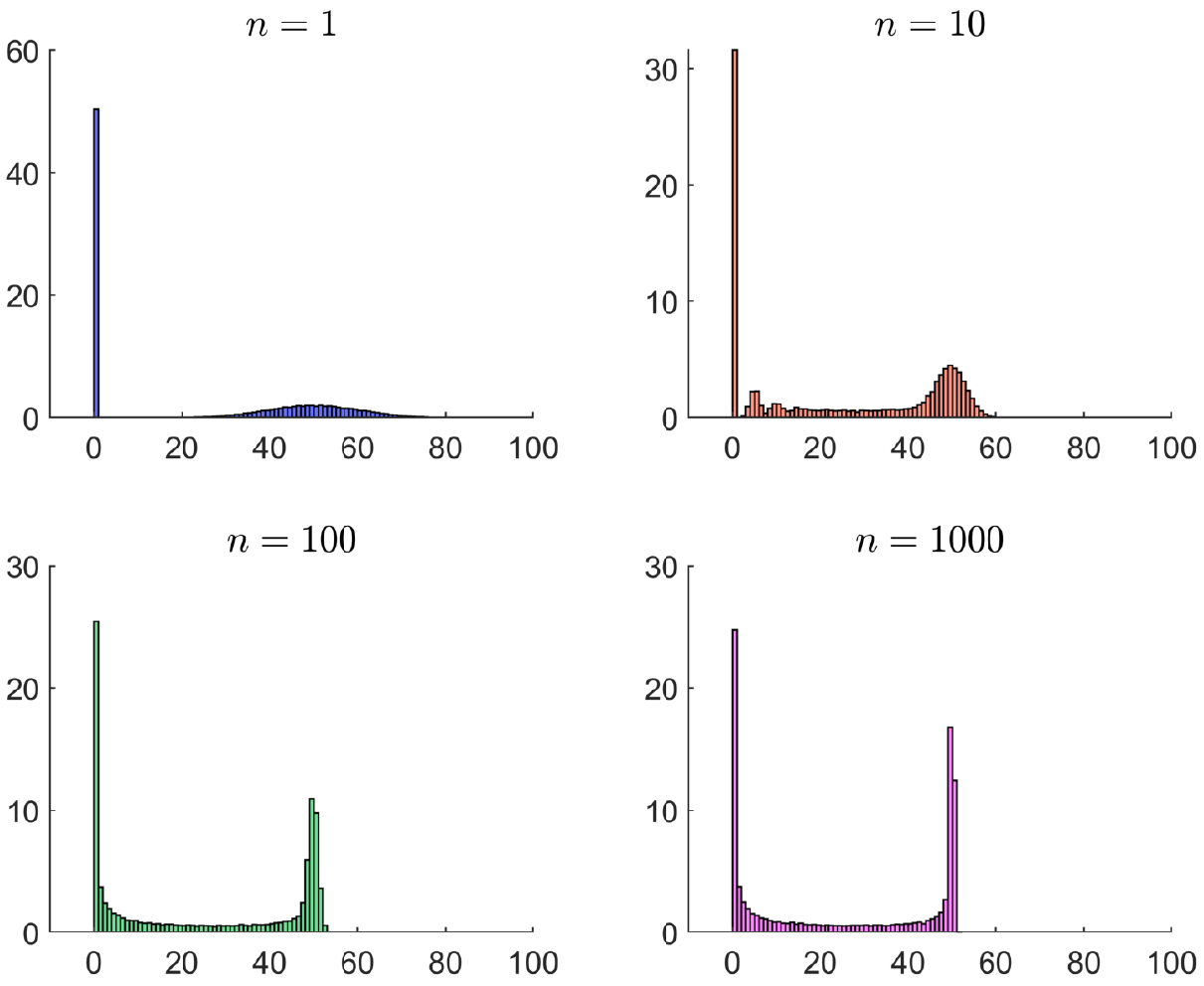}
\vspace*{50pt}
\end{figure}

\begin{table}[tbph]
\centering
\caption{Calibrated Pearson correlation (Normal copula, $\mathcal{H}\left( \omega \right)=1/20$)}
\label{tab:correl2c}
\begin{tabular}{c|ccccc}
\hline
& & & & & \\[-2ex]
\multirow{2}{*}{$\hat{\sigma}\left( \frequency\right)$}
         & \multicolumn{5}{c}{$\overline{\frequency}$} \\[0.75ex]
         & ${\TsV}10.0\%$ & ${\TsV}20.0\%$ &      $25.0\%$ & $30.0\%$ & $40.0\%$ \\
& & &  \\[-2ex]  \hline
$10.0\%$ &           &           &           &           &           \\
$20.0\%$ & $39.88\%$ & $24.58\%$ &           &           &           \\
$30.0\%$ & $50.00\%$ & $42.83\%$ & $38.88\%$ & $35.70\%$ & $31.70\%$ \\
$40.0\%$ &           & $50.00\%$ & $49.20\%$ & $47.77\%$ & $45.30\%$ \\
\hline
\end{tabular}
\end{table}

\begin{figure}[tbph]
\centering
\caption{Dependogram of the bivariate Normal copula}
\label{fig:correl5}
\includegraphics[width = \figurewidth, height = \figureheight]{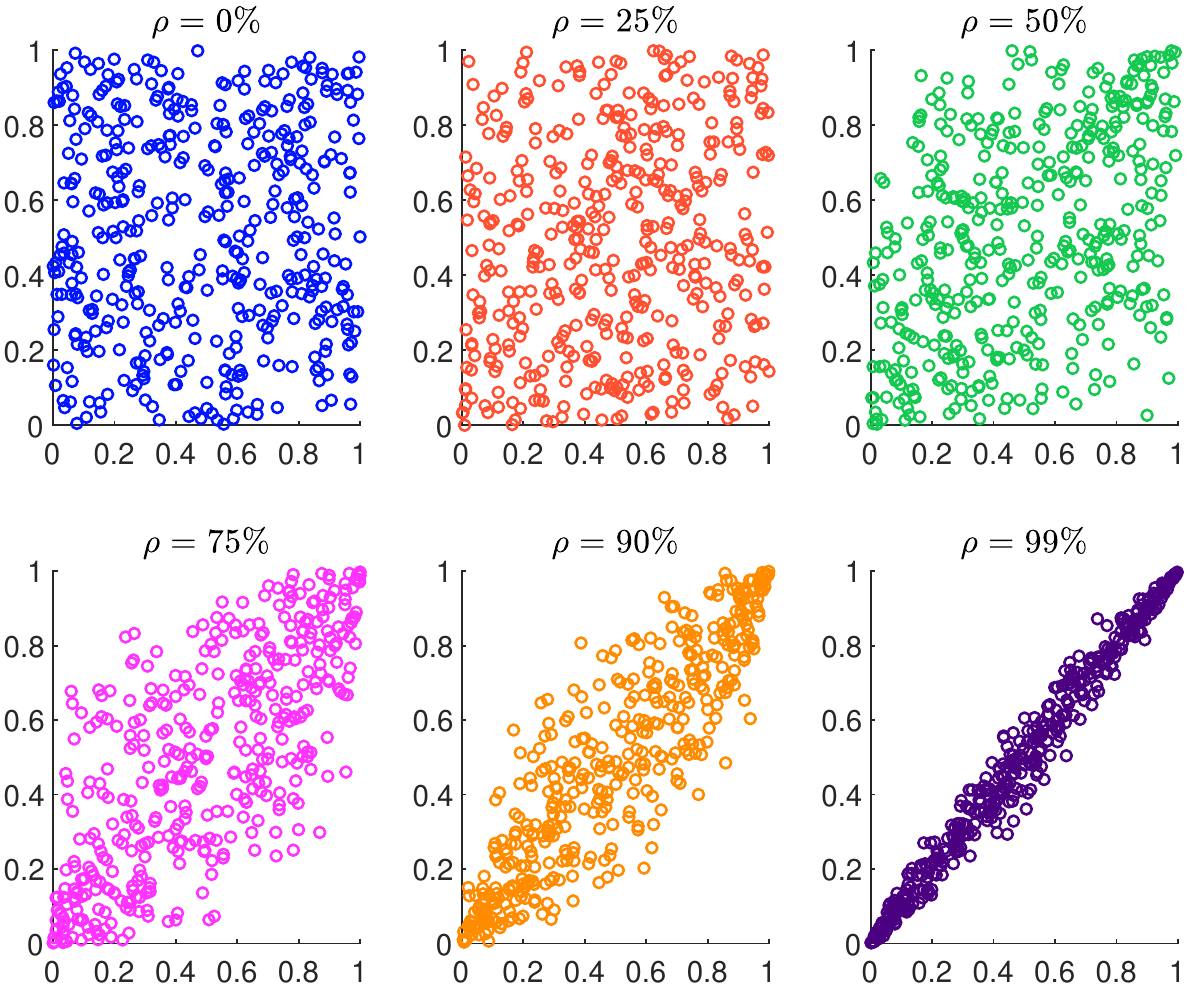}
\end{figure}

\begin{figure}[tbph]
\centering
\caption{Dependogram of redemption rate frequencies for equity funds}
\label{fig:correl3c}
\includegraphics[width = \figurewidth, height = \figureheight]{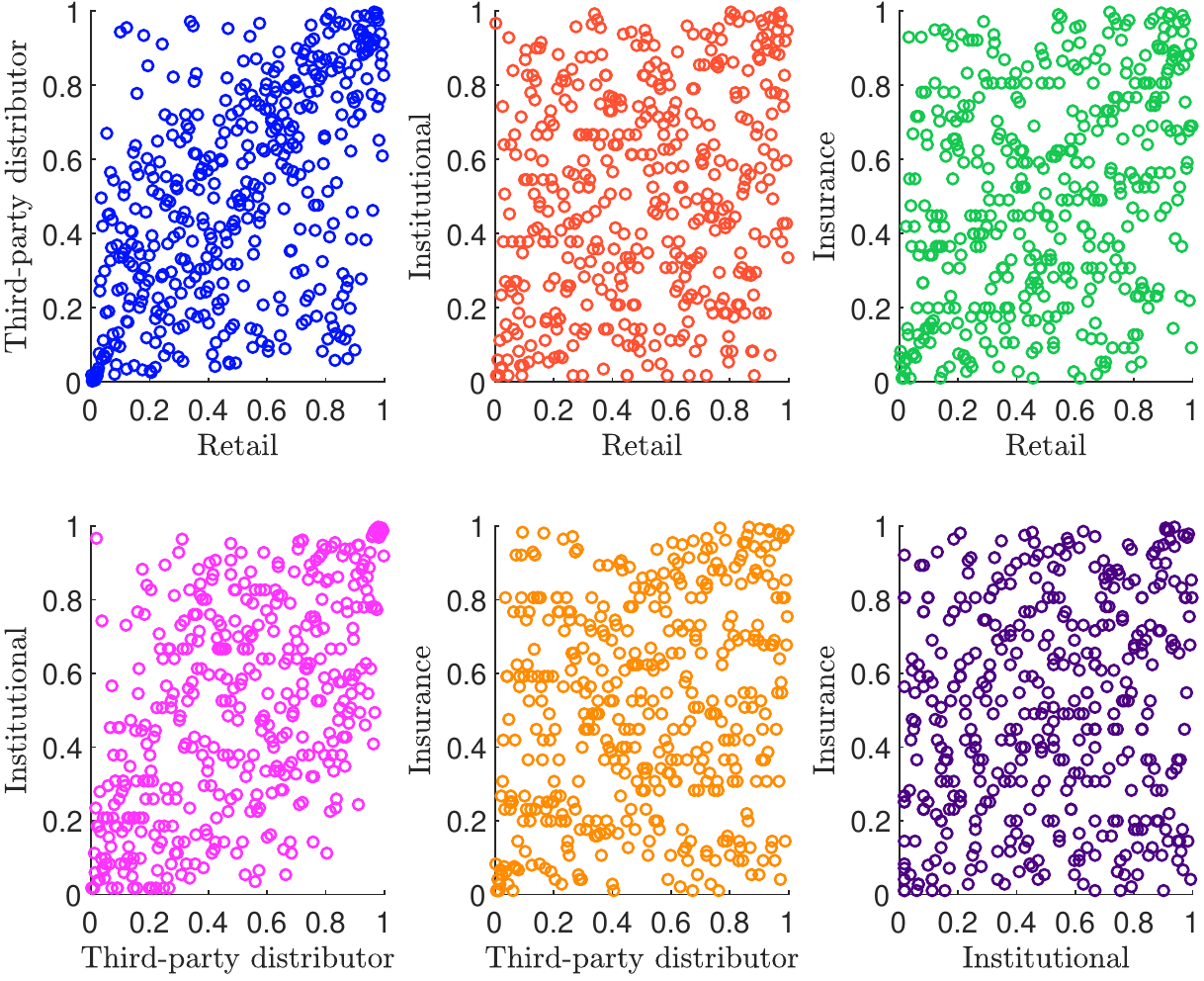}
\end{figure}

\begin{figure}[tbph]
\centering
\caption{Histogram of the weekly redemption rate in \% with respect to the autocorrelation $\rho _{\mathrm{time}}$
($\tilde{p} = 50\%, \tilde{\mu} = 50\%, \tilde{\sigma} = 10\%, \rho = 50\%, n = 10$)}
\label{fig:autocor1c}
\includegraphics[width = \figurewidth, height = \figureheight]{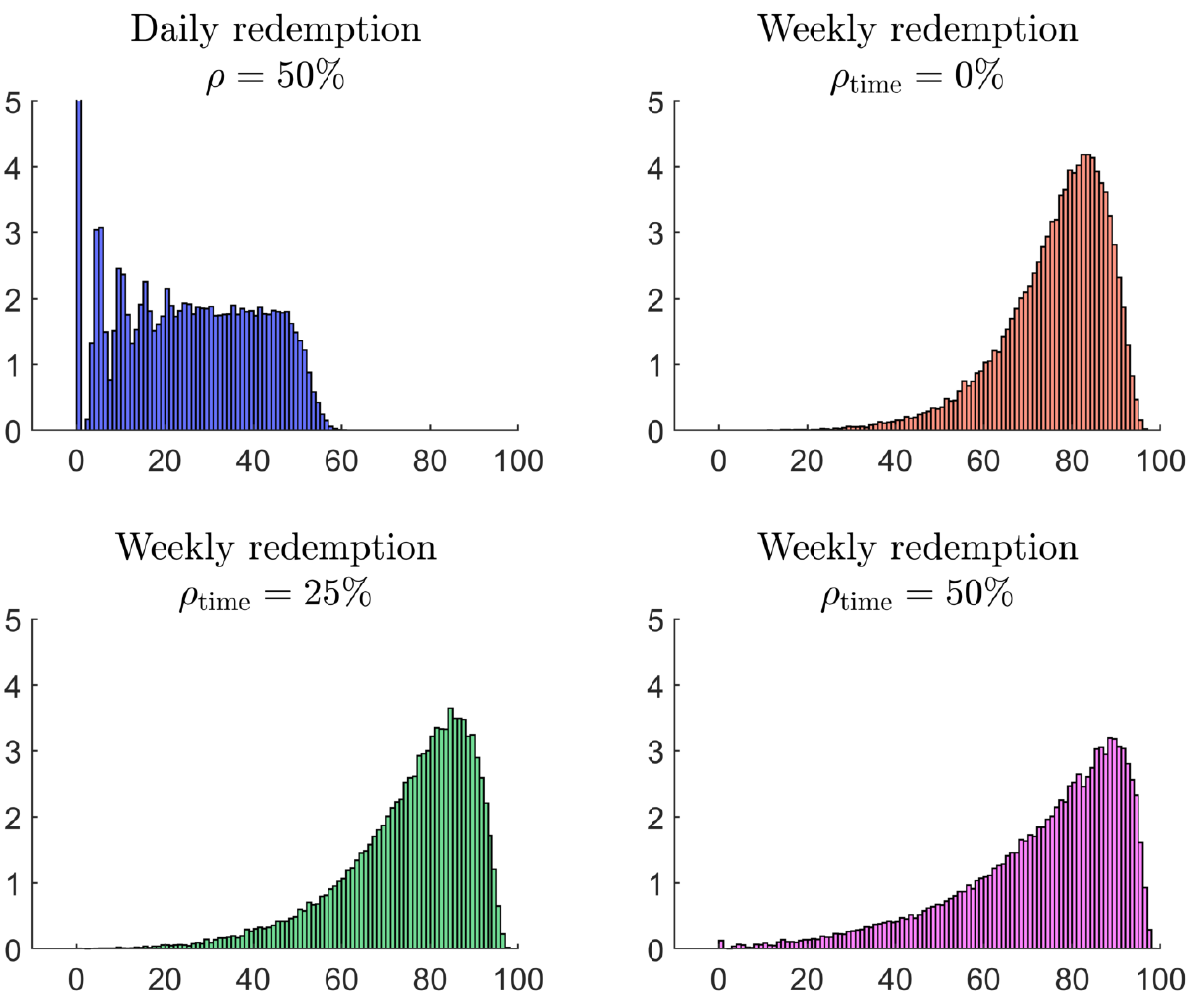}
\end{figure}

\begin{table}[tbph]
\centering
\caption{Autocorrelation of the redemption frequency in \%}
\label{tab:time5-p}
\begin{tabular}{lcccc}
\hline
                        & Balanced & Bond & Equity & Money market  \\ \hline
Institutional           &          $26.3^{**}$ &          $12.9^{**}$ & ${\TsV}3.8{\TsVIII}$ &     ${\TsVIII}33.9^{**}$ \\
Insurance               &          $39.8^{**}$ &          $10.5^{**}$ & ${\TsV}1.8{\TsVIII}$ &     ${\TsVIII}16.9^{**}$ \\
Retail                  & ${\TsV}7.9{\TsVIII}$ & ${\TsV}9.8{\TsVIII}$ &          $25.2^{**}$ &          $-0.1{\TsVIII}$ \\
Third-party distributor &          $15.0^{**}$ &          $32.5^{**}$ &          $42.4^{**}$ &     ${\TsVIII}13.9^{**}$ \\ \hline
\end{tabular}
\end{table}

\begin{table}[tbph]
\centering
\caption{Autocorrelation of the redemption severity in \%}
\label{tab:time5-mu}
\begin{tabular}{lcccc}
\hline
                        & Balanced & Bond & Equity & Money market  \\ \hline
Institutional           &     ${\TsIII}$$16.9^{**}$ & ${\TsV}2.0{\TsVIII}$ & ${\TsV}6.1{\TsVIII}$ & $21.4^{**}$ \\
Insurance               &           $-1.1{\TsVIII}$ & ${\TsV}8.4{\TsVIII}$ & ${\TsV}8.5{\TsVIII}$ & $18.3^{**}$ \\
Retail                  &     ${\TsIII}$$13.5^{**}$ & ${\TsV}3.1{\TsVIII}$ &          $10.1^{**}$ & $12.5^{**}$ \\
Third-party distributor & ${\TsVIII}$$1.6{\TsVIII}$ &          $13.4^{**}$ &     ${\TsV}9.9^{**}$ & $21.3^{**}$ \\ \hline
\end{tabular}
\end{table}


\begin{table}[tbph]
\centering
\caption{Coefficient of determination $\mathfrak{R}_{c}^{2}$ in \% ---
$\redemption\left(t\right) = \beta_0 + \beta_1 \frequency\left(t\right) + u\left(t\right)$}
\label{tab:time7b-p}
\begin{tabular}{lcccc}
\hline
                        & Balanced & Bond & Equity & Money market  \\ \hline
Institutional           & ${\TsV}9.2$ & $45.2$ & $59.1$ & $55.1$ \\
Insurance               & ${\TsV}2.8$ & $18.4$ & $22.2$ & $53.3$ \\
Retail                  &      $68.2$ & $61.9$ & $60.1$ & $55.2$ \\
Third-party distributor &      $51.8$ & $66.4$ & $54.2$ & $64.7$ \\ \hline
\end{tabular}
\end{table}

\begin{table}
\centering
\caption{Coefficient of determination $\mathfrak{R}_{c}^{2}$ in \% ---
$\redemption\left(t\right) = \beta_0 + \beta_1 \redemption^{\star}\left(t\right) + u\left(t\right)$}
\label{tab:time7b-mu}
\begin{tabular}{lcccc}
\hline
                        & Balanced & Bond & Equity & Money market  \\ \hline
Institutional           & $88.1$ & $78.3$ & $51.3$ & $93.2$ \\
Insurance               & $99.2$ & $85.3$ & $85.4$ & $94.4$ \\
Retail                  & $88.6$ & $93.2$ & $99.1$ & $89.4$ \\
Third-party distributor & $96.3$ & $95.9$ & $95.6$ & $98.0$ \\ \hline
\end{tabular}
\end{table}

\begin{table}
\centering
\caption{Coefficient of determination $\mathfrak{R}_{c}^{2}$ in \% ---
$\redemption\left(t\right) = \beta_0 + \beta_1 \frequency\left(t\right)
+ \beta_2 \redemption^{\star}\left(t\right) + u\left(t\right)$}
\label{tab:time7b-p-mu}
\begin{tabular}{lcccc}
\hline
                        & Balanced & Bond & Equity & Money market  \\ \hline
Institutional           & $89.0$ & $86.8$ & $84.0$ & $96.4$ \\
Insurance               & $99.3$ & $87.2$ & $88.2$ & $97.1$ \\
Retail                  & $96.2$ & $97.3$ & $99.7$ & $95.7$ \\
Third-party distributor & $98.4$ & $98.2$ & $97.6$ & $98.9$ \\ \hline
\end{tabular}
\end{table}

\begin{table}
\centering
\caption{Coefficient of determination $\mathfrak{R}^2_c$ in \% --- Equation (\ref{eq:factor2}), one-week time horizon}
\label{tab:factor1b}
\begin{tabular}{lcccc}
\hline
                        & Balanced & Bond & Equity & Money market  \\ \hline
Institutional           & $0.3$ & $0.7$ & $1.0$ & $1.4$ \\
Insurance               & $0.2$ & $0.5$ & $1.4$ & $2.3$ \\
Retail                  & $0.8$ & $2.3$ & $0.6$ & $0.3$ \\
Third-party distributor & $0.8$ & $0.8$ & $1.2$ & $3.8$ \\ \hline
\end{tabular}
\end{table}

\begin{table}
\centering
\caption{Coefficient of determination $\mathfrak{R}^2_c$ in \% --- Equation (\ref{eq:factor2}), two-week time horizon}
\label{tab:factor1c}
\begin{tabular}{lcccc}
\hline
                        & Balanced & Bond & Equity & Money market  \\ \hline
Institutional           & $1.3$ & $0.7$ & $2.8$ & $2.8$ \\
Insurance               & $0.1$ & $0.3$ & $1.5$ & $5.1$ \\
Retail                  & $2.3$ & $2.0$ & $0.8$ & $0.9$ \\
Third-party distributor & $1.1$ & $2.1$ & $1.5$ & $3.7$ \\ \hline
\end{tabular}
\end{table}

\begin{figure}
\centering
\caption{Relationship between redemption rate and two-week stock returns (equity category)}
\label{fig:factor2}
\includegraphics[width = \figurewidth, height = \figureheight]{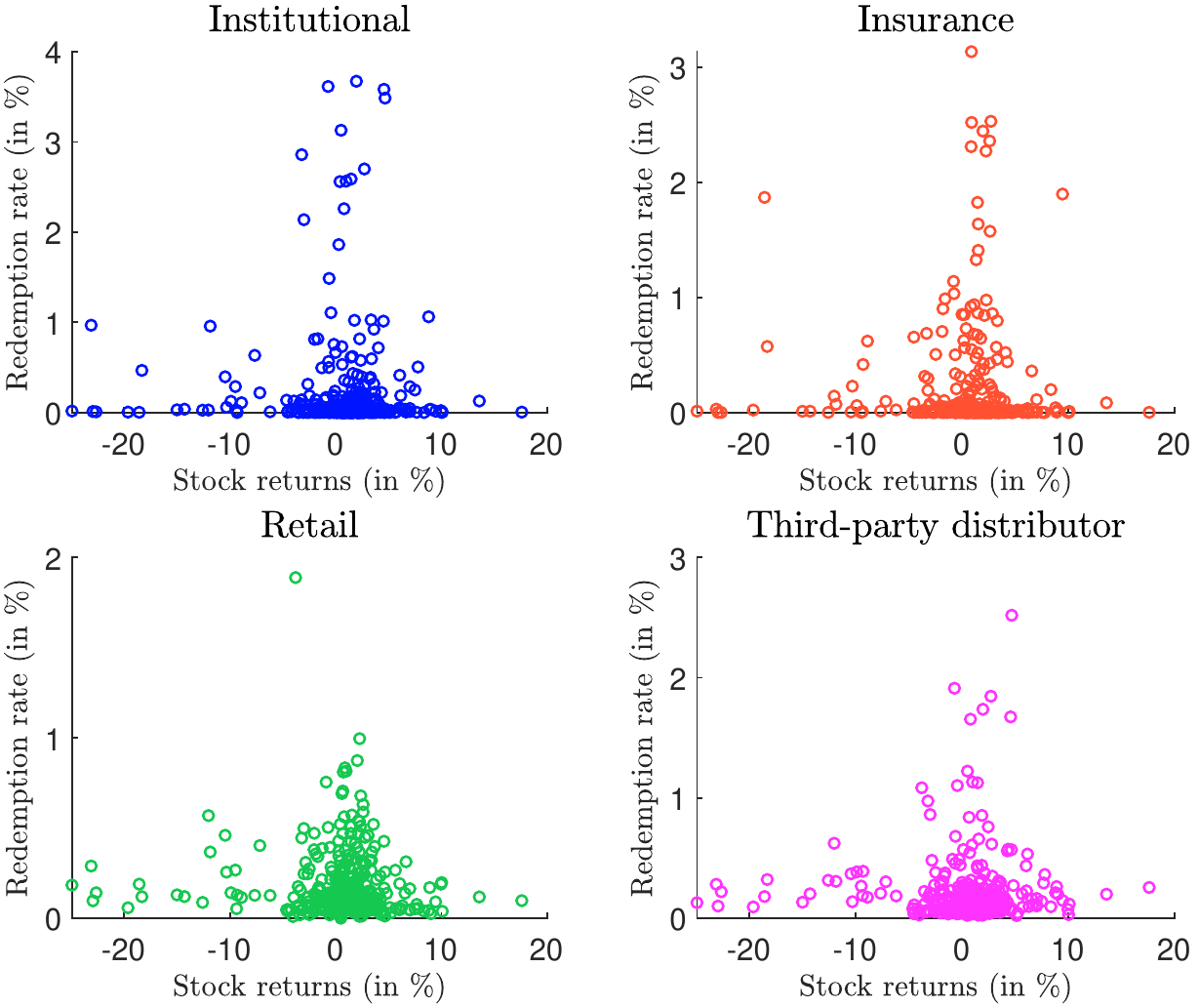}
\end{figure}

\ifResearchVersion

\clearpage

\tableofcontents

\fi

\end{document}